\documentclass[12pt]{report}
\usepackage{amsmath,authblk}
\usepackage[utf8]{inputenc}
\usepackage[T1]{fontenc}
\usepackage{amssymb}
\usepackage{dsfont}
\usepackage{graphicx, subcaption}
\usepackage{listings}
\usepackage[dvipsnames]{xcolor}
\usepackage[colorlinks]{hyperref}
\usepackage{float}
\usepackage{pgfplots}
\usepackage{bm}

\usepackage{physics}
\usepackage{slashed}
\pgfplotsset{compat=1.18}
\usepackage{amsmath,amssymb,graphicx,hyperref}
\usepackage{qcircuit}
\usepackage{tikz}
\usetikzlibrary{decorations.pathmorphing}
\usepackage{caption}
\usepackage{pgfplots}
\pgfplotsset{compat=1.18} 
\usepackage{siunitx} 
\usepackage{wrapfig, appendix, simplewick}
\AtBeginEnvironment{subappendices}{%
\chapter*{Appendix}
\addcontentsline{toc}{chapter}{Appendices}
\counterwithin{figure}{section}
\counterwithin{table}{section}
}

\usepackage{booktabs, enumitem}
\usepackage{tkz-berge}
\usetikzlibrary{arrows.meta}
\usetikzlibrary{positioning}

\usepackage[textwidth=15cm, left=2cm, right=2cm, top=2cm, bottom=2cm]{geometry}

\newcommand{\qft}{\widehat{\text{QFT}}}
\newcommand{\iqft}{\widehat{\text{IQFT}}}

\lstdefinelanguage{Mathematica}{
  morekeywords={ClearAll,Module,Table,Exp,Conjugate,MatrixExp,DiagonalMatrix,IdentityMatrix,
  Eigensystem,FindPeaks,Mean,Range,Plot,Abs,ComplexExpand,Re,Im,N,ArgMax,Subdivide,
  ListLinePlot,Transpose,ConjugateTranspose,Position,Ordering,Select,Between,HannWindow},
  sensitive=true,
  morecomment=[l]{(*},
  morecomment=[s][\color{gray}]{(*}{*)},
  morestring=[b]"
}
\lstdefinestyle{mma}{
  language=Mathematica,
  basicstyle=\ttfamily\small,
  keywordstyle=\color{blue!60!black}\bfseries,
  commentstyle=\itshape\color{green!40!black},
  stringstyle=\color{orange!50!black},
  showstringspaces=false,
  frame=single,
  framerule=0.3pt,
  rulecolor=\color{black!30},
  backgroundcolor=\color{black!2},
  numbers=left,
  numberstyle=\tiny\color{black!40},
  numbersep=8pt,
  xleftmargin=12pt,
  breaklines=true,
  columns=fullflexible,
  keepspaces=true,
  upquote=true
}
\newenvironment{diagram}
{
\begin{tikzpicture}[baseline = (current bounding box.center),every node/.style={scale=0.85},scale=1]
}
{
\end{tikzpicture}
}

\tikzset{
  pics/applyTransferRight/.style n args={3}{
    code={
      \draw (-0.5,0)   -- (0,0);
      \draw (-0.5,-3)  -- (0,-3);

      \filldraw[fill = green!30, rounded corners] (0,0.5) rectangle (1,-0.5);
      \node at (0.5,0) {#1};

      \draw (0.5,-0.5) -- (0.5,-2.5);

      \filldraw[fill = blue!30, rounded corners] (0,-2.5) rectangle (1,-3.5);
      \node at (0.5,-3) {#3};

      \draw (1,0)   edge[out=0,in=90]   (2,-1);
      \filldraw[fill = orange!30] (2,-1.5) circle (0.5);
      \node at (2,-1.5) {#2};
      \draw (2,-2) edge[out=-90,in=0] (1,-3);

      \coordinate (-leftTop)    at (-0.5,0);
      \coordinate (-leftBottom) at (-0.5,-3);
      \coordinate (-topEast)    at (1,0);
      \coordinate (-botEast)    at (1,-3);
      \coordinate (-topKnee)    at (2,-1);     
      \coordinate (-botKnee)    at (2,-2);     
      \coordinate (-circle)     at (2,-1.5);   
      \coordinate (-centerTop)  at (0.5,0);
      \coordinate (-centerBot)  at (0.5,-3);
    }
  }
}
\usetikzlibrary{positioning, calc}

\tikzset{
  pics/applyTransferLeft/.style n args={3}{
    code={
      \filldraw[fill = red!30] (0,0) circle (.5);
      \node at (0,0) {#2};

      \draw (0,0.5) edge[out=90,in=180] (1,1.5);
      \draw (0,-0.5) edge[out=270,in=180] (1,-1.5);

      \filldraw[fill = green!30, rounded corners] (1,2) rectangle (2,1);
      \node at (1.5,1.5) {#3};

      \filldraw[fill = blue!30, rounded corners] (1,-1) rectangle (2,-2);
      \node at (1.5,-1.5) {#1};

      \draw (2,1.5) -- (2.5,1.5);
      \draw (2,-1.5) -- (2.5,-1.5);

      \draw (1.5,1) -- (1.5,-1);

      \coordinate (-center) at (0,0);
      \coordinate (-top) at (2.5,1.5);
      \coordinate (-bottom) at (2.5,-1.5);
      \coordinate (-left) at (-0.5,0);
    }
  }
}

\tikzset{
  pics/MPStens/.style n args={1}{
    code={
      \draw (-0.5,0) -- (0,0);
      \filldraw[fill = green!30, rounded corners] (0, -0.5) rectangle (1,0.5);
      \node at (0.5,0) {#1};
      \draw (1,0) -- (1.5,0);
      \draw (0.5,-0.5) -- (0.5,-1);

      \coordinate (-west) at (0,0);
      \coordinate (-east) at (1,0);
      \coordinate (-down) at (0.5,-1);
      \coordinate (-up)   at (0.5,0.5);
    }
  }
}

\tikzset{
  pics/MpsTens/.style n args={1}{
    code={
      \draw (-0.5,-0.5) -- (0,-0.5);
      \filldraw[fill=blue!30, rounded corners] (0,0) rectangle (1,-1);
      \node at (0.5,-0.5) {#1};
      \draw (0.5,0.0) -- (0.5,0.5);
      \draw (1,-0.5) -- (1.5,-0.5);

      \coordinate (-west) at (0,-0.5);
      \coordinate (-east) at (1,-0.5);
      \coordinate (-down) at (0.5,-1.5);
      \coordinate (-up)   at (0.5,0);
    }
  }
}

\tikzset{
  pics/Matrix/.style n args={1}{
    code={
      \filldraw[fill=red!30] (0,0) circle (.5);
      \node at (0,0) {#1};
      \draw (-1,0)--(-.5,0);
      \draw ( .5,0)--( 1,0);

      \coordinate (-west) at (-1,0);
      \coordinate (-east) at (1,0);
      \coordinate (-center) at (0,0);
    }
  }
}

\tikzset{
  pics/drawMatrixLeft/.style n args={1}{
    code={
      \filldraw[fill=red!30] (0,0) circle (.5);
      \node at (0,0) {#1};
      \draw (0,0.5) edge[out=90,in=180] (1,1.5);
      \draw (0,-0.5) edge[out=270,in=180] (1,-1.5);

      \coordinate (-top) at (1,1.5);
      \coordinate (-bottom) at (1,-1.5);
      \coordinate (-center) at (0,0);
    }
  }
}

\tikzset{
  pics/drawMatrixRight/.style n args={1}{
    code={
      \draw (2,1.5) edge[out=0,in=90] (3,0.5);
      \filldraw[fill = orange!30] (3,0) circle (.5);
      \node at (3,0) {#1};
      \draw (3,-0.5) edge[out=-90,in=0] (2,-1.5);

      \coordinate (-top) at (2,1.5);
      \coordinate (-bottom) at (2,-1.5);
      \coordinate (-center) at (3,0);
    }
  }
}
\tikzset{
  pics/tensor/.style n args={1}{ 
    code={
      \draw (-0.5,0) -- (0,0);
      \filldraw[rounded corners, fill=green!30] (0,-0.5) rectangle (1,0.5);
      \node at (0.5,0) {#1};
      \draw (1,0) -- (1.5,0);
      \draw (0.5,-0.5) -- (0.5,-1);

      \coordinate (-west) at (0,0);
      \coordinate (-east) at (1,0);
      \coordinate (-down) at (0.5,-0.5);
      \coordinate (-up)   at (0.5, 0.5);
    }
  },
  MPStensor/.style = {insert path={
    \pic {tensor={#1}};
  }}
}
\tikzset{
  pics/tensorUp/.style n args={1}{ 
    code={
      \draw (-0.5,0) -- (0,0);
      \filldraw[rounded corners, fill=green!30] (0,-0.5) rectangle (1,0.5);
      \node at (0.5,0) {#1};
      \draw (1,0) -- (1.5,0);
      \draw (0.5,0.5) -- (0.5,1.0);

      \coordinate (-west) at (0,0);
      \coordinate (-east) at (1,0);
      \coordinate (-up)   at (0.5,0.5);
      \coordinate (-down) at (0.5,-0.5); 
    }
  },
  MPStensorUp/.style = {insert path={ \pic {tensorUp={#1}} }}
}

\definecolor{codegreen}{rgb}{0,0.6,0}
\definecolor{codegray}{rgb}{0.5,0.5,0.5}
\definecolor{codepurple}{rgb}{0.58,0,0.82}
\definecolor{backcolour}{rgb}{0.95,0.95,0.92}

\lstdefinestyle{mystyle}{
    backgroundcolor=\color{backcolour},
    commentstyle=\color{codegreen},
    keywordstyle=\color{magenta},
    numberstyle=\tiny\color{codegray},
    stringstyle=\color{codepurple},
    basicstyle=\ttfamily\footnotesize,
    breakatwhitespace=false,
    breaklines=true,
    captionpos=b,
    keepspaces=true,
    numbers=left,
    numbersep=5pt,
    showspaces=false,
    showstringspaces=false,
    showtabs=false,
    tabsize=2
}

\lstdefinestyle{pythonstyle}{
    style=mystyle,
    language=Python,
}

\lstdefinelanguage{Julia}{
  morekeywords={
    abstract,break,case,catch,const,continue,do,else,elseif,end,export,
    false,for,function,immutable,import,importall,if,in,macro,module,
    otherwise,quote,return,switch,true,try,type,typealias,using,while,
    struct,mutable,where
  },
  sensitive=true,
  morecomment=[l]\#,
  morecomment=[n]{\#=}{=\#},
  morestring=[b]",
}

\lstdefinestyle{juliastyle}{
    style=mystyle,
    language=Julia,
    basicstyle       = \ttfamily\footnotesize,
    keywordstyle     = \bfseries\color{blue},
    stringstyle      = \color{magenta},
    commentstyle     = \color{ForestGreen},
    showstringspaces = false,
    literate=
      {ϕ}{{$\phi$}}1
      {π}{{$\pi$}}1
      {ℂ}{{$\mathbb{C}$}}1
      {ψ}{{$\psi$}}1
      {←}{{$\leftarrow$}}1
      {μ}{{$\mu$}}1
      {λ}{{$\lambda$}}1
      {⊗}{{$\otimes$}}1
}

\usepackage[colorlinks]{hyperref}

\usepackage[sorting=none]{biblatex}
\title{\bf Lectures on Quantum Field Theory \\on a Quantum Computer}
\author[1,2]{Aninda Sinha}
\affil[1]{Centre for High Energy Physics, Indian Institute of Science, Bangalore, India.}
\affil[2]{Department of Physics and Astronomy, University of Calgary, Canada. }
\author[1,3]{Ujjwal Basumatary}
\affil[3]{Raman Research Institute, Light and Matter Physics, Bangalore, India. }

\date{\today}

\begin{document}
\maketitle
\begin{abstract}
The lecture notes cover the basics of quantum computing methods for quantum field theory applications. No detailed knowledge of either quantum computing or quantum field theory is assumed and we have attempted to keep the material at a pedagogical level. We review the anharmonic oscillator, using which we develop a hands-on treatment of certain interesting QFTs in $1+1D$: $\phi^4$
 theory, Ising field theory, and the Schwinger model. We review quantum computing essentials as well as tensor network techniques. The latter form an essential part for quantum computing benchmarking. Some error modeling on QISKIT is also done in the hope of anticipating runs on NISQ devices.  \\

 These lecture notes are the expanded version of a one semester course taught by AS during August-November 2025 at the Indian Institute of Science and TA-ed by UB. The programs written for this course are available in a GitHub repository. 
\end{abstract}


\tableofcontents

\chapter{Overview}

\section{Why these lectures}

Feynman famously said \cite{feynman}:
{\it Nature isn’t classical, dammit, and if you want to make a simulation of nature, you’d better make it quantum mechanical…and by golly it’s a wonderful problem, because it doesn’t look so easy.}

Many countries all over the world have launched missions to harness the power of the ``quantum world." The time is apt to push this power to the boundaries of our knowledge. Many fundamental unanswered questions in theoretical physics rely on the framework of quantum field theory (QFT). The younger generation, submerged by the growing crescendo surrounding the excitement promised by the ``new quantum age," needs to be made aware of these fundamental questions. They need modern tools to explore these unchartered waters. 

The aim of these pedagogical lectures is to gently introduce (and hopefully inspire) students to the fascinating world of quantum field theory and quantum computing. While some knowledge of the basics \cite{tong, bl, schwartz, shankar, nc, laflamme} will help, it is our attempt to keep the lectures as pedagogical as possible. For a recent lecture course of the somewhat advanced topics of gauge theories on a quantum computer, see \cite{davoudi}. Our intention is to keep these lectures at a more introductory level so that students after a course on quantum mechanics should find them accessible and then take on \cite{davoudi}. 

We feel that delving into quantum field theory and thinking about the subject keeping a quantum computer in mind, is a wonderful way to remain (relevant and) motivated. It is an example of the ``doesn't look so easy" that Feynman was referring to. The seminal work of \cite{JLP2011ScalarArXiv} introduced this problem and since then a lot has happened. We wish to give a flavour of these developments. Paraphrasing John Preskill \cite{pressquote}: {\it Ken Wilson invented the renormalization group by thinking about quantum field theory on a classical computer. May be we will learn something about quantum gravity by thinking about quantum field theory on a quantum computer.}

In our opinion, the time is ripe to expand the vocabulary of a typical quantum field theorist to include the language of quantum computing. Time and again, such cross-fertilization has led to fresh new insights and even major breakthroughs (like the RG paradigm). It is important to start early---that is our motto for these lectures; to help get going fast! Our goal is not to be exhaustive, but rather to introduce enough background material to enable the reader to embark on this wonderful journey.

\section{Background expected}
\subsection{Physics}
We expect the reader to be familiar with the harmonic oscillator, perturbation theory techniques for the anharmonic oscillator and scattering in quantum mechanics. While some familiarity with the classical Ising model may help, we will review the basics when the time comes. For the Schwinger and Thirring model parts, familiarity with the Dirac equation will help although we will review the necessary mathematics. 

For further reading on the basics, we recommend \cite{bl, tong, shankar} followed by \cite{schwartz}. Lattice  models in 1+1D are discussed extensively in condensed matter textbooks like \cite{sachdev, shankar}. The bosonized Schwinger and Thirring models are discussed in \cite{shankar}.

\subsection{Quantum computing and MPS}
We will introduce the necessary quantum computing aspects. However, some familiarity with the circuit model of quantum computing \cite{nc} may be useful. For further reading on the basics, we recommend \cite{nc, laflamme}. A recent exhaustive review on quantum algorithms is \cite{childs}.

For an introduction to basic matrix product state (MPS) techniques, we refer the reader to Ref.~\cite{Schollw_ck_2011}, and to Ref.~\cite{Hauschild_2018} for a brief overview of \texttt{TeNPy}. For infinite MPS and tangent-space methods, we recommend Refs.~\cite{Vanderstraeten_2019,Haegeman_2013}. The simulations in this work were carried out primarily using the Julia packages \texttt{MPSKit.jl} and \texttt{TensorKit.jl}. Their documentation is largely self-contained; in cases where certain implementation details were not immediately clear, we have provided explicit scripts and examples in our accompanying GitHub repository.

\subsection{Programming}
We will use \texttt{Qiskit}, Python, Mathematica and modern tensor network techniques. Familiarity with none of these is a prerequisite although some experience in programming will definitely help. The part with Mathematica can be completely omitted if the reader wishes to. However, in our opinion this is a user-friendly symbolic manipulation package, worth investing some time in. For QISKIT, python and tensor network techniques, we will introduce the reader to these as and when necessary. We use the \texttt{Julia} libraries \texttt{MPSKit.jl} and \texttt{TensorKit.jl} for our MPS simulations and have included some short code snippets to ease the reader into this new programming language. The GitHub repository will have commented scripts and notebooks to complement these lectures. For relevant introductory material that we recommend as companion reading, see \cite{combarro1, combarro2}.  

The computers we have used are a) MacBook Pro 16GB RAM, M1 processor, b) Acer Nitro 5 -- 51545 (2020) and c) Intel 40 core workstation, 128 GB RAM. We will indicate the runtimes on these when they differ widely. {\it We will not present results from runs on actual quantum hardware.}

\section{Plan}
The lectures are divided into 4 parts. 
\begin{enumerate}
\item PHYSICS: In this part, we review and introduce the necessary background physics material. We start with the anharmonic oscillator, discuss some simple nonperturbative techniques, backing them up with the classic instanton methods. We review scattering in quantum mechanics and the semi-classical WKB technique. Then we consider a lattice of anharmonic oscillators. When nearest neighbour spring couplings are introduced, in the continuum limit, this model becomes the standard $1+1D$ $\phi^4$ scalar field theory. This discussion is followed by projecting the lattice anharmonic oscillator's double well form onto the lowest excited states. The continuum limit of this theory is the Ising field theory. The Ising field theory is probably the simplest interesting situation where one can study scattering on a quantum computer and ask questions regarding particle production. The final physics chapter deals with the bosonized Schwinger and Thirring models. Both of these are cases where the $\phi^4$ potential in the anharmonic oscillator is replaced by $\cos \phi$ potential---the so-called Sine-Gordon models. We limit ourselves to $1+1D$ only. 

\item QUANTUM COMPUTING ESSENTIALS: In this part we introduce the necessary quantum computing aspects to simulate the physics models considered in the first part. The main focus of these simulations are to find the ground state in the interacting theory, build excitations on top of it and then scatter them. The necessary quantum algorithms include the quantum Fourier transform ($\widehat{QFT}$ to distinguish from quantum field theory QFT), quantum phase estimation (QPE), the Hadamard test and Linear Combination of Unitaries (LCU). The recent technique of using W-state adiabatic state preparation will be reviewed in a pedagogical way. 

\item TN/MPS ESSENTIALS: Quantum computing techniques are expected to be superior to any classical computing techniques, when entanglement is large. For modest amount of entanglement, there are quantum inspired classical techniques called Tensor-Networks.  To complement and benchmark the quantum computing side, we will also talk about Tensor Network techniques like MPS (Matrix Product States) and algorithms like TEBD (Time Evolving Block Decimation), DMRG (Density Matrix Renormalization Group) and Time-Dependent Variational Principle (TDVP). These can be thought of as quantum inspired classical algorithms and feature in the forefront of research on modern numerical methods in quantum theory.  

\item RESULTS: We present the results separately. The focus is on getting the spectrum, building single particle states nonperturbatively and scattering them in order to study particle production. In some instances, we have also incorporated noise models using \texttt{Qiskit} to show what can be expected on real runs on the IBMq platform. The GitHub repository hosts all our programs. Sometimes we have given snippets of the programs to explain essential steps. 
\end{enumerate}

{\bf Reading guide:} At first pass we would recommend that the details in part III are skimmed over/omitted. One could start with part II, followed by chapters 2,3,4 and then have a look at the corresponding results chapters. 

\section{Use of chatGPT5}
\subsection{Reviewing}
 We used chatGPT5 as an auxiliary tool for literature review and preliminary coding. All results and derivations presented here were independently verified and, where necessary, we rewrote programs and proofs by hand. Our experience has been a mixed bag. Here are our observations. 

\begin{itemize}
    \item For well-known basic Physics like the anharmonic oscillator, we found that chatGPT5 was very useful. It was able to review standard material, point at correct, authoritative references as well as perform basic numerical checks in \texttt{Python}.
    \item When it came to the quantum field theory aspects, its performance was less stellar. It required a lot of prompting and the material that is presented in these lectures is an outcome of many laborious hours of connecting the logical dots, albeit with good help from it. In many instances, its output to queries seemed to us like an eager student wanting to show-off! At this stage, basic calculations were not always reliable and needed quite a few cross-checks.
    \item When it came to reviewing quantum computing literature, it was helpful but again up to a point. Sometimes it needed prompting to check its confident facts against what was actually presented in the papers. Breaking down the logical flow into understandable pieces took many hours of careful prompting and to and fro. 
\end{itemize}
\subsection{Programming}
When it came to programming, again our experience was a mixed bag, more on the negative side.
\begin{itemize}
    \item For basic programming help, it was reasonably good. It can possibly serve as a good reference for \texttt{Python} APIs. Most of the Mathematica programs were either very buggy or unnecessarily convoluted \footnote{Towards the end of our writing, we tried chatGPT5.1 which seems to be somewhat better!}. It tends to incorrectly use functions that are not available in a niche package's public API. Do not expect it to write reliable long programs: choose your prompts carefully, with lots of inbuilt checks to help debug. 
    \item For somewhat advanced coding, we were mostly left very frustrated. Despite several hours of effort, we were unable to generate tensor network programs with its help and it seemed unfamiliar with modern Qiskit API. It confidently gave us Jupyter notebooks, riddled with numerous errors. Some we were able to debug with its own help, but most of the time we were left tearing our hair! Eventually, we resorted to writing our own programs for most of the work. Results presented in the final section are obtained from scripts written entirely on our own.
    \item Warning: To share an instance, when prompted to help us write MPS/TN based programs, it started by giving us programs using \texttt{QiskitAerMPS}, which never completed a run on any of our machines. When pursuaded further, it confidently supplied buggy \texttt{TeNPy} programs which crashed the Macbook. Our own attempts were more reliable and we have shared these on a GitHub repository.
\end{itemize}

All responsibility for the content of these notes, including any parts drafted with AI assistance, rests with the listed authors.

The programs written for our lectures are on \href{https://github.com/ujjwalbasumatary/qftonqc}{GitHub} and the repository is being maintained by UB. Final term papers were offered and we have shared some of these on the GitHub repository.

\section*{Acknowledgments} We thank the students of HE-381, IISc for bearing with us during this `experimental' course and for very useful feedback. We also thank Faizan Bhat, Apoorva Patel and Barry Sanders for feedback and discussions and Raghav Jha for useful correspondence. UB would like to thank Lukas Devos for clarifications on the usage of \texttt{MPSKit.jl} and \texttt{TensorKit.jl}. AS acknowledges support from a Quantum Horizons Alberta chair professorship. 

\begin{figure}[h]
  \centering
  \includegraphics[width=0.7\linewidth]{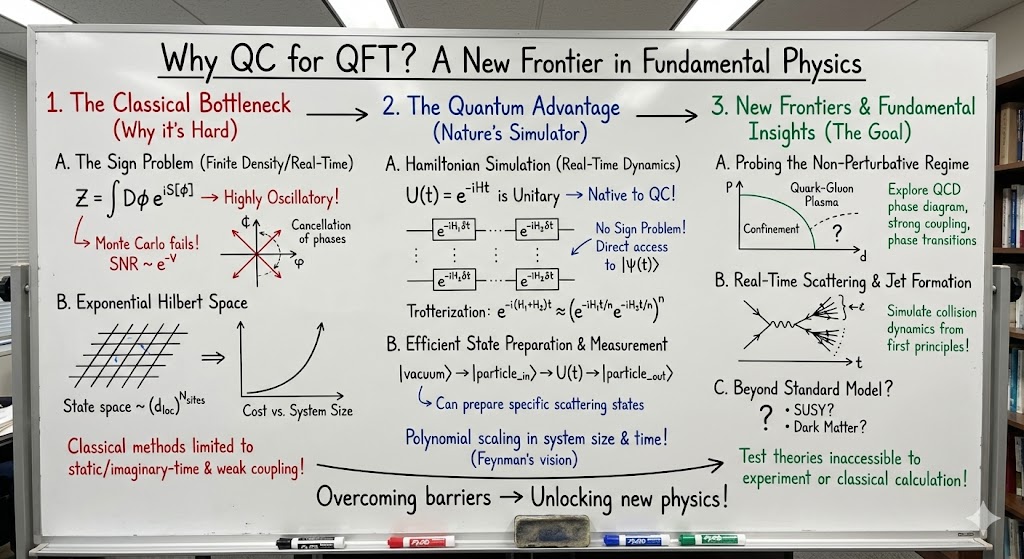}
  \caption{Google Gemini 3 pro's summary of these lectures. Courtesy: Chayanka Kakati.}
\end{figure}

\newpage
\part{Physics background}




\chapter{The anharmonic oscillator}
\section{Introduction}

In these lectures, the quartic harmonic oscillator (and its generalization to the $\cos$ potential) will form the bedrock for all the QFT models we consider. Standard perturbation theory teaches us an important lesson: the series diverges factorially, with coefficients that grow like $n!$ so that the expansion is only asymptotic.  The divergence is symptomatic of nonperturbative effects---instantons in the double well and barrier–penetration saddles more generally---which contribute terms of the schematic form $e^{-S/g}$ multiplied by trans-series in the coupling.\footnote{A \emph{trans-series} is a generalized asymptotic expansion that includes both perturbative and nonperturbative sectors, schematically
\[
f(g)\;\sim\;\sum_{k=0}^\infty C_k\,e^{-kA/g}\sum_{n=0}^\infty a_{k,n}\,g^n\,.
\]
Such structures arise very naturally in semiclassical and resurgent analyses, where exponentially small instanton effects supplement divergent perturbation series.}
In parallel with these analytic structures, there is an accompanying numerical story: eigenvalues and eigenvectors can be obtained essentially exactly using shooting methods, spectral techniques, or high–order Bender–Wu recursions. For pedagogical purposes, this combination is perfect: in a single Hamiltonian one sees how perturbation theory, semiclassics and numerics talk to each other and where each of them succeeds or fails \cite{BenderWu1969, BenderWu1973, coleman, ZinnJustinQFT, DelabaerePham1999, DunneUnsal2014}.

For us, the anharmonic oscillator plays two closely related roles. First, it builds intuition that we will reuse in field theory. The usual single–well anharmonic oscillator is the warm–up for the $\phi^4$ field theory. The double–well version is the warm–up for the Ising field theory. Generalizing the potential from $x^4$ to $\cos x$ connects directly with the bosonized Schwinger and Thirring models. These are precisely the three field theories that we will study in these lectures, and it is helpful to have their ``baby versions'' under complete control. Second, the oscillator provides a precise benchmarking ground for quantum–computing workflows. Because spectra, correlators, tunnelling splittings and real–time observables are all known to high accuracy, we can validate ground–state preparation, real–time evolution and spectroscopy pipelines before moving on to latticized field theories. Variational eigensolvers can be tested in a situation where the exact answers are unambiguous; phase–estimation and Fourier–spectroscopy approaches can be checked against known line positions and widths; and adiabatic ramps across avoided crossings in the double–well can be compared to textbook Landau–Zener and instanton expectations. In short, starting with the anharmonic oscillator allows us to build and calibrate our entire toolbox---perturbative, semiclassical, numerical and algorithmic---on a problem that is both nontrivial and completely under control, so that when the same broad themes reappear later on, we are better prepared for any potential surprises \cite{Peruzzo2014, McClean2016, Kandala2017, Kitaev1995}.

\section{The model}
Let us start with the usual simple harmonic oscillator with the Hamiltonian 
$$
H=\frac{p^2}{2}+\frac{x^2}{2}\,.
$$
Quantizing gives us
\begin{eqnarray}
    \hat x&=&\frac{1}{\sqrt{2}}(a^\dagger+a)\,,\\
    \hat p&=&\frac{i}{\sqrt{2}}(a^\dagger-a)\,,
\end{eqnarray}
with $[a,a^\dagger]=1$ giving $\hat H=(a^\dagger a+\frac{1}{2})$. The energy eigenstates $|n\rangle$ with $n\in \mathbb{Z}_{\geq 0}$ satisfy
\begin{eqnarray}
  \hat H|n\rangle=(n+\frac12)|n\rangle\,,\quad \hat a|n\rangle=\sqrt{n}|n-1\rangle\,,\quad \hat a^\dagger |n\rangle=\sqrt{n+1}|n+1\rangle\,.  
\end{eqnarray}
With these conventions in place, we move to the anharmonic oscillator. 
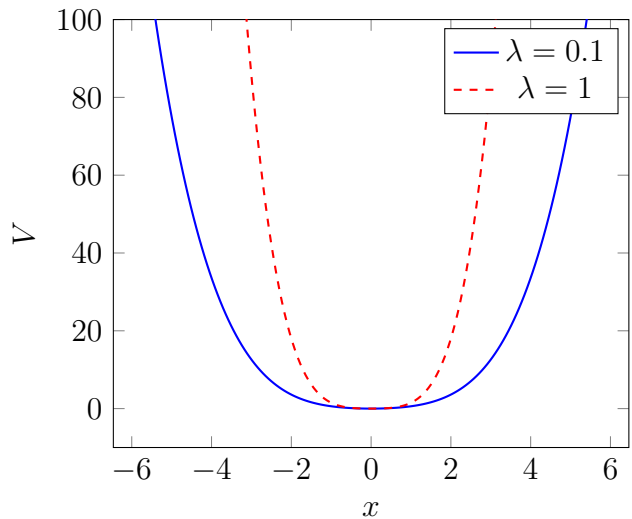
\begin{figure}[hbt]
\begin{center}
\begin{tikzpicture}
  \begin{axis}[xlabel={$x$}, ylabel={$V$}, ymax=100, legend style={cells={align=left}}]
    \addplot[thick, blue] [domain=-10:10, samples=300, thick]  {x^2/2+0.1*x^4};
    \addlegendentry{$\lambda=0.1$}
    \addplot[thick, red, dashed] [domain=-10:10, samples=300, thick] {x^2/2+x^4};
    \addlegendentry{$\lambda=1$}
  \end{axis}
\end{tikzpicture}
\end{center}
\caption{The anharmonic potential.}
\end{figure}

Consider
$$
H=\frac{p^2}{2}+\frac{x^2}{2}+\lambda x^4\,.
$$
To find the eigen-spectrum numerically, we can find a matrix representation of $H$ in the SHO basis and perform exact diagonalization. Then we check for convergence of the eigenvalues as we keep increasing the size of the matrix. The 4-decimal place converged results are shown in table \ref{tab:anharmonic-combined}.

\vskip 1cm

\begin{table}[h]
\centering
\caption{Eigenvalues and level spacings for $H=\frac{p^2}{2}+\frac{x^2}{2}+\lambda x^4$ with $\hbar=\omega=m=1$.}
\label{tab:anharmonic-combined}
\begin{tabular}{@{}l r r r r@{}}
\toprule
 & $\lambda=0.1$ & $\Delta E$ & $\lambda=1$ & $\Delta E$ \\
\midrule
E$_{0}$ & 0.5591 & 1.2104 & 0.8038 & 1.9341 \\
E$_{1}$ & 1.7695 & 1.3691 & 2.7379 & 2.4414 \\
E$_{2}$ & 3.1386 & 1.4903 & 5.1793 & 2.7631 \\
E$_{3}$ & 4.6289 & 1.5914 & 7.9424 & 3.0212 \\
E$_{4}$ & 6.2203 & 1.6795 & 10.9636 & 3.2395 \\
E$_{5}$ & 7.8998 & 1.7580 & 14.2031 & 3.4309 \\
E$_{6}$ & 9.6578 & 1.8295 & 17.6340 & 3.6024 \\
E$_{7}$ & 11.4873 & 1.8952 & 21.2364 & 3.7585 \\
E$_{8}$ & 13.3825 & 1.9561 & 24.9949 & 3.9024 \\
E$_{9}$ & 15.3386 & -- & 28.8973 & -- \\
\bottomrule
\end{tabular}
\end{table}

 \paragraph{Perturbative expansion.}  For comparison, let us review perturbative results. 
Using perturbation theory we find
\begin{align}
E_0(\lambda) &= \tfrac{1}{2} + \tfrac{3}{4} \lambda - \tfrac{21}{8} \lambda^2 + \mathcal O(\lambda^3),\\
E_1(\lambda) &= \tfrac{3}{2} + \tfrac{15}{4} \lambda - \tfrac{405}{8} \lambda^2 + \mathcal O(\lambda^3).
\end{align}
The coefficients grow factorially at high order. From table \ref{tab:anharmonic-combined}  we have for $\lambda=0.1$, the approximate values
$$E_0\approx 0.559\,,\quad E_1\approx 1.769.$$
Plugging $\lambda=0.1$ in the perturbative expressions, we get $0.549$ and $1.369$ respectively. While $E_0$ agrees, $E_1$ is off. This is suggestive that $\lambda=0.1$ is not small enough and simple-minded perturbation theory is not of much use!

\begin{figure}[hbt]
  \centering
  \includegraphics[width=0.8\textwidth]{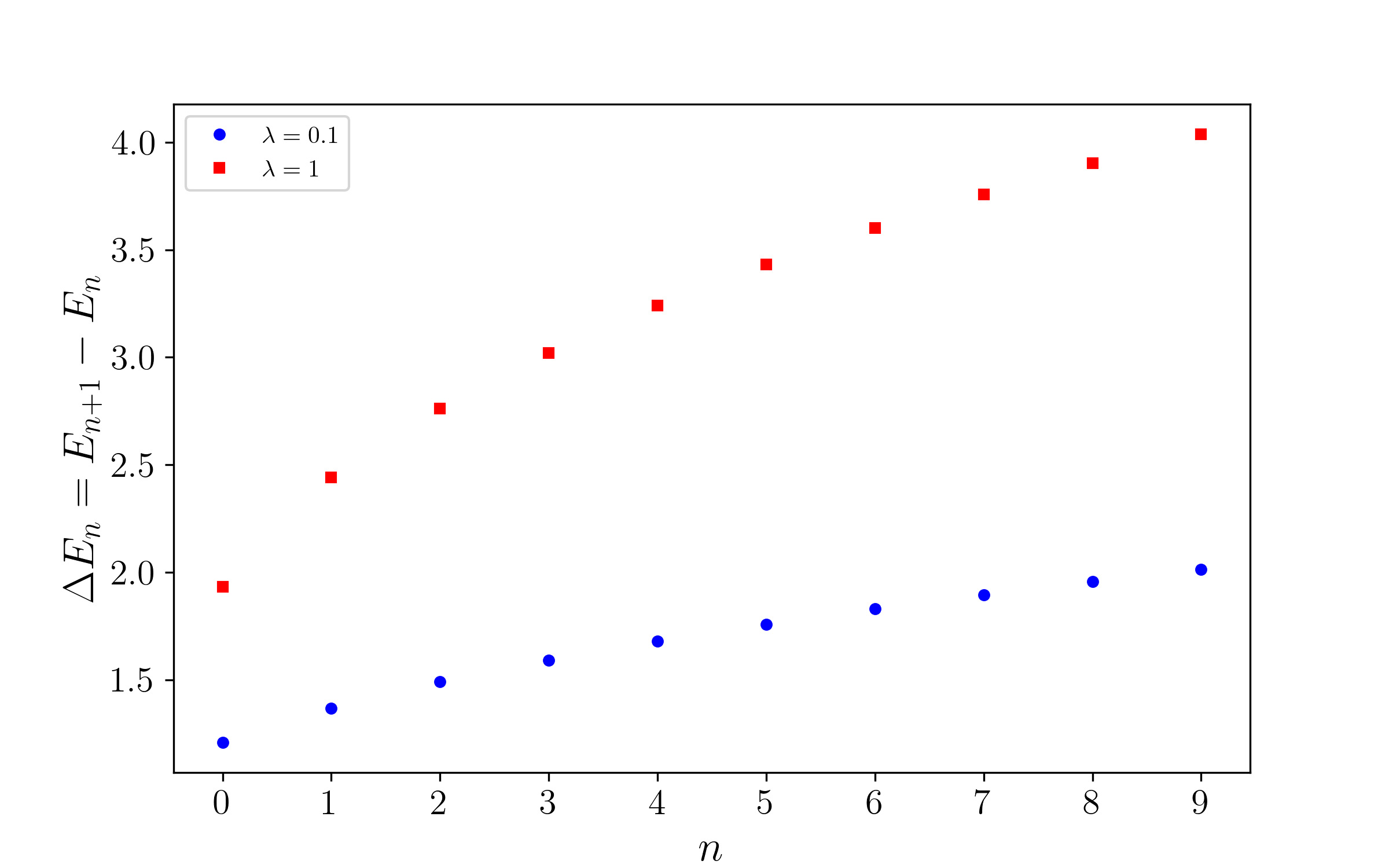}
  \caption{Level spacings $\Delta E_n$ for $H=\tfrac{p^2}{2}+\tfrac{x^2}{2}+\lambda x^4$.}
  \label{fig:deltaE-anharmonic}
\end{figure}

The level spacings are plotted in fig.\ref{fig:deltaE-anharmonic}. The increase in level-spacing is significant. One cannot hope to have a qubit if the level spacings are constant. If the level spacings become bigger then the low lying modes will separate out. This helps in having qubits in realistic physical systems, especially if the gap between the first two levels from the rest is high. 

By the way, an \emph{increasing} level spacing with excitation number is not an exotic curiosity but a common feature of systems with an effective positive quartic correction to the restoring potential. In cold atoms confined by optical traps or lattices, the potential is approximately harmonic near the center, while the Gaussian intensity profile of the beams generates quartic corrections away from the minimum; as a result the spacings \(\Delta E_n\) grow with \(n\). Trapped ions in Paul or Penning devices show the same trend when imperfections in the RF/DC fields add quartic terms beyond ideal quadrupole confinement, producing a hardening anharmonicity and larger vibrational gaps at higher quanta. In crystalline solids, the expansion of interatomic potentials about equilibrium includes an \(x^4\) term; at large amplitudes phonon modes harden, and in the quantized picture the spacing between vibrational levels increases with the vibrational quantum number. Mesoscopic electron traps such as quantum dots often realize an electrostatic confinement that is harmonic to leading order with quartic corrections set by gate geometry, leading to excitation spectra that curve upward in precisely the way expected for \(\lambda>0\). The same physics underlies nonlinear mechanical and optical oscillators operating in the Duffing \emph{hardening} regime: a positive quartic restoring term produces an amplitude-dependent frequency increase, which in the quantum description appears as rising \(\Delta E_n\). Even in superconducting circuits one can engineer this behavior—although transmons are weakly \emph{softening}, other architectures (for example, fluxonium at appropriate flux bias or Josephson arrays with tailored inductive/capacitive networks) realize effective positive-quartic potentials and therefore exhibit level spacings that increase with excitation\footnote{The strict $\cos x$ type potential gives decreasing level spacings.}.

\begin{center}
\begin{tikzpicture}
  \begin{axis}[
    width=8cm, height=6cm,
    xlabel={\(x\)}, ylabel={\(y\)},
    grid=both, legend pos=north east]
    \addplot[domain=0:6, samples=300, thick] {(1-exp(1-x))^2};
    \addlegendentry{$Morse$}
  \end{axis}
\end{tikzpicture}
\end{center}
On the other hand, in certain diatomic molecules, the level spacings in fact {\it decreases} with increasing level. Modeling this is harder but can be done using the so-called Morse potential. This is given by 
$$V_{Morse}=D_e(1-\exp(-x+x_0))^2\,.$$


\section{Double well}

\begin{figure}[H]
\begin{center}
\begin{tikzpicture}
  \begin{axis}[xlabel={$x$}, ylabel={$V$}, legend style={cells={align=left}}, ymin=-1, ymax=2]
    \addplot[thick, blue] [domain=-3:3,  samples=300, thick]  {-x^2/2+0.1*x^4};
    \addlegendentry{$\lambda=0.1$}
    \addplot[thick, red, dashed] [domain=-3:3, samples=300, thick] {-x^2/2+x^4};
    \addlegendentry{$\lambda=1$}
  \end{axis}
\end{tikzpicture}
\end{center}
\end{figure}
Now we turn to the symmetric double-well potential. First let's write it as
$$H=\frac{p^2}{2}-\frac{x^2}{2}+\lambda x^4\,.$$
For this the eigenspectrum can be computed similarly. We have the benchmarking table and level-spacing plots below. Notice that unlike the previous case, $E_0, E_1$ are similar in magnitude for $\lambda=0.1$. There is rich physics behind this, which we will explain next. 
\begin{table}[h]
\centering
\caption{Eigenvalues and level spacings for $H=\tfrac{p^2}{2}-\tfrac{x^2}{2}+\lambda x^4$ with $\hbar=\omega=m=1$.}
\label{tab:inverted-anharmonic-combined}
\begin{tabular}{@{}l r r r r@{}}
\toprule
 & $\lambda=0.1$ & $\Delta E$ & $\lambda=1$ & $\Delta E$ \\
\midrule
E$_{0}$ & -0.1541 & 0.2969 & 0.5148 & 1.5058 \\
E$_{1}$ & 0.1428 & 0.8674 & 2.0206 & 2.1705 \\
E$_{2}$ & 1.0102 & 0.9389 & 4.1911 & 2.5144 \\
E$_{3}$ & 1.9491 & 1.1095 & 6.7055 & 2.7961 \\
E$_{4}$ & 3.0586 & 1.2301 & 9.5016 & 3.0305 \\
E$_{5}$ & 4.2887 & 1.3352 & 12.5321 & 3.2342 \\
E$_{6}$ & 5.6239 & 1.4278 & 15.7663 & 3.4156 \\
E$_{7}$ & 7.0517 & 1.5108 & 19.1819 & 3.5798 \\
E$_{8}$ & 8.5625 & 1.5865 & 22.7617 & 3.7305 \\
E$_{9}$ & 10.1490 & -- & 26.4922 & -- \\
\bottomrule
\end{tabular}
\end{table}

\begin{figure}[h]
  \centering
  \includegraphics[width=0.6\textwidth]{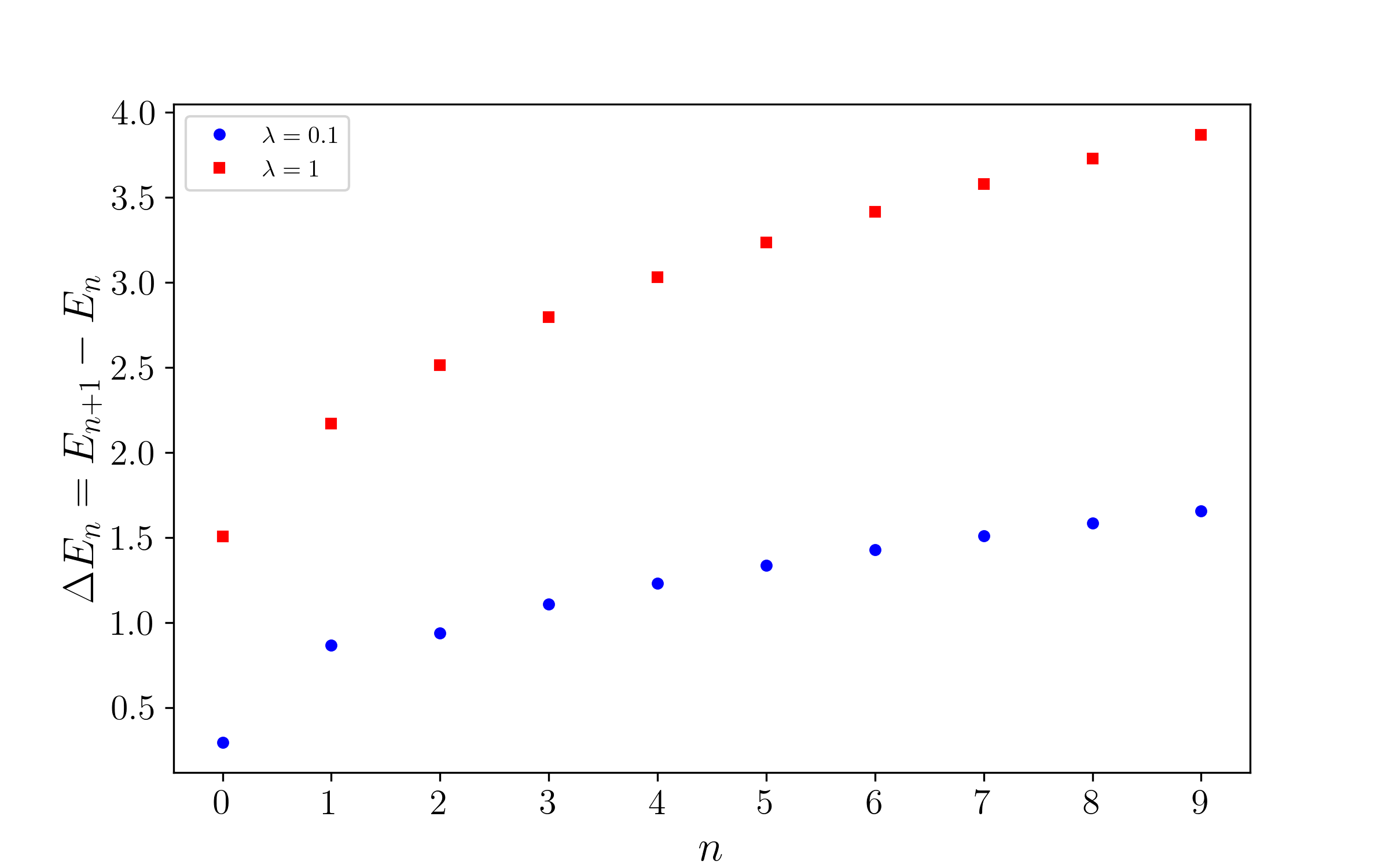}
  \caption{Level spacings $\Delta E_n$ for $H=\frac{p^2}{2}-\frac{x^2}{2}+\lambda x^4$.}
  \label{fig:deltaE-inverted}
\end{figure}

The symmetric quartic double-well potential
\begin{equation}
V(x) = \lambda \,(x^2 - a^2)^2
\end{equation}
is the canonical example of a quantum-mechanical system where semiclassical, nonperturbative effects are essential. The two classical minima at $x=\pm a$ each support an approximately harmonic ground state. However, quantum tunneling through the central barrier mixes the states localized in the left and right wells, producing symmetric and antisymmetric combinations as the true energy eigenstates. This mixing generates an exponentially small level splitting between the ground doublet. 

Perturbation theory around either minimum captures the harmonic oscillator corrections but entirely misses the splitting: no finite-order perturbative expansion can couple the two wells. The appropriate tool is the semiclassical instanton method, which evaluates the Euclidean path integral by including classical Euclidean solutions connecting the two minima. In what follows we review some of the physics behind this.

\section{Localized states and exact propagators}

Let us denote by $|L\rangle$ and $|R\rangle$ wavefunctions localized near $-a$ and $+a$, respectively. These are not exact eigenstates of the Hamiltonian but provide a useful approximate basis. The true eigenstates of definite parity are
\begin{equation}
|+\rangle = \frac{|L\rangle + |R\rangle}{\sqrt{2(1+S)}}, \qquad
|-\rangle = \frac{|L\rangle - |R\rangle}{\sqrt{2(1-S)}},
\end{equation}
where $S=\langle L|R\rangle$ is the small overlap.

The Euclidean propagators defined via
\begin{equation}
K_{LL}(T) = \langle L| e^{-HT} |L\rangle, \qquad
K_{LR}(T) = \langle L| e^{-HT} |R\rangle
\end{equation}
can be expressed in terms of the low-lying eigenstates. Neglecting higher excitations,
\begin{align}
K_{LL}(T) &= \frac{1+S}{2}\,e^{-E_+ T} + \frac{1-S}{2}\,e^{-E_- T}, \\
K_{LR}(T) &= \frac{1+S}{2}\,e^{-E_+ T} - \frac{1-S}{2}\,e^{-E_- T}.
\end{align}
Adding and subtracting gives
\begin{equation}
K_{LL}(T) \pm K_{LR}(T) = (1\pm S)\, e^{-E_{\pm} T}.
\end{equation}
Thus by computing the Euclidean propagators between localized states, one can directly read off the symmetric and antisymmetric energies $E_{\pm}$.

\section{Euclidean path integral and instantons}

\begin{figure}[h]
\centering
\begin{tikzpicture}[scale=0.8]
\draw[->] (-3,0) -- (3,0) node[right] {$x$};
\draw[->] (0,-0.2) -- (0,3) node[above] {$V(x)$};
\draw[thick,blue, domain=-3.7:3.7,smooth,variable=\x] plot ({\x},{0.1*(\x*\x-2*2)^2/4});
\node at (-2,0) [below] {$-a$};
\node at (2,0) [below] {$+a$};
\node at (0,1.2) {$V(0)=\lambda a^4$};
\end{tikzpicture}
\caption{The symmetric quartic double well. The instanton connects $-a$ to $+a$ in Euclidean time.}
\end{figure}
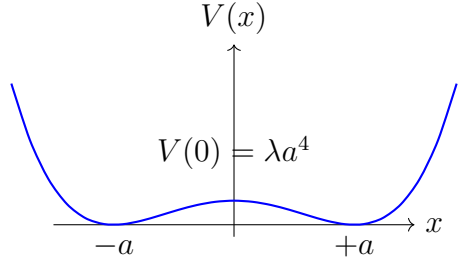

The Euclidean path integral representation is
\begin{equation}
K_{LR}(T) = \int_{x(0)=-a}^{x(T)=+a}\! \mathcal{D}x(\tau)\, e^{-S_E[x]/\hbar}, \qquad
S_E[x] = \int_0^T d\tau \left[\tfrac12 \dot{x}^2 + V(x)\right].
\end{equation}
Saddle points of $S_E$ dominate in the semiclassical limit. The Euler--Lagrange equation is
\begin{equation}
\ddot{x} = V'(x).
\end{equation}
Besides the trivial vacua at $\pm a$, there exist finite-action solutions interpolating between them. These are the \emph{instantons}---a terminology attributed to 't Hooft. 

The strategy to derive them is as follows. Notice that 
\begin{equation}
    \frac12 \dot{x}^2+\lambda(x^2-a^2)^2=\frac12(\dot x+\sqrt{2\lambda}(x^2-a^2))^2-\sqrt{2\lambda}\dot x (x^2-a^2)\,.
\end{equation}
Therefore, for the action to be minimum (saddle) we must have
\begin{equation}\label{BPS}
    \dot x+\sqrt{2\lambda}(x^2-a^2)=0\,.
\end{equation}
This also reads:
\begin{equation}
    \frac{dx}{d\tau}=\sqrt{2V(x)}\,.
\end{equation}
[Since $-a<x<a$, we choose this branch.]
This equation eq.(\ref{BPS}) is easy to solve. Explicitly, one finds
\begin{equation}
x_{\text{inst}}(\tau) = a\, \tanh\!\left(\frac{\omega}{2}(\tau-\tau_0)\right), \qquad \omega = \sqrt{V''(a)} = 2\sqrt{2\lambda}\,a,
\end{equation}
with center $\tau_0$ (integration constant). 

\begin{center}
    
\begin{tikzpicture}
  \begin{axis}[xlabel={$x$}, ylabel={$V$}, legend style={cells={align=left}}, ymin=-2, ymax=2]
    \addplot[thick, blue] [domain=-7:7,  samples=300, thick]  {tanh(x)};
    \addlegendentry{kink}
    \addplot[thick, red, dashed] [domain=-7:7, samples=300, thick] {-tanh(x)};
    \addlegendentry{anti-kink}
  \end{axis}
\end{tikzpicture}

\end{center}

The action of a single instanton is
\begin{equation}
S_0 = \int_0^T d\tau 2 V(x) =\int_{-a}^a \sqrt{2V(x)}\,dx = \frac{4}{3}\sqrt{2\lambda}\,a^3.
\end{equation}
There is another solution (which we leave has homework): 

\begin{equation}
x_{\text{inst}}(\tau) = -a\, \tanh\!\left(\frac{\omega}{2}(\tau-\tau_0)\right), \qquad \omega = \sqrt{V''(a)} = 2\sqrt{2\lambda}\,a,
\end{equation}
which corresponds to $K_{RL}$ which is called the anti-instanton.
The instanton has a characteristic width $\Delta\tau \sim 1/\omega$ in Euclidean time. The parameter $\tau_0$ is a collective coordinate: translating the instanton does not change its action, and the path integral includes an integration $\int d\tau_0$ over its position.

\section{The dilute instanton gas approximation}

A path contributing to $K_{LL}(T)$ or $K_{LR}(T)$ may contain multiple instantons and anti-instantons. For instance the amplitude for $L\rightarrow R$ tunneling should also include the possiblities that $L\rightarrow R$ followed by $R\rightarrow L$ and again $L\rightarrow R$ events (i.e., an odd number of tunnelling events). Each tunneling event carries the system across the barrier in a Euclidean interval of width $\sim 1/\omega$. If these events are widely separated, their actions add approximately: $S \approx n S_0$. This is the dilute instanton gas approximation (DIGA).

For a configuration with $n$ tunneling events, each contributes a factor $K e^{-S_0}$, where $K$ is the one-loop determinant prefactor. Each event has a free center $\tau_i$, and integrating over its position in $[0,T]$ gives $\int_0^T d\tau_i$. For $n$ events this yields $\int_0^T d\tau_1\cdots d\tau_n$. To avoid overcounting permutations of indistinguishable instantons, one divides by $n!$. Thus the contribution is
\begin{equation}
\frac{1}{n!}\,\big(K T e^{-S_0}\big)^n.
\end{equation}

Summing over all even $n$ (returning to the initial well) produces
\begin{equation}
K_{LL}(T) \;\propto\; e^{-E_{\rm pert} T}\cosh\!\big(K T e^{-S_0}\big),
\end{equation}
while summing over all odd $n$ (ending in the opposite well) gives
\begin{equation}
K_{LR}(T) \;\propto\; e^{-E_{\rm pert} T}\sinh\!\big(K T e^{-S_0}\big).
\end{equation}
Here $E_{\rm pert}$ denotes the perturbative vacuum energy around a single minimum. In the dilute instanton gas approximation the tunneling amplitude always comes multiplied by a factor of the form $e^{-E_{\rm pert}T}$. This contribution originates from the fact that, away from the localized instanton core of width $\mathcal{O}(\omega^{-1})$, the system spends essentially the entire Euclidean time interval $T$ sitting in one of the perturbative vacua. The corresponding path integral over fluctuations around a well gives precisely the perturbative ground state energy $E_{\rm pert}$: the classical potential at the minimum, the Gaussian zero--point energy coming from the fluctuation determinant, and higher--order loop corrections. Since this background contribution scales linearly with $T$, it factors out as $e^{-E_{\rm pert}T}$ in all instanton sectors, while the instanton itself only supplies a finite action $S_0$ and a prefactor independent of $T$.

Adding and subtracting again, one finds
\begin{equation}
K_{LL}(T) \pm K_{LR}(T) \;\propto\; e^{-E_{\rm pert} T} \exp\!\left(\pm K T e^{-S_0}\right),
\end{equation}
so that the energies of the symmetric and antisymmetric states are
\begin{equation}
E_{\pm} = E_{\rm pert} \mp K e^{-S_0} + \cdots,
\end{equation}
and the ground-state splitting is
\begin{equation}
\Delta = E_{-}-E_{+} = 2K e^{-S_0}\left[1+\mathcal{O}(\hbar)+\mathcal{O}(e^{-S_0})\right].
\end{equation}

Explicitly for $H=p^2/2-x^2/2+\lambda x^4$ and the lowest states\footnote{$E_{2n+1}-E_{2n}\propto e^{-S_0}$, with the $n$-dependence in the prefactor.}, we have
\begin{equation}
    \Delta\approx \frac{1.34}{\sqrt{\lambda}}\exp(-\frac{0.236}{\lambda})\,.
\end{equation}
This means that $|+\rangle$ is the lowest energy state, which is a symmetric combination of $L$ and $R$ vacua. Thus, there is ``restoration of broken symmetry" (where either the $L$ or the $R$ is the vacua) due to non-perturbative effects. 
For $\lambda=0.1$ this predicts $\Delta\approx 0.4$ while the numerical answer is $0.3$. For $\lambda=0.7$ this predicts $\Delta\approx 1.14$ while numerics gives $1.28$. The agreement is expected to improve on including multi-instanton contributions. What we have added are just the one instanton case weighted by $e^{-S_0}$ and have ignored terms like $e^{-2S_0}$ and higher.

The sharp benchmark targets for a quantum computer are to produce:
\begin{eqnarray}
    \Delta=0.3 \qquad \lambda=0.1\\
    \Delta=1.28 \qquad \lambda=0.7
\end{eqnarray}

\begin{figure}[hbt]
\centering
\begin{tikzpicture}
\pgfmathsetmacro{\a}{2}     
\pgfmathsetmacro{\w}{2}     
\pgfmathsetmacro{\tL}{2.0}  
\pgfmathsetmacro{\tR}{6.0}  
\pgfmathsetmacro{\dt}{1/\w} 

\begin{axis}[
  axis lines=middle,
  xlabel={$\tau$}, ylabel={$x(\tau)$},
  xmin=0, xmax=8, ymin=-2.5, ymax=2.5,
  width=0.85\textwidth, height=5.2cm,
  ticks=none, clip=false
]

\addplot[thick,red,domain=0:8,samples=400]
  { \a * (exp((\w/2)*(x-\tL)) - exp(-(\w/2)*(x-\tL)))
        /(exp((\w/2)*(x-\tL)) + exp(-(\w/2)*(x-\tL))) };

\addplot[thick,blue,domain=0:8,samples=400]
  { -\a * (exp((\w/2)*(x-\tR)) - exp(-(\w/2)*(x-\tR)))
         /(exp((\w/2)*(x-\tR)) + exp(-(\w/2)*(x-\tR))) };

\addplot[dashed] coordinates {(\tL-\dt,-2.4) (\tL-\dt,2.4)};
\addplot[dashed] coordinates {(\tL+\dt,-2.4) (\tL+\dt,2.4)};
\addplot[<->] coordinates {(\tL-\dt,-2.1) (\tL+\dt,-2.1)};
\node[anchor=north] at (axis cs:\tL,-2.1) {$\;\sim\,\tfrac{1}{\omega}$};
\node[anchor=north] at (axis cs:\tL,-2.75) {instanton center $\tau_0=\tL$};

\addplot[dashed] coordinates {(\tR-\dt,-2.4) (\tR-\dt,2.4)};
\addplot[dashed] coordinates {(\tR+\dt,-2.4) (\tR+\dt,2.4)};
\addplot[<->] coordinates {(\tR-\dt,2.1) (\tR+\dt,2.1)};
\node[anchor=south] at (axis cs:\tR,2.1) {$\;\sim\,\tfrac{1}{\omega}$};
\node[anchor=south] at (axis cs:\tR,2.75) {anti-instanton center $\tau_0=\tR$};

\end{axis}
\end{tikzpicture}
\caption{A dilute instanton–anti-instanton pair. Each kink (and anti-kink) switches between vacua over an Euclidean-time window of width $\sim 1/\omega$, shown by dashed lines at $\tau_0\pm 1/\omega$ and a double arrow. Widely separated means the two cores do not overlap: $|\tau_0^{(\text{I})}-\tau_0^{(\text{AI})}|\gg 1/\omega$.}
\end{figure}
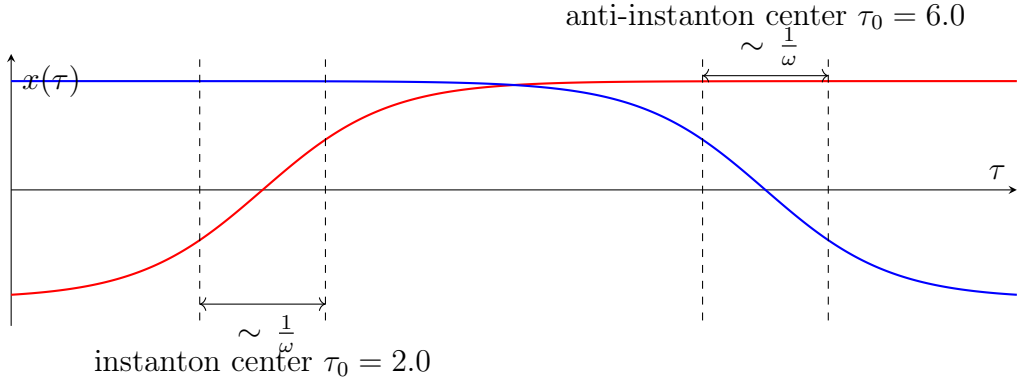

An important clarification: In the dilute instanton gas approximation, configurations with $n$ tunnelling events (instantons or anti--instantons) are described by paths that remain in a perturbative vacuum for almost all of the Euclidean interval $[0,T]$, with $n$ well--separated localized transitions of width $\mathcal{O}(\omega^{-1})$. Each instanton possesses a collective coordinate $\tau_i$ corresponding to the freedom of translating its center in Euclidean time. The path integral over such $n$--instanton configurations therefore contains an integration over all possible choices of these centers:
\[
\int_0^T d\tau_1 \int_0^T d\tau_2 \cdots \int_0^T d\tau_n \, .
\]
However, because the instantons are indistinguishable objects, different orderings of the same set of centers $\{\tau_1,\dots,\tau_n\}$ represent the \emph{same physical configuration}. To avoid overcounting, one restricts to an ordered domain $\tau_1<\tau_2<\cdots<\tau_n$, which gives
\[
\int_{0<\tau_1<\tau_2<\cdots<\tau_n<T} d\tau_1 \cdots d\tau_n
= \frac{T^n}{n!}.
\]
Thus the contribution of the $n$--instanton sector is proportional to $(K T e^{-S_0})^n/n!$, where $K$ is the one--instanton fluctuation prefactor. In this way DIGA does not assume the events are simultaneous; rather, it sums over all possible positions of the instantons, with the $1/n!$ factor arising from the indistinguishability of their centres in the path integral.


\section{Beyond the dilute gas}

If an instanton and anti-instanton occur close together, their cores overlap and the action is less than $2S_0$. These configurations are important for understanding the analytic structure of perturbation theory: they generate ambiguities in the so-called Borel resummation of divergent series, which are cancelled by the perturbative sector in the framework of resurgence. However, in the dilute limit $K T e^{-S_0}\ll 1$, the measure for such close pairs is suppressed, and they do not affect the leading real tunneling splitting. The dilute instanton gas approximation therefore provides the leading-order semiclassical prediction for the level splitting, with corrections systematically organized as loop effects around the instanton and multi-instanton contributions.

After this discussion of the spectrum of the anharmonic oscillator, we turn to the dynamics of scattering---the central theme of these lectures is to study scattering in QFT, especially probing inelasticity. 

\section{Conventional scattering: a recap}

\begin{figure}[h]
\centering
\begin{tikzpicture}[scale=0.9]
  \draw[->] (-5,0) -- (5.2,0) node[right] {$x$};
  \draw[->] (0,0) -- (0,3) node[above] {$V(x)$};


  \draw[thick] (-5,0) -- (-3,0);
  \draw[thick] (-2,0) -- (2,0);
  \draw[thick] (3,0) -- (5,0);

  \filldraw[fill=blue!20,draw=black,thick] (-3,0) rectangle (-2,2.5);
  \filldraw[fill=blue!20,draw=black,thick] (2,0) rectangle (3,2.5);

  \draw[dashed] (-3,0)  (-4,.8) node[below] {$-w/2-b$};
  \draw[dashed] (-2,0)  (-1.5,0.8) node[below] {$-w/2$};
  \draw[dashed] (0,0)  (0,-0.1) node[below] {$0$};
  \draw[dashed] (2,0)  (1.5,0.8) node[below] {$w/2$};
  \draw[dashed] (3,0)  (4,0.8) node[below] {$w/2+b$};

  \node at (-2.5,2.8) {$V_b$};
  \node at (2.5,2.8) {$V_b$};

  \draw[<->] (-3,-0.9) -- (-2,-0.9); \node at (-2.5,-1.2) {$b$};
  \draw[<->] (-2,-0.9) -- (2,-0.9);  \node at (0,-1.2) {$w$};
  \draw[<->] (2,-0.9) -- (3,-0.9);   \node at (2.5,-1.2) {$b$};
\end{tikzpicture}
\caption{Square double–barrier potential with barrier height $V_b$, barrier width $b$, and well width $w$.}
\label{fig:potential}
\end{figure}
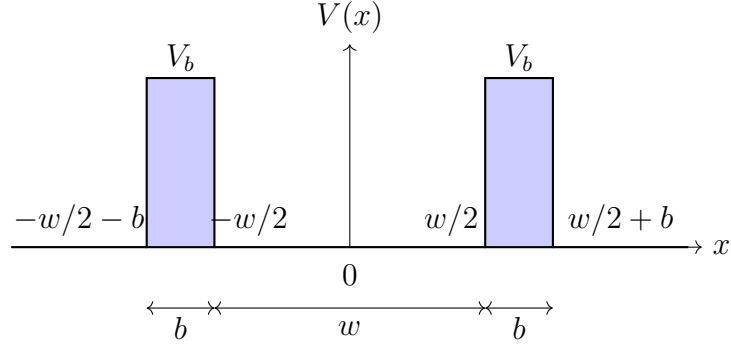

We consider one--dimensional stationary scattering in the potential shown in Fig.~\ref{fig:potential}, consisting of two identical rectangular barriers of height $V_b$ and width $b$, separated by a central well of width $w$.

We work with the time--independent Schr\"odinger equation in natural units $\hbar=1$, $2m=1$:
\begin{equation}
  -\psi''(x)+V(x)\psi(x)=E\,\psi(x),
\end{equation}
so that for free propagation $E=k^2$. 


In each region of constant potential, the solutions are superpositions of right-- and left--moving plane waves:
\begin{align}
\psi(x) &= A^+ e^{ikx} + A^- e^{-ikx}, \quad \text{if } V(x)=0, \; k=\sqrt{E},\\
\psi(x) &= B^+ e^{iqx} + B^- e^{-iqx}, \quad \text{if } V(x)=V_b, \; q=\sqrt{E-V_b}.
\end{align}
For $E<V_b$, $q=i\kappa$ with $\kappa=\sqrt{V_b-E}$, so the barrier solutions are real exponentials.


At an abrupt interface between two regions with wave numbers $k_a$ (left) and $k_b$ (right), continuity of $\psi$ and $\psi'$ gives
\begin{equation}
\begin{pmatrix}B^+\\ B^-\end{pmatrix}
=
\frac{1}{2}\begin{pmatrix}
1+\tfrac{k_b}{k_a} & 1-\tfrac{k_b}{k_a}\\[4pt]
1-\tfrac{k_b}{k_a} & 1+\tfrac{k_b}{k_a}
\end{pmatrix}
\begin{pmatrix}A^+\\ A^-\end{pmatrix}.
\end{equation}
This $2\times 2$ \emph{interface matrix} $S_{a\to b}$ encodes matching conditions.


Propagation through a uniform region of width $d$ and wave number $k$ multiplies the amplitudes by a diagonal phase matrix
\begin{equation}
P(k,d)=
\begin{pmatrix} e^{ikd} & 0 \\ 0 & e^{-ikd}\end{pmatrix}.
\end{equation}


Concatenating the interfaces and propagations, the total transfer matrix across the double--barrier structure is
\begin{equation}
M(E)=S_{k\to q}\,P(q,b)\,S_{q\to k}\,P(k,w)\,S_{k\to q}\,P(q,b)\,S_{q\to k}.
\end{equation}
This relates the incoming/outgoing amplitudes in the left lead $(A_L^+,A_L^-)$ to those in the right lead $(A_R^+,A_R^-)$:
\begin{equation}
\begin{pmatrix}A_R^+\\ A_R^-\end{pmatrix}
= M(E)\begin{pmatrix}A_L^+\\ A_L^-\end{pmatrix}.
\end{equation}

\subsection{Scattering amplitudes}

For left incidence, we choose
\begin{equation}
\binom{A_L^+}{A_L^-}=\binom{1}{r},\qquad
\binom{A_R^+}{A_R^-}=\binom{t}{0},
\end{equation}
so that there is a unit incoming wave from the left, reflection amplitude $r$, and transmission amplitude $t$ to the right. Inserting into the transfer relation gives
\begin{equation}
r(E)=-\frac{M_{21}(E)}{M_{22}(E)},\qquad
t(E)=\frac{1}{M_{22}(E)}.
\end{equation}
The reflection and transmission probabilities are $R(E)=|r(E)|^2$, $T(E)=|t(E)|^2$, with $R+T=1$.

\subsection{Resonances and Wigner time delay}

Let the incident state be a narrowband packet centered at $E_0$:
\[
\Psi_{\rm in}(x,t)=\int dE\,a(E)\,e^{i(kx-Et)},\qquad a(E)\ \text{peaked at }E_0.
\]
The transmitted packet is
\[
\Psi_{\rm tr}(x,t)=\int dE\,a(E)\,t(E)\,e^{i(kx-Et)},\qquad t(E)=|t(E)|e^{i\phi(E)}.
\]
Expand the phase $\phi(E)\approx \phi(E_0)+\phi'(E_0)(E-E_0)$ and take $|t(E)|$ slowly varying across the packet. A stationary-phase (or envelope) analysis shows the peak of the transmitted packet is shifted in time by
\begin{equation}
\boxed{\ \Delta t \;=\; \dv{\phi(E)}{E}\Big|_{E_0}\ \equiv\ \tau(E_0)\ ,\ }
\end{equation}
the \emph{Wigner time delay} for the transmission channel. Intuitively, near a quasi--bound level the wave dwells inside the structure, generating a rapidly varying scattering phase $\phi(E)$; the slope $\dv{\phi}{E}$ measures that extra dwell time relative to free motion.

\paragraph{Single isolated resonance.}
Near an isolated resonance at $E_r$ with total width $\Gamma=\Gamma_L+\Gamma_R$, the transmission amplitude takes the Breit--Wigner form (up to a slowly varying background phase $\theta$):
\begin{equation}
t(E)\ \approx\ e^{i\theta(E)}\,\frac{\sqrt{\Gamma_L\Gamma_R}}{E-E_r+i\Gamma/2}.
\end{equation}
Then
\begin{equation}
\boxed{\
\tau(E)=\dv{E}\arg t(E)\ \approx\ \frac{\Gamma/2}{(E-E_r)^2+(\Gamma/2)^2}\ ,\qquad
\tau(E_r)=\frac{2}{\Gamma},}
\end{equation}
a Lorentzian peak of height $2/\Gamma$ centered at $E_r$. On the real axis one also observes a $\sim\pi$ phase jump in $\phi(E)$ across the resonance and a near--unity peak in $T(E)$ (for symmetric barriers).
Formally, resonances are the poles of the scattering matrix (or $t$) in the \emph{lower} half complex--energy plane:
\[
E_\star=E_r-i\Gamma/2.
\]

\subsection*{Resonance condition in the transfer--matrix language}
Since $t(E)=1/M_{22}(E)$ for identical leads and real $V(x)$, the complex resonance energies are the zeros of $M_{22}$:
\begin{equation}
\boxed{\ M_{22}(E_\star)=0,\qquad E_\star=E_r-i\Gamma/2.\ }
\end{equation}

\section{Resonances in the inverted double well}

There is another case of the anharmonic oscillator which is of great interest. Consider
\begin{eqnarray}
H=\frac{p^2}{2}+\frac{x^2}{2}-\lambda x^4\,.
\end{eqnarray}
This gives rise to a double-hump or inverted double-well potential. This is instructive for two reasons:
\begin{itemize}
    \item A transmon qubit is a Josephson junction in parallel with a capacitor. In the transmon qubit models, the Hamiltonian is 
    \begin{equation}
        H=-E_J \cos \phi+\frac{Q^2}{2C}\,.
    \end{equation}
    Here $E_J, C$ are constants and $\phi, Q$ are conjugate variables. Expanding the $\cos\phi$ term we get a Hamiltonian that looks like the anharmonic oscillator with $\lambda<0$. Thus, the anharmonic oscillator with $\lambda<0$ is an approximation to this case. Here the energy level spacings in fact decrease with level number (we will need the $\cos\phi$ potential to show this).
    \item Our focus will be resonances. As we will examine below, a simple and instructive “scattering–without–asymptotics” example is the anharmonic oscillator with $\lambda<0$.
\end{itemize}
For convenience, we will consider a slight rewriting of the Hamiltonian for the
\emph{inverted} double–well
\begin{equation}
  H \;=\; \frac{p^2}{2m} \;-\; \lambda\,(x^2-a^2)^2,
  \qquad (m=\hbar=1 \text{ in what follows}).
\end{equation}
The “barrier tops” sit at \(E=0\), while the pocket at the origin has depth
\(V(0)=-\lambda a^4\).
Near \(x=0\) the pocket is approximately harmonic with
\(\omega_0=\sqrt{V''(0)}=2a\sqrt{\lambda}\).

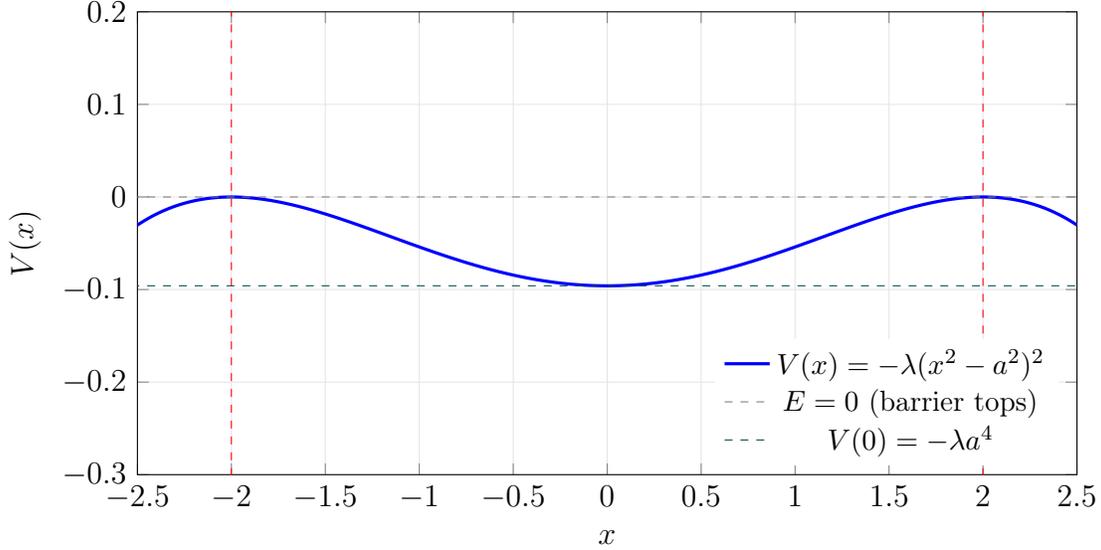
\begin{figure}[h]
  \centering
  \begin{tikzpicture}
    \begin{axis}[
      width=0.8\textwidth, height=0.44\textwidth,
      xlabel={$x$}, ylabel={$V(x)$},
      xmin=-2.5, xmax=2.5, ymax=0.2, ymin=-0.3,
      domain=-2.5:2.5, samples=400,
      grid=both, grid style={gray!20},
      legend style={draw=none, at={(0.98,0.02)}, anchor=south east, font=\small}
    ]
      \pgfmathsetmacro{\A}{2.0}
      \pgfmathsetmacro{\Lam}{0.006}
      \pgfmathsetmacro{\Vzero}{-\Lam*\A*\A*\A*\A}
      \addplot[blue, very thick] {-\Lam*(x^2-\A^2)^2};
      \addlegendentry{$V(x)=-\lambda(x^2-a^2)^2$}
      \addplot[gray, dashed] coordinates {(-6.5,0) (6.5,0)};
      \addlegendentry{$E=0$ (barrier tops)}
      \addplot[teal!60!black, dashed] coordinates {(-6.5,\Vzero) (6.5,\Vzero)};
      \addlegendentry{$V(0)=-\lambda a^4$}
      \addplot[red, dashed] coordinates {(-\A,-4) (-\A,0.5)};
      \addplot[red, dashed] coordinates {(\A,-4) (\A,0.5)};
      \node[red!70!black] at (axis cs:-\A,0.35) {$-a$};
      \node[red!70!black] at (axis cs:\A,0.35) {$+a$};
    \end{axis}
  \end{tikzpicture}
  \caption{Inverted double well with \(a=2\), \(\lambda=0.006\).
  The pocket bottom is at \(V(0)=-\lambda a^4\), the barrier tops at \(E=0\).}
  \label{fig:idw-potential}
\end{figure}

\paragraph{Why conventional scattering is ill-posed here.}
As \(|x|\!\to\!\infty\), \(V(x)\sim -\lambda x^4\to -\infty\). There are no
asymptotically free regions supporting plane waves, so the usual \(S\)-matrix
with in/out states at \(t\to\pm\infty\) is not defined. This does \emph{not}
kill all physics: localized wave packets \(\psi(x,0)\) near the pocket do
escape and evolve in time in a highly structured way. The correct language is
\emph{resonances and transients}, not asymptotic phase shifts.

\paragraph{Physics picture: why resonances \emph{aren’t} ordinary eigenvalues.}
For a closed, Hermitian Hamiltonian $H$, the stationary Schrödinger equation
$H\psi=E\psi$ has only \emph{real} $E$. Time evolution is unitary:
$e^{-iHt}$ just rotates phases; the norm of a state is just unity.

So where does a resonance come from? From a \textbf{metastable trap} that
\emph{leaks probability to infinity}. Prepare a wave packet $\psi(0)$
localized near a pocket/barrier. The packet lingers, then leaks out as an
\emph{outgoing} wave. If leakage is slow and featureless, the pocket
amplitude shows a long exponential window,
\[
A(t)=\mel{\psi(0)}{e^{-iHt}}{\psi(0)} \;\approx\; Z\,e^{-iE_R t}\,e^{-\Gamma t/2}
\quad\text{(intermediate times).}
\]
Here $E_R$ is the “resonance energy” and $\Gamma$ is the \emph{leakage rate}
(lifetime $\tau=1/\Gamma$). This exponential behaviour is the physics definition
of a resonance.

\smallskip
\noindent
\textbf{Where it lives mathematically.} The Laplace/Fourier transform of $A(t)$
is (up to $i$) a matrix element of the {\it resolvent} $1/(z-H)$,
\[
\tilde A(z)\;=\;\int_0^\infty\!dt\,e^{izt}A(t)\;\propto\;\mel{\psi}{(z-H)^{-1}}{\psi}
=:\,F_\psi(z).
\]
An exponential $e^{-iE_R t}e^{-\Gamma t/2}$ transforms into a \emph{simple pole}
at $z_\star=E_R-\tfrac{i}{2}\Gamma$.
But for a self-adjoint $H$, $F_\psi(z)$ is what is called a Herglotz/Nevanlinna function, which is
\emph{analytic off the real axis on the physical sheet}; it cannot have poles with
$\Im z\neq 0$ there. The way out is standard in scattering: \emph{analytically continue}
$F_\psi(z)$ across the continuum cut to the \textbf{second sheet}. On that sheet the same
complex point $z_\star$ is allowed to be a pole. That pole is the resonance.


\section{WKB estimate of the resonance energy and width}

Consider
\[
V(x)=-\lambda(x^2-a^2)^2,\qquad (m=\hbar=1).
\]
For a resonance with energy $E\in\big(V(0),\,0\big)$ (i.e.\ $-\,\lambda a^4<E<0$),
there are four turning points $\pm x_1(E)$ (inner) and $\pm x_2(E)$ (outer), defined by
$E=V(x)$:
\[
|x|=x_{1,2}(E)=\sqrt{a^2 \mp \sqrt{\frac{-E}{\lambda}}}\,,
\qquad \varepsilon\equiv -E>0.
\]
The motion is classically allowed in $|x|<x_1$ (pocket) and in $|x|>x_2$ (escape
region), separated by a forbidden “barrier” $x_1<|x|<x_2$.

\paragraph{Step 1: quantize the real part $E_R$ (inner Bohr–Sommerfeld).}
Ignoring the leakage, the pocket supports quasi-bound levels given at leading WKB by
\[
\oint_{-x_1}^{x_1} p(x;E)\,dx \;=\; 2\pi\Big(n+\tfrac12\Big),\qquad
p(x;E)=\sqrt{2\,[E-V(x)]}.
\]
For a quick estimate near the pocket bottom one may use the local harmonic
approximation $V(x)\approx V(0)+\tfrac12\omega_0^2 x^2$ with
$\omega_0=\sqrt{V''(0)}=2a\sqrt{\lambda}$, but for shape resonances close to
the barrier top the integral is best \emph{evaluated numerically} at the $E$
you will eventually identify as $E_R$.

\paragraph{Step 2: tunneling through the barrier (WKB action).}
Define the barrier action
\[
S_b(E)\;=\;\int_{x_1}^{x_2} \kappa(x;E)\,dx,\qquad
\kappa(x;E)=\sqrt{2\,[V(x)-E]} \quad (x_1< x < x_2).
\]
For our quartic one can write $S_b$ in a single integral with elementary limits.
Set $y=x^2$, $dy=2x\,dx$, and $\varepsilon=-E>0$; then
\[
S_b(E)\;=\;\frac{\varepsilon}{\sqrt{2\lambda}}\int_{-1}^{1}
\frac{\sqrt{1-s^2}}{\sqrt{\,a^2+\alpha s\,}}\;ds,
\qquad \alpha=\sqrt{\frac{\varepsilon}{\lambda}}.
\]
This integral is elementary to evaluate numerically.  Close to the barrier top
($\varepsilon\ll \lambda a^4$) a simple expansion gives the handy estimate
\[
S_b(E) \;=\; \frac{\pi\,\varepsilon}{2\,a\,\sqrt{2\lambda}}
\Big[\,1+\frac{3}{32}\frac{\varepsilon}{\lambda a^4}+ \mathcal O(\varepsilon^2)\Big].
\]

\paragraph{Step 3: “attempt rate” (classical oscillation in the pocket).}
The classical oscillation period in the pocket at energy $E$ is
\[
T_{\rm in}(E)\;=\;\sqrt{2}\int_{-x_1}^{x_1}\frac{dx}{\sqrt{E-V(x)}} \,,
\qquad
\omega_{\rm in}(E)=\frac{2\pi}{T_{\rm in}(E)}.
\]
(Equivalently $T_{\rm in}(E)=\partial S_{\rm in}/\partial E$ with
$S_{\rm in}(E)=\oint p\,dx$.)  Physically, the packet hits a barrier once every
half period, so the total number of “escape attempts” per unit time at the two
sides is $\omega_{\rm in}(E)/\pi$.

\paragraph{Step 4: Gamow formula for the width.}
Each attempt transmits with probability $T(E)\approx e^{-2S_b(E)}$ (WKB).
Hence the decay probability per unit time is
\[
W(E)\;\approx\;\frac{\omega_{\rm in}(E)}{\pi}\,e^{-2S_b(E)}\,,
\]
and the resonance width is (with $\hbar=1$)
\[
\boxed{\;\Gamma(E)\;\approx\;\frac{\omega_{\rm in}(E)}{\pi}\,e^{-2S_b(E)}\; }.
\]
Evaluated at the (real) quantized energy $E_R$ from Step 1 this gives the
leading Gamow width $\Gamma\equiv\Gamma(E_R)$.  In practice one computes
$E_R$ and $\Gamma$ self-consistently: $E_R$ from the pocket quantization and
$\Gamma$ from the barrier action and pocket frequency at that $E_R$.

\subsection*{Numerical WKB estimate}

We take
\[
V(x)=-\lambda(x^2-a^2)^2,\qquad (m=\hbar=1),\qquad
a=2,\ \lambda=0.006.
\]
Then
\[
V(0)=-\lambda a^4=-0.096,\qquad
\omega_0=\sqrt{V''(0)}=2a\sqrt{\lambda}\approx 0.30984.
\]

\paragraph{Step 1: fix the resonance energy \(E_R\) (inner pocket level).}
For the lowest even quasi–bound state we place the center in the pocket band
\(V(0)<E_R<0\).
A crude inner quantization (harmonic pocket corrected by the shallow well) puts
\[
E_R^{\text{(top)}}\approx -0.037 \quad(\text{i.e. }E_R\text{ measured from the barrier top }E=0).
\]
This is consistent with the time–domain peak we observe in numerics (an exercise for you to attempt).

\paragraph{Step 2: turning points and barrier action.}
Turning points solve \(V(x)=E_R\).
With \(\varepsilon=-E_R>0\) and \(s=\sqrt{\varepsilon/\lambda}\),
\[
x_{1,2}(E_R)=\sqrt{\,a^2 \mp s\,},\qquad
x_1\simeq 1.23,\ \ x_2\simeq 2.55.
\]
The WKB barrier action is
\[
S_b(E_R)=\int_{x_1}^{x_2}\!\sqrt{2\,[\,V(x)-E_R\,]}\,dx \;\approx\; 0.277.
\]
(verify using Mathematica)

\paragraph{Step 3: “attempt rate” in the pocket.}
The classical oscillation period in the pocket at energy \(E_R\) is
\[
T_{\rm in}(E_R)=\sqrt{2}\int_{-x_1}^{x_1}\frac{dx}{\sqrt{E_R-V(x)}}\;\approx\; 12.02,
\]
so one encounter with a given barrier occurs roughly once per period.
Hence the \emph{per–barrier} attempt rate is \(1/T_{\rm in}\approx 0.0832\).
(Equivalently, the angular frequency \(2\pi/T_{\rm in}\approx 0.523\); some texts
count both sides and use \(2/T_{\rm in}\). We quote the single–barrier convention
to match our numerics below.)

\paragraph{Step 4: Gamow width.}
With WKB transmission probability \(T(E_R)\approx e^{-2S_b}\),
\[
\Gamma_{\rm WKB}\;\approx\;\frac{1}{T_{\rm in}(E_R)}\,e^{-2S_b(E_R)}.
\]
Numerically,
\[
e^{-2S_b}\approx e^{-0.554}\approx 0.574,\qquad
\Gamma_{\rm WKB}\approx 0.0832\times 0.574\approx 0.048.
\]

\[
\boxed{\,E_R^{\text{(top)}}\approx -0.037,\qquad \Gamma_{\rm WKB}\approx 4.8\times 10^{-2}\,}
\]

\medskip
In this chapter, we reviewed various aspects of the anharmonic oscillator, which will be useful for our studies of quantum field theory. But before we dive into QFT, there is another important piece of physics that we need to review: the Adiabatic Theorem. This will be needed when we prepare scattering states in the presence of interactions.


\medskip

\chapter{The Adiabatic Theorem}
A key tool from quantum mechanics that we will make use of when discussing scattering is the adiabatic theorem. This is why we thought it was prudent to include this in a separate short chapter\footnote{Apart from standard textbooks \cite{sakurai}, Wikipedia is a good source for this.}. When we talk about scattering we will need to build the initial scattering state. In a free theory, this is in principle easy since we rely on Gaussian techniques which are exact in this case. However, for an interacting theory, the procedure we will follow will rely on ramping up from the free theory to the interacting theory by means of the adiabatic theorem.

\noindent\textbf{Setup and statement.}
Let $H(t)$ be a Hamiltonian that varies smoothly on $t\in[0,T]$. Assume (i) a discrete, nondegenerate spectrum with instantaneous eigenpairs
\[
H(t)\,\ket{n(t)}=E_n(t)\ket{n(t)},\qquad \braket{m(t)}{n(t)}=\delta_{mn},
\]
(ii) a finite minimum gap $g(t)=\min_{m\neq n}|E_m(t)-E_n(t)|>0$, and
(iii) sufficiently slow variation of $H(t)$ (quantified below).
If $\ket{\psi(0)}=\ket{n(0)}$, then for all $t\in[0,T]$ the solution of the Schr\"odinger equation
\[
i\hbar\,\partial_t\ket{\psi(t)}=H(t)\ket{\psi(t)}
\]
remains in the corresponding instantaneous eigenspace:
\begin{equation}\label{berry}
\ket{\psi(t)} \simeq e^{i\theta_n(t)}\,e^{i\gamma_n(t)}\ket{n(t)},
\end{equation}
where the \emph{dynamical phase} and \emph{geometric (Berry) phase} are
\[
\theta_n(t)=-\frac{1}{\hbar}\int_0^{t}E_n(t')\,dt',\qquad
\gamma_n(t)=i\int_0^{t}\!\braket{n(t')}{\dot n(t')}dt'.
\]
The approximation error is $O(\varepsilon)$ with $\varepsilon$ set by the adiabatic small parameter below.

\medskip
Expand the state in the instantaneous eigenbasis:
\[
\ket{\psi(t)}=\sum_k c_k(t)\ket{k(t)}.
\]
Insert into the Schr\"odinger equation and project with $\bra{m(t)}$:
\begin{equation}
i\hbar\,\dot c_m(t)+i\hbar\sum_n c_n(t)\,\braket{m(t)}{\dot n(t)}
=E_m(t)\,c_m(t).
\label{eq:cm-eq}
\end{equation}
Differentiate $H(t)\ket{n(t)}=E_n(t)\ket{n(t)}$ and project onto $\bra{m(t)}$ to eliminate the nonadiabatic couplings $\braket{m}{\dot n}$ for $m\neq n$:
\begin{equation}
\braket{m(t)}{\dot n(t)}=
-\frac{\bra{m(t)}\dot H(t)\ket{n(t)}}{E_m(t)-E_n(t)}\qquad(m\neq n).
\label{eq:coupling}
\end{equation}
Equation~\eqref{eq:cm-eq} becomes
\begin{equation}
\dot c_m(t)+\Bigl(\tfrac{i}{\hbar}E_m(t)+\braket{m}{\dot m}\Bigr)c_m(t)
=\sum_{n\neq m}\frac{\bra{m}\dot H\ket{n}}{E_m-E_n}\,c_n(t).
\label{eq:cm-master}
\end{equation}
If the right-hand side can be neglected (slow drive and finite gaps), then
\[
\dot c_m(t)=\Bigl(-\tfrac{i}{\hbar}E_m(t)-\braket{m}{\dot m}\Bigr)c_m(t),
\]
which integrates to
\[
c_m(t)=c_m(0)\,\exp\!\Bigl[i\theta_m(t)\Bigr]\exp\!\Bigl[i\gamma_m(t)\Bigr].
\]
Thus $|c_m(t)|^2=|c_m(0)|^2$ and an initial eigenstate stays in the corresponding instantaneous eigenstate up to the phases above.

\medskip
\noindent\textbf{Adiabatic condition (nondegenerate case).}
A standard sufficient condition controlling the neglected terms in \eqref{eq:cm-master} is
\begin{equation}
\max_{t\in[0,T]}\;\max_{m\neq n}\;
\frac{\big|\bra{m(t)}\dot H(t)\ket{n(t)}\big|}{\hbar\,|E_m(t)-E_n(t)|^2}
\;\equiv\;\varepsilon\;\ll\;1.
\label{eq:adiabatic-condition2}
\end{equation}
Equivalently, when one introduces a scaled time $s=t/T$ with $H(t)=H(s)$,
nonadiabatic transition amplitudes decay like $O(1/T)$ provided the gap stays finite.

\medskip
\noindent\textbf{Remarks.}
\begin{itemize}
\item The Berry phase $\gamma_n$ is real since $\frac{d}{dt}\braket{n}{n}=0\Rightarrow
2\,\mathrm{Re}\,\braket{n}{\dot n}=0$.
\item More refined theorems relax gap assumptions and give explicit error bounds; here we keep the classical form used in physics applications.
\end{itemize}

\section{The remainder term in the Adiabatic theorem}


We have
\begin{equation}\label{main}
    \dot{c}_m(t)+\left(\frac{i}{\hbar} E_m(t)+\langle m(t)|\dot{m}(t)\rangle\right)c_m(t)=\sum_{n\neq m}\frac{
    \langle m(t)|\dot H|n(t)\rangle
    }{\Delta_{mn}} c_n(t)\,,
\end{equation}
where $\Delta_{mn}=E_m-E_n$. Write
\begin{equation}
    c_m(t)=d_m(t) e^{i\theta_m(t)}e^{i\gamma_m(t)}\,,
\end{equation}
where $\theta_m,\gamma_m$ were introduced in eq.(\ref{berry}). Then eq.(\ref{main}) becomes 
\begin{equation}
    \dot{d}_m(t)=\sum_{n\neq m}\frac{
    \langle m(t)|\dot H|n(t)\rangle
    }{\Delta_{mn}} d_n(t) e^{i(\theta_n(t)-\theta_m(t))}e^{i(\gamma_n(t)-\gamma_m(t))}\,.
\end{equation}
Integrating both sides between $0,T$ we have
\begin{equation}
    d_m(T)-d_m(0)=\sum_{n\neq m}\int_{0}^T dt \ \frac{
    \langle m(t)|\dot H|n(t)\rangle
    }{\Delta_{mn}} d_n(t) e^{i(\theta_n(t)-\theta_m(t))}e^{i(\gamma_n(t)-\gamma_m(t))}
\end{equation}
Now let $s=t/T$ and write $H(t)=\widetilde H(s)$; then $\dot H=(1/T)\,\partial_s \widetilde H$ and
\[
d_m(T)-d_m(0)\;=\;\int_0^1\!ds\;
\sum_{n\neq m}\frac{\langle m(s)|\partial_s \widetilde H(s)|n(s)\rangle}{\Delta_{mn}(s)}d_n(s) e^{iT(\theta_n(s)-\theta_m(s))}e^{iT(\gamma_n(s)-\gamma_m(s))}.
\]
One can invoke the Riemann-Lebesgue lemma to say that the rhs goes to 0. But without invoking that, let us see how far we can proceed.

Applying the triangle inequality we get
\begin{equation}
    |d_m(T)-d_m(0)|\leq \int_0^1\!ds\;\sum_{n\neq m}
|\frac{\langle m(s)|\partial_s \widetilde H(s)|n(s)\rangle}{\Delta_{mn}(s)}d_n(s)|\,.
\end{equation}
Now the maximum value of $|d_n|$ is 1. So replacing on the rhs we have
\begin{equation}
    |d_m(T)-d_m(0)|\leq \int_0^1\!ds\;\sum_{n\neq m}
|\frac{\langle m(s)|\partial_s \widetilde H(s)|n(s)\rangle}{\Delta_{mn}(s)}|\,.
\end{equation}

This is the best we can do with these rudimentary tools. The bare minimum requirement from here that must be obeyed is that each term in the summand must be small and that the sum over $n$ converges. Only then, we can claim that $|d_m(T)-d_m(0)|\rightarrow 0$ as needed by the adiabatic approximation to hold.

\paragraph{Conditions which may work for a negligible remainder.}
\begin{enumerate}
\item \textbf{Gap condition:} $\Delta_{\min}>0$ on $[0,1]$ (no crossings with the target level).
\item \textbf{Smooth slow drive:} $H(t)=\widetilde H(t/T)$ with bounded $\partial_s \widetilde H$ are finite.
\item \textbf{Small matrix elements:} $|\frac{\langle m(s)|\partial_s \widetilde H(s)|n(s)\rangle}{\Delta_{mn}(s)}|\ll 1$ when $s\in [0,1]$ for all $m\neq n$.
\end{enumerate}

\section{Time-ordered evolution in ASP}
\label{sec:asp-time-ordered}

In an adiabatic ramp we let the Hamiltonian vary smoothly along the parameter
$s=t/T\in[0,1]$. The Schr\"odinger equation reads
\begin{equation}
i\,\partial_s\,|\psi(s)\rangle \;=\; T\,H(s)\,|\psi(s)\rangle,
\end{equation}
whose exact solution is the \emph{time-ordered} (parameter-ordered) exponential
\begin{equation}
\label{eq:timeordered}
|\psi(s)\rangle \;=\; \mathcal T\exp\!\Big(-\,i\,T\!\int_0^{s}\!du\;H(u)\Big)\,|\psi(0)\rangle.
\end{equation}
Only when $[H(u),H(v)]=0$ for all $u,v$ can the ordering symbol be dropped and
\eqref{eq:timeordered} collapse to a single exponential. 

\medskip

\noindent\textbf{Product formula:}
Discretize the ramp into $M$ slices of width $\Delta s=1/M$ (so $\Delta t=T/M$), pick
$s_k=(k-\tfrac{1}{2})\Delta s$, and freeze the Hamiltonian on each slice:
\begin{equation}
U_k \;\equiv\; \exp\!\big(-\,i\,\Delta t\,H(s_k)\big),\qquad
U(1)\;\approx\; U_M U_{M-1}\cdots U_1.
\end{equation}
As $\Delta t\to0$ this ordered product converges to \eqref{eq:timeordered}. The leading
corrections are governed by commutators $[H(s),H(s')]$; in the Magnus/Dyson
expansion the first nontrivial term is
\(
\Omega_2= -\tfrac{T^2}{2}\!\int_0^{s}\!\!\int_0^{u}\!du\,dv\,[H(u),H(v)].
\)

\medskip

\noindent\textbf{Within each slice: Strang splitting.}
If $H(s)=H_{\rm on}(s)+H_{\rm bond}(s)$ we can approximate $U_k$ by the symmetric
(second-order) Trotter step
\begin{equation}
\label{eq:strang}
U_k \;=\; e^{-\,i\frac{\Delta t}{2}H_{\rm on}(s_k)}\;
          e^{-\,i\,\Delta t\,H_{\rm bond}(s_k)}\;
          e^{-\,i\frac{\Delta t}{2}H_{\rm on}(s_k)}
\;+\;O(\Delta t^{3}),
\end{equation}
which yields a \emph{global} error $O(\Delta t^{2})$ over the full ramp when $H(s)$ is
smooth and commutators are bounded. We will discuss higher-order Suzuki-Trotter formulas in the next part. 

\section{Two benchmark examples of the adiabatic theorem}

For a Hamiltonian $H(s)$ with $s=t/T\in[0,1]$ and instantaneous eigenstates
$H(s)\ket{n(s)} = E_n(s)\ket{n(s)}$, the simplest non–degenerate adiabatic 
theorem states that if $E_0(s)$ is separated from the rest of the spectrum 
by a gap $\Delta(s) = E_1(s)-E_0(s) > 0$ and $H(s)$ is sufficiently smooth, 
then a system prepared in $\ket{0(0)}$ at $s=0$ remains in the instantaneous 
ground state $\ket{0(s)}$ up to a phase, provided the total runtime $T$ is large
compared to
\begin{equation}
 T_{\text{ad}} \sim 
 \max_{s\in[0,1]} 
 \frac{|\bra{1(s)}\partial_s H(s)\ket{0(s)}|}
      {\Delta(s)^2}\,.
 \label{eq:adiabatic-condition}
\end{equation}
The two examples below benchmark this statement: first in a situation where the
assumptions are satisfied, and then in a case where they are explicitly violated
by a level crossing.

\subsubsection*{Example 1: adiabatic preparation of the anharmonic oscillator ground state}

Consider, in units with $\hbar=m=1$, the family of Hamiltonians
\begin{equation}
 H(s) = (1-s)\,H_0 + s\,H_1,\qquad s=\frac{t}{T}\in[0,1],
\end{equation}
with
\begin{equation}
 H_0 = \frac{p^2}{2} + \frac{\omega^2 x^2}{2},\qquad
 H_1 = \frac{p^2}{2} + \frac{\omega^2 x^2}{2} + \lambda x^4,
\end{equation}
for some fixed $\omega$ and $\lambda>0$.  For all $s\in[0,1]$ the ground state
is non–degenerate and the gap $\Delta(s)=E_1(s)-E_0(s)$ remains finite.

For numerical illustration, truncate to the first $N_{\max}$ harmonic oscillator
eigenstates of $H_0$ and represent $H(s)$ as an $N_{\max}\times N_{\max}$ matrix.
For each $s$ on a grid, diagonalise to obtain $E_n(s)$ and $\ket{n(s)}$.  Two
basic diagnostics are then:

\begin{itemize}
 \item The \emph{instantaneous spectrum} $E_n(s)$ (at least $n=0,1,2$) as a
   function of $s$.  This shows explicitly that the gap $\Delta(s)$ never closes.
 \item The \emph{final ground–state fidelity} as a function of total runtime $T$:
   starting from the ground state $\ket{\psi(0)}=\ket{0(0)}$ of $H_0$, solve 
   the time–dependent Schr\"odinger equation
   \begin{equation}
     i\frac{\partial}{\partial t}\ket{\psi(t)} = H\!\left(\frac{t}{T}\right)\ket{\psi(t)},
   \end{equation}
   and at $t=T$ compute
   \begin{equation}
     F(T) = \left|\braket{0(1)}{\psi(T)}\right|^2,
   \end{equation}
   where $\ket{0(1)}$ is the exact ground state of $H_1$ from static diagonalisation.
\end{itemize}

A simple explicit choice is $\omega=1$, $\lambda=0.1$, $N_{\max}= 7$ and
a uniform time discretisation.  One typically finds that $F(T)$ rises rapidly
towards $1$ once $T$ exceeds the adiabatic timescale suggested by
\eqref{eq:adiabatic-condition}.

\begin{figure}[t]
\centering
\includegraphics[width=0.47\textwidth]{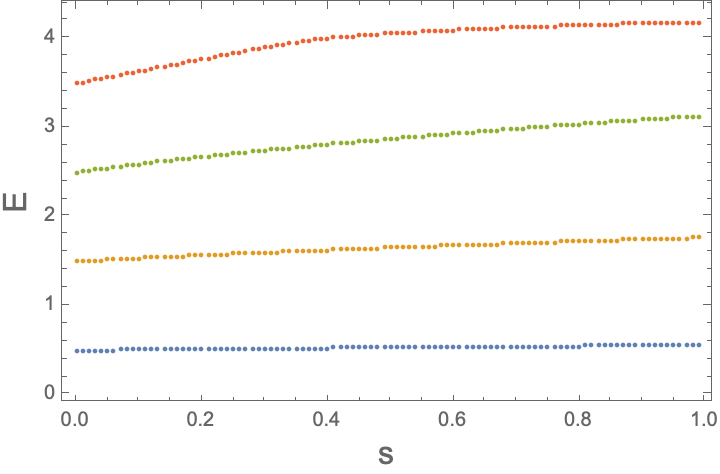}\hfill
\includegraphics[width=0.47\textwidth]{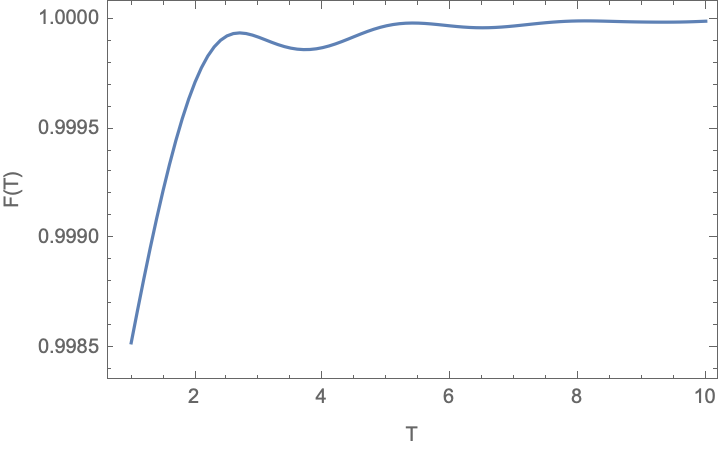}
\caption{\textbf{Adiabatic preparation of the anharmonic oscillator ground state.}
Left: lowest few instantaneous energies $E_n(s)$ of $H(s)$, showing a finite
gap $\Delta(s)$ throughout the evolution. Right: final ground–state fidelity
$F(T)$ versus total runtime $T$, demonstrating convergence to the true ground
state for sufficiently slow evolution. Plots were made using Mathematica.}
\label{fig:aho-adiabatic}
\end{figure}

Figure~\ref{fig:aho-adiabatic} thus gives a concrete benchmark where the
adiabatic theorem works exactly as advertised: as $T$ increases, the prepared
state rapidly approaches the true ground state of the anharmonic oscillator.

\subsubsection*{Example 2: breakdown at a level crossing}

To show how the adiabatic theorem can fail when its hypotheses are violated,
it is useful to consider the simplest possible two–level system.  Take
\begin{equation}
 H(s) = (2s-1)\,\sigma_z,\qquad s=\frac{t}{T}\in[0,1],
 \label{eq:crossing-H}
\end{equation}
where $\sigma_z$ is the usual Pauli matrix.  The instantaneous eigenvalues are
\begin{equation}
 E_{\pm}(s) = \pm\,(2s-1),
\end{equation}
with corresponding eigenstates $\ket{0}$ and $\ket{1}$ (eigenstates of $\sigma_z$).
At $s=0$ we have $H(0)=-\sigma_z$, so the ground state is $\ket{0}$; at $s=1$
we have $H(1)=+\sigma_z$, so the ground state is $\ket{1}$.  Crucially, the
gap
\begin{equation}
 \Delta(s) = E_{+}(s)-E_{-}(s) = 2|2s-1|
\end{equation}
\emph{closes} at $s^\ast=\tfrac12$, where the two levels cross exactly.

Because $H(s)$ at different times all commute,
$[H(s),H(s')]=0$, the time–evolution operator is just
\begin{equation}
 U(T) = \exp\!\left(-i\int_0^T \mathrm{d}t\,H\!\left(\frac{t}{T}\right)\right)
 = \exp\!\left(-i\phi(T)\,\sigma_z\right)
\end{equation}
for some real phase $\phi(T)$ independent of the initial state.  If we start in
the ground state at $s=0$,
\begin{equation}
 \ket{\psi(0)} = \ket{0},
\end{equation}
then at $t=T$ we simply have
\begin{equation}
 \ket{\psi(T)} = e^{-i\phi(T)}\ket{0}.
\end{equation}
On the other hand, the instantaneous ground state of $H(1)$ is $\ket{1}$, which
is orthogonal to $\ket{0}$.  The final ground–state fidelity is therefore
\begin{equation}
 F_{\text{cross}}(T) = \left|\braket{1}{\psi(T)}\right|^2
 = \left|\braket{1}{0}\right|^2 = 0,
\end{equation}
\emph{for any runtime $T$}, no matter how slowly we vary the Hamiltonian.

\begin{figure}[t]
\centering
\includegraphics[width=0.47\textwidth]{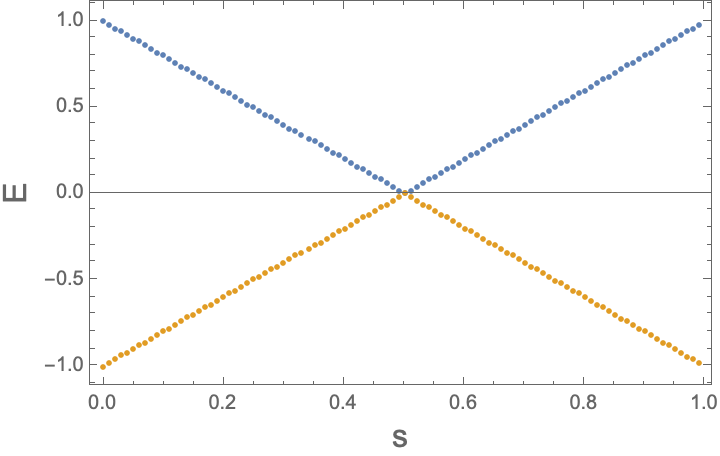}\hfill
\includegraphics[width=0.47\textwidth]{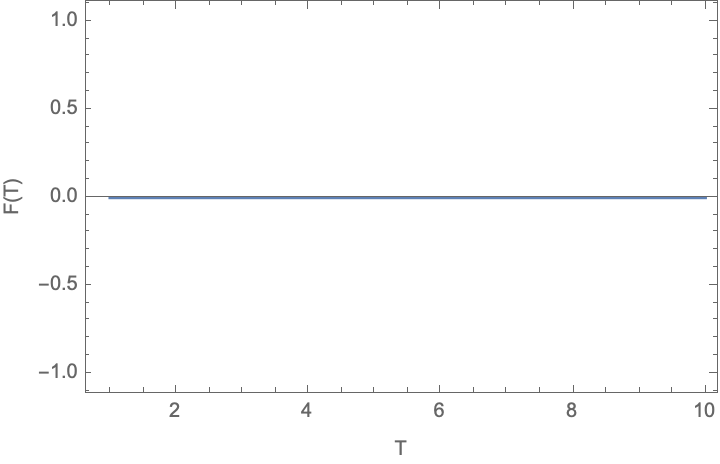}
\caption{\textbf{Failure of adiabatic following at a level crossing.}
Left: instantaneous energies $E_{\pm}(s)$ of the two–level system
\eqref{eq:crossing-H}, displaying an exact level crossing at $s=\tfrac12$ where
the gap closes. Right: Fidelity vs $T$ is always 0 as explained in the text.}
\label{fig:crossing-failure}
\end{figure}

This example cleanly highlights the role of the spectral gap in the
adiabatic theorem.  If one perturbs \eqref{eq:crossing-H} to introduce a small
avoided crossing, for example
\begin{equation}
 H_{\text{LZ}}(s) = (2s-1)\,\sigma_z + \varepsilon\,\sigma_x,
\end{equation}
the gap no longer closes.  In that case, the evolution reduces to the familiar
Landau--Zener problem: the excitation probability decays exponentially with $T$
for fixed $\varepsilon$, in accordance with the adiabatic condition
\eqref{eq:adiabatic-condition}.


\chapter{The $\phi^4$ theory in 1+1 dimensions}

\section{\texorpdfstring{$\phi^4$}{phi4} in \(1+1\) dimensions}

The canonical example of quantum field theory that is taught in virtually every course is the \(\phi^4\) theory. We will take this theory up, restricting to \(1+1\) dimensions, and show how one can think of it as a lattice generalization of the anharmonic oscillator that we learnt in the first chapter. The background material reviewed in this chapter is by no means exhaustive and focuses only on what is necessary to understand the quantum–computing applications or what we feel may be interesting future research to undertake on a quantum computer.

Concretely, we study a real scalar field with the Lagrangian (schematic for now):
\[
\mathcal L \;=\; \tfrac12 (\partial\phi)^2 \;+\; \tfrac12 \mu_0^2\,\phi^2 \;+\; \tfrac{g_0}{4!}\,\phi^4 \;+\; \delta \mathcal L,
\]
where \(\delta\mathcal L\) contains counterterms chosen in a convenient renormalization scheme and $\mu_0, g_0$ being the bare parameters (we will elaborate on this). In \(1+1\) dimensions the field \(\phi\) is classically dimensionless while the coupling \(\lambda\) has mass dimension two, so the theory is \emph{super-renormalizable}---this means only the vacuum energy and mass require renormalizations, and wave-function renormalization is at most finite. The model enjoys a \(\mathbb Z_2:\phi\mapsto-\phi\) symmetry and exhibits two phases separated by what is called an Ising–universality critical point: a symmetric phase with a single massive particle, and a broken phase with kink and antikink excitations whose quantum fluctuations generate a tower of mesonic bound states. This minimal setting already contains the ingredients we will need later--—mass gaps, scattering amplitudes, resonances and multi-particle thresholds.

On a spatial lattice with spacing \(a\), the Hamiltonian becomes a chain of coupled anharmonic oscillators,
\[
H \;=\; a\sum_{n}\!\left[\tfrac12 \pi_n^2 \;+\; \tfrac12 (\nabla_a \phi_n)^2 \;+\; \tfrac12 \mu_0^2\,\phi_n^2 \;+\; \tfrac{g_0}{4!}\,\phi_n^4\right] \;+\; \text{c.t.},
\]
so that every site inherits precisely the single-particle physics of the quartic oscillator while the gradient term couples sites. This “oscillator-of-oscillators’’ viewpoint makes \(\phi^4_{1+1}\) the natural bridge from Chapter~2 to field theory. 

From the perspective of quantum algorithms, \(\phi^4\) is also the historically pivotal testing ground. The seminal work of Jordan, Lee and Preskill (JLP) \cite{JLP2012Science} gave the first polynomial-time quantum algorithms for preparing scattering states and estimating cross sections in scalar quantum field theories, with explicit resource analyses at weak and strong coupling. This work also elaborated on adiabatic state preparation and how to measure nonperturbative 2-2 scattering amplitudes. How would we benchmark such results? On the nonperturbative physics side, Hamiltonian-truncation studies in \(1+1\) dimensions provide precise benchmarks for spectra and matrix elements across both phases, giving us comparison points against which to validate quantum-computing outputs in small volumes before scaling up. 

\paragraph*{Suggested references.}
For quantum-algorithm foundations and resource estimates for scalar QFT scattering, see \cite{JLP2011ScalarArXiv, JLP2012Science}. Extensions to fermionic field theories and complexity-theory refinements appear in \cite{JLP2014Fermions, JLP2018BQP}. For nonperturbative benchmarks of \(\phi^4_{1+1}\) via Hamiltonian truncation, see \cite{RychkovVitale2014, RychkovVitale2015, HVVR}. For constructive field theory and textbook background on \(\phi^4\) in two dimensions, see \cite{GlimmJaffe1968, SimonPphi2, zinnjustin}.

\section{From many anharmonic oscillators to scalar $\phi^4$}
\label{sec:aho-to-phi4}

\begin{figure}[H]
\centering
\begin{tikzpicture}[
    x=1.4cm, y=1.4cm,
    site/.style={circle, draw=blue!60, fill=blue!30, minimum size=5pt, inner sep=0pt},
    spring/.style={red, very thick, decorate, decoration={coil, segment length=3.3mm, amplitude=1.4pt}}
  ]

  \node at (-0.6,0) {$\cdots\cdots$};
  \node at (7.2,0) {$\cdots\cdots$};

  \node[site] (s0) at (0,0) {};
  \node[site] (s1) at (1.3,0) {};
  \node[site] (s2) at (2.6,0) {};
  \node[site] (s3) at (3.9,0) {};
  \node[site] (s4) at (5.2,0) {};
  \node[site] (s5) at (6.5,0) {};

  \node[below=3pt] at (s0) {$j-2$};
  \node[below=3pt] at (s1) {$j-1$};
  \node[below=3pt] at (s2) {$j$};
  \node[below=3pt] at (s3) {$j+1$};
  \node[below=3pt] at (s4) {$j+2$};
  \node[below=3pt] at (s5) {$j+3$};

  \draw[spring] (s0) -- (s1);
  \draw[spring] (s1) -- (s2);
  \draw[spring] (s2) -- (s3);
  \draw[spring] (s3) -- (s4);
  \draw[spring] (s4) -- (s5);

  \foreach \sitename/\lab in {s0/{j-2}, s1/{j-1}, s2/{j}, s3/{j+1}, s4/{j+2}, s5/{j+3}} {

    \begin{scope}[shift={(\sitename)}, yshift=0.5cm]
      \draw[thin] (-0.15,0.00) -- (0.15,0.00);
      \draw[thin] (-0.15,0.10) -- (0.15,0.10);
      \draw[thin] (-0.15,0.19) -- (0.15,0.19);
      \draw[thin] (-0.15,0.27) -- (0.15,0.27);
    \end{scope}

    \node[above=23pt] at (\sitename) {$n_{\lab}$};
  }

\end{tikzpicture}
\caption{Lattice of anharmonic oscillators. Each blue site at position
$j, j\pm1,\dots$ hosts a local anharmonic oscillator with energy levels
labelled by $n_j, n_{j+1},\dots$. Nearest--neighbour couplings are indicated
by red springs, representing the discrete gradient term.}
\label{fig:lattice-AHO-cartoon}
\end{figure}
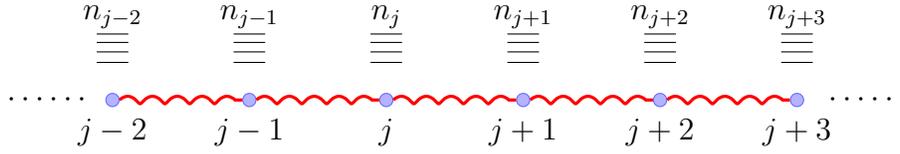

\noindent\textbf{Motivation.} We have mastered the single-site anharmonic oscillator (AHO). The natural next step is a \emph{lattice of AHOs}. If we simply stack $N$ copies, nothing talks to anything: $H=\sum_j H_{\rm AHO}(\phi_j,\pi_j)$. To make physics, we couple neighbors. The simplest choice is a bilinear spring:
\[
H \;=\; \sum_{j=1}^{N}\!\Big[\tfrac12\pi_j^2 + V(\phi_j)\Big]
\;+\; \frac{\kappa}{2}\sum_{j=1}^{N}(\phi_{j+1}-\phi_j)^2,
\qquad V(\phi)=\tfrac12\mu_0^2\phi^2+g_0\,\phi^4.
\]
The cartoon in fig.(\ref{fig:lattice-AHO-cartoon}) depicts a chain of AHOs (sites) joined by springs (the gradient term).


\subsection{Continuum limit: why this \emph{is} scalar $\phi^4$}

In order to get the continuum limit we write $x_j= j a$ and map 
\begin{equation}
    \phi_j\rightarrow \sqrt{a}\phi(x_j)\,,\qquad \pi_j \rightarrow \sqrt{a}\pi(x_j)\,,\qquad \qquad \sum_j \to \frac{1}{a}\int dx,
\end{equation}
which ensure that 
\begin{equation}
    [\phi_j, \pi_k]=\delta_{jk}\longrightarrow [\phi(x_j),\pi(x_k)]=i \delta(x_j-x_k)\,.
\end{equation}
This leads to 
\begin{equation}
    \sum_{j=1}^{N}\!\Big[\tfrac12\pi_j^2 + V(\phi_j)\Big]\longrightarrow \int dx \left[\frac12 \pi^2+\frac12 \mu_0^2 \phi^2 +g_0 a \phi^4\right]\,.
\end{equation}
Finally, $\phi_{j+1}-\phi_j = a\,\partial_x {\sqrt{a}}\phi + \frac{a^2}{2}\partial_x^2 {\sqrt{a}}\phi + \cdots$. Then
\[
\sum_j (\phi_{j+1}-\phi_j)^2 \;\to\; a^2\int dx\,  (\partial_x\phi)^2 \;+\; \mathcal O(a^3).
\]
After parameter redefinitions $g_0 a \mapsto g, \kappa a^2\rightarrow v^2$, the leading continuum Hamiltonian density is
\[
\mathcal H \;=\; \tfrac12 \pi^2 + \tfrac{v^2}{2}(\partial_x\phi)^2 + \tfrac12 \mu_0^2 \phi^2 + g\,\phi^4 \;+\; \underbrace{c_4 (\partial_x^2\phi)^2 + \cdots}_{\text{irrelevant in $1{+}1$D}}.
\]
A quick clarification on nomenclature. The Hamiltonian density has units of $ML^{-1}$. In natural units where $c=1, \hbar=1$, this is just $M^2$. $\partial_x$ has dimensions of $M$. Thus, $\phi$ is dimensionless if $\kappa$ is dimensionless and $\mu_0$ has dimensions of mass. This means $g$ has dimensions of $M^2$ while $c_4$ has dimensions of $M^{-2}$. Operators in the Hamiltonian which have negative mass dimensions are called ``irrelevant" as their importance diminishes at low energies. Conversely, operators with positive mass dimensions are called ``relevant" and zero mass dimensions are called ``marginal."

Summary: a chain of AHOs with the simplest spring becomes $\phi^4$ plus higher-derivative corrections; the latter are irrelevant at long distances, so the infrared theory is the familiar scalar $\phi^4$.

\subsection{Usual mode expansion}
In standard approaches to QFT, we expand the operators $\phi(x), \Pi(x)$ at the fixed time $t=0$ as follows:
\begin{eqnarray}
    \phi(x)&=&\int_{-\infty}^\infty \frac{dp}{2\pi\sqrt{2|p|}}\left[a(p)e^{i p x}+a^\dagger(p)e^{-i p x}\right]\,,\\
    \Pi(x)&=&\int_{-\infty}^\infty \frac{dp |p|}{2\pi\sqrt{2|p|}}\left[-ia(p)e^{i p x}+ia^\dagger(p)e^{-i p x}\right]\,,\\
\end{eqnarray}
Then 
$$ [a(p),a^\dagger(p')]=2\pi \delta(p-p')$$
ensures that $[\phi[x],\Pi(y)]=i\delta(x-y)$ is satisfied. This choice of normalization also gives a simple form for the Hamiltonian in the free theory in terms of $a,a^\dagger$:
\begin{equation}
    H=\int_{-\infty}^\infty \frac{dp}{2\pi} a^\dagger(p) a(p) |p|\,.
\end{equation}
The vacuum is defined as the state $|0\rangle$ which satisfies $a(p)|0\rangle=0$ for all $p$.

\paragraph{Single and multi-particle states:} Single particle states are defined as 
\begin{equation}
    |p\rangle=a^\dagger(p)|0\rangle\,.
\end{equation}
Similarly, multiparticle states are defined via:
\begin{equation}
|p_1,p_2,\cdots p_n\rangle=a^\dagger(p_1)a^\dagger(p_2)\cdots a^\dagger(p_n)|0\rangle.
\end{equation}
These are all in the free theory. The challenge is to find them in the interacting theory. This is where the adiabatic theorem helps.

\section{Adiabatic preparation}
\label{sec:asp-aho-lattice}

We will apply the adiabatic state preparation (ASP) protocol to start from the strictly solvable product ground state at zero coupling and turn on the nearest–neighbor interaction slowly to let correlations and entanglement build up.

\medskip

\noindent\textbf{Model and conventions.}
We keep the lattice Hamiltonian introduced above, now viewed as a one–parameter family:
\begin{equation}
\label{eq:aho-chain-asp}
H(\kappa)\;=\;\sum_{j=1}^{N}\!\Big[\tfrac12\,\pi_j^2 + V(\phi_j)\Big]
\;+\;\frac{\kappa}{2}\sum_{j=1}^{N-1}\big(\phi_{j+1}-\phi_j\big)^2,
\qquad
V(\phi)=\tfrac12\,\mu_0^{\,2}\,\phi^2 + g_0\,\phi^4.
\end{equation}
At $\kappa=0$ the ground state is the direct product of single–site AHO ground states, exactly known and easy to prepare. Throughout, $\mu_0^{\,2}$ denotes the coefficient of the quadratic term (so the code variable \verb|mu0| corresponds to $\mu_0^{\,2}$), and $g_0$ is the onsite quartic strength. No continuum limit or field redefinitions are invoked in this section.

\medskip

\noindent\textbf{Adiabatic path and schedule.}
ASP varies the coupling as $\kappa:0\to\kappa_{\mathrm f}$ over a total time $T$, for example by setting $H(t)=H(\kappa(s))$ with $s=t/T\in[0,1]$. A smooth schedule with endpoint flattening,
\[
\dot\kappa(0)=\dot\kappa(1)=0,
\]
suppresses boundary terms in the adiabatic remainder (see the “adiabatic remainder” subsection for the $1/T$ vs.\ $1/T^2$ scaling). For any finite chain the many–body gap never strictly closes; nonetheless, near critical regimes larger $T$ is needed to keep diabatic errors small.

\medskip

\noindent\textbf{What to measure along the ramp.}
The cleanest, renormalization–free witness here is the half–chain entanglement entropy,
\begin{equation}
S_{1/2}(\kappa)\;=\;-\mathrm{Tr}\,\rho_{[1..N/2]}(\kappa)\,\ln\rho_{[1..N/2]}(\kappa),
\end{equation}
which is exactly zero at $\kappa=0$ (product state), then rises as correlations spread, and saturates once the correlation length is shorter than the system size (area law away from criticality). In part 2, $S_{1/2}(\kappa)$ is computed with a minimal matrix–product–state (MPS) implementation.

\begin{figure}[hbt]
  \centering
  \includegraphics[width=0.6\linewidth]{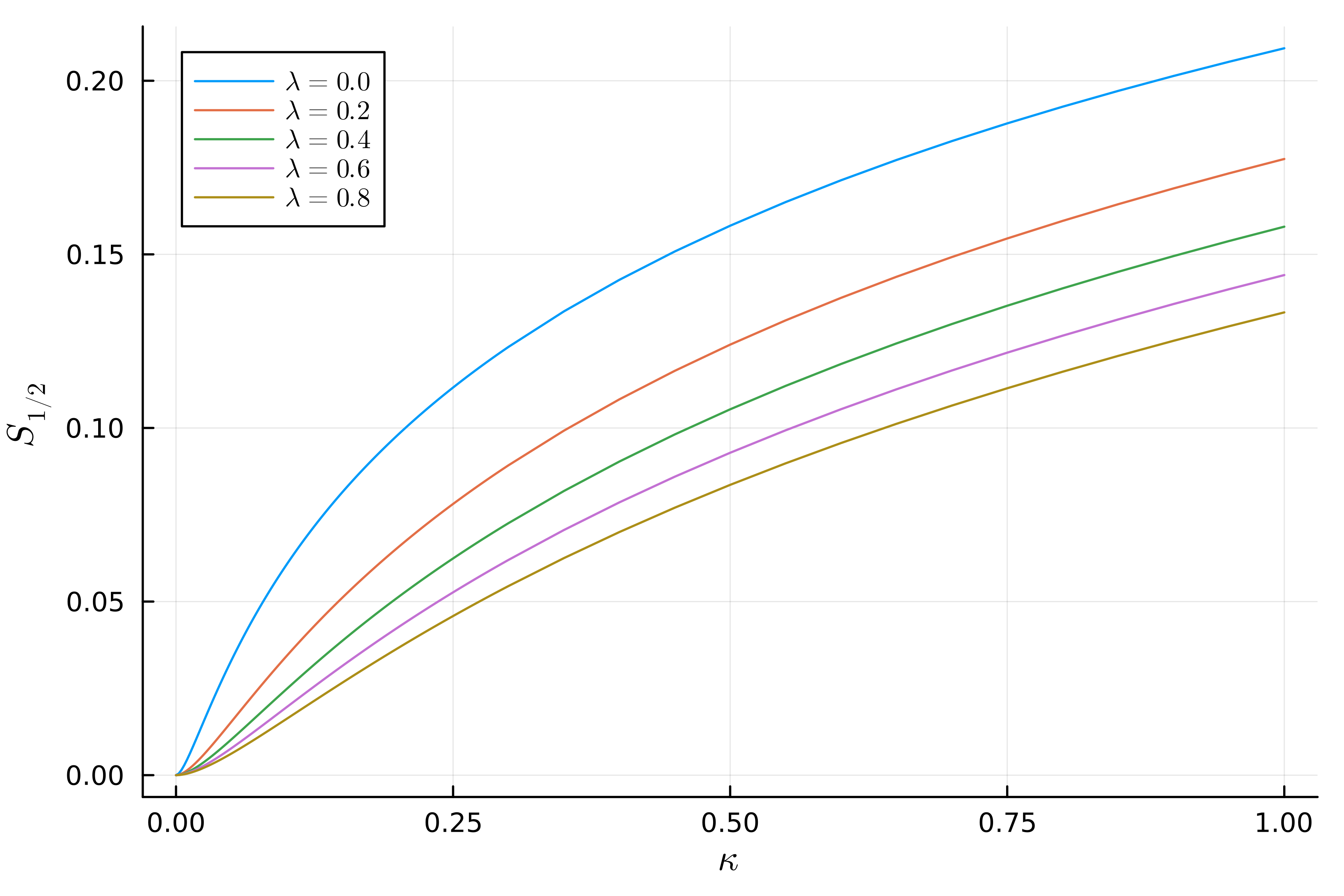}
  \caption{Half-chain entanglement entropy plotted as a function of the gradient coupling $\kappa$. The local Hilbert space was truncated at $16$ modes and the maximum bond dimension is $\chi_{\text{max}} = 20$. The ground state was obtained with the VUMPS algorithm.}\label{EEmps}
\end{figure}

\medskip

\noindent\textbf{What to expect}: 
For small~$\kappa$, the first nontrivial physics is inherited from a single AHO matrix element: the dipole $f_1=\langle 1|\phi|0\rangle_{\rm AHO}$. The one–particle hopping scale is proportional to $\kappa\,|f_1|^2$, so softening the onsite problem (smaller $\mu_0^{\,2}$ or, at fixed $\mu_0^{\,2}$, smaller effective AHO gap) enhances the growth of $S_{1/2}$, while increasing $g_0$ at fixed~$\kappa$ typically suppresses it by enlarging single–site gaps and shrinking~$f_1$. Thus the $\kappa$–ramp provides a direct, nonperturbative bridge from “many decoupled AHOs” to an interacting many–body ground state with extended correlations, without invoking any continuum language. The figure \ref{EEmps} validates this intuition. For smallish $\kappa$, larger $\lambda$'s give smaller entanglement since the entanglement does not spread as much across the chain.

\section{Physical parameters}
How do we define the physical mass of a particle by considering the lattice perspective? In the relativistic limit we have the relation
\begin{equation}
    E^2=p^2+m^2\,.
\end{equation}
This is what we will leverage. First we need to introduce momentum on the lattice. This is easily done by doing a discrete Fourier transform:
\begin{equation} \phi_j=\tfrac{1}{\sqrt{N}}\sum_q e^{iqja}\,\varphi_q\,,\qquad  q=\tfrac{2\pi}{Na}n, \quad n=0,\dots,N{-}1\,.\end{equation}
Using
\begin{equation}
    \sum_{j=1}^N e^{i(q+q')j a}=N \delta_{q+q',0}\,,
\end{equation}
and introducing the notation $|b_q|^2=b_q b_{-q}$, we find that the quadratic part of the Hamiltonian can be written as:
\[
H_2=\frac12\sum_q\Big(|\pi_q|^2+\omega_q^2\,|\varphi_q|^2\Big),\qquad
\omega_q^2 \;=\; \mu_0^2+2\kappa \big(1-\cos qa\big).
\]
Notice from here that in the $a\rightarrow 0$ limit, $$\omega_q^2\rightarrow \mu_0^2+\underbrace{(\kappa a^2)}_{v^2} q^2, $$ which is the relativistic form we expect in the continuum. Let us define $E_0$ to be the ground-state energy (with $\lambda$ turned on) and $E_1(p)$ be the energy of the lowest odd $Z_2$ excitation at momentum $p$. Then we should define the physical mass $m_{phys}$ as:
\begin{equation}
    m_{phys}=E_1(p=0)-E_0\,.
\end{equation}
A consistency check is that for low values of $p$, we should have $E_1(p)-E_0\approx \sqrt{m_{phys}^2+p^2}$. Now notice that in introducing a momentum, we were forced to introduce the lattice spacing into the story. The total length of the space that we are considering is $L=N a$. The correlation length between excitations on dimensional grounds goes like $1/m_{phys}$. We need $L\gg 1/m_{phys}$. This needs to be ensured for consistency. We know how to use the adiabatic theorem to get the true ground state with $\lambda\neq 0$ for the AHO. For the $\phi^4$ case, we adopt the same route. We initiate all $j$-sites to be in the ground state and then use ASP. 

\subsection{One particle state}
In a free scalar quantum field theory, we defined a 1-particle state at momentum $p$ to be
\begin{equation}
    |p\rangle=a^\dagger(p)|0\rangle\,.
\end{equation}
Now 
\begin{equation}
    a^\dagger(p)=\frac{1}{\sqrt{N}}\sum_{j=1}^N e^{i p x_j} a_j^\dagger\,,
\end{equation}
which gives 
\begin{equation}
    |p\rangle=\frac{1}{\sqrt{N}}\sum_{j=1}^N e^{i p x_j} a_j^\dagger |0\rangle=\frac{1}{\sqrt{N}}\sum_{j=1}^N e^{i p x_j} |0\cdots \underbrace{1}_{j'th} \cdots 0\rangle\,,
\end{equation}
where in the $j'th$ term, $j$'th lattice site is in the first excited state and all the other sites are in the ground state. Hence, in the free theory, the single particle state is a linear superposition in the position basis as exhibited above. Now that we have the single particle state in the free theory, we can use our friend the adiabatic theorem to perform ASP and get the true single particle state in the interacting theory.

\subsection{Bare parameters vs physical parameters}
On the lattice we started off with the following parameters:
\begin{equation}
    N, \mu_0, g_0, \kappa\,.
\end{equation}
For convenience let's set $\kappa=1$, to admit a relativistic continuum limit. We introduced the lattice spacing by considering the DFT. We have a box of size $L=N a$. Let's keep the size of the box fixed by fixing $N, a$. Then we are left with the ``bare" parameters $\mu_0, g_0$. In the discussion above we introduced the physical mass $m_{phys}$ in terms of $E_1(0), E_0$ which are $\mu_0,g_0$ dependent. It makes ample physical sense to say that we are given a particular $m_{phys}$ and we want to run our simulations for this value. This gives us 1 condition for 2 bare parameters $\mu_0, g_0$. We can specify another physical condition by giving the value of the 2-2 scattering amplitude at some centre-of-mass momentum $p_0$. Using these to specifications, we can tune $\mu_0, g_0$ for our choice of $N, a$. Then we can rerun for different $a$'s to approach the continuum limit.

\section{What if the coupling is \emph{not} just nearest neighbor?}
Write the most general finite-range, translation-invariant quadratic form
\[
H_{\rm nn} \;=\; \frac12 \sum_{r=1}^{R} \eta_r \sum_j (\phi_{j+r}-\phi_j)^2,
\qquad \eta_r\ge 0.
\]
Fourier transforming gives a lattice “Laplacian”
\(
\omega^2(k) = \mu^2 + 4\sum_{r}\eta_r \sin^2(\tfrac{rk}{2}).
\)
\begin{itemize}[leftmargin=1.2em]
  \item \textbf{Same universality (generic case).} If $\eta_r$ are short-ranged and positive, the small-$k$ expansion is
  \[
  \omega^2(k)=\mu^2 + v^2 k^2 + \alpha_4 k^4 + \cdots,
  \]
  so the continuum limit is the \emph{same} $\phi^4$ up to irrelevant $(\partial_x^2\phi)^2$ terms and a rescaled $v$. Emergent Lorentz symmetry (linear dispersion) appears once you tune to the critical line ($\mu^2\to \mu_c^2$).
\end{itemize}

\subsection{Lifshitz theories}
Consider adding a next to nearest neighbor coupling:
\begin{equation}
    \frac{\kappa}{2}(\phi_{j+1}-\phi_j)^2+\frac{\tilde \kappa}{2}(\phi_{j+2}-\phi_j)^2\,.
\end{equation}
In the continuum limit this leads to 
\begin{equation}
 \left[ \frac{\kappa}{2}(a\partial_x\phi(x)+ \frac{a^2}{2}\partial_x^2\phi(x))^2+ \frac{\tilde \kappa}{2}(2a\partial_x\phi(x)+2 a^2\partial_x^2\phi(x))^2\right]{\bigg |}_{x=x_j}
\end{equation}
Expanding and dropping the total derivatives $\partial_x \phi\partial_x^2 \phi=1/2\partial_x(\partial_x \phi\partial_x \phi$), and using $\partial_x \phi \partial_x^3\phi=\partial_x(\partial_x \phi \partial_x^2 \phi)-(\partial_x^2\phi)^2$ we can choose:
\begin{equation}
    \tilde\kappa=-\frac{1}{4}\kappa\,,
\end{equation}
to be left with $(\partial_x^2\phi)^2$ as the leading term. This is the Lifshitz Hamiltonian as the dispersion relation will correspond to $E\sim k^2$.

\section{The $g_0\gg 1$ case}
\label{sec:small-kappa}

What happens when $g_0\gg 1$? Is there another way to set up the problem more efficiently? This question was briefly addressed in \cite{Nishiyama2001}.

We start with our usual lattice Hamiltonian and do the following rescaling:  $\phi_j\rightarrow g_0^{-1/6}\phi_j$, $\pi_j\rightarrow g_0^{1/6}\pi_j$ we get
\begin{equation}
    g_0^{1/3} H=\sum_j \left[\frac12 \pi_j^2+\phi_j^4\right]+\frac{\kappa}{g_0^{2/3}}\left[\frac12 (\phi_{j+1}-\phi_j)^2\right]
\end{equation}
With this rescaling, we have reorganized the perturbative expansion in a non-standard way, where we can now treat the quadratic term as the perturbation. Even when we simulate on a quantum computer, we can do ASP in this manner as the first part of the Hamiltonian we can handle using approximate techniques. 

A key point to note is that the perturbation is now on the gradient term which was responsible for generating entanglement across the chain. The first piece is a local piece. As such in the limit $g_0\gg 1$, we have the situation where there is no entanglement! 
Let $\{|n\rangle\}_{n\ge0}$ be the single-site AHO eigenstates of $h_{\rm AHO}=\tfrac12 p^2+\tfrac12 m_0^2 x^2+g x^4$ with energies $\varepsilon_n$.
We use tensor products $\bigotimes_j |n_j\rangle$ as our unperturbed basis.


\section{Scattering on a quantum computer}
\label{sec:qc-scattering}

In this section, we will discuss some of the issues that pertains to studying scattering on actual quantum hardware or quantum simulators.
We outline two complementary ways to extract elastic \(2\!\to\!2\) phase shifts for the \(1{+}1\)D lattice \(\phi^4\) chain directly on a quantum processor. Both methods share the same state-preparation backbone: \emph{prepare exact free-theory eigenstates and adiabatically dress them} to the interacting theory at the target couplings. Working at fixed lattice \((N,\kappa,\mu_0^2,g_0)\) and boundary condition (periodic for sharp crystal momentum, or open for parity-resolved standing waves), one first calibrates the single–particle dispersion \(E_1(k)\), then proceeds with either a time–domain scattering experiment or a finite–volume spectroscopic inversion.

\paragraph{Common preparation (adiabatic dressing from the free theory).}
At \(g_0=0\) the chain is quadratic. Prepare the exact free vacuum \(|0_{\mathrm{free}}\rangle\). A single free one–particle plane wave at momentum \(k\) is \(a_k^\dagger|0_{\mathrm{free}}\rangle\), where \(a_k^\dagger=\tfrac{1}{\sqrt N}\sum_j e^{ikx_j} a_j^\dagger\) is the normal–mode raiser. For two particles at total momentum \(P=0\), use \(a_k^\dagger a_{-k}^\dagger|0_{\mathrm{free}}\rangle\) or a narrow superposition around \(k\). With the state prepared, turn on interactions along a smooth schedule \(s\in[0,1]\mapsto g_0(s)\) (and optionally co–tune \(\mu_0^2(s)\)) using a boundary–flattened ramp so that leakage scales as \(O(1/T^2)\). Momentum and \(\mathbb Z_2\) parity keep the evolution within the intended symmetry sector, adiabatically mapping free states to their interacting counterparts.

\subsection*{A. Wave–packet scattering in real time}
Prepare two narrow, counter–propagating one–particle packets centered at \(\pm k_0\) and localized far apart. Dress them adiabatically to the interacting theory at \(g_\star\), then evolve in \emph{real time} under \(H(g_\star)\) long enough for the packets to separate again after a single collision. The phase shift is read off in two equivalent ways. One can measure a Wigner time delay by comparing to a “free” reference evolution that reproduces the measured \(E_1(k)\); the peak of the cross–overlap \(C(\Delta t)=\langle\Psi^{\rm free}(t{+}\Delta t)|\Psi(t)\rangle\) yields \(\tau(k_0)=2\,\tfrac{d\delta}{dE}\big|_{k_0}\), which integrates locally to \(\delta(k_0)\). Alternatively, one can project the outgoing right–moving packet onto the corresponding freely propagated packet and take the complex argument of the overlap; for a sufficiently narrow packet this phase equals approximately \(2\,\delta(k_0)\). Both readouts use standard Hadamard–test primitives with a single clean ancilla controlling short segments of the same Trotterized evolution used for state preparation.

\subsection*{B. Finite–volume spectroscopy and the quantization condition}
Instead of scattering, prepare symmetry–projected few–body states and measure energy levels. First tabulate the single–particle dispersion \(E_1(k)\) by dressing \(a_k^\dagger|0_{\mathrm{free}}\rangle\) and estimating \(E_1(k)-E_0\) via a short real– or imaginary–time autocorrelator. Then prepare a dressed two–particle, \(P=0\), even–parity state and extract the lowest two–body level \(E_2(L)\) either by a brief imaginary–time filter (single–exponential slope), a short real–time autocorrelation (single–frequency fit), or a shallow variational deflation pass. In the elastic regime one solves \(E_2(L)=2\,E_1(k)\) for the relative momentum \(k\), and substitutes into the finite–volume quantization rule. For periodic boundaries,
\[
kL+2\,\delta(k)=2\pi n \qquad (n\in\mathbb Z),
\]
while for hard walls the even/odd standing–wave branches read \(kL+2\,\delta_+(k)=(2n{+}1)\pi\) and \(kL+2\,\delta_-(k)=2n\pi\).

\section{Unitarity in two--particle scattering: elastic vs.\ inelastic}
\label{unitarity}
In this section, we will review the constraints arising from unitarity on the 2-2 scattering amplitude. 
\paragraph{Setup and normalization.}
Work in the center-of-mass (CM) basis of two-particle states
\[
|p\rangle_{2}\;\equiv\;|p,-p\rangle,\qquad
\langle p'|p\rangle_{2}\;=\;(2\pi)\,2E_p\,\delta(p'-p),
\]
with \(E_p=\sqrt{m^2+p^2}\). Translation invariance implies momentum conservation, so two-to-two matrix elements in the CM basis are diagonal in \(p\).

\paragraph{Start from operator unitarity.}
Unitarity is
\[
S S^\dagger=\mathbf{1}.
\]
Sandwiching between CM states and inserting a resolution of the identity over a complete set of out-states \(\{\ket{k}_{2}\}\cup \{\ket{X}_{\neq 2}\}\) (two–particle and all multi–particle sectors):
\begin{align}
\langle q|S S^\dagger|p\rangle_{2}
&=\sum_{k}\langle q|S|k\rangle_{2}\,{}_{2}\langle k|S^\dagger|p\rangle
\;+\;\sum_{X\neq 2}\langle q|S|X\rangle\langle X|S^\dagger|p\rangle \nonumber\\
&=\langle q|p\rangle_{2}\;=\;(2\pi)\,2E_p\,\delta(q-p). \label{eq:unitarity-master}
\end{align}

\paragraph{Parameterizing the elastic matrix element.}
By CM momentum conservation the elastic \(2\to2\) matrix element is diagonal:
\[
{}_{2}\langle q|S|p\rangle_{2} \;=\;(2\pi)\,2E_p\,\delta(q-p)\;S_2(p),
\]
which defines the (single ``partial wave'' in \(1{+}1\)D) two–body \(S\)-eigenvalue \(S_2(p)\).

\subsection*{Purely elastic scattering}
If no inelastic channels are kinematically open, the second sum in \eqref{eq:unitarity-master} vanishes:
\[
\sum_{X\neq 2}\langle q|S|X\rangle\langle X|S^\dagger|p\rangle=0,
\]
 and \eqref{eq:unitarity-master} reduces to \footnote{Obviously we need an identity like $\sum_k (2\pi)^2 (2 E_p)^2 \delta(k-p)\delta(k-q)=(2\pi)(2E_p)\delta(p-q)$. Figure out where this comes from!}
\[
(2\pi)\,2E_p\,\delta(q-p)\;|S_2(p)|^2 \;=\; (2\pi)\,2E_p\,\delta(q-p).
\]
Hence
\[
|S_2(p)|=1 \quad\Longrightarrow\quad S_2(p)=e^{2i\delta(p)},
\]
with a real phase shift \(\delta(p)\) (time-reversal and real analyticity ensure \(\delta\in\mathbb{R}\)). Therefore,
\[
\boxed{\;{}_{2}\langle q|S|p\rangle_{2}^{\text{(elastic)}}=(2\pi)\,2E_p\,\delta(q-p)\;e^{2i\delta(p)}\;}
\]
i.e.\ the two–body matrix element is a pure phase multiplying the momentum delta.

It is an apt place to introduce an important nomenclature. 
In scattering theory, a \emph{threshold} is the minimal centre--of--mass energy at which a given reaction channel first becomes kinematically allowed. Below this energy there is simply not enough energy to create the particles in the final state, and the corresponding cross section vanishes. For example, if a new particle of mass $m_2$ can first appear in the inelastic process
\[
A_1 + A_1 \;\to\; A_1 + A_2\,,
\]
then the inelastic threshold is
\[
E_{\rm cm}^{\text{(th)}} = 2 m_1 + m_2\,,
\]
and only for $E_{\rm cm} > E_{\rm cm}^{\text{(th)}}$ can this channel contribute to the scattering.

\subsection*{Inelastic scattering}
When channels with \(n\ge 3\) final particles are open, the second term in \eqref{eq:unitarity-master} is nonzero. Using the elastic parameterization above, unitarity gives
\begin{align}
(2\pi)\,2E_p\,\delta(q-p)\;|S_2(p)|^2
\;+\;\sum_{X\neq 2}\langle q|S|X\rangle\langle X|S^\dagger|p\rangle
\;=\;(2\pi)\,2E_p\,\delta(q-p).
\end{align}
Rotational (here: parity) and translational invariance force the inelastic sum to be proportional to \((2\pi)\,2E_p\,\delta(q-p)\). Defining the \emph{inelasticity} (absorption) parameter \(\eta(p)\in[0,1]\) by
\[
|S_2(p)|\equiv \eta(p),
\qquad
1-\eta^2(p)\;=\;\frac{1}{(2\pi)\,2E_p}\sum_{X\neq 2}\langle p|S^\dagger|X\rangle\langle X|S|p\rangle_{2},
\]
we may write the most general solution as
\[
S_2(p)=\eta(p)\,e^{2i\delta(p)} ,
\]
with \(\delta(p)\in\mathbb{R}\). Consequently,
\[
\boxed{\;{}_{2}\langle q|S|p\rangle_{2}^{\text{(inelastic)}}=(2\pi)\,2E_p\,\delta(q-p)\;\eta(p)\,e^{2i\delta(p)},\qquad 0\le \eta(p)\le 1\;}
\]
and probability conservation becomes
\[
1-\eta^2(p)
=\sum_{n\ge 3}\int d\Phi_n\;\big|\mathcal{M}_{2\to n}(p\to \{k_i\})\big|^2,
\]
i.e.\ the loss from the elastic amplitude equals the total inclusive probability into all inelastic channels (the \(1{+}1\)D optical theorem in this single–channel setting).
 In \(1{+}1\)D there is only one independent ``partial wave'' in the CM frame; the above \(S_2(p)\) coincides with that single eigenvalue. For identical particles (bosons/fermions) the exchange symmetry is already baked into the definition of the two–body sector; the form \(S_2=\eta e^{2i\delta}\) still holds.


\chapter{Ising field theory}
\section{Introduction}

The Ising field theory occupies a particularly appealing intersection of conceptual simplicity and structural complexity. On the one hand, it is arguably the simplest nontrivial relativistic quantum field theory in $1+1$ dimensions: the scaling limit of the transverse-field Ising chain gives a conformal field theory with central charge $c=\tfrac12$.\footnote{If you are not familiar with conformal field theory or integrability, you may safely treat phrases like ``Ising CFT'', ``central charge'', and ``$E_8$'' as labels for now. What we will actually use is only that: (i) there is a `free--fermion line' in the space of couplings along which scattering is purely elastic and exactly solvable; and (ii) there is a special ``magnetic line'' with a discrete spectrum of massive particles. The rest of this chapter will work directly with the spin chain and the continuum kinematics developed earlier.} Perturbing this critical point by the two relevant operators of the Ising CFT, the energy density $\epsilon$ and the spin operator $\sigma$, produces a massive continuum with surprisingly rich dynamics. At special loci in coupling space, the model becomes integrable and admits an exact S--matrix with factorized, purely elastic scattering; along the famous magnetic line the spectrum is organized by the exceptional Lie algebra $E_8$, and the corresponding mass ratios have even been observed in real materials tuned close to an Ising quantum critical point \cite{BPZ, MussardoBook, SmirnovBook, Zamolodchikov1989, Coldea2010}. If we move just a bit away from the strictly integrable regime, the very same theory already displays many features typically linked to much more intricate systems: confinement of kinks into meson-like bound states, resonances with finite lifetimes, and inelastic production of multi-particle states above threshold. Thus, within a single model we can interpolate between exactly solvable dynamics and genuinely difficult quantum behaviour, while still defining continuum observables—such as masses, phase shifts, and inelasticities—in a clear, field-theoretic framework.

From the perspective of these lectures, this makes the Ising field theory an ideal playground for quantum computation. The difficult questions for classical computers are precisely those involving real--time, out--of--equilibrium dynamics at intermediate and high energies. If we launch localized wave packets, scatter them, and ask how entanglement grows during and after the collision, or how spectral weight is distributed between elastic and inelastic channels, standard tensor--network techniques eventually  run into trouble: the bond dimension required to track the state typically grows rapidly with time. Truncated--space approaches built on conformal field theory (such as TCSA) give us controlled spectra and matrix elements at low energies, but they struggle once multiple channels open and resonant, nonperturbative physics takes over. None of this is a failure of the methods; it is a reflection of the physics. Scattering in a strongly--coupled, nonintegrable quantum field theory generates complexity. A quantum computer, by directly implementing the unitary time evolution in the full Hilbert space, is designed to operate precisely in this regime.

It is therefore quite remarkable that controlled ``field--theory experiments'' are now being carried out on real quantum processors. Recent work \cite{Farrell2025WStates} has shown that even noisy digital hardware can be used to extract the confined meson spectrum of the longitudinal--field Ising chain via quench spectroscopy, by converting time--series data into rest--energy peaks that agree with nonperturbative expectations. More directly for our purposes, gate--based simulations with $\mathcal{O}(10^2)$ qubits have already implemented genuine scattering experiments in the Ising field theory: wave packets are prepared, made to collide, and the outgoing state is analysed to study the opening of inelastic channels, and the probability of multi--particle production as a function of collision energy. These results do not yet constitute demonstrations of quantum advantage ``beyond all doubt,'' but they do mark an important qualitative milestone: they probe inherently real-time, strongly coupled observables formulated in continuum field theory, which are famously costly for classical algorithms to simulate at similar spacetime volumes \cite{Lamb:2023eyr, FarrellPRD2024, Jha2025IFT}.

Seen from a broader physics viewpoint, these developments also point towards the quantum simulation of gauge theories---something that we will examine in the next chapter. Kink confinement in the longitudinal--field Ising model is a close cousin of string formation in gauge theories; mapping out meson towers and their dissolution under nonintegrable perturbations is a minimalist rehearsal for hadron physics. Likewise, understanding how elastic probability is depleted into many--body continua in this simple setting prepares us to ask---and eventually answer---the analogous questions in more complicated arenas. In summary, there are two main reasons to investigate the Ising field theory on a quantum computer: it provides an ideal framework for refining quantum simulation techniques, and it offers exactly the kind of physics from which we can draw insights that will be crucial once we move on to gauge theories and more complex models.

\paragraph{Suggested references:} For an overview of integrable QFT and statistical models see \cite{MussardoBook}. E\(_8\) scattering and mass ratios were originally studied in \cite{Zamolodchikov1989}; form-factor technology is reviewed in \cite{SmirnovBook, KarowskiWeisz}. Finite-volume effects are studied in \cite{YangYang1969, Luescher1986}. Confinement and meson spectra in the nonintegrable Ising regime are discussed in \cite{FonsecaZamo2003, Rutkevich2010}. CFT finite-size methods are in \cite{BPZ, CardyFS}. Experimental signatures of E\(_8\) in quasi-1D magnets provide useful physical context \cite{Coldea2010}.

\section{From \texorpdfstring{$\phi^4$}{phi4} to the Ising chain}


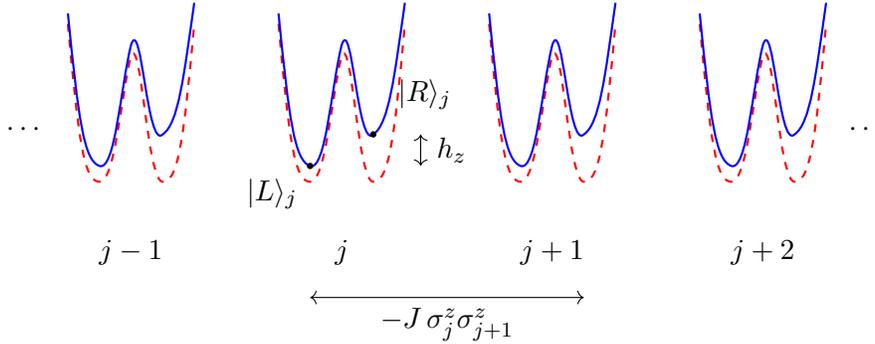
\begin{figure}[H]
\centering
\begin{tikzpicture}[
    x=1.4cm,
    y=1.4cm,
    every node/.style={font=\small}
  ]

  \node at (-3,0.1) {$\cdots$};
  \node at (5,0.1) {$\cdots$};

  \foreach \x/\lab in {-2/{j-1}, 0/{j}, 2/{j+1}, 4/{j+2}} {

    \begin{scope}[shift={(\x,0)}]
      \draw[dashed,thick, red]
        (-0.6,1.1)
          .. controls (-0.5,0.0)  and (-0.45,-0.4) .. (-0.30,-0.4)
          .. controls (-0.15,-0.4) and (-0.10,0.4)  .. (0.0,0.8)
          .. controls (0.10,1.0)  and (0.15,-0.4) .. (0.30,-0.4)
          .. controls (0.45,-0.4) and (0.50,0.0)  .. (0.6,1.1);

      \draw[thick, blue]
        (-0.6,1.2)
          .. controls (-0.5,0.2)  and (-0.45,-0.2) .. (-0.30,-0.25)
          .. controls (-0.15,-0.30) and (-0.10,0.5)  .. (0.0,0.9)
          .. controls (0.10,1.2)  and (0.15,-0.1)  .. (0.30,0.05)
          .. controls (0.45,0.15) and (0.50,0.6)   .. (0.6,1.3);

      \node[below=6pt] at (0,-0.7) {$\lab$};
    \end{scope}
  }

  \begin{scope}[shift={(0,0)}]
    \coordinate (Lmin) at (-0.30,-0.25);
    \coordinate (Rmin) at ( 0.30, 0.05);

    \fill (Lmin) circle (1.2pt);
    \fill (Rmin) circle (1.2pt);
    \node[below left=1pt]  at (Lmin) {$|L\rangle_j$};
    \node[above right=7.5pt] at (Rmin) {$|R\rangle_j$};

    \draw[<->] (0.75,-0.25) -- (0.75,0.05);
    \node[right=2pt] at (0.75,-0.10) {$h_z$};
  \end{scope}

  \draw[<->] (-0.3,-1.5) -- (2.3,-1.5);
  \node[below] at (1.0,-1.5) {$-J\,\sigma^z_j\sigma^z_{j+1}$};

\end{tikzpicture}
\caption{Ising spin chain from anharmonic oscillators. Each lattice site carries
a projected double-well potential: the dashed curve is the symmetric well
($h_z=0$), while the solid curve includes a longitudinal tilt $h_z$, biasing
the local states $|L\rangle_j, |R\rangle_j$ with $\sigma^z_j=\pm1$. A typical
nearest--neighbour coupling $-J\,\sigma^z_j\sigma^z_{j+1}$ is indicated between
two central sites. Intra-site tunneling, controlled by the transverse field
$h_x$, mixes $|L\rangle_j$ and $|R\rangle_j$ but is not shown explicitly to
avoid clutter.}
\label{fig:ising-double-well-tilt}
\end{figure}

In the previous chapter, we built a lattice scalar theory by placing an anharmonic oscillator on every site,
\begin{equation}
H_{\phi^4}^{\text{lat}}
= \sum_{j}\!\Big[\tfrac{1}{2}\Pi_j^2 + U(\phi_j)\Big]
\;+\; \frac{\kappa}{2a^2}\sum_{j}(\phi_{j+1}-\phi_j)^2,
\qquad
U(\phi)=\frac{m^2}{2}\phi^2+\frac{\lambda}{4}\phi^4,
\label{eq:lattice-phi4}
\end{equation}
with lattice spacing \(a\), kinetic coupling \(\kappa\) for gradients, and on–site double–well when \(m^2<0\) (spontaneous \(Z_2\) breaking). Let \(v\equiv \sqrt{|m^2|/\lambda}\) denote the well minima; in the deep–well regime the local spectrum at each site consists of two low–lying states localized near \(\phi=\pm v\), split by a small tunneling amplitude through the barrier.

The key observation is that, at energies well below the local barrier, each site is effectively a \emph{two–level system} \cite{degennes, blinczeks, levitskii}. Let \(\{|L\rangle_j,|R\rangle_j\}\) be wavefunctions localized in the left/right well. Form symmetric/antisymmetric combinations
\[
|+\rangle_j=\frac{|L\rangle_j+|R\rangle_j}{\sqrt{2}},
\qquad
|-\rangle_j=\frac{|L\rangle_j-|R\rangle_j}{\sqrt{2}},
\]
and use them as a qubit basis. We now define Pauli operators by their action in this two–dimensional subspace:
\[
\sigma_j^z\,|L\rangle_j=+|L\rangle_j,\quad \sigma_j^z\,|R\rangle_j=-|R\rangle_j,
\qquad
\sigma_j^x\,|L\rangle_j=|R\rangle_j,\quad \sigma_j^x\,|R\rangle_j=|L\rangle_j.
\]
(Equivalently, \(\sigma_j^z\) measures the sign of \(\phi_j\) and \(\sigma_j^x\) flips between the two wells.)

Projecting the \emph{on–site} part of \eqref{eq:lattice-phi4} onto this subspace produces two familiar terms. First, quantum tunneling between \(\pm v\) gives a level splitting \(\Delta\), which appears as a transverse field:
\begin{equation}
\Big[\tfrac{1}{2}\Pi_j^2 + U(\phi_j)\Big]\;\longrightarrow\; -\frac{\Delta}{2}\,\sigma_j^x \;+\; \text{const}.
\label{eq:tunnel-to-sigmax}
\end{equation}
Second, a small \emph{tilt} of the double well (e.g.\ by adding a linear term \(\epsilon\,\phi\) or by asymmetrically shifting \(m^2\)) biases one well over the other. In the two–level subspace this is a longitudinal field,
\begin{equation}
\epsilon\,\phi_j \;\longrightarrow\; -\frac{\varepsilon}{2}\,\sigma_j^z,
\qquad
\varepsilon \propto \epsilon\,\langle L|\phi|L\rangle - \epsilon\,\langle R|\phi|R\rangle \simeq 2\epsilon v.
\label{eq:tilt-to-sigmaz}
\end{equation}

The \emph{inter–site} coupling originates from the discrete gradient term. In the deep–well limit the low–energy configurations have \(\phi_j\approx v\, s_j\) with Ising variables \(s_j=\pm1\) (the sign of the local displacement). Then
\[
(\phi_{j+1}-\phi_j)^2 \approx v^2(s_{j+1}-s_j)^2 = 2v^2\big(1 - s_js_{j+1}\big).
\]
Up to an overall constant, this favours aligned neighbours and projects to a nearest–neighbour Ising coupling:
\begin{equation}
\frac{\kappa}{2a^2}\,(\phi_{j+1}-\phi_j)^2
\;\longrightarrow\;
-\;J\,\sigma_j^z \sigma_{j+1}^z \;+\; \text{const},
\qquad
J \propto \frac{\kappa v^2}{a^2},
\label{eq:grad-to-zz}
\end{equation}
where the proportionality absorbs scheme–dependent form factors from the projection. Putting \eqref{eq:tunnel-to-sigmax}–\eqref{eq:grad-to-zz} together we obtain precisely the transverse–field Ising model with a possible longitudinal bias,
\begin{equation}
H_{\text{eff}} \;=\; -J \sum_j \sigma_j^z\sigma_{j+1}^z \;-\; h_x \sum_j \sigma_j^x \;-\; h_z \sum_j \sigma_j^z \;+\; \text{const},
\qquad
h_x \sim \tfrac{\Delta}{2},\;\; h_z \sim \tfrac{\varepsilon}{2}.
\label{eq:tfim-from-phi4}
\end{equation}
The absence of a same–site \(\sigma_j^2\) term is automatic: \(\sigma_j^2=\mathbf{1}\) in the projected two–level Hilbert space. Thus, \emph{replacing the local coordinate \(\phi_j\) by its two lowest well states} is precisely the step that takes us from a lattice \(\phi^4\) theory to a spin chain with \(\sigma^z\sigma^z\) couplings.

This construction makes the continuum connection transparent. Near its quantum critical point \((h_z=0,\; h_x/J=1)\), the transverse–field Ising chain flows to the \(1{+}1\)–dimensional Ising conformal field theory with central charge \(c=\tfrac{1}{2}\). On the other hand, the continuum Landau–Ginzburg effective theory for the Ising universality class is a real scalar field with a \(Z_2\)-symmetric \(\phi^4\) potential. In other words, the same \(Z_2\)–invariant quartic theory we developed from a lattice of anharmonic oscillators is the coarse–grained field theory of the Ising chain near criticality, while the deep–well projection of the lattice \(\phi^4\) recovers the Ising spins themselves. The two descriptions are therefore \emph{dual perspectives on the same long–distance physics}: \(\phi\) is the coarse–grained order parameter; \(\sigma^z\) is its two–state remnant upon discretization and projection.

The dictionary extends to excitations and will be useful later. The classical kink of \(\phi^4\) interpolating between \(-v\) and \(+v\) maps to a \emph{domain wall} between opposite \(\sigma^z\) domains. A small longitudinal field \(h_z\) tilts one well, confining kink–antikink pairs into ``mesons''; this has a precise counterpart in both languages: a linear string tension for \(\sigma^z\) domain walls and a confining potential for \(\phi\) kinks. Finally, at \(h_z=0\) the low–energy fermionic quasiparticles of the transverse–field chain are the familiar Bogoliubov modes of the quadratic (free) \(\phi\) fluctuations around the symmetric point, completing the bridge between the operator content on both sides.

For our purposes, the takeaway is practical. We can think of the AHO chain with a deep double–well as the \emph{parent} model: keeping the full local Hilbert space and weak gradients yields the lattice \(\phi^4\) of Eq.~\eqref{eq:lattice-phi4}; projecting each site to its lowest doublet and keeping nearest–neighbor gradients yields the TFIM \eqref{eq:tfim-from-phi4}. The quantum–computing discussions we had for \(\phi^4\) (wave–packet preparation, vacuum subtraction, two–particle scattering) thus carry over seamlessly to Ising, with the simple replacement \(\phi_j\mapsto \sigma_j^z\) and a transverse \(\sigma^x\) term governing local tunneling.

\section{Ising chain: definitions, notation}

We work on a ring of \(N\) spin\(-\tfrac12\) degrees of freedom (sites labeled \(j=1,\dots,N\)) with lattice spacing \(a\) and periodic boundary conditions. The Hamiltonian family we will study is
\begin{equation}
\label{eq:ising-xyz}
H(J,h_x,h_z)
\;=\;
-\,J\sum_{j=1}^{N}\sigma^z_j\,\sigma^z_{j+1}
\;-\;h_x\sum_{j=1}^{N}\sigma^x_j
\;-\;h_z\sum_{j=1}^{N}\sigma^z_j,
\qquad
\sigma^\alpha_{N+1}\equiv\sigma^\alpha_{1},
\end{equation}
with \(J>0\) the ferromagnetic Ising coupling, \(h_x\) the transverse field, and \(h_z\) the longitudinal field. Throughout, \(\sigma^\alpha_j\) denotes a Pauli matrix \(\sigma^\alpha\) (\(\alpha=x,y,z\)) acting \emph{only} on site \(j\).

\paragraph{Notation and operator placement.}
When we write \(\sigma^\alpha_j\) we mean the tensor product
\[
\sigma^\alpha_j
\;\equiv\;
\mathbb{I}^{\otimes (j-1)} \otimes \sigma^\alpha \otimes \mathbb{I}^{\otimes (N-j)},
\]
so every local term in \eqref{eq:ising-xyz} acts nontrivially on its indicated site(s) and as the identity elsewhere. In particular, the nearest--neighbour Ising interaction is the sum of two-site operators
\(\sigma^z_j\sigma^z_{j+1}
=\big(\mathbb{I}^{\otimes (j-1)}\otimes\sigma^z\otimes\sigma^z\otimes\mathbb{I}^{\otimes (N-j-1)}\big)\),
wrapped periodically by \(\sigma^\alpha_{N+1}\equiv\sigma^\alpha_1\).

\subsection{The solvable limit \(H(J,0,0)\).}
Setting \(h_x=h_z=0\) gives
\begin{equation}
\label{eq:classical-ising-H}
H(J,0,0)
\,=\,
-\,J\sum_{j=1}^{N}\sigma^z_j\,\sigma^z_{j+1},
\end{equation}
the quantum operator whose eigenbasis is the simultaneous \(\sigma^z\)-eigenbasis. Equivalently, \eqref{eq:classical-ising-H} is the energy functional of the \emph{classical} one-dimensional Ising model with spins \(s_j=\pm 1\) via \(\sigma^z_j|s_j\rangle=s_j|s_j\rangle\):
\[
E[s_1,\dots,s_N] \;=\; -\,J\sum_{j=1}^{N} s_j\,s_{j+1}.
\]
This limit is “solvable’’ in two complementary senses:

\emph{(i) As a quantum Hamiltonian:} it is already diagonal in the computational basis. Every product state \(|s_1\rangle\otimes\cdots\otimes|s_N\rangle\) with \(s_j=\pm 1\) is an eigenstate, and the energy depends only on the number of \emph{domain walls} (bonds with \(s_j\neq s_{j+1}\)). Writing \(N_{\rm dw}\) for that number,
\[
E \;=\; -J\sum_j s_js_{j+1}
\;=\; -J\big[(N-N_{\rm dw}) - N_{\rm dw}\big]
\;=\; -JN \;+\; 2J\,N_{\rm dw}.
\]
Thus the two fully aligned ferromagnets \(|\uparrow\uparrow\cdots\uparrow\rangle\) and \(|\downarrow\downarrow\cdots\downarrow\rangle\) are degenerate ground states with \(E_0=-JN\); each domain wall costs an energy \(2J\). Excitations are therefore freely moving domain walls (kinks) at this level, with no interactions between them.

\emph{(ii) As a classical statistical model:} the finite-temperature partition function \(Z=\sum_{\{s\}}e^{-\beta E[\{s\}]}\) is exactly computable by a \(2\times 2\) transfer matrix. Defining
\[
T \;=\; 
\begin{pmatrix}
e^{\beta J} & e^{-\beta J} \\
e^{-\beta J} & e^{\beta J}
\end{pmatrix},
\quad
\lambda_{\pm} \;=\; e^{\beta J}\pm e^{-\beta J},
\]
one has \(Z=\lambda_+^{\,N}+\lambda_-^{\,N}\), and in the thermodynamic limit the free energy density is \(f = -\beta^{-1}\log\lambda_+\). All thermodynamic quantities follow directly; for example, the zero-field two-point function decays as \(\langle s_0 s_r\rangle=(\tanh \beta J)^r\), so the correlation length is \(\xi^{-1}=-\log\tanh(\beta J)\). In one dimension there is no finite-temperature phase transition (no spontaneous magnetization at \(T>0\)), a fact that historically motivated the search for and eventual discovery of nontrivial critical behavior in two dimensions.

\paragraph{Historical context and importance.}
The Ising model originated as a minimal theory of cooperative magnetism. The one-dimensional case was solved early on by Ising and showed \emph{no} finite-\(T\) transition, whereas the two-dimensional model famously \emph{does} order below a critical temperature and was solved exactly by Onsager, inaugurating modern critical phenomena. For us, the \(H(J,0,0)\) limit plays a concrete pedagogical role: it anchors the notion of ferromagnetic order and domain-wall excitations in a setting where the spectrum is transparent. Turning on \(h_x\) later will produce quantum fluctuations that delocalize those domain walls (and, at criticality, lead to a relativistic scaling limit), while a longitudinal field \(h_z\) breaks the \(\mathbb{Z}_2\) symmetry and ultimately enables inelastic scattering in the Ising field theory studied on quantum hardware. We will approach those deformations step by step, using only what we need for the scattering experiments.

\subsection{The solvable limit \(H(J,h_x,0)\)}

We now turn on a transverse field and set the longitudinal field to zero, giving the Transverse Field Ising Model (TFIM)
\begin{equation}
\label{eq:tfim}
H(J,h_x,0)
\;=\;
-\,J\sum_{j=1}^{N}\sigma^z_j\,\sigma^z_{j+1}
\;-\;h_x\sum_{j=1}^{N}\sigma^x_j,
\qquad
\sigma^\alpha_{N+1}\equiv\sigma^\alpha_{1}.
\end{equation}
As before, \(\sigma^\alpha_j\) means a Pauli \(\sigma^\alpha\) acting on site \(j\) and the identity elsewhere. The model is translation invariant and has a global \(\mathbb{Z}_2\) spin-flip symmetry generated by \(P=\prod_{j}\sigma^x_j\), under which \(\sigma^z_j\!\mapsto\!-\sigma^z_j\). This is the standard transverse-field Ising model (TFIM).

At zero temperature there are two quantum phases separated by a continuous quantum phase transition. For small transverse field, \(g\equiv h_x/J<1\), the ground state is ferromagnetically ordered along \(z\) (two nearly degenerate vacua on a ring, split only by nonperturbative tunneling at finite \(N\)). For large transverse field \(g>1\), the ground state is a unique paramagnet polarized along \(x\). Exactly at \(g=1\) the system is critical with emergent relativistic invariance and dynamical exponent \footnote{If the dispersion relation is $\omega=k^z$, then $z$ here is defined to be the dynamical exponent.} \(z=1\). The scaling limit at this critical point is the Ising conformal field theory with central charge \(c=\tfrac12\). This is the quantum \(1{+}1\)D counterpart of the \(2\)D classical Ising critical point via the standard Trotter (quantum–to–classical) mapping.

The TFIM  is exactly solvable: one diagonalizes it by a sequence of transforms (Fourier, Jordan–Wigner, Bogoliubov). Without reproducing that machinery here, we record the outputs we will need later. The elementary excitations are free Majorana fermions with single-particle dispersion
\begin{equation}
\label{eq:dispersion}
\varepsilon(k)\;=\;2J\,\sqrt{1+g^{2}-2g\cos(ka)}\,,
\end{equation}
so the spectral gap is
\begin{equation}
\label{eq:gap}
m\;\equiv\;\varepsilon(0)\;=\;2J\,|1-g|\,,
\end{equation}
and the group velocity is \(v(k)=\partial_k\varepsilon(k)\); at criticality the characteristic velocity is \(v_\ast=2Ja\). The equal-time correlations and order parameters are known in closed form: in the ordered phase \(g<1\) the spontaneous magnetization along \(z\) is \(M_z=(1-g^2)^{1/8}\), while the correlation length diverges near criticality as \(\xi\sim|1-g|^{-1}\). The critical exponents are the Ising ones, \(\beta=\tfrac18\), \(\nu=1\), \(\eta=\tfrac14\), consistent with the \(c=\tfrac12\) CFT. In the scaling limit the finite-size spectrum on a circle of length \(L=Na\) organizes into conformal towers with level spacings \(\sim (2\pi v_\ast/L)\), a fact we will use as a diagnostic when we calibrate dispersion and gap on finite rings.

Two remarks are especially relevant for the scattering agenda later on. First, the \(h_z=0\) chain is \emph{integrable} and its continuum limit is a free Majorana theory: multi-particle scattering is purely elastic and factorized, and in the fermionic description the two-body \(S\)-matrix is just a sign from Fermi statistics. As a result, wave-packet collisions at \(h_z=0\) exhibit vanishing inelastic production and zero Wigner time delay; this provides an ideal hardware baseline. Second, even away from strict criticality the long-wavelength sector is well captured by relativistic kinematics with the mass \(m\) in \eqref{eq:gap} and the lattice-measured \(\varepsilon(k)\) in \eqref{eq:dispersion}, so our later conversion between lattice energies \(E\), momenta \(p\), and rapidities \(\theta\) will be quantitatively controlled once we have calibrated \(\varepsilon(k)\) directly on the device.

Historically, the TFIM in one dimension was among the first quantum spin chains to be solved exactly. Lieb, Schultz, and Mattis established the fermionization framework for \(XY\)-type chains; Pfeuty specialized it to the Ising case and derived the spectrum, gap, and order parameters; Barouch and McCoy obtained exact space–time correlation functions and scaling forms; and later developments connected the quantum critical point to the Ising minimal model of conformal field theory. Standard references include \cite{LSM1961, Pfeuty1970, BarouchMcCoy1971, Sachdev2011, Cardy1996}. 

\subsection{Turning on a longitudinal field: the nonintegrable Ising chain \(H(J,h_x,h_z)\)}

We now let both fields act:
\begin{equation}
\label{eq:ising-full}
H(J,h_x,h_z)
\,=\,
-\,J\sum_{j=1}^{N}\sigma^z_j\,\sigma^z_{j+1}
\;-\;h_x\sum_{j=1}^{N}\sigma^x_j
\;-\;h_z\sum_{j=1}^{N}\sigma^z_j,
\qquad
\sigma^\alpha_{N+1}\equiv\sigma^\alpha_{1}.
\end{equation}
As in the preceding sections, \(\sigma^\alpha_j\) denotes a Pauli operator acting on site \(j\) and the identity elsewhere. Translation invariance remains intact for all \((J,h_x,h_z)\). The global \(\mathbb{Z}_2\) spin-flip generated by \(P=\prod_j\sigma^x_j\) is a symmetry at \(h_z=0\) and is \emph{explicitly broken} as soon as \(h_z\neq 0\).

Two structural changes follow immediately from the longitudinal field. First, integrability is lost in the lattice model away from special lines: the exactly solvable free–fermion structure at \(h_z=0\) no longer applies once \(h_z\neq 0\). Second, in the ferromagnetic regime the longitudinal field lifts the degeneracy of the two Ising vacua and produces a \emph{linear confining force} between domain walls (kinks): a pair of kinks that delimit a flipped domain now pay an energy proportional to their separation, so the asymptotic excitations are not free kinks but rather a discrete tower of \emph{meson-like} kink–antikink bound states. In the weak-field limit their energies form a characteristic almost equally spaced set near the two-kink threshold, a hallmark of linear confinement in one dimension (see, e.g., \cite{McCoyWu1978, Rutkevich2005}).

The continuum interpretation sharpens these statements. Near the quantum critical point of the transverse-field chain (\(h_z=0\), \(h_x/J=1\)), the long-wavelength limit is the Ising conformal field theory with central charge \(c=\tfrac12\). Moving away from criticality corresponds to perturbing the CFT by its two relevant primaries, the energy density \(\varepsilon\) (even under \(\mathbb{Z}_2\)) and the spin \(\sigma\) (odd). In field-theory language one writes the Ising Field Theory (IFT)\footnote{Here $\varepsilon$ is the local energy density and $\sigma$ the spin/order-parameter operator; the numbers $\Delta_\varepsilon=1$, $\Delta_\sigma=1/8$ are their scaling dimensions in the critical theory. We won’t need any further CFT machinery; we only use that $\tau$ and $h$ correspond to tuning $h_x$ and $h_z$ away from criticality.}
\[
\mathcal{L}_{\mathrm{IFT}}
=\mathcal{L}_{\mathrm{Ising\ CFT}}
\;+\;\tau\!\int\!\varepsilon(x)\,d^2x
\;+\;h\!\int\!\sigma(x)\,d^2x,
\qquad
(\,\Delta_\varepsilon=1,\ \Delta_\sigma=\tfrac18\,),
\]
with couplings \(\tau\propto h_x/J-1\) and \(h\propto h_z/J\). Two distinguished “integrable corridors’’ are then visible in this plane. Along \(h=0\) the theory is a \emph{free massive Majorana} field: scattering is elastic and factorized, and the two–body \(S\)-matrix is just a sign. Along \(\tau=0\), i.e. at the critical transverse field with a longitudinal perturbation, Zamolodchikov discovered an \emph{integrable} massive theory whose exact spectrum contains eight stable particles with mass ratios fixed by the exceptional Lie algebra \(E_8\); the two–body amplitudes are again purely elastic and exactly known \cite{Zamolodchikov1989}. The generic case with \(\tau\neq 0\) and \(h\neq 0\) is \emph{nonintegrable}: elastic scattering persists below the first inelastic threshold, but as soon as sufficient energy is available, \emph{particle production} \(2\to 4,6,\dots\) occurs. It is precisely in this nonintegrable IFT regime that modern experiments and simulations test real–time inelastic processes; the celebrated neutron–scattering observation of the \(E_8\) mass pattern in a quasi–one–dimensional magnet \cite{Coldea2010} provides a complementary view of the integrable edge of this physics.

For our purposes this landscape serves three roles. It provides a clean baseline at \(h_z=0\), where kink excitations map to free fermions and wave–packet collisions exhibit vanishing time delay and no inelasticity. It supplies a conceptually sharp integrable checkpoint at \(\tau=0\) with \(h_z\neq 0\) (the \(E_8\) theory), where one again expects purely elastic behavior but with a rich multiplet of masses. And, most importantly, it explains why turning on \emph{both} deformations produces inelastic channels and long–lived mesonic resonances: this is the regime in which wave–packet collisions probe \(\delta(E)\), Wigner time delays, and energy flow into multi–particle sectors—the observables we will extract in the results section. Modern field–theory and scattering reviews place this nonintegrable deformation of the Ising CFT within a broader program of “integrable–plus–perturbations’’ where exact data (masses, form factors) along integrable rays constrain and organize the physics away from them \cite{MussardoBook, DelfinoMussardo1995, FonsecaZamo2003}.

To summarize this section: 
In the $(\tau,h)$ plane of the Ising field theory it is useful to keep three special
corridors in mind:
\begin{itemize}
  \item The \emph{free Majorana line} $h=0$ ($h_z=0$ on the lattice), where the
  TFIM is integrable and the continuum limit describes free fermions with purely
  elastic, factorized scattering.
  \item The \emph{$E_8$ line} $\tau=0$ (critical transverse field with a longitudinal
  perturbation), where the exact spectrum contains eight stable particles with
  fixed mass ratios and scattering is again purely elastic and integrable.
  \item The \emph{generic nonintegrable regime} with $\tau\neq 0$ and $h\neq 0$,
  where integrability is broken and above threshold one has genuine inelastic
  processes $2\to 4,6,\dots$.
\end{itemize}
The scattering experiments that were emulated on
quantum hardware \cite{Farrell2025WStates} are deliberately carried out in this third regime: they choose
$(J,h_x,h_z)$ so that the continuum IFT has \emph{two} stable massive particles
with masses $m_1<m_2$ but is otherwise nonintegrable, and then study inelastic
processes such as $A_1+A_1\to A_1+A_2$ above threshold. The $E_8$ line plays the
role of a nearby integrable ``anchor'' for our intuition, but it is \emph{not} the
point actually realized in the hardware runs.

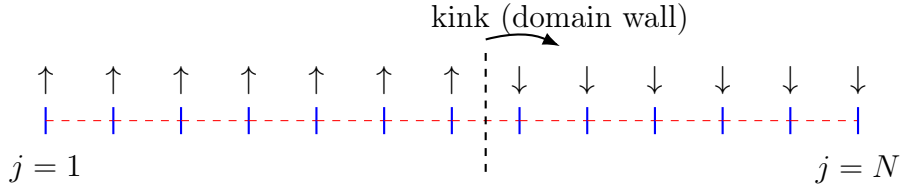
\begin{figure}[t]
\centering
\begin{tikzpicture}[x=0.9cm,y=0.9cm,>=Latex]
  \tikzstyle{site}=[inner sep=0pt, minimum width=0.6cm]
  \foreach \j in {0,...,12}{
    \node[site] (s\j) at (\j,0) {};
    \draw[blue, thick] (\j,0.2) -- (\j,-0.2);
    \ifnum\j<12\relax
      \draw[red,dashed] (\j,0) -- (\numexpr\j+1\relax,0);
    \fi
  }
  \foreach \j in {0,...,6}{
    \node at (\j,0.6) {$\uparrow$};
  }
  \foreach \j in {7,...,12}{
    \node at (\j,0.6) {$\downarrow$};
  }
  \draw[dashed,thick] (6.5,1.0) -- (6.5,-0.8);
  \draw[->,thick] (6.5,1.0) ++(0,0.2) to[bend left=20] (7.6,1.1);
  \node[above] at (7.6,1.1) {kink (domain wall)};
  \node[below] at (0,-0.35) {$j=1$};
  \node[below] at (12,-0.35) {$j=N$};
\end{tikzpicture}
\caption{A single domain wall (kink) separating ferromagnetic regions. In the classical/diagonal limit \(H(J,0,0)\), each kink costs energy \(2J\).}
\label{fig:kink}
\end{figure}

\begin{figure}[hbt]
\centering
\begin{tikzpicture}[x=0.9cm,y=0.9cm,>=Latex]
  \tikzstyle{site}=[inner sep=0pt, minimum width=0.6cm]
  \foreach \j in {0,...,14}{
    \node[site] (s\j) at (\j,0) {};
    \draw[blue,thick] (\j,0.2) -- (\j,-0.2);
    \ifnum\j<14\relax
      \draw[red, dashed] (\j,0) -- (\numexpr\j+1\relax,0);
    \fi
  }
  \foreach \j in {0,...,4}{
    \node at (\j,0.6) {$\uparrow$};
  }
  \foreach \j in {5,...,9}{
    \node at (\j,0.6) {$\downarrow$};
  }
  \foreach \j in {10,...,14}{
    \node at (\j,0.6) {$\uparrow$};
  }
  \draw[dashed,thick] (4.5,1.0) -- (4.5,-0.8);
  \draw[dashed,thick] (9.5,1.0) -- (9.5,-0.8);
  \node[above] at (4.5,1.1) {kink};
  \node[above] at (9.5,1.1) {antikink};
  \draw[decorate,decoration={brace,amplitude=6pt}] (4.5,-0.6) -- (9.5,-0.6);
  \node[below] at (7,-1.0) {separation \(\ell\)};
  \draw[->] (7,0.9) -- (7,0.2);
\end{tikzpicture}
\caption{A kink–antikink pair encloses a flipped domain of length \(\ell\). With a longitudinal field \(h_z\neq 0\), each flipped spin raises the energy by \(\sim 2h_z\), producing a linear “string” potential \(V(\ell)\simeq 2m_{\rm kink}+\sigma\,\ell\) with \(\sigma\propto h_z/a\).}
\label{fig:kinkantikink}
\end{figure}
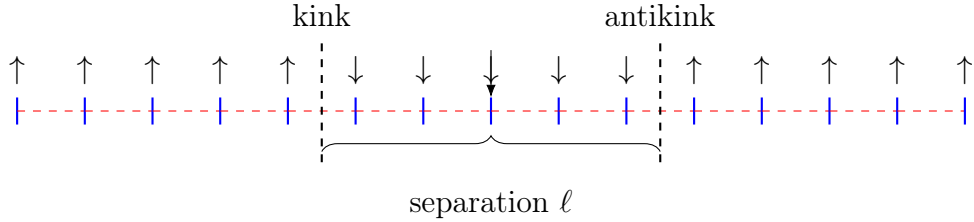

\begin{figure}[t]
\centering
\begin{tikzpicture}[x=1cm,y=1cm]
  \draw[->] (-0.1,0) -- (6.2,0) node[below] {$\ell$};
  \draw[->] (0,-0.1) -- (0,4.2) node[left] {$V(\ell)$};

  \draw[thick] (0,0.8) -- (6,3.8);
  \node[anchor=east] at (0,0.8) {$2m_{\mathrm{kink}}$};
  \node at (4.6,3.3) {string tension $\sigma$};

  \draw[thick, blue] (0.6,1.3) -- (2.2,1.3); \node[right] at (2.25,1.3) {$E_0$};
  \draw[thick, blue] (0.6,2.1) -- (2.2,2.1); \node[right] at (2.25,2.1) {$E_1$};
  \draw[thick, blue] (0.6,2.8) -- (2.2,2.8); \node[right] at (2.25,2.8) {$E_2$};
  \draw[thick, blue] (0.6,3.4) -- (2.2,3.4); \node[right] at (2.25,3.4) {$E_3$};

  \begin{scope}[shift={(0.3,-1.3)},scale=0.6]
    \foreach \j in {0,...,12} {
      \draw[red,dashed] (\j,0) -- (\j+1,0);
    }
    \foreach \j in {0,...,4}  { \node at (\j,0.5) {$\uparrow$}; }
    \foreach \j in {5,...,7}  { \node at (\j,0.5) {$\downarrow$}; }
    \foreach \j in {8,...,12} { \node at (\j,0.5) {$\uparrow$}; }
    \draw[dashed] (4.5,0.9) -- (4.5,-0.6);
    \draw[dashed] (7.5,0.9) -- (7.5,-0.6);
    \draw[<->] (4.5,-0.6) -- (7.5,-0.6);
    \node[below] at (6,-0.9) {$\ell$};
  \end{scope}
\end{tikzpicture}
\caption{In a longitudinal field, a kink–antikink pair feels a linearly rising potential $V(\ell)\simeq 2m_{\mathrm{kink}}+\sigma\,\ell$, producing a discrete tower of confined “meson” bound states $E_n$.}
\label{fig:mesons}
\end{figure}
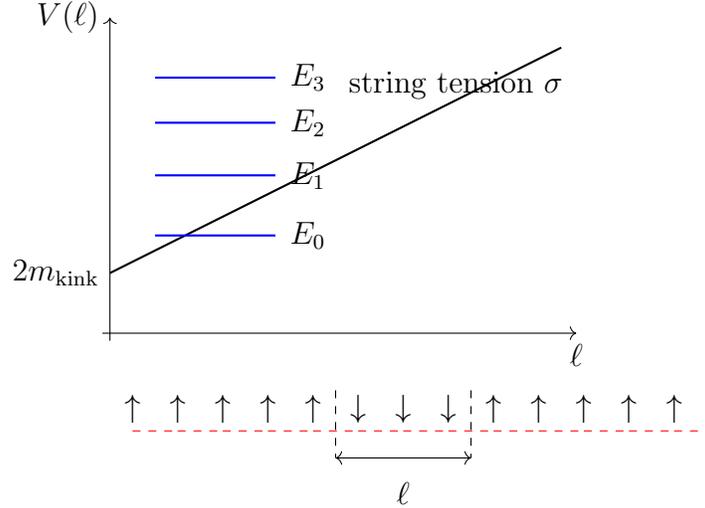

\section{The \(S\)-matrix for Ising field theory}

For a recap of 1+1D two-particle scattering kinematics and elastic vs inelastic unitarity, see section \ref{unitarity}.

\paragraph{Continuum kinematics.}
We have in 1+1D 
\begin{equation}
    E^2=p^2+m^2\,.
\end{equation}
These are solved by the parametrization in terms of the rapidity $\theta$:
\begin{equation}
E=m\cosh\theta,\qquad p=m\sinh\theta,\qquad s=4m^2\cosh^2(\theta/2).
\end{equation}
There is no scattering angle; two-body kinematics is one-dimensional up to particle exchange.

Our goal is to describe what we mean by “scattering’’ in the quantum Ising field theory (IFT) language, in exactly the form that will be used later for wave–packet collisions and for extracting time delays and inelasticity. 

\paragraph{Asymptotic particles and states.}
In the scaling regime of the Ising chain, the long–wavelength excitations are massive relativistic particles with mass \(m\). In the center–of–mass (c.m.) frame, two incoming particles carry momenta \(\pm p\) and energies \(E/2\) each. An \emph{in} state is prepared by sending two well–separated wave packets towards each other; an \emph{out} state is what you have long after the collision, when the outgoing packets are again well separated. The scattering operator \(S\) maps the in–basis to the out–basis,
\[
|\text{out}\rangle \;=\; S\,|\text{in}\rangle,
\]
and encodes the amplitudes for all possible outcomes. More details about how we prepare the scattering states is given in the next part of these lectures. For now we note the following in order to connect with \cite{Farrell2025WStates}:
\begin{enumerate}
  \item We work with periodic boundary conditions and diagonalize $H(J,h_x,h_z)$ in
  sectors of fixed lattice momentum $k = 2\pi n/L$.
  \item For each $k$, order the eigenvalues $E_\alpha(k)$ and identify the lowest
  one--particle branches $E_1(k), E_2(k),\dots$ by continuity in $k$ and by their
  quantum numbers. In the regime relevant for \cite{Farrell2025WStates} there are two
  isolated single--particle branches separated from the multi--particle continuum.
  \item Define the finite--volume mass gaps
  \[
    m_a(L) = E_a(k{=}0,L) - E_0(L),\qquad a=1,2,
  \]
  where $E_0(L)$ is the true ground state energy on the ring. In a massive
  1+1D theory these converge exponentially in $L$, so one fits
  $m_a(L) = m_a + A_a e^{-L/\xi}$ to extract the infinite--volume masses $m_1,m_2$.
  \item The functions $E_a(k)$ then play the role of the lattice dispersion relations
  for the continuum particles $A_a$, and $m_a$ set the continuum mass scale that
  appears in our kinematic formulas $E^2 = p^2 + m_a^2$.
\end{enumerate}

In \cite{Farrell2025WStates} this entire calibration step is performed
classically (via ED/DMRG/MPS) and the resulting $m_1,m_2,E_a(k)$ are used to
choose the incoming momenta and collision energies for the quantum circuits.
The quantum computer is then used purely for the genuinely real--time, strongly
entangled scattering dynamics.

\paragraph{Elastic \(2\!\to\!2\) in \(1{+}1\)D and the phase shift.}
Below the first inelastic threshold (i.e.\ for total c.m.\ energy \(2m\le E<4m\)), the only process available is elastic \(2\!\to\!2\) scattering. In \(1{+}1\) dimensions with identical particles, kinematics leaves a single invariant: the rapidity difference \(2\theta\). Unitarity and translation invariance then imply that the entire two–body process is characterized by a single complex number of unit modulus,
\[
S(2\theta)\;=\;e^{\,i\,\delta(2\theta)} ,
\]
whose phase \(\delta(2\theta)\) is the \emph{elastic phase shift}. In a free (integrable) Majorana theory—our \(h_z=0\) baseline—the two–body \(S\) is just a sign from fermionic exchange, so \(\delta\) is a constant (no energy dependence) and wave packets suffer no time delay.

\paragraph{Above threshold and inelasticity.}
Once the c.m.\ energy exceeds the first production threshold (\(E\ge 4m\) for IFT without additional bound states), channels like \(2\!\to\!4\) open. The two–body subspace no longer evolves unitarily by itself; a convenient parameterization is
\[
S_{2\to 2}(2\theta)\;=\;\eta(2\theta)\,e^{\,i\,\delta(2\theta)}, \qquad 0\le \eta\le 1,
\]
where \(\eta(2\theta)\) is the \emph{inelasticity parameter}. Probability conservation says that the missing weight is precisely the total inelastic probability,
\[
P_{\mathrm{inel}}(E)\;=\;1-\big|S_{2\to 2}(2\theta)\big|^{2}\;=\;1-\eta(E)^{2}.
\]
In the purely elastic window one has \(\eta=1\); as inelastic channels open, \(\eta\) decreases from unity.

\paragraph{What wave packets actually measure: Wigner time delay.}
Real experiments and our quantum–hardware runs use \emph{packets}, not plane waves. For a narrow packet peaked at energy \(E\) the peak of the transmitted/reflected packet is shifted in time by
\[
\Delta t(E)\;=\;\frac{\partial}{\partial E}\,\delta\!\left(2\theta(E)\right),
\]
the \emph{Wigner time delay}. Thus the energy–dependence of the phase shift is directly observable as an advance or delay in the packet’s centroid. We will use this relation verbatim when we analyze collisions.

\paragraph{How spectra encode the same information (finite volume).}
On a ring of length \(L\) the two–body c.m.\ momenta are quantized. In the elastic window, imposing single–valuedness of the two–particle wave function yields the Bethe–Yang condition (see the $\phi^4$ discussion)
\[
p(E)\,L\;+\;\delta\!\left(2\theta(E)\right)\;=\;2\pi\,n,\qquad n\in\mathbb{Z}.
\]
Given two–body energies \(E_2(L)\) at several \(L\)’s (measured on hardware or computed classically), one can invert this relation to obtain \(\delta(2\theta)\). In the inelastic regime the simple one–channel form breaks down, and spectral methods must be generalized to coupled channels; in our work we instead diagnose inelasticity directly from late–time observables (sector weights and energy–density tracks).

\subsection{Conventions-in-brief}

We set \(\hbar=c=1\) and keep the lattice spacing \(a\) explicit when it clarifies dimensions. States are normalized so that one-particle plane waves on a ring of length \(L=Na\),
\[
|p\rangle=\frac{1}{\sqrt{N}}\sum_{j=1}^N e^{ipaj}\,|0\cdots 1_j\cdots 0\rangle,\qquad p=\frac{2\pi n}{L},
\]
obey \(\langle p'|p\rangle=\delta_{p,p'}\). In the scaling regime we convert between \((E,p)\) and rapidity \(\theta\) by \(E=2m\cosh\theta\), \(p=m\sinh\theta\); this is the only relativistic dictionary we will use later. The \emph{phase shift} \(\delta(2\theta)\) is defined by \(S(2\theta)=e^{i\delta(2\theta)}\) in the elastic window; the \emph{Wigner time delay} is \(\Delta t(E)=\partial_E\delta(2\theta(E))\).

In part 2, following \cite{Farrell2025WStates}, we will outline an efficient way to prepare scattering states using the so-called W-state.

\chapter{Schwinger and Thirring models}

\section{Introduction}

Our treatment of the Thirring and Schwinger models will follow the  bosonization approach, in the spirit of \cite{shankar}. We will discuss current algebra, normal ordering and short-distance regularization rather than the path-integral formalism or the staggered fermion picture. In this language the kinetic term $\bar\psi i\slashed\partial \psi$ maps to a free Gaussian field, the vector current $j^\mu$ maps to the dual of $\partial_\mu\phi$, and the mass term $\bar\psi\psi$ maps to a vertex operator $\cos(\sqrt{4\pi}\,\phi)$. We begin by fixing normalizations using the regulated two-point function and by proving the ``Gaussian identity'' for vertex correlators. We then bosonize the chiral kinetic term using a symmetric point-splitting prescription \cite{shankar}.

With this dictionary in hand, the massive Thirring model becomes a sine--Gordon (SG) theory after a simple field rescaling, precisely as in the classic analyses of Coleman and Mandelstam \cite{Coleman1975,Mandelstam1975}. In our conventions
\[
\mathcal L_{TH}
=\bar\psi(i\slashed\partial-m)\psi-\frac{g}{2}\,j_\mu j^\mu
\quad\longleftrightarrow\quad
\frac{1}{2}\Big(1+\frac{g}{\pi}\Big)(\partial\phi)^2-\frac{m}{\pi\alpha}\cos\!\big(\sqrt{4\pi}\,\phi\big)\,,
\]
so the Thirring coupling simply renormalizes the boson stiffness, while the fermion mass becomes the SG cosine. This equivalence is nonperturbative at the level of the continuum theory and will be our organising principle for ``single-particle'' physics: in the neutral sector the lightest excitation is the so-called first SG breather $B_1$, whose elastic $2\to2$ scattering amplitude is known in closed form in the integrable regime. When we extract finite-volume phase shifts from two-body levels, we can benchmark the neutral channel against the exact SG $B_1$--$B_1$ S-matrix and the associated breather mass relations \cite{Zamolodchikov1977,ZZ1979}.

The Schwinger model is equally transparent in bosonic variables. In the massless case, the theory is equivalent to a free massive scalar with $\mu=e/\sqrt{\pi}$, generated by the axial anomaly \cite{Schwinger1962}. Turning on a fermion mass adds a cosine potential on top of the quadratic term that comes from integrating out the electric field, leading to a local bosonic Lagrangian with both $\mu^2\phi^2$ and $\cos(\sqrt{4\pi}\,\phi)$ pieces. Coleman’s analysis makes clear how $\theta$-vacua, confinement and string breaking appear in this setting \cite{Coleman1976}. In this chapter, we derive this mapping explicitly---again using the same regulator and normal-ordering conventions as above---so that the massive Schwinger model appears as ``sine--Gordon deformed by a quadratic mass''. This gives us two calibration lines for dynamics: at $m=0$ the model is free ($S=1$), while at $e=0$ it reduces to integrable SG, allowing us to reuse the Thirring/SG benchmarks.

For lattice and hardware implementations we will only keep the minimal essentials. Rather than the Kogut--Susskind Hamiltonian formulation \cite{KogutSusskind1975,KogutRMP1979}, we will rely on the bosonized formalism \footnote{Two student term papers uploaded on our GitHub repository examine the staggered fermion approach.}. In our opinion, this makes the connection with the previous discussions on $\phi^4$ and Ising field theory natural and avoids the introduction of technical complications. However, from the perspective of quantum simulations, the staggered fermion approach has its advantages \cite{FarrellPRD2024}. 

\paragraph{Suggested references.}
For the original solution of the Schwinger model and its massive generalization, see Schwinger’s classic work~\cite{Schwinger1962} and Coleman’s detailed analysis~\cite{Coleman1976}, together with his treatment of the equivalence between the massive Thirring and sine-Gordon models~\cite{Coleman1975} and Mandelstam’s complementary construction of soliton operators~\cite{Mandelstam1975}; a broader pedagogical perspective is given in Coleman’s collected lectures~\cite{ColemanAspects} and in Shankar’s condensed-matter QFT text~\cite{shankar}. The Hamiltonian formulation of lattice gauge theories and their connection to spin systems are nicely reviewed in Kogut and Susskind~\cite{KogutSusskind1975} and in Kogut’s RMP article~\cite{KogutRMP1979}. For modern numerical and tensor-network approaches to the Schwinger and Thirring models, see the reviews by Ba{\~n}uls and Cichy~\cite{banuls}, Ba{\~n}uls \emph{et al.}~\cite{BanulsEtAl2019PoS}, and the SciPost lecture notes by Emonts and Zohar~\cite{EmontsZohar2020}. Experimental and quantum-computing realizations of the Schwinger model include the early trapped-ion experiment of Martinez \emph{et al.}~\cite{Martinez2016} and more recent large-scale or high-energy simulations in circuit QED and digital quantum computers~\cite{BelyanskyPRL2024,FarrellPRD2024,angel,pufu2,sakamoto}, which provide useful benchmarks and motivation for the algorithms discussed in this chapter.

\section{From the transmon AHO to a lattice cosine scalar}
A single transmon is a weakly-anharmonic oscillator with compact phase $\hat\varphi$:
\begin{equation}
H_{\rm trans}
= 4E_C(\hat n-n_g)^2 \;-\; E_J\cos\hat\varphi,\qquad [\hat\varphi,\hat n]=i.
\label{eq:transmon}
\end{equation}
In the phase basis, $4E_C\hat n^2$ acts as a kinetic term. An \emph{array} (capacitive/inductive couplings) reduces to a 1D field $\varphi_x$ with conjugate $\Pi_x$:
\begin{equation}
H
= \sum_{x=1}^L \Big[
\frac{1}{2C_{\rm eff}}\Pi_x^2
+ \frac{K}{2}(\varphi_{x+1}-\varphi_x)^2
- E_J\cos(\varphi_x-\varphi_0)
\Big].
\label{eq:cosine-chain}
\end{equation}
Expanding about the bias setpoint $\varphi_0$:
\(
- E_J\cos(\varphi_x-\varphi_0)
= -E_J+\tfrac{E_J}{2}\delta\varphi_x^2-\tfrac{E_J}{24}\delta\varphi_x^4+\cdots,
\)
so the AHO/$\phi^4$ story is the small-excursion limit of a \emph{compact} cosine theory. This theory is related to the much studied Sine-Gordon theory which will feature in different avatars in our discussion below.


For the rest of this chapter, we will closely follow \cite{shankar}: we will consider the Schwinger and Thirring models by starting with the fermionic oscillator and putting it on a lattice, similar to how we put the anharmonic oscillator on the lattice and coupled them to get the kinetic gradient term. Then we will rely on a technique called {\it Bosonization} to find equivalence between these models and Sine-Gordon like models.

\section{Dirac fermion and the lattice}
Recall that the usual `bosonic' simple harmonic oscillator has the Hamiltonian $H=a^\dagger a+\frac{1}{2}$. To introduce fermions, we need to replace $a,a^\dagger$ with anticommuting objects.
The fermionic oscillator is made of anticommuting `Grassmann' variables. These are defined as:
\begin{equation}
    \{\psi, \psi\}=0\,.
\end{equation}
The notation is $\{a,b\}=a b+b a$. $a,b$ are Grassmann numbers and should not be mentally replaced by usual numbers. 
Thus we cannot write $\psi^2$ as it is zero. Let us introduce $\psi^\dagger$ such that
\begin{equation}
    \{\psi^\dagger,\psi\}=1\,.
\end{equation}
Let all other anticommutators vanish. The Hamiltonian is 
\begin{equation}
    H=\Omega_0 \psi^\dagger \psi\,.
\end{equation}
The Hilbert space is very simple. If we define the ground state by $|0\rangle$ such that $\psi|0\rangle=0$, then clearly $H|0\rangle=0$. We define $|1\rangle=\psi^\dagger |0\rangle$. There are no other states in the Hilbert space except $|0\rangle, |1\rangle$. $H|1\rangle=\Omega_0 |1\rangle.$

Now we move to the lattice. We introduce the fermionic oscillator at each site such that:
\begin{equation}
\{\psi_j^\dagger, \psi_{j'}\}=\delta_{jj'}\,.
\end{equation}
All other anticommutation relations vanish.
Then we use nearest neighbour coupling. The lattice Hamiltonian will have two kinds of terms--- $H_{o.s.}=\psi^\dagger_j \psi_j$ representing the on-site Hamiltonian and $\psi^\dagger_{j}\psi_{j+1}$ representing the nearest neighbour coupling. We can also consider $\psi_{j}\psi_{j+1}$ but for now let's ignore them.

\begin{figure}[t]
  \centering
  \includegraphics[width=0.85\linewidth]{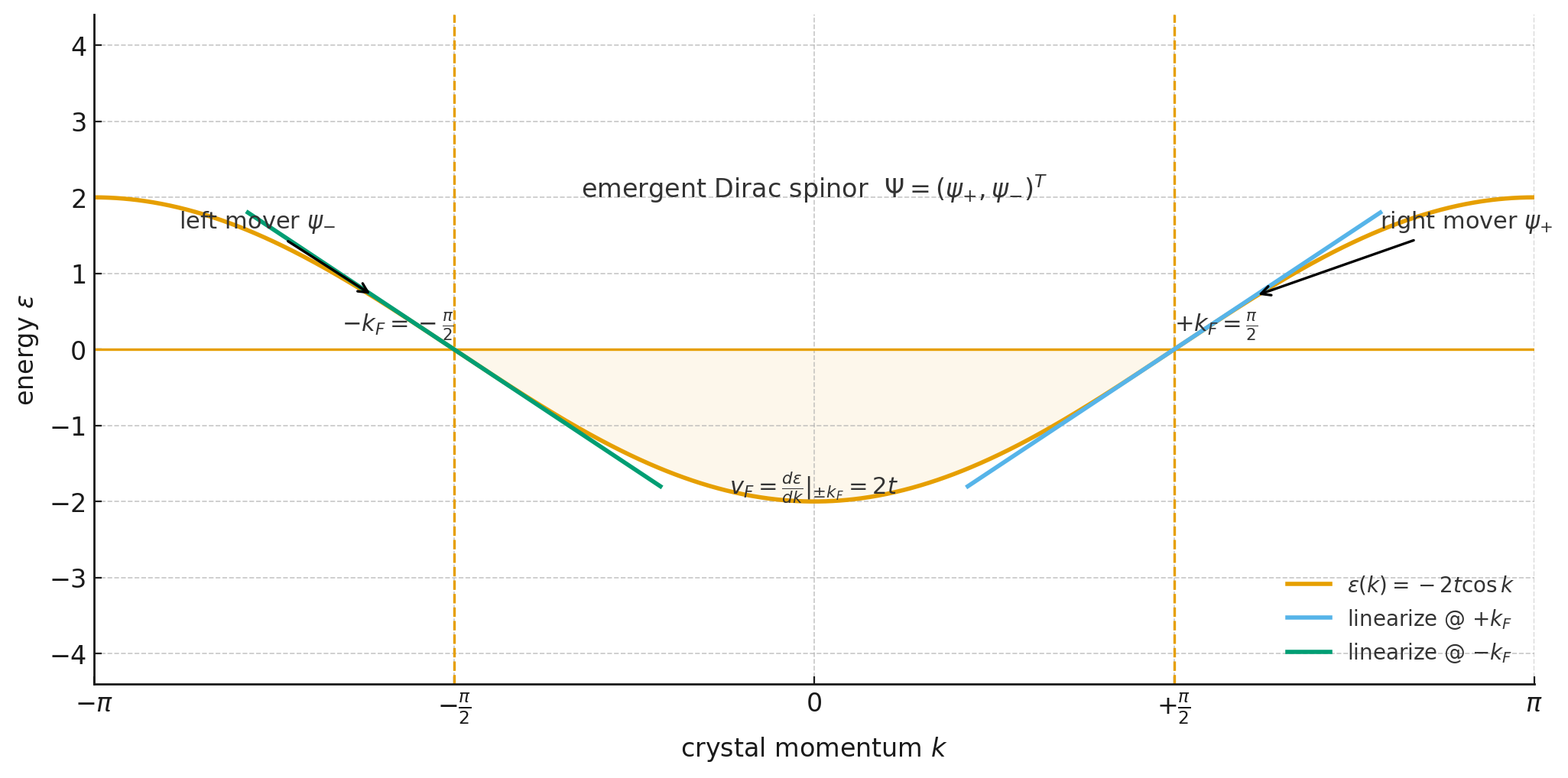}
  \caption{Emergent Dirac fermion from two nonrelativistic chiral patches. 
  The tight-binding dispersion $\varepsilon(k)=-2t\cos k$ at half filling has Fermi points at $k=\pm \pi/2$. 
  Linearizing near these points defines right/left movers $\psi_{\pm}$ with $v_F=2t$; packing $(\psi_{+},\psi_{-})^T$ yields the $1{+}1$D Dirac spinor.}
  \label{fig:dirac-from-fermions}
\end{figure}

Next we need the momentum states:
\begin{equation}\label{psij}
    \psi_j=\int_{-\pi}^\pi \frac{dk}{2\pi} \psi(k) e^{ikj}\,,\quad  \psi^\dagger_j=\int_{-\pi}^\pi \frac{dk}{2\pi} \psi^\dagger(k) e^{-ikj}\,.
\end{equation}
The inverses read:
\begin{equation}
    \psi(k)=\sum_j e^{-ikj}\psi_j\,,\quad \psi^\dagger(k)=\sum_j e^{ikj}\psi^\dagger_j\,.
\end{equation}
These equations imply
\begin{equation}
    \int_{-\pi}^\pi \frac{dk}{2\pi}e^{ik(j-j')}=\delta_{jj'}\,,\quad \sum_j e^{i(k-k')j}=2\pi \delta(k-k')\,.
\end{equation}
Then
\begin{eqnarray}
    H_0&\equiv& -\frac12\left(\sum_j \psi^\dagger_{j+1} \psi_j-\psi_{j+1}\psi^\dagger_j\right)\,,\\
    &=&-\frac12\sum_j\int_{-\pi}^\pi\int_{-\pi}^\pi \frac{dk}{2\pi}\frac{dk'}{2\pi}\left[\psi^\dagger(k)\psi(k')e^{-i(k-k')j}e^{-ik} +\psi^\dagger(k)\psi(k')e^{-i(k-k')j}e^{ik}\right]\,,\\
    &=&-\frac12\int_{-\pi}^\pi\int_{-\pi}^\pi \frac{dk}{2\pi}\frac{dk'}{2\pi}\left[\psi^\dagger(k)\psi(k')2\pi\delta(k-k')e^{-ik} +\psi^\dagger(k)\psi(k')2\pi \delta(k-k')e^{ik}\right]\,,\\
    &=&-\int_{-\pi}^\pi \frac{dk}{2\pi} \psi^\dagger(k)\psi(k)\cos(k)\,.
\end{eqnarray}
Now since $\cos(k)\approx 1-k^2/2$, this form is non-relativistic. Between $k\in(-\pi/2,\pi/2)$, $H_0$ will give negative eigenvalues. This is dealt with the concept of the Dirac sea where $|k|\leq k_F=\pi/2$, the negative energy states are filled. We get the concept of a Fermi surface, which in this case are the points $k=\pm \pi/2$. Near $k=\pi/2$, we have $-\cos k=-\pi/2+k$ while near $k=-\pi/2$, $-\cos k=-\pi/2-k$. We can discard the constant terms (or absorb them into the on-site quadratic piece, giving rise to a mass term). If we consider $\psi(k)$ to have support near the Fermi surface, then 
\begin{equation}\label{H0f}
    H_0\approx \int \frac{dp}{2\pi} \psi_+^\dagger(p) \, p\,  \psi_+(p)+\int \frac{dp}{2\pi} \psi_-^\dagger(p) \,(- p)\,  \psi_-(p)\,,
\end{equation}
where we have introduced the notation $\psi_\pm$ to indicate that these correspond to $\psi(k)$ with support near $k=\pm \pi/2$. The situation is depicted in fig.(\ref{fig:dirac-from-fermions}). So two ``patches'' of the nonrelativistic fermion together make up the relativistic Dirac fermion.

This form is identical to the Hamiltonian of a free relativistic Dirac fermion in 1+1D. We can rewrite eq.(\ref{psij}) as:
\begin{eqnarray}
    \psi_\pm(x)=\int_{-\infty}^\infty \frac{dp}{2\pi} \psi_\pm(k)e^{ipx} e^{-\frac{1}{2}\alpha|p|}\theta(\pm p)\,.
\end{eqnarray}
Here $\alpha>0$ is a regulator that also is reminder of the fact that we have support near the Fermi surface. The step function is needed since $p\sim \pm\pi/2$ is for $\psi_{\pm}$.
To make contact with relativistic expressions, we will need to send $\alpha\rightarrow 0$ as its presence with the $|p|$ factor breaks Lorentz invariance. The Dirac equation in question reads:

\begin{equation}
    i \partial_t \psi=H \psi
\end{equation}
where $H=\sigma_3 p+\sigma_2m$. For the massless case, $H=H_0$ (defined above) is obviously diagonal. We can call the upper component of $\psi$ as $\psi_+$ and lower one as $\psi_-$. It turns out that following the standard procedure of canonical quantization gives precisely the operator form in eq.(\ref{H0f}). $\psi_\pm(x,t)\propto e^{i p (x\mp t)}$ respectively.
We can now compute the equal time correlation function (expectation value in the ground state):
\begin{eqnarray}
    \langle \psi_+(x)\psi_+^\dagger(0)\rangle=\int_{-\infty}^\infty \frac{dp}{2\pi}e^{-\frac12\alpha|p|} \int_{-\infty}^\infty \frac{dq}{2\pi} e^{-\frac12 \alpha|q|} \langle \psi_+(p)\psi_+^\dagger(q)\rangle\,.
\end{eqnarray}
We compute the expectation value as follows:
\begin{eqnarray}
    \langle \psi_+(p)\psi_+^\dagger(q)\rangle&=&\sum_{j,j'} e^{iqj-ipj'}\langle \psi_{+j'}\psi_{+j}^\dagger\rangle\,,\\
    &=&\sum_{j,j'} e^{iqj-ipj'}\delta{jj'}=2\pi\delta(p-q)\theta(q)\,.
\end{eqnarray}
Inserting this inside the correlator we have
\begin{equation} \label{ferm1}
    \langle \psi_+(x)\psi_+^\dagger(0)\rangle=\frac{1}{2\pi(\alpha-i x)}\,.
\end{equation}
Similarly we find:
\begin{eqnarray} \label{ferm2}
     \langle \psi_\pm(x)\psi_\pm^\dagger(0)\rangle&=&\pm\frac{1}{2\pi(\pm\alpha-i x)}\,,\\
     \langle \psi^\dagger_\pm(0)\psi_\pm(x)\rangle&=&\mp\frac{1}{2\pi(\mp\alpha-i x)}\,.
\end{eqnarray}
The Dirac fermion (2 component) is
\begin{equation}
    \psi=\begin{pmatrix} \psi_+ \\ \psi_- \end{pmatrix}\,.
\end{equation}
Using this we note:
\begin{equation}
    j_0=\psi^\dagger\psi=\psi^\dagger_+\psi_++\psi_-^\dagger\psi_-\,,\quad j_1=\psi^\dagger \sigma_3 \psi=\psi^\dagger_+\psi_+-\psi_-^\dagger\psi_-\,,
\end{equation}
and 
\begin{equation}
    \bar\psi \psi=\psi^\dagger \sigma_2\psi=-i \psi_+^\dagger \psi_-+i \psi_-^\dagger \psi_+\,.
\end{equation}

\section{Bosonization}
The remarkable result in 1+1D is that there is an equivalence between certain fermionic theories and bosonic theories. This goes by the name of Bosonization. We can motivate this by studying and comparing correlation functions in the free fermionic theories and free bosonic theories. 
Let's consider the Hamiltonian for a massless scalar field in 1+1D given by
\begin{eqnarray}
    H_B=\frac12 \int dx \left(\Pi(x)^2+(\partial_x\phi)^2\right)\,.
\end{eqnarray}
Like in the fermionic case, we want the Hamiltonian to be of the form:
\begin{eqnarray}
    H_B=\int_{-\infty}^\infty \frac{dp}{2\pi} a^\dagger(p)a(p) |p|\,,
\end{eqnarray}
where 
\begin{eqnarray}
    [a(p),a^\dagger(p')]=2\pi \delta(p-p')\,.
\end{eqnarray}
These are obeyed if 
\begin{eqnarray}
    \phi(x)&=& \int_{-\infty}^\infty \frac{dp}{2\pi \sqrt{2|p|}}\left(a(p)e^{ipx}+a^\dagger(p)e^{-ipx}\right)\,,\\
    \Pi(x)&=& \int_{-\infty}^\infty \frac{dp |p|}{2\pi \sqrt{2|p|}}\left(-i a(p)e^{ipx}+i a^\dagger(p)e^{-ipx}\right)\,.
\end{eqnarray}
As in the fermionic case we will stick in an additional convergence factor of $e^{-\frac12 \alpha |p|}$ in each of these integrals. After this we introduce the analogs of the $\psi_\pm$:
\begin{eqnarray}
    \phi_\pm(x)=\pm \int_0^{\pm \infty} \frac{dp}{2\pi \sqrt{2|p|}}\left(e^{ipx}a(p)+e^{-ipx}a^\dagger(p)\right)e^{-\frac12 \alpha|p|}\,,
\end{eqnarray}
so that $\phi(x)=\phi_+(x)+\phi_-(x)$, which obey $[\phi_+(x),\phi_-(y)]=i/4$,
using which one can show 
\begin{eqnarray}
    G_\pm(x)&=&\langle \phi_\pm(x)\phi_\pm(0)-\phi^2_\pm(0)\rangle=\frac{1}{4\pi}\ln \frac{\alpha}{\alpha\mp i x}\,,\\
    G(x)&=&\langle \phi(x)\phi(0)-\phi^2(0)\rangle=\frac{1}{4\pi}\ln \frac{\alpha^2}{\alpha^2+x^2}\,.
\end{eqnarray}
The $\ln$ form makes it suggestive that if exponentials of $\phi$ were involved then one could presumably recover the fermion correlators. This is in fact what happens. First we note
\begin{eqnarray}
    G_\beta(x)\equiv \langle e^{i\beta\phi(x)}e^{-i\beta\phi(0)}\rangle=\left(\frac{\alpha^2}{\alpha^2+x^2}\right)^{\frac{\beta^2}{4\pi}}\,.
\end{eqnarray}
This is proved as follows:
\subsection*{BCH/normal-ordering proof of \texorpdfstring{$e^{A}e^{B}$}{e^A e^B} identity}

\paragraph{Setup.}
Let $A$ and $B$ be operators \emph{linear} in free-boson creation/annihilation modes.
Decompose
\[
A=A^{(+)}+A^{(-)},\qquad B=B^{(+)}+B^{(-)},
\]
where $A^{(+)}$ ($B^{(+)}$) contains only creation operators and $A^{(-)}$ ($B^{(-)}$) only annihilation operators, so that
\(
[A^{(+)},B^{(+)}]=[A^{(-)},B^{(-)}]=0
\)
and all commutators that appear below are $c$-numbers.
Normal ordering is defined by moving all $(+)$ parts to the left of all $(-)$ parts:
\[
: e^{A} : \;\equiv\; e^{A^{(+)}} e^{A^{(-)}} ,\qquad
: e^{B} : \;\equiv\; e^{B^{(+)}} e^{B^{(-)}} .
\]
Define the \emph{contraction} (a $c$-number) by either equivalent expression
\[
\langle A B\rangle \;\equiv\; [A^{(-)},B^{(+)}] \;=\; \langle 0| A B |0\rangle .
\]

\paragraph{BCH lemma for central commutators.}
If $[X,Y]$ is a $c$-number, then
\begin{equation}
e^{X+Y}=e^{X}e^{Y}e^{-\frac12[X,Y]},\qquad
e^{X}e^{Y}=e^{Y}e^{X}e^{[X,Y]} .
\label{eq:BCH-central}
\end{equation}

\paragraph{Lemma 1 (single exponential).}
\[
e^{A} \;=\; :e^{A}:\; \exp\!\Big(\tfrac12\langle A A\rangle\Big).
\]
\emph{Proof.} Apply \eqref{eq:BCH-central} with $X=A^{(+)}$, $Y=A^{(-)}$:
\[
e^{A}=e^{A^{(+)}+A^{(-)}}=e^{A^{(+)}}e^{A^{(-)}}e^{-\frac12[A^{(+)},A^{(-)}]}
= :e^{A}:\; e^{+\frac12\langle A A\rangle},
\]
since $[A^{(+)},A^{(-)}]=-\,\langle A A\rangle$. \hfill$\square$

\paragraph{Lemma 2 (product of normal-ordered exponentials).}
\[
: e^{A} :\; : e^{B} : \;=\; : e^{A+B} :\; \exp\!\big(\langle A B\rangle\big).
\]
\emph{Proof.} Using the definitions and \eqref{eq:BCH-central},
\[
: e^{A} :\; : e^{B} :
= e^{A^{(+)}} e^{A^{(-)}} e^{B^{(+)}} e^{B^{(-)}}
= e^{A^{(+)}} e^{B^{(+)}} e^{A^{(-)}} e^{B^{(-)}} e^{[A^{(-)},B^{(+)}]}
= : e^{A+B} :\; e^{\langle A B\rangle}.
\]
\hfill$\square$

\paragraph{Theorem}\footnote{This is set as an exercise in \cite{shankar}; here, we prove it for you!}
\begin{equation}
\boxed{%
e^{A} e^{B}
= :e^{A+B}:\; \exp\!\Big(\langle A B\rangle+\tfrac12\langle A^2\rangle+\tfrac12\langle B^2\rangle\Big)}
\label{eq:ShankarIdentity}
\end{equation}
\emph{Proof.} Combine Lemma~1 on each factor with Lemma~2:
\[
\begin{aligned}
e^{A}e^{B}
&=\big(:e^{A}:\, e^{\tfrac12\langle A^2\rangle}\big)\,
  \big(:e^{B}:\, e^{\tfrac12\langle B^2\rangle}\big) \\
&= \big(:e^{A}:\, :e^{B}:\big)\,
   \exp\!\Big(\tfrac12\langle A^2\rangle+\tfrac12\langle B^2\rangle\Big) \\
&= :e^{A+B}:\, \exp\!\Big(\langle A B\rangle+\tfrac12\langle A^2\rangle+\tfrac12\langle B^2\rangle\Big).
\end{aligned}
\]
\hfill$\square$

\paragraph{Corollary (Gaussian identity for vertex operators).}
Taking $A=i\beta\,\phi(x)$ and $B=-i\beta\,\phi(0)$ for a free boson (so all contractions are $c$-numbers),
\[
\big\langle e^{\,i\beta\phi(x)} e^{-\,i\beta\phi(0)} \big\rangle
= \exp\!\Big(-\tfrac{\beta^2}{2}\,\langle\big(\phi(x)-\phi(0)\big)^2\rangle\Big),
\]
which is the Gaussian identity used in bosonization.

We similarly have
\begin{eqnarray}
    G^\pm_\beta(x)\equiv \langle e^{i\beta\phi_\pm(x)}e^{-i\beta\phi_\pm(0)}\rangle=\left(\frac{\alpha}{\alpha\mp i x}\right)^{\frac{\beta^2}{4\pi}}\,.
\end{eqnarray}

This finally leads to the Bosonization dictionary:
\begin{eqnarray}
    \psi_\pm(x)=\frac{1}{\sqrt{2\pi\alpha}} e^{\pm i \sqrt{4\pi}\phi_\pm(x)}\,,
\end{eqnarray}
where it is to be understood that this dictionary {\it only} holds inside correlation functions where the cutoffs on the two sides ($\alpha$) are exactly the same and correlators are calculated in the Fermi vacuum and bosonic vacuum in the corresponding cases. 

Using bosonization, we can reproduce eq.(\ref{ferm1}) and eq.(\ref{ferm2}) for instance. The most useful applications in our case rely on some identities, which we prove below.
\begin{eqnarray}
    \bar\psi\psi&=&-i \psi^\dagger_+\psi_-+i \psi^\dagger_- \psi_+\,,\\
    &=& -\frac{1}{\pi\alpha}\cos(\sqrt{4\pi}\phi)\,,
\end{eqnarray}
where we have used the BCH lemma for $[A,[A,B]]=[B,[A,B]]=0$: $e^{A+B}=e^A e^B e^{-\frac12 [A,B]}$. Similarly we note that $\bar\psi i\sigma_3\psi=\sin\sqrt{4\pi}\phi/(\pi\alpha)$. There are two more important identities that follow from bosonization and careful treatment of coincidence points (point splitting). The first one is:
\begin{equation}
    :\psi_\pm^\dagger(x)\psi_+(x):=\frac{1}{\sqrt{\pi}}\partial_x \phi_\pm\,,
\end{equation}
using which we have
\begin{eqnarray}
    j_\mu=\frac{\epsilon_\mu^{\ \ \nu}}{\sqrt{\pi}}\partial_\nu\phi
\end{eqnarray}
where $\epsilon_{\mu\nu}$ is the 2D Levi-ci-vita tensor. Explicitly we have $j_0=\partial_x\phi/\sqrt{\pi}$ and $j_1=\partial_x(\phi_+-\phi_-)/\sqrt{\pi}=-\Pi/\sqrt{\pi}$. We further have using point-splitting
\begin{eqnarray}
    (\bar\psi\psi)^2&=&-\frac{1}{\pi}(\partial_x\phi)^2+\cdots\,,\\
    (\bar\psi i\sigma_3\psi)^2&=& -\frac{1}{\pi}(\partial_x\phi)^2+\cdots\,,
\end{eqnarray}
where in the $\cdots$ we have omitted non-renormalizable terms. We focused on displaying the results for $t=0$. The result for general $t$ is to replace $(\partial_x\phi)^2\rightarrow (\partial_x\phi)^2+(\partial_t\phi)^2$ in Euclidean space. The fermion kinetic term maps to the bosonic kinetic term \footnote{This is set as an exercise in \cite{shankar}.}. To see this we proceed as follows:


\subsection*{Bosonizing the chiral kinetic term (symmetric point-splitting)}

We work with chiral spinless fermions $\psi_{+}(x)$ (right-mover) and $\psi_{-}(x)$ (left-mover).
Define the \emph{symmetrized} kinetic-density operator for the $+$ chirality by point-splitting:
\begin{equation}
\label{eq:Hplus-def}
\mathcal{H}_{+}(x)
\equiv
-\frac{i}{2}\lim_{\varepsilon\to 0}
\frac{\psi_{+}^\dagger\!\left(x+\tfrac{\varepsilon}{2}\right)\psi_{+}\!\left(x-\tfrac{\varepsilon}{2}\right)
      -\psi_{+}^\dagger\!\left(x-\tfrac{\varepsilon}{2}\right)\psi_{+}\!\left(x+\tfrac{\varepsilon}{2}\right)}{\varepsilon},
\end{equation}
with normal-ordering understood (vacuum piece subtracted).

Use the chiral-boson representation:
\begin{equation}
\psi_{+}(x)=\frac{1}{\sqrt{2\pi\alpha}}\,\!e^{\,i\sqrt{4\pi}\,\phi_{+}(x)}\!\,,\qquad
\langle \phi_{+}(x)\phi_{+}(0)\rangle=-\frac{1}{4\pi}\ln(\alpha - i x),
\end{equation}
and the normal-ordering/BCH identity (proved earlier)
\[
e^{A}e^{B} = :e^{A+B}:\exp\!\Big(\langle AB\rangle + \tfrac12\langle A^2\rangle + \tfrac12\langle B^2\rangle\Big).
\]
With $A=-i\sqrt{4\pi}\,\phi_{+}(x+\tfrac{\varepsilon}{2})$ and $B=+i\sqrt{4\pi}\,\phi_{+}(x-\tfrac{\varepsilon}{2})$ this gives
\begin{align}
\psi_{+}^\dagger\!\left(x+\tfrac{\varepsilon}{2}\right)\psi_{+}\!\left(x-\tfrac{\varepsilon}{2}\right)
&= \frac{1}{2\pi}\,\frac{1}{\alpha - i\varepsilon}\;
:\!\exp\!\big(-i\sqrt{4\pi}\,\Delta_{+}(x;\varepsilon)\big)\!:\,,\\
\psi_{+}^\dagger\!\left(x-\tfrac{\varepsilon}{2}\right)\psi_{+}\!\left(x+\tfrac{\varepsilon}{2}\right)
&= \frac{1}{2\pi}\,\frac{1}{\alpha + i\varepsilon}\;
:\!\exp\!\big(+i\sqrt{4\pi}\,\Delta_{+}(x;\varepsilon)\big)\!:\,,
\end{align}
where the symmetric difference is
\begin{equation}
\Delta_{+}(x;\varepsilon)
\equiv \phi_{+}\!\left(x+\tfrac{\varepsilon}{2}\right)-\phi_{+}\!\left(x-\tfrac{\varepsilon}{2}\right)
= \varepsilon\,\partial_x\phi_{+}(x)+\frac{\varepsilon^{3}}{24}\,\partial_x^{3}\phi_{+}(x)+\mathcal{O}(\varepsilon^{5}).
\end{equation}

\paragraph{Expansion to quadratic order in $\varepsilon$.}
Since $\Delta_{+}=\mathcal{O}(\varepsilon)$, expand
\[
:\!e^{\mp i\sqrt{4\pi}\,\Delta_{+}}\!:\;=\;1 \mp i\sqrt{4\pi}\,\Delta_{+} - 2\pi\,\Delta_{+}^{\,2} + \mathcal{O}(\varepsilon^{3}),
\qquad
\frac{1}{\alpha \mp i\varepsilon}=\frac{\alpha \pm i\varepsilon}{\alpha^{2}+\varepsilon^{2}}.
\]
Insert these into \eqref{eq:Hplus-def}. The $c$-number vacuum term and the total-derivative term $\propto \partial_x\phi_{+}$ drop (normal ordering and boundary conditions). The finite operator comes from the $\Delta_{+}^{\,2}$ piece:
\begin{equation}
\mathcal{H}_{+}(x)
= \frac{1}{4\pi}\,:\!(\partial_x\phi_{+})^{2}\!:\,.
\label{eq:Hplus-final}
\end{equation}

\paragraph{Left mover ($-$ chirality).}
Repeating the same steps for $\psi_{-}$, with
\(
\psi_{-}(x)=\frac{1}{\sqrt{2\pi\alpha}}\,:\!e^{-\,i\sqrt{4\pi}\,\phi_{-}(x)}\!:
\)
and
\(
\langle \phi_{-}(x)\phi_{-}(0)\rangle=-\frac{1}{4\pi}\ln(\alpha + i x),
\)
one finds
\begin{equation}
\mathcal{H}_{-}(x)=\frac{1}{4\pi}\,:\!(\partial_x\phi_{-})^{2}\!:\,.
\label{eq:Hminus-final}
\end{equation}

\paragraph{Full kinetic Hamiltonian.}
\begin{equation}
\boxed{%
- i \!\int\!dx\;\Big(:\psi_{+}^\dagger\partial_x\psi_{+}:\;-\;:\psi_{-}^\dagger\partial_x\psi_{-}:\Big)
\;=\;\frac{1}{4\pi}\!\int\!dx\;\Big[ :(\partial_x\phi_{+})^{2}:\;+\;:(\partial_x\phi_{-})^{2}:\Big]. }
\end{equation}
Equivalently, in nonchiral fields $\phi=\phi_{+}+\phi_{-}$ and $\theta=\phi_{+}-\phi_{-}$,
\[
\mathcal{H}_{0}=\frac{1}{2\pi}\int dx\,\big[(\partial_x\phi)^{2}+(\partial_x\theta)^{2}\big]=\frac{1}{2\pi}\int dx\,\big[(\partial_x\phi)^{2}+ \Pi^{2}\big].
\]

\section{Thirring model bosonized}

Using the identities derived above, we find
\begin{eqnarray}
    L_{TH}=\bar\psi (i\slashed{\partial}-m)\psi-\frac{g}{2} j_\mu j^\mu\,,
\end{eqnarray}
bosonizes to 
\begin{eqnarray}
  \boxed{  L^B_{TH}=\frac{1}{2}(1+\frac{g}{\pi})(\partial\phi)^2-\frac{m}{\pi\alpha}\cos\sqrt{4\pi}\phi\,.}
\end{eqnarray}
This model is the Sine-Gordon model. 
Note that if the fermion was massless, we get a free theory in the bosonized version!

\section{Schwinger model bosonized}


\paragraph{Starting point.}
With metric $\eta_{\mu\nu}=\mathrm{diag}(+,-)$ and Levi–Civita $\epsilon^{01}=+1$, take
\begin{equation}
\mathcal L_{\text{Sch}}=
-\frac{1}{4}F_{\mu\nu}F^{\mu\nu}
+\bar\psi\,(i\slashed{\partial}-e\slashed{A}-m)\psi,
\qquad
j^\mu\equiv \bar\psi\gamma^\mu\psi.
\end{equation}

\paragraph{Bosonization dictionary.}
For the scalar $\phi$ (with velocity set to $1$),
\begin{equation}
:\bar\psi\,i\slashed{\partial}\psi:\;\longleftrightarrow\;\tfrac12(\partial_\mu\phi)^2,
\qquad
j^\mu\;\longleftrightarrow\;\frac{1}{\sqrt{\pi}}\,\epsilon^{\mu\nu}\partial_\nu\phi,
\qquad
:\bar\psi\psi:\;\longleftrightarrow\;-\frac{1}{\pi\alpha}\cos\!\big(\sqrt{4\pi}\,\phi\big),
\end{equation}
where $\alpha$ is a short-distance cutoff. (This is the same normalization you used for the Thirring mapping.)

\paragraph{Gauge coupling in bosonic variables.}
The minimal coupling term becomes
\begin{equation}
e\,A_\mu j^\mu\;\longleftrightarrow\;\frac{e}{\sqrt{\pi}}\,\epsilon^{\mu\nu}A_\mu\,\partial_\nu\phi
=\frac{e}{\sqrt{\pi}}\big(A_0\partial_x\phi-A_1\partial_t\phi\big)
= \,\frac{e}{\sqrt{\pi}}\,\phi\,E \;+\; \partial_\mu(\cdots),
\end{equation}
after an integration by parts, where $E\equiv F_{01}=\partial_t A_1-\partial_x A_0$ is the electric field in $1{+}1$\,D.

The Maxwell term in this signature is
\begin{equation}
-\frac{1}{4}F_{\mu\nu}F^{\mu\nu}
=+\frac{1}{2}E^2.
\end{equation}

\paragraph{Integrating out the gauge field (completing the square).}
The $A_\mu$-dependent part of the Lagrangian is
\begin{equation}
\mathcal L_A \;=\; \frac{1}{2}E^2 \;-\; \frac{e}{\sqrt{\pi}}\,\phi\,E
= \frac{1}{2}\Big(E-\frac{e}{\sqrt{\pi}}\phi\Big)^2 \;-\; \frac{1}{2}\,\frac{e^2}{\pi}\,\phi^2.
\end{equation}
Since $E$ is nondynamical in $1{+}1$\,D, integrating it out sets $E=\tfrac{e}{\sqrt{\pi}}\phi$ and drops the perfect square, leaving the local mass term
\(
-\tfrac{1}{2}\,\tfrac{e^2}{\pi}\,\phi^2.
\)

\paragraph{Result (massive Schwinger model).}
Putting everything together,
\begin{equation}
\boxed{%
\mathcal L_{\text{Sch}}^{\;B}
=\frac{1}{2}(\partial_\mu\phi)^2
-\frac{1}{2}\,\frac{e^2}{\pi}\,\phi^2
-\frac{m}{\pi\alpha}\,\cos\!\big(\sqrt{4\pi}\,\phi\big)\,.
}
\end{equation}
In the massless limit $m\to 0$ this is a free massive scalar with
\(
\mu^2=\tfrac{e^2}{\pi}
\)
(the standard Schwinger mass).

\paragraph{ \texorpdfstring{$\theta$}{theta}-angle.}
Including a topological term
\(
\mathcal L_\theta=\frac{e\,\theta}{2\pi}\,\epsilon^{\mu\nu}F_{\mu\nu}
=\frac{e\,\theta}{\pi}\,E
\)
shifts the square as
\(
E-\frac{e}{\sqrt{\pi}}\phi \;\to\; E-\frac{e}{\sqrt{\pi}}\phi-\frac{e\,\theta}{\pi},
\)
which is equivalent to shifting the cosine:
\begin{equation}
\mathcal L_{\text{Sch}}^{\;B}(\theta)
=\frac{1}{2}(\partial\phi)^2
-\frac{1}{2}\,\frac{e^2}{\pi}\,\phi^2
-\frac{m}{\pi\alpha}\,\cos\!\big(\sqrt{4\pi}\,\phi-\theta\big).
\end{equation}
The $\theta=\pi$ is called the deconfined phase (where the degrees of freedom are deconfined quarks and antiquarks---here quark and antiquark refer to the fundamental fermion exciations) while $\theta=0$ is the confined phase (where the degrees of freedom are quark-antiquark bound states, i.e., mesons). We elaborate on the slightly confusing ``confinement''/``deconfinement'' terminology below.

\subsection{Phases of the theory}

\begin{figure}[t]
\centering
\begin{tikzpicture}[>=stealth,scale=1.1]

\draw[->] (0,0) -- (6.5,0) node[right] {$\theta$};
\draw[->] (0,0) -- (0,3.8) node[above] {$m/g$};

\draw (0,0) -- (0,-0.08) node[below] {$0$};
\draw (3,0) -- (3,-0.08) node[below] {$\pi$};
\draw (6,0) -- (6,-0.08) node[below] {$2\pi$};

\draw (0,1.6) -- (-0.08,1.6) node[left] {$m_c/g \approx 0.33$};

\draw[thick,blue] (3,1.6) -- (3,3.8);
\node[rotate=90] at (3.35,3.0) {\small 1st-order};

\fill[red] (3,1.6) circle (1.8pt);
\node[above left] at (3,1.6) {\small Ising critical point};

\node[align=center] at (3,0.8)
  {\small no phase transition as a function of $\theta$\\[-1mm]
   \small for $m/g < m_c/g$};


\end{tikzpicture}
\caption{Schematic phase diagram of the Schwinger model in the $m/g$--$\theta$ plane \cite{angel}.
Physics is periodic in $\theta$ with period $2\pi$, so only one period is shown.
A line of first-order transitions occurs at $\theta=\pi$ for $m/g > m_c/g \approx 0.33$,
ending at a second-order critical point at $m_c/g$. Below $m_c/g$ there is no
phase transition as $\theta$ is varied; throughout the phase diagram the Schwinger model has a massive gauge boson
and exhibits screening/confinement of static charges; there is no Coulomb
(massless-photon) phase.  The only phase transition separates a
$\mathsf{CP}$-symmetric confining regime from a $\mathsf{CP}$-violating
confining regime near $\theta=\pi$.} \label{schphases}
\end{figure}
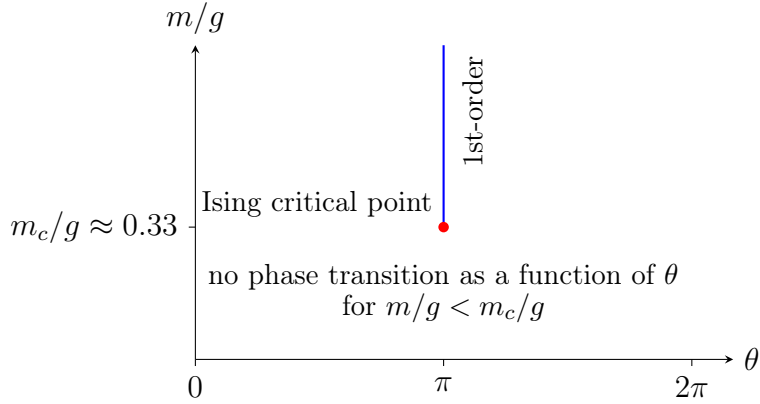

In our conventions the massive Schwinger model is defined in $(1+1)$
dimensions by
\[
\mathcal{L}_{\rm Sch} \;=\; -\frac14 F_{\mu\nu}F^{\mu\nu}
\;+\; \bar\psi\bigl(i\gamma^\mu D_\mu - m\bigr)\psi
\;+\; \frac{e\,\theta}{2\pi}\,\epsilon^{\mu\nu}\partial_\mu A_\nu\,,
\qquad
D_\mu = \partial_\mu - i e A_\mu\,,
\]
with metric $\eta_{\mu\nu} = \mathrm{diag}(+,-)$ and
$\epsilon^{01}=+1$.  For $\theta\neq\pi$ the ground state is unique.
Fractionally charged external probes experience a linear potential,
while integer charges are completely screened by dynamical pairs; in
either case the physical spectrum consists of a single massive
pseudoscalar “meson” (the Schwinger boson).  In the chiral limit its
mass is $M_\gamma = e/\sqrt{\pi}$ and it remains nonzero for $m\neq
0$, so there is no phase transition as a function of $m/e$ away from
$\theta\simeq\pi$.  The interesting phase structure arises near
$\theta=\pi$, where the theory has a $\mathsf{CP}$ symmetry.  For
small $m/e$ this $\mathsf{CP}$ symmetry is unbroken and there is again
a unique vacuum, but for sufficiently large $m/e$ the ground state
enters the Dashen phase and becomes doubly degenerate with opposite
signs of the electric field and of the pseudoscalar condensate
$\langle\bar\psi i\gamma^5\psi\rangle$, signalling spontaneous
$\mathsf{CP}$ violation.  These two regimes are separated by a phase
transition line in the $(m/e,\theta)$ plane, expected to terminate at
an Ising-type critical point near $\theta=\pi$ \cite{Jentsch:2021trr}, so that the phase
diagram---see fig.~\ref{schphases}---consists, in the static-charge
sense, of a single confining/screening phase with either unbroken or
spontaneously broken $\mathsf{CP}$ symmetry rather than a Coulomb
(deconfined) phase.  In the $\mathsf{CP}$-violating region near
$\theta=\pi$ the spectrum contains kink excitations that behave as
“half-asymptotic” charged particles; following Coleman \cite{Coleman1976} and
Shankar--Murthy \cite{shankarmurthy} this regime is sometimes loosely described as
deconfined, but the gauge boson remains massive and static
fractional charges are still linearly confined.

\subsection{Charges}

\paragraph{Charge in the Thirring (and SG) description.}
With our bosonization conventions for the massive Thirring model,
\[
j^\mu=\bar\psi\gamma^\mu\psi\quad\longleftrightarrow\quad
\frac{1}{\sqrt{\pi}}\,\epsilon^{\mu\nu}\partial_\nu\phi,
\]
so the U(1) (fermion-number) charge in an interval $[x_1,x_2]$ is
\[
Q([x_1,x_2])=\int_{x_1}^{x_2}\!dx\,j^0(x)
=\frac{1}{\sqrt{\pi}}\big[\phi(x_2)-\phi(x_1)\big].
\]
For localized configurations this gives the total charge
\(
Q=\frac{1}{\sqrt{\pi}}\,[\phi(+\infty)-\phi(-\infty)]\in\mathbb{Z},
\)
so an elementary fermion corresponds to a kink with $\Delta\phi=\sqrt{\pi}$, while neutral excitations (e.g.\ breathers in the SG language) have $Q=0$.
After rescaling to the canonically normalized SG field $\varphi=\sqrt{1+\tfrac{g}{\pi}}\,\phi$ with cosine $\cos(\beta\varphi)$ and
\(
\beta^2=\frac{4\pi}{1+g/\pi},
\)
the current reads
\[
j^\mu\;\longleftrightarrow\;\frac{\beta}{2\pi}\,\epsilon^{\mu\nu}\partial_\nu\varphi,
\quad\Rightarrow\quad
Q=\frac{\beta}{2\pi}\,[\varphi(+\infty)-\varphi(-\infty)].
\]
Equivalently, the vacuum minima are spaced by $\Delta\varphi=\tfrac{2\pi}{\beta}$, and a topological soliton (the Thirring fermion) carries $Q=\pm1$ corresponding to $\Delta\varphi=\pm\tfrac{2\pi}{\beta}$ (or $\Delta\phi=\pm\sqrt{\pi}$).

\paragraph{Charge in the bosonized Schwinger model.}
The discussion here is similar. With our conventions
\(j^\mu=\frac{1}{\sqrt{\pi}}\epsilon^{\mu\nu}\partial_\nu\phi\)
(and compactification \(\phi\sim\phi+\sqrt{\pi}\) since \(\beta=\sqrt{4\pi}\)),
the dynamical U(1) charge in an interval \([x_1,x_2]\) is
\[
Q_{\rm dyn}([x_1,x_2])=\int_{x_1}^{x_2}\!dx\,j^0(x)
=\frac{1}{\sqrt{\pi}}\big[\phi(x_2)-\phi(x_1)\big].
\]
For localized configurations this gives the total dynamical charge
\(
Q_{\rm dyn}=\frac{1}{\sqrt{\pi}}\big[\phi(+\infty)-\phi(-\infty)\big]\in\mathbb{Z},
\)
so an elementary fermion corresponds to a kink with
\(\Delta\phi=\sqrt{\pi}\).
Static external sources \(\rho_{\rm ext}(x)=\sum_a Q_a\,\delta(x-x_a)\)
enter Gauss’ law as
\(
\partial_x\!\big(E-\tfrac{e}{\sqrt{\pi}}\phi\big)=e\,\rho_{\rm ext},
\)
implying that across a point charge \(Q_a\) the field steps by
\(\Delta\phi=\sqrt{\pi}Q_a\).
Equivalently, the physical charge enclosed in \([x_1,x_2]\) is
\[
Q_{\rm phys}([x_1,x_2])=Q_{\rm dyn}([x_1,x_2])+\!\int_{x_1}^{x_2}\!dx\,\rho_{\rm ext}(x)
=\frac{1}{e}\,\big[E(x_2)-E(x_1)\big],
\]
so the jump of \(\phi\) (hence of \(E=\tfrac{e}{\sqrt{\pi}}\phi\) at \(\theta=0\))
directly measures the enclosed charge.

\section{Scattering}

\subsection*{Massive Thirring: what is being scattered and how to benchmark}

Via Coleman’s mapping, the massive Thirring model
\[
\mathcal L_{TH}=\bar\psi(i\slashed{\partial}-m)\psi-\frac{g}{2}\,j_\mu j^\mu
\]
is equivalent to sine–Gordon (SG) after a field rescaling, with coupling
\[
\beta^{2}=\frac{4\pi}{1+g/\pi},\qquad
\xi\equiv\frac{\beta^{2}}{8\pi-\beta^{2}}=\frac{1}{1+2g/\pi}.
\]
In this correspondence the Thirring fermion carries the SG topological charge (soliton), while the neutral single-particle excitations in the $Q{=}0$ sector are the SG breathers. The lightest such state $B_{1}$ plays the role of the “scalar particle” one would scatter in analogy with the $\phi^{4}$ case. In the attractive SG regime ($\beta^{2}<8\pi$, equivalently $g>-\pi/2$) the breather tower has masses
\[
m_{n}=2M\,\sin\!\Big(\frac{n\pi\xi}{2}\Big),\qquad n=1,2,\dots,\Big\lfloor \tfrac1{\xi}\Big\rfloor,
\]
where $M$ is the soliton mass and $m_{B_1}\equiv m_1=2M\sin(\pi\xi/2)$.

The benchmark for $2\to2$ scattering in this neutral sector is the exactly known elastic $B_{1}$–$B_{1}$ S–matrix,
\[
S_{B_1B_1}(\theta_R)
= \frac{\sinh\theta_R+i\sin(\pi\xi)}{\sinh\theta_R-i\sin(\pi\xi)}
= e^{2i\delta_{B_1B_1}(\theta)},
\qquad
\delta_{B_1B_1}(\theta_R)=\arctan\!\frac{\sin(\pi\xi)}{\sinh\theta_R},
\]
with rapidity difference $\theta_R$. 

\paragraph{Rapidity difference in $S_{B_1B_1}(\theta_R)$.}
Here we use the notation $\theta_R$ to emphasise that it is \emph{not} the topological $\theta$–angle. In $1{+}1$ dimensions each on–shell momentum is parametrized by a rapidity $\vartheta$,
\[
p^\mu=(E,p)=(m\cosh\vartheta,\;m\sinh\vartheta).
\]
For two identical particles (here the first breather $B_1$) Lorentz invariance implies that the $2\!\to\!2$ S–matrix depends only on the rapidity \emph{difference}
\[
\theta \equiv \vartheta_1-\vartheta_2.
\]
It fixes the center–of–mass energy via
\[
s=(p_1+p_2)^2=2m^2\big(1+\cosh\theta_R\big)=4m^2\cosh^2\!\frac{\theta_R}{2},
\]
so in the COM frame $k=m\sinh(\theta_R/2)$ and $E_{\rm pair}=2m\cosh(\theta_R/2)$. The elastic amplitude is then written as
\[
S_{B_1B_1}(\theta_R)=e^{2i\delta(\theta_R)},
\]
with $\delta(\theta_R)$ the phase shift as a function of the rapidity difference.

In practice one prepares two $B_{1}$ wavepackets and extracts the phase shift from the asymptotic time delay, or—more conveniently on a lattice/in finite volume—from two-body energy levels. In $1{+}1$ dimensions the finite-volume quantization in the center-of-mass frame reads
\[
k L + 2\,\delta(k)=2\pi n,\qquad
E=2\sqrt{m_{B_1}^{2}+k^{2}},\qquad n\in\mathbb{Z},
\]
so measuring $E(L)$ gives $k(L)$ and hence $\delta(k)$. Agreement with the closed form above as a function of $\theta=2\,\mathrm{arcsinh}(k/m_{B_1})$ and the coupling (via $\xi$) is the nonperturbative benchmark for your pipeline. This test is directly comparable to the neutral-scalar scattering we perform in $\phi^{4}$, with the added bonus that here the exact answer is known for all energies (below inelastic thresholds to higher breathers, if present).

\subsection*{Massive Schwinger: what is being scattered and how to benchmark}

In our conventions the massive Schwinger model,
\[
\mathcal L_{\rm Sch}=-\frac14 F_{\mu\nu}F^{\mu\nu}+\bar\psi(i\slashed{\partial}+e\slashed A-m)\psi,
\]
bosonizes to
\[
\mathcal L_{\rm Sch}^{\,B}=\frac12(\partial\phi)^{2}-\frac12\,\frac{e^{2}}{\pi}\,\phi^{2}-\frac{m}{\pi\alpha}\cos\!\big(\sqrt{4\pi}\,\phi-\theta\big).
\]
At $m=0$ the theory is a free neutral boson of mass $\mu=e/\sqrt{\pi}$, so two-particle scattering is trivial ($S\equiv 1$). This provides a clean “null test” of your scattering extraction (wavepacket collisions should show no time delay; finite-volume levels should follow the free quantization condition). At $e=0$ the model reduces to pure sine–Gordon, so we recover integrability and can reuse the $B_{1}$–$B_{1}$ benchmark described above, with the breather spectrum and S–matrix determined solely by the cosine coupling.

For generic $(e,m)$ the bosonic theory is SG deformed by a quadratic mass term and is not integrable. The appropriate “single particle” to scatter is the lightest neutral excitation (the Schwinger boson continuously deformed by $m\neq 0$). At low energies and small $m$ one may expand the cosine to read off the induced mass shift and an effective quartic coupling for $\phi$, which yields a controlled perturbative prediction for the phase shift $\delta(k)$ in the Born/one-loop approximation. This furnishes a quantitative benchmark curve near threshold. As the energy increases, inelastic channels can open (production of heavier excitations if present), and the S–matrix ceases to be purely elastic; in finite volume this shows up as avoided crossings and volume dependence beyond the simple quantization condition $kL+2\delta(k)=2\pi n$. Nonperturbative benchmarks are then provided by comparing our extracted masses and phase-shift trends against established continuum methods (e.g. TCSA/DMRG/FRG studies of the massive Schwinger model), while retaining the two exactly solvable edges, $m=0$ (free) and $e=0$ (integrable SG), as calibration lines.

In summary, in the Thirring case one should interpret the neutral “single particle’’ as the first breather $B_{1}$ and benchmark the full energy-dependent phase shift against the exact $S_{B_1B_1}(\theta)$. In the Schwinger case the single particle is the lightest neutral boson; one benchmarks first against the two integrable/free limits and then against perturbation theory at low energy and nonperturbative numerical results away from those limits, using the same finite-volume phase-shift extraction workflow as in $\phi^{4}$.

\subsection{Adiabatic preparation of mesons in the confined Schwinger model}

In the bosonized description at $\theta=0$ the massive Schwinger model reads
\begin{equation}
    \mathcal L = \frac12 (\partial_\mu\phi)^2 - \frac12\,\mu^2 \phi^2
    - \kappa m_f \cos\!\big(\sqrt{4\pi}\,\phi\big),
    \qquad \mu^2 = \frac{e^2}{\pi}\,,
\end{equation}
where $m_f$ is the bare fermion mass and $\mu=e/\sqrt{\pi}$ is the Schwinger mass. At $m_f=0$ the theory is just a free massive boson of mass $\mu$, while for $m_f\neq0$ the cosine term confines the underlying fermions into neutral ``mesons''. In the confined regime the lightest excitations are gapped, stable mesons; they can be viewed as deformations of the free boson particle at $m_f=0$.

On the lattice we take the bosonized Hamiltonian in momentum space,
\begin{equation}
    H_{\text{free}} = \sum_p \omega_p\, a_p^\dagger a_p + E_0^{\text{free}}, \qquad
    \omega_p^2 = \mu^2 + \hat p^{\,2}, \quad
    \hat p \equiv \frac{2}{a}\sin\frac{p a}{2},
\end{equation}
with vacuum $|0\rangle$ and one–particle states $|1_p\rangle = a_p^\dagger|0\rangle$. We now switch on the cosine interaction via an adiabatic path
\begin{equation}
    H(s) = H_{\text{free}} + s\,V_{\cos}, \qquad
    V_{\cos} = -\,\kappa m_f \sum_j a\,\cos\!\big(\sqrt{4\pi}\,\phi_j\big), \qquad s\in[0,1].
\end{equation}
The parameter $s$ plays the role of ``time'' in parameter space: at $s=0$ we have the free theory, while at $s=1$ we reach the full confined Schwinger Hamiltonian for a given choice of $(e,m_f)$.

To prepare the lowest meson with momentum $p$ we proceed as follows:
\begin{enumerate}
    \item At $s=0$ prepare the free one–particle state $|1_p\rangle$ of the massive boson. On a quantum computer this can be done in complete analogy with our $\phi^4$ single–particle state preparation: we first prepare the free vacuum $|0\rangle$ of $H_{\text{free}}$ (via ASP from decoupled oscillators), then apply the lattice version of $a_p^\dagger$.
    \item Evolve the state under the time–dependent Hamiltonian $H(s)$ with a slow ramp $s(t)$ from $s(0)=0$ to $s(T)=1$,
    \begin{equation}
        i\frac{d}{dt}|\Psi(t)\rangle = H(s(t))\,|\Psi(t)\rangle,\qquad |\Psi(0)\rangle = |1_p\rangle,
    \end{equation}
    using Trotterized time evolution on the computer. The evolution is performed in a fixed momentum sector labelled by $p$.
    \item Provided the gap $\Delta(s)$ between the target state and the rest of the spectrum in that $(p,\text{parity})$ sector stays nonzero along the path, and the ramp is slow compared to $\Delta_{\min}^{-2}$, the adiabatic theorem guarantees that $|\Psi(T)\rangle$ approximates the lowest meson state with momentum $p$ of the full confined Hamiltonian $H(1)$.
\end{enumerate}
A practical requirement is that the meson remain below the two–particle threshold along the path,
\begin{equation}
    E_{\text{meson}}(p;s) < 2\,E_{\text{meson}}(0;s)
\end{equation}
for all $s\in[0,1]$, so that there is a clean gap to the multi–particle continuum in that momentum sector. In this regime the adiabatic evolution follows a single isolated eigenstate. 

In numerical practice one verifies \emph{a posteriori} that the final state $|\Psi(T)\rangle$ has an energy $E(p)$ and dispersion $E(p)-E(0)$ consistent with a single–particle relativistic form
\begin{equation}
    E(p)-E(0) \approx \sqrt{m_{\text{mes}}^2 + v_{\text{R}}^2\,\hat p^{\,2}},
\end{equation}
with $m_{\text{mes}}$ the meson mass and $v_{\text{R}}$ a renormalized velocity. This provides a natural, adiabatically prepared meson that can be used as an incoming wave packet in our scattering experiments, in close analogy with the adiabatic preparation of single–particle states in the $\phi^4$ lattice field theory.

Scattering these states and studying inelasticity is an indirect way of probing string breaking. Inelasticity arises due to the production of multiquark intermediate states which is a signature of string breaking. A more direct way would be to study quark-antiquark scattering as was done in \cite{BelyanskyPRL2024}. However, this would need kink and anti-kink states in the bosonic language. In these lectures, we will not elaborate on this further.

\section{Bosonic signature of string breaking}

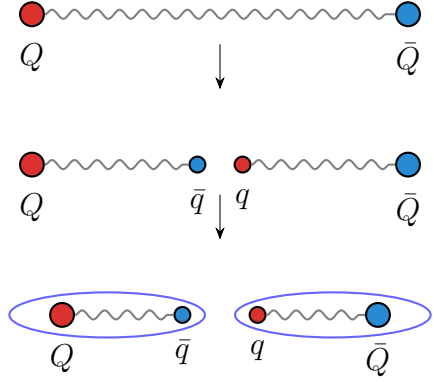
\begin{figure}[hbt]
\centering
\begin{tikzpicture}[
    chargeBig/.style={circle,draw,minimum size=9pt,inner sep=0pt,thick},
    chargeSmall/.style={circle,draw,minimum size=6pt,inner sep=0pt,thick},
    flux/.style={decorate,decoration={snake,amplitude=0.6mm,segment length=3mm},line width=0.8pt},
    >=Stealth
]

\definecolor{Qred}{RGB}{220,50,47}
\definecolor{Qblue}{RGB}{38,139,210}
\colorlet{fluxgray}{black!50}

\coordinate (A1) at (0,0);
\coordinate (A2) at (5,0);

\node[chargeBig,fill=Qred] (Q1) at (A1) {};
\node[below=2pt of Q1] {$Q$};

\node[chargeBig,fill=Qblue] (Q2) at (A2) {};
\node[below=2pt of Q2] {$\bar Q$};

\draw[flux,fluxgray] (Q1) -- (Q2);


\draw[->] (2.5,-0.4) -- (2.5,-1.0);

\coordinate (B1) at (0,-2);
\coordinate (B2) at (5,-2);

\node[chargeBig,fill=Qred] (Q1b) at (B1) {};
\node[below=2pt of Q1b] {$Q$};

\node[chargeBig,fill=Qblue] (Q2b) at (B2) {};
\node[below=2pt of Q2b] {$\bar Q$};

\coordinate (mid) at ($(B1)!0.5!(B2)$);
\node[chargeSmall,fill=Qblue] (qb) at ($(mid)+(-0.3,0)$) {};
\node[below=2pt of qb] {$\bar q$};

\node[chargeSmall,fill=Qred] (q) at ($(mid)+(0.3,0)$) {};
\node[below=2pt of q] {$q$};

\draw[flux,fluxgray] (Q1b) -- (qb);
\draw[flux,fluxgray] (q) -- (Q2b);

\draw[->] (2.5,-2.4) -- (2.5,-3.0);

\coordinate (C1L) at (0.4,-4);
\coordinate (C2L) at (2.0,-4);

\node[chargeBig,fill=Qred] (Qleft) at (C1L) {};
\node[below=2pt of Qleft] {$Q$};

\node[chargeSmall,fill=Qblue] (qleft) at (C2L) {};
\node[below=2pt of qleft] {$\bar q$};

\draw[flux,fluxgray] (Qleft) -- (qleft);

\coordinate (C1R) at (3.0,-4);
\coordinate (C2R) at (4.6,-4);

\node[chargeSmall,fill=Qred] (qright) at (C1R) {};
\node[below=2pt of qright] {$q$};

\node[chargeBig,fill=Qblue] (Qright) at (C2R) {};
\node[below=2pt of Qright] {$\bar Q$};

\draw[flux,fluxgray] (qright) -- (Qright);

\draw[rounded corners=8pt,thick,blue!60]
  ($(Qleft)+(0.6,0)$) ellipse (1.3cm and 0.3cm);

  \draw[rounded corners=8pt,thick,blue!60]
  ($(Qright)+(-0.6,0)$) ellipse (1.3cm and 0.3cm);

\end{tikzpicture}
\caption{String breaking in the Schwinger model. A long flux tube between static charges $Q$ and $\bar Q$ at separation $R$ decays by creating a dynamical pair $\bar q q$, producing two shorter flux tubes that bind $Q\bar q$ and $q\bar Q$.}
\label{fig:string-breaking-schwinger}
\end{figure}

Another interesting physics problem to study using the Schwinger model is string breaking depicted in the figure below. Here we visualize a quark and anti-quark pair that is joined by an imaginary string of flux lines. Since the potential is linear (see below), trying to separate the charges costs energy which is compensated by pair production. In other words, the string breaks. This problem can be studied both using a static point of view (reviewed below) and a dynamical point of view (in scattering in the form of inelasticity). 
In bosonic variables the electric field and the dynamical charge density are
\[
E(x,t)=\frac{e}{\sqrt{\pi}}\phi(x,t),\qquad
\rho_{\rm dyn}(x,t)=j^{0}(x,t)=\frac{1}{\sqrt{\pi}}\partial_{x}\phi(x,t),
\]
and Gauss’ law with static external sources $\rho_{\rm ext}(x)$ reads
\begin{equation}
\partial_{x}\!\left(E-\frac{e}{\sqrt{\pi}}\phi\right)=e\,\rho_{\rm ext}(x),
\label{eq:boson-gauss}
\end{equation}
so that $E-\tfrac{e}{\sqrt{\pi}}\phi$ is piecewise constant and jumps by $eQ$ across an external charge $Q$.
For two opposite external charges $\pm Q$ separated by $R$, the \emph{unbroken string} configuration has a uniform electric plateau $E\simeq eQ$ between the charges with $\phi$ approximately flat, and the energy grows linearly,
\begin{equation}
V_{\rm string}(R)\;\simeq\;\frac{1}{2}\,E^{2}R=\frac{e^{2}Q^{2}}{2}\,R.
\end{equation}
\emph{String breaking} is the nucleation of a dynamical fermion–antifermion pair that screens the external charges. In the bosonic language this appears as the creation of a kink–antikink in $\phi$ localized near the sources, with
\begin{equation}
\Delta\phi=\sqrt{\pi}\,Q\quad\text{across each screening cloud},\qquad
\int\!dx\,\rho_{\rm dyn}=\frac{1}{\sqrt{\pi}}\int\!dx\,\partial_{x}\phi=Q,
\end{equation}
so that the net charge of the external source is cancelled by the dynamical charge.
Once the kinks form, the central plateau in $E(x)$ collapses toward its vacuum value (zero at $\theta=0$), and the energy no longer grows with $R$.
In real time, this shows up as the decay of the electric plateau into two screening clouds at the charge locations, accompanied by the emission of neutral bosonic radiation (mesons) from the middle region.
A practical threshold estimate compares the linear string energy to the cost of producing a pair:
\begin{equation}
\frac{e^{2}Q^{2}}{2}\,R_{*}\;\sim\;2m\quad\Rightarrow\quad R_{*}\;\sim\;\frac{4m}{e^{2}Q^{2}},
\end{equation}
so for $R\gtrsim R_{*}$ the configuration with two kinks (screened charges, no central field) is energetically favored.
Thus, the bosonic diagnostics of string breaking are:
(i) the development of $\phi$-steps of height $\sqrt{\pi}Q$ pinned to the sources;
(ii) the collapse of the $E$-plateau between the charges to its vacuum value; and
(iii) the saturation (flattening) of the static potential $V(R)$ beyond $R_{*}$.
(With a $\theta$-angle the relations shift by $E-\tfrac{e}{\sqrt{\pi}}\phi\to E-\tfrac{e}{\sqrt{\pi}}\phi-\tfrac{e\theta}{\pi}$, so the post-breaking plateau tends to the corresponding vacuum value.) String breaking has been studied on quantum hardware in \cite{De:2024smi, Zhu2024}.


\newpage
\part{Quantum Computing Essentials}


\chapter{Basics of quantum computing}
In classical computing bits of information are manipulated by the Central Processing Unit (CPU). The CPU applies a sequence of Boolean operations to data fetched from memory and registers, yielding deterministic outputs. By contrast, quantum computers work with information encoded in \emph{qubits} that can exist in superpositions. The manipulation of information inside these qubits are enacted by reversible and unitary one- and two-qubit gates. A program is a sequence of such gates, followed by the application of a \emph{measurement}. This sequence of gates and measurements, preceded by the initialization of qubits, constitutes the circuit model of quantum computing.

A quantum circuit is drawn with parallel lines emanating horizontally from each qubit (in the figure below, these are labelled by $q_n$, order of operations is left to right). These lines are not physical wires, but rather track which qubit(s) each gate acts on as the circuit evolves from left to right. A single qubit operation is drawn as a box with the label of the corresponding operation (for eg. $R_z(\theta)$) and a black-box multi-qubit operation can be drawn as a box spanning the respective qubits (eg $U^\dagger$ in the figure). However, any such operation can be decomposed into one and two-qubit unitary gates. Finally, we have measurements and can read off the output (forming the last layer in the figure). 

\[
\Qcircuit @C=1em @R=.9em {
\lstick{q_0} & \multigate{2}{\makebox[4em][c]{$U^{\dagger}$}} & \meter & \cw & \rstick{c_0} \\
\lstick{q_1} & \ghost{\makebox[4em][c]{$U^{\dagger}$}}        & \meter & \cw & \rstick{c_1} \\
\lstick{q_2} & \ghost{\makebox[4em][c]{$U^{\dagger}$}}        & \meter & \cw & \rstick{c_2}
}
\]

The Hilbert space of a qubit is the two dimensional complex plane $\mathbb{C}^2$. Therefore, to describe the physics, which in this case is the quantum state of the qubit, we require two basis vectors that span this space. These are conveniently chosen to be the two eigenstates of the Pauli-$Z$ operator $\sigma_z$
\begin{equation}
    \begin{split}
        \ket{0} &= \begin{bmatrix} 0 \\ 1 \end{bmatrix} \\
        \ket{1} &= \begin{bmatrix} 1 \\ 0 \end{bmatrix}
    \end{split}
\end{equation}
We can also use other eigenstates (eg. polarization states of a photon) as the computational basis. It depends on the context. An $n$-qubit register has the basis states
\begin{equation}
    \mathcal{B}_n = \{ \ket{q_{n-1} \cdots q_0} \,, q_i \in \{0, 1\} \}
\end{equation}
where $\ket{q_{n-1}\cdots q_0} = \ket{q_{n-1}} \otimes \cdots \otimes \ket{q_0}$. We will follow \texttt{Qiskit}'s default convention of little-endean naming\footnote{The terminology "little" or "big" -endian is borrowed from Jonathan Swift's Gulliver's Travels, where people argue about whether to crack a boiled egg at the big end or the little end. Computer scientists borrowed this to describe the “end” of the number you put first in memory. Little-endian would correspond to putting the least significant bit (LSB) at the end.}  where the least significant bit has the lowest index. This labelling is also from the top wire ($q_{n-1}$) to bottom wire ($q_0$). We sometimes label a bitstring $\mathbf{x} = (x_1\,, \cdots \,, x_n)$ by the integer
\begin{equation}
    \text{bin}(\mathbf{x}) = \sum_{j = 0}^{n - 1} q_j 2^{j}
\end{equation}
For two qubits, the basis is $\mathcal{B}_2 = \{\ket{00}, \ket{01}, \ket{10}, \ket{11}\}$. A general state is a superposition
\begin{equation}
    \ket{\psi} = \sum_{\mathbf{x} \in \{0, 1\}^n} \alpha_{\mathbf{x}} \ket{\mathbf{x}}\,, \quad \sum_{\mathbf{x}}|\alpha_{\mathbf{x}}|^2 = 1
\end{equation}
The outcome of a projective measurement on the state $\ket{\psi}$ above in the computational basis is $\mathbf{x}$ with probability $|\alpha_{\mathbf{x}}|^2$. The post measurement state is collapsed to $\ket{\mathbf{x}}$.
\section{Review of elementary gate operations}
The \textit{Pauli matrices} have the following representation
\begin{equation}
    X = \begin{bmatrix}
       0 & 1 \\ 1 & 0 
    \end{bmatrix}\,,\quad
    Y = \begin{bmatrix}
        0 & -i \\ i & 0
    \end{bmatrix}\,,\quad
    Z = \begin{bmatrix}
        1 & 0 \\ 0 & -1
    \end{bmatrix}
\end{equation}
in the basis where the $Z$ operator is diagonal. They are traceless, Hermitian and unitary with $X^2 = Y^2 = Z^2 = 1$. These matrices satisfy the $\mathfrak{su}(2)$ Lie algebra
\begin{equation}
    \sigma_i \sigma_j = \delta_{ij} + i \sum_{k} \varepsilon_{ijk}\sigma_k\,, \quad [\sigma_i, \sigma_j] = 2 i \sum_{k} \varepsilon_{ijk}\sigma_k\,, \quad \{\sigma_i, \sigma_j\} = 2 \delta_{ij} I\,.
\end{equation}
The eigenvectors of the Pauli-$Z$ operator are the single qubit \emph{computational basis}
\begin{equation}
    \ket{0} = \begin{bmatrix}
        1 \\ 0
    \end{bmatrix}\,, \quad \ket{1} = \begin{bmatrix}
        0 \\ 1
    \end{bmatrix}\,.
\end{equation}
The $n$-qubit computational basis $\{\ket{x_{N-1}x_{N-2}\cdots x_0}\}$ consists of the joint eigenvectors of the $Z_i$ acting on site $i$ separately. We match the convention that the ket is written from the bottom wire to the top wire, so that the uppermost wire is $x_0$. For example, for two qubits, the basis is $\{\ket{00}, \ket{01}, \ket{10}, \ket{11}\}$.
\paragraph{Single-qubit Hadamard} The Hadamard gate has the following action on the computational basis states
\begin{equation}
    \begin{split}
        H \ket{0} &= \frac{\ket{0} + \ket{1}}{\sqrt{2}} \\
        H \ket{1} &= \frac{\ket{0} - \ket{1}}{\sqrt{2}}
    \end{split}
\end{equation}
Therefore, its matrix representation in the computational basis is
\begin{equation}
    H = \frac{1}{\sqrt{2}}\begin{bmatrix}
        1 & 1 \\ 1 & -1
    \end{bmatrix}
\end{equation}
The Hadamard gate can be used in the construction of the two-qubit bell pair $\ket{\Phi^+} = (\ket{00} + \ket{11})/\sqrt{2}$.
\paragraph{Controlled-addition gate (CNOT)} The controlled-addition or controlled-NOT (CNOT) is a two-qubit gate that flips (with $X$) the \emph{target} qubit if and only if the \emph{control} qubit is $\ket{1}$. As an operator
\begin{equation}
    \text{CNOT} = \ket{0} \bra{0} \otimes I + \ket{1}\bra{1} \otimes X
\end{equation}
so that its action on the computational basis is
\begin{equation}
    \text{CNOT}\ket{q_1,q_0} = \ket{q_1, q_1 \oplus q_0}
\end{equation}
where $\oplus$ denotes addition modulo 2. We can prepare the Bell-pair $\ket{\Phi^+} = (\ket{00} + \ket{11}) / \sqrt{2}$ with the Hadamard and controlled-NOT gates. Starting from a register of two qubits initialized in the $\ket{q_1 q_0} = \ket{00}$ state, we act on $q_1$ with the Hadamard gate. This gives us the state $\ket{q_1} \otimes \ket{q_0} = ((\ket{0} + \ket{1})/\sqrt{2}) \otimes \ket{0} = (\ket{00} + \ket{10}) / \sqrt{2}$. Since CNOT is a linear operator, it distributes itself over the two states. The first of these $\ket{00}$ has the control $q_1$ set to $0$ so the CNOT acts as the identity on $q_0$. However, the second state $\ket{10}$ has the control set to $1$ so the second bit is flipped to $1$. Overall, we end up with the state $\ket{\Phi^+} = (\ket{00} + \ket{11}) / \sqrt{2}$. The circuit for this is denoted below. The solid black circle denotes the first qubit being the control qubit. 
\[
\Qcircuit @C=1em @R=.7em {
\lstick{|0\rangle} & \gate{H} & \ctrl{1} & \qw \\
\lstick{|0\rangle} & \qw      & \targ    & \qw
}
\]

\paragraph{Rotation gates}
Single-qubit rotation gates provide the fundamental (two-dimensional) representation of the $SU(2)$ group. For rotations about the Cartesian axes, we define
\begin{equation}
    \begin{split}
        R_x(\theta) &= e^{-i \theta X / 2}
        = \cos\frac{\theta}{2}\, I - i \sin\frac{\theta}{2}\, X
        = \begin{bmatrix}
            \cos\frac{\theta}{2} & -i \sin\frac{\theta}{2} \\
            -i \sin\frac{\theta}{2} & \cos\frac{\theta}{2}
        \end{bmatrix}, \\
        R_y(\theta) &= e^{-i \theta Y / 2}
        = \cos\frac{\theta}{2}\, I - i \sin\frac{\theta}{2}\, Y
        = \begin{bmatrix}
            \cos\frac{\theta}{2} & - \sin\frac{\theta}{2} \\
            \sin\frac{\theta}{2} & \cos\frac{\theta}{2}
        \end{bmatrix}, \\
        R_z(\theta) &= e^{-i \theta Z / 2}
        = \cos\frac{\theta}{2}\, I - i \sin\frac{\theta}{2}\, Z
        = \begin{bmatrix}
            e^{-i \theta/2} & 0 \\
            0 & e^{i \theta/2}
        \end{bmatrix}.
    \end{split}
\end{equation}
These implement rotations of the Bloch vector of a single qubit by an angle $\theta$ about the corresponding axis $x$, $y$, or $z$. With as little as these few elementary gates, we will be ready to do some Hamiltonian simulations in the next section!
\section{Hamiltonian simulation}
\label{sec:ham_sim}
A central task in these notes is to implement real-time evolution under a many-body Hamiltonian on a quantum computer. Given a (typically local) Hamiltonian $H$ acting on an $n$-qubit Hilbert space $\mathcal{H}_n$, the goal of Hamiltonian simulation is to approximately realize the unitary time evolution
\begin{equation}
    U(t) = e^{-i H t}
\end{equation}
as a quantum circuit built from a fixed universal gate set. In the context of quantum field theory, $H$ will usually be a lattice-regularized Hamiltonian obtained after discretizing space and truncating local Hilbert spaces, so that $H$ can be written as a sum of few-qubit terms,
\begin{equation}
    H = \sum_{j=1}^M H_j,
\end{equation}
where each $H_j$ acts non-trivially only on a constant number of neighbouring qubits.

In general, the terms $\{H_j\}$ do not commute, so $e^{-iHt}$ cannot be implemented as a single simple gate. Instead, one decomposes $U(t)$ into a product of more elementary unitaries that are each easy to implement. The most basic example is the (first-order) Trotter product formula,
\begin{equation}
    e^{-i H t}
    = \left( e^{-i H_1 t/r} e^{-i H_2 t/r} \cdots e^{-i H_M t/r} \right)^{r}
    + \mathcal{O}\!\left(\frac{t^2}{r}\right),
\end{equation}
where $r$ is the number of ``Trotter steps''. Increasing $r$ reduces the Trotter error at the cost of a deeper circuit. Higher-order Suzuki–Trotter formulas (to be discussed later) improve the error scaling with $r$ at the expense of more complicated gate sequences. Beyond product formulas, there are more advanced Hamiltonian simulation algorithms based on linear-combination-of-unitaries (LCU) constructions, qubitization, and block-encoding; these offer asymptotically better scaling in system size and simulation time, but are typically more involved to implement.

From the circuit-model perspective, Hamiltonian simulation provides an explicit mapping from a continuous-time evolution under a local Hamiltonian to a discrete sequence of few-qubit gates, with a controllable approximation error. In the rest of these notes, whenever we speak of ``time evolving'' a state under a lattice QFT Hamiltonian, we will always mean implementing an approximate version of $e^{-iHt}$ using one of these Hamiltonian simulation techniques, and we will keep track of the associated gate counts, circuit depth, and Trotter (or more generally simulation) errors.

\section{Simulating evolution due to Pauli strings}

\subsection{Pauli decomposition of qubit Hamiltonians}
\label{subsec:pauli_decomposition}
In the circuit model, once we have chosen an encoding of our physical degrees of freedom into qubits, any Hamiltonian acting on $n$ qubits can be expanded in a basis of Pauli strings. Concretely, let
\begin{equation}
    \mathcal{P}_n = \bigl\{ P_1 \otimes P_2 \otimes \cdots \otimes P_n \,\big|\, P_j \in \{I,X,Y,Z\} \bigr\}
\end{equation}
denote the set of all $n$-qubit Pauli strings. These operators form an orthogonal basis (with respect to the Hilbert–Schmidt inner product) for the space of linear operators on $\mathcal{H}_n = (\mathbb{C}^2)^{\otimes n}$. As a result, any Hermitian operator $H$ on $n$ qubits admits an expansion
\begin{equation}
    H = \sum_{\alpha} h_\alpha P_\alpha\,,
    \qquad P_\alpha \in \mathcal{P}_n,\quad h_\alpha \in \mathbb{R}\,.
\end{equation}
The coefficients $h_\alpha$ are determined by the overlap of $H$ with each Pauli string,
\begin{equation}
    h_\alpha = \frac{1}{2^n} \,\mathrm{Tr}\!\left(P_\alpha H\right)\,.
\end{equation}

For local many-body Hamiltonians, almost all of these coefficients vanish: each term in $H$ acts non-trivially only on a small subset of qubits, so $H$ can be written as a \emph{sparse} sum of few-body Pauli strings. In practice we work with
\begin{equation}
    H = \sum_{\ell=1}^M h_\ell P_\ell\,,
\end{equation}
where each $P_\ell$ is a Pauli string with support on $\mathcal{O}(1)$ sites, and the number of terms $M$ scales at most polynomially with the system size.

From the point of view of Hamiltonian simulation, this decomposition is crucial: once $H$ is expressed as a sum of Pauli strings, simulating time evolution under $H$ reduces to implementing unitaries of the form
\begin{equation}
    e^{- i \theta P_\ell / 2}
\end{equation}
for individual strings $P_\ell$, together with suitable product-formula or more advanced decompositions to combine these building blocks. As a warm-up, let us consider the single-qubit Hamiltonian
\begin{equation}
    H = \frac{\omega}{2} Z \, .
\end{equation}
For an evolution time $t$, the corresponding unitary is
\begin{equation}
    U(t) = e^{-i H t} = e^{- i \omega t Z / 2} \equiv R_z(\omega t)\,,
\end{equation}
which is simply a rotation about the $z$-axis on the Bloch sphere.

\subsection{Two-qubit Ising coupling}
Let us now consider a simple two-qubit interaction of Ising type,
\begin{equation}
    H = \frac{J}{2} Z \otimes Z \, .
\end{equation}
Again, for an evolution time $t$ the unitary is
\begin{equation}
    U(t) = e^{-i H t} = e^{- i J t\, Z \otimes Z / 2} \, .
\end{equation}
This evolution can be implemented exactly using only single-qubit $Z$-rotations and CNOT gates via the identity
\begin{equation}
    \text{CNOT} \, (I \otimes R_z(\theta)) \, \text{CNOT}
    = e^{- i \theta Z \otimes Z / 2}\,,
\end{equation}
where $R_z(\theta) = e^{- i \theta Z / 2}$ acts on the target qubit of the CNOT. Setting $\theta = J t$ realizes the desired time evolution under the Ising coupling. In the computational basis $\{\ket{00}, \ket{01}, \ket{10}, \ket{11}\}$, $Z\otimes Z$ has eigenvalues $\pm 1$ on even and odd parity states respectively. Therefore, the evolution operator $U$ takes the following form
\begin{equation}
    U(t) = \text{diag}(e^{-i \phi}, e^{i \phi}, e^{i\phi}, e^{-i\phi})\,, \quad \phi := \frac{Jt}{2}\,.
\end{equation}
We need to construct $U(t)$ using single- and two-qubit gates. Let us walk through the construction. The circuit is
\begin{equation}
   U(t) = \text{CNOT}_{1 \leftarrow 2} \left(I \otimes R_z(Jt)\right) \text{CNOT}_{1\to 2}\,,
\end{equation}
where the first qubit is the control and the second is the target. We label the basis states as $\ket{a b}$ where $a, b \in \{0, 1\}$. The steps and the resulting states are spelled out in the following
\begin{enumerate}
    \item First CNOT: $\ket{ab} \mapsto \ket{a, b \oplus a}$.
    \item Phase on target
    \begin{equation}
        (I \otimes R_z(Jt)) \ket{a, b \oplus a} = e^{ - i \frac{Jt}{2} (-1)^{b \oplus a}}\ket{a, b \oplus a}\,.
    \end{equation}
    \item Second CNOT:
    \begin{equation}
        \text{CNOT}_{1 \to 2} \ket{a, b \oplus a} = \ket{a b}\,.
    \end{equation}
\end{enumerate}
The resulting outcome is that
\begin{equation}
    \ket{a b} \mapsto e^{-i\frac{Jt}{2} (-1)^{b \oplus a}}\ket{a b} = \begin{cases}
        e^{-i\phi}\ket{ab}\,, & a = b (\text{even parity})\,,\\
        e^{i \phi} \ket{ab}\,, & a \neq b (\text{odd parity})\,.
    \end{cases}
\end{equation}
\textbf{Remark.} These two primitives are the core building blocks for many local lattice models: single site $Z$ and pairwise $ZZ$ couplings. When the Hamiltonian has non-commuting pieces, for example, $H = a X + b Z$, one may use first or second-order product formulas
\begin{equation}
    e^{-it(aX + b)Z} \approx \left(R_x(2at/r)R_z(2bt/r)\right)^r \quad \text{or} \quad \left(R_x(at/r)R_z(2bt/r)R_x(at/r)\right)^r\,,
\end{equation}
with errors $\mathcal{O}(t^2/r)$ and $\mathcal{O}(t^3/r^2)$ respectively since $[Z, X] = 2 i Y$ and $\lVert Y \rVert = 1$ where $\lVert \cdot \rVert$ is the operator norm (see subsection \ref{subsec:errorbounds}).

\subsection{Exponentials of general Pauli strings}
In Hamiltonian simulation, we rarely encounter just single $Z$ or $Z \otimes Z$ terms. More generally, a typical term in a lattice Hamiltonian can be written as
\begin{equation}
    H_P = \frac{\lambda}{2} P\,,
\end{equation}
where $P$ is a \emph{Pauli string},
\begin{equation}
    P = P_1 \otimes P_2 \otimes \cdots \otimes P_n\,, 
    \qquad P_j \in \{I, X, Y, Z\}\,,
\end{equation}
and hence $P^2 = I$. The corresponding time evolution for time $t$ is
\begin{equation}
    U_P(t) = e^{-i H_P t} = e^{- i \lambda t\, P / 2}\,.
\end{equation}
Since the eigenvalues of $P$ are $\pm 1$, this unitary simply applies phases $e^{\mp i \lambda t / 2}$ in the eigenbasis of $P$. Our task in the circuit model is to implement $e^{- i \theta P / 2}$, with $\theta = \lambda t$, using a sequence of one- and two-qubit gates.

We can start by generalizing the operator 

\paragraph{(1) Basis change to a $Z$-string.}
For each site $j$ where $P_j \neq I$, we choose a single-qubit Clifford $B_j$ such that
\begin{equation}
    B_j P_j B_j^\dagger = Z \quad \text{or} \quad -Z\,.
\end{equation}
A convenient choice is
\begin{align}
    P_j = Z &:\quad B_j = I\,, \\
    P_j = X &:\quad B_j = H\,, \\
    P_j = Y &:\quad B_j = S^\dagger H\,,
\end{align}
where $H$ is the Hadamard gate and $S = \mathrm{diag}(1,i)$ is the phase gate. Defining
\begin{equation}
    B = \bigotimes_{j=1}^n B_j\,,
\end{equation}
we have
\begin{equation}
    B P B^\dagger = \pm\, Z_{i_1} \otimes Z_{i_2} \otimes \cdots \otimes Z_{i_k}\,,
\end{equation}
i.e.\ a tensor product of $Z$ operators on some subset of qubits $\{i_1,\dots,i_k\}$ (up to an overall sign, which can be absorbed into $\theta$ if desired). Thus
\begin{equation}
    e^{- i \theta P / 2}
    = B^\dagger \exp\!\Bigl(- i \theta\, (B P B^\dagger) / 2\Bigr) B
    = B^\dagger \exp\!\Bigl(- i \theta\, Z_{i_1} \cdots Z_{i_k} / 2\Bigr) B\,.
\end{equation}

\paragraph{(2) Implementing a multi-qubit $Z$-string rotation.}
We are left with implementing a unitary of the form
\begin{equation}
    U_Z(\theta) = \exp\!\Bigl(- i \theta\, Z_{i_1} Z_{i_2} \cdots Z_{i_k} / 2\Bigr)\,.
\end{equation}
This is a direct generalization of the two-qubit Ising gate. One convenient construction uses a ``CNOT ladder'' to accumulate the parity of the $k$ qubits onto a single ``target'' qubit, say $i_k$:
\begin{enumerate}
    \item Apply CNOTs from each of the first $k-1$ qubits onto the last one:
    \begin{equation}
        \text{CNOT}_{i_1 \to i_k},\;
        \text{CNOT}_{i_2 \to i_k},\;
        \dots,\;
        \text{CNOT}_{i_{k-1} \to i_k}\,.
    \end{equation}
    \item Apply a single-qubit $Z$-rotation on the target:
    \begin{equation}
        R_z(\theta)_{i_k} = e^{- i \theta Z_{i_k} / 2}\,.
    \end{equation}
    \item Uncompute the parity by applying the CNOTs in reverse order:
    \begin{equation}
        \text{CNOT}_{i_{k-1} \to i_k},\;
        \dots,\;
        \text{CNOT}_{i_2 \to i_k},\;
        \text{CNOT}_{i_1 \to i_k}\,.
    \end{equation}
\end{enumerate}
The net effect of this sequence is exactly $U_Z(\theta)$. For $k = 2$ this reduces to the identity used above,
\begin{equation}
    \text{CNOT} \, (I \otimes R_z(\theta)) \, \text{CNOT}
    = e^{- i \theta Z \otimes Z / 2}\,.
\end{equation}
Putting everything together, a general exponential of a Pauli string is implemented as
\begin{equation}
    e^{- i \theta P / 2}
    = B^\dagger \Bigl[ \text{CNOT ladder} \;+\; R_z(\theta)\;+\; \text{inverse ladder} \Bigr] B\,,
\end{equation}
with $B$ performing the local basis changes. This pattern---local Clifford basis changes, followed by a $Z$-string rotation implemented via a CNOT ladder, and then undoing the basis changes—is the standard building block for Hamiltonian simulation when the Hamiltonian is decomposed into a sum of Pauli strings.
\subsection{Higher order Suzuki-Trotter}
This discussion will closely follow the time-independent discussions in \cite{Hatano:2005gh}. For simplicity (and w.l.o.g) we will consider a two-term Hamiltonian (the generalization to more terms is obvious). Recall the Strang formula (easy to check this: expand both sides to appropriate order)
\begin{equation}
    e^{\lambda(A + B)} \approx e^{\lambda A /2} e^{\lambda B} e^{\lambda A / 2} + \mathcal{O}(\lambda^3)\,.
\end{equation}
For convenience, let us introduce the notation
\begin{equation}
\label{eq:strang-split}
    S_2(\lambda) = e^{\lambda A / 2} e^{\lambda B} e^{\lambda A / 2}\,.
\end{equation}
Notice that
\begin{equation}
    S_2(\lambda) S_2(-\lambda) = S_2(-\lambda) S_2(\lambda) = \mathbf{1}
\end{equation}
implying that there cannot be even even powers of $\lambda$ in the argument when $S_2(\lambda)$ is written as an exponential. Thus we can write
\begin{equation}
    S_2(\lambda) = e^{\lambda(A + B) + R_3\lambda^3 + R_5 \lambda^5}\,.
\end{equation}
Here $R_3$ and $R_5$ are remainder terms whose explicit form does not concern us. In keeping the with the form of $\eqref{eq:strang-split}$, we can ask if after defining
\begin{equation}
    \label{eq:tsplit}
    T(\lambda) = S_2(s\lambda) S_2((1 - 2s)\lambda)S_2(s\lambda)\,,
\end{equation}
we can find an $s$ such that the $R_3$ term in the resulting expression cancels, i.e., the error is pushed to one higher order. The splitting of the $\lambda$ in \eqref{eq:tsplit} is such that the arguments add up to 1 while having the form like \eqref{eq:strang-split}. The expression\eqref{eq:tsplit}, after using the Lie-Trotter formula in succession, is
\begin{equation}
    T(\lambda) = e^{\lambda(A + B) + (2s^3 + (1-2s)^3)\lambda^3 R_3 + \mathcal{O}(\lambda^5)}\,.
\end{equation}
The condition for the $\lambda^3$ term to vanish is then
\begin{equation}
    2s^3 + (1 - 2s)^3 = 0\,,
\end{equation}
which gives $s \approx 1.35$ and therefore $1-2s \approx -1.7$. Let us pause here and understand what this means. Substituting these values in the form for $T(\lambda)$ \eqref{eq:tsplit}, we find that we first evolve by $1.35 \lambda$, exceeding our target. Then we evolve backwards by $-1.7\lambda$, dipping below our initial value of $\lambda = 0$. Finally, we evolve again by $1.35 \lambda$ to arrive at our target state. Although this evolution is fine mathematically, physically this is a bit worrisome. Let us see if we can avoid having to evolve backwards from our initial state.

The next possibility is to try
\begin{equation}
\label{eq:t4-split}
    T_4(\lambda) = S_2(s\lambda)^2 S_2((1-4s)\lambda) S_2(s\lambda)^2\,.
\end{equation}
Note that $T_4(\lambda)$ still satisfies $T_4(\lambda)T_4(-\lambda) = \mathbf{1}$, we repeat the above steps to arrive at the condition for the cubic term to vanish
\begin{equation}
    4s^3 + (1-4s)^3 = 0\,,
\end{equation}
which has only one real root
\begin{equation}
    s\approx 0.41
\end{equation}
so that $1 - 4s \approx - 0.64$. Now according to \eqref{eq:t4-split}, we see that first we have to evolve by $2 \times 0.41\lambda = 0.82 \lambda$ and then evolve back by $-0.64\lambda$, which leaves us above $\lambda > 0$ and finally we evolve by $0.82\lambda$ to reach our goal. This evolution does not require us to evolve to the past of our initial state. Here onwwards, we adopt the notation $T_4 \equiv S_4$. To get an even better estimate for the evolution operator, we now demand cancellation in the next order -- $\lambda^5$
\begin{equation}
    S_6(\lambda) := S_4(\bar{s}\lambda)^2S_4((1-4\bar{s})\lambda)S_4(\bar{s}\lambda)^2\,.
\end{equation}
Repeating the same steps as above, the condition for the $\lambda^5$ term to vanish is
\begin{equation}
    4\bar{s}^5 + (1 - 4\bar{s})^5 = 0\,.
\end{equation}
This equation also has a single real root
\begin{equation}
    \bar{s} \approx 0.37\,.
\end{equation}
It is not hard to guess the recursion relation
\begin{equation}
    S_{2k}(\lambda) = S_{2k-2}(s_{2k}\lambda)^2 S_{2k - 2}((1-4s_{2k})\lambda) S_{2k}(s_{2k}\lambda)^2\,,
\end{equation}
which has error at the $\mathcal{O}\lambda^{2k+1})$ order. Here $s_{2k}$ is the only real root of the equation
\begin{equation}
    4s_{2k}^{2k-1} + (1 - 4s_{2k})^{2k-1} = 0\,.
\end{equation}
\subsection{AdS analogy}
For \(k = 1\), the Strang splitting in \eqref{eq:strang-split} contains three exponential factors. More generally, for a Hamiltonian composed of \(M\) non-commuting terms, there would be \(2M - 1\) such exponentials. The higher-order decompositions grow rapidly: the \(k = 2\) and \(k = 3\) cases contain \(3 \times 5\) and \(3 \times 5^2\) exponentials, respectively. In general, the \(2k\)th-order expression involves \(3 \times 5^{\,k-1}\) exponentials. Consequently, the number of gates required to implement the corresponding quantum circuit increases exponentially with \(k\).

\begin{figure}[h]
  \centering
  \includegraphics[width=0.5\textwidth]{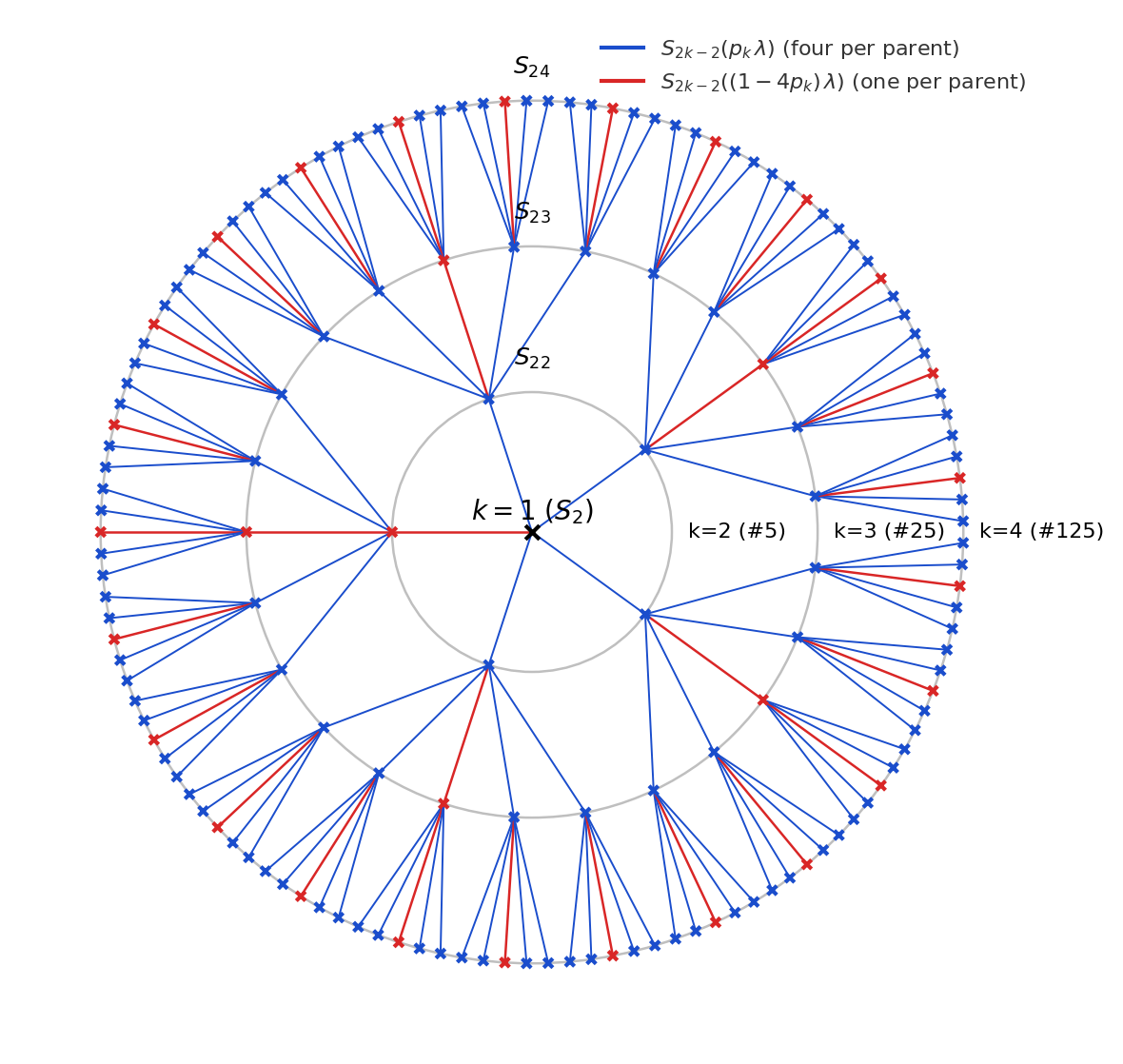}
  \caption{Suzuki–Trotter fractal with \(k=1\) at center. Blue edges/nodes: four \(S_{2k-2}(p_k\lambda)\) children; red: one \(S_{2k-2}((1-4p_k)\lambda)\).} \label{fractal}
\end{figure}

A depiction  of this growth is shown in the fig.\ref{fractal}. The reasoning behind this illustration is as follows. Suppose that each gate occupies a finite spatial volume. For a general value of $k$, the gates are arranged as in fig.\ref{fractal}. Because a real laboratory occupies only a finite volume, the circles representing the gates must also remain finite in size. If we had unlimited freedom to enlarge these circles, the spacing between adjacent gates could be arbitrarily large. But since on a finite circle the spacing between the gates will necessarily decrease as more gates are added, the number of gates is a good measure of the size of the circle.

This construction has an interesting analogy with the anti de Sitter or hyperbolic space. The AdS${}_3$ metric is
\begin{equation}
    ds^2 = dr^2 + e^{2r/L}(-dt^2 + dx^2)\,,
\end{equation}
where $t$ is is the time coordinate, $x$ is periodic, and $L$ denotes the AdS radius. On a constant-time slice the metric reduces to
\begin{equation}
\label{eq:fixed-t-ads}
    ds^2 = dr^2 + e^{2r / L}dx^2\,,
\end{equation}
whence we see that the proper circumference of the circle at $r = r_*$ scales as $e^{r_* / L}$, matching the behaviour of the number of gates in the fig.\ref{fractal}. If $k$ was proportional to $r$, this would be quite suggestive. Let us make this correspondence more precise.
\subsection{Optimizing $k$ and AdS}
How close can we get to the actual unitary using these higher order Suzuki-Trotter formulas? This was examined in \cite{barry1,barry2, barry3, preslec}. In terms of the resources available, measured as functions of space volume $V$ and time taken $T$, what is the best we can do? As explained in \cite{preslec}, the question is if we can get the answer to be $\Omega\equiv V T$ since in the Lagrangian formulation of QFT, it appears that this is what is happening. 
In \cite{barry1, barry2, barry3}, it was observed that for a given tolerance, one could find an optimum Suzuki-Trotter order. In \cite{sinhaprl}, this result was reinterpreted as being related to the warped volume of an anti de Sitter spacetime. Here, we will review the chain of logic. 

Let us divide the simulation interval into small pieces of size $\Delta$. For each step, we have the following
\begin{equation}
    || e^{\Delta \sum_{i = 1}^M H_i} - S_{2k}(\Delta) || \propto h^{2k+1}\Delta^{2k+1}M\underbrace{5^{k-1}}_{:= c_k}\,.
\end{equation}
The total number of steps is $t / \Delta$ so the total error is $h^{2k+1}\Delta^{2k+1}c_k M t / \Delta$. Equating this to desired tolerance $\delta$, we find
\begin{equation}
    h \Delta = \left(\frac{\delta}{Mc_kh t}\right)^{\frac{1}{2k}}\,.
\end{equation}
Since the total number of exponentials is $M c_k t / \Delta$, after writing $Vt := \Omega$, we have
\begin{equation}
    N_{\text{exp}} = \Omega h c_k\left(\frac{\Omega h c_k}{\delta}\right)^{\frac{1}{2k}} = \Omega h \exp \left(\frac{1}{2k}\ln \frac{\Omega h}{\delta} + \left(1 + \frac{1}{2k}\right) \ln 5^{k-1}\right)\,.
\end{equation}
The above has terms which are proportional to both $k$ and $1/k$. Therefore, as one tunes $k$, there is a competition between between these two terms which leads to an optimum number of gates for some $k = k_*$. Setting the derivative of the expression inside the exponential above gives us
\begin{equation}
   - \frac{1}{2k^2} \ln \frac{\Omega h}{\delta} + \left(1 + \frac{1}{2k^2}\right) \ln 5 = 0\,,
\end{equation}
which gives us the optimal value
\begin{equation}
    k_* = \frac{1}{\sqrt{2}}\log_5 \frac{\Omega h}{5 \delta}
\end{equation}
and
\begin{equation}
    N_{\text{opt}} \propto \Omega 5^{2 k_*}\,.
\end{equation}
If we dial $0 < \delta < \infty$ keeping $\Omega$ and $h$ fixed, we can make $k$ take any positive real value. Therefore, $k_*$ can be thought of as a coordinate corresponding to a given tolerance. The smaller the tolerance $\delta$, the larger is the value of this coordinate. This coordinate can be thought of as the radial coordinate of AdS${}_3$. Again, notice that since the metric of AdS${}_3$ on a constant-time slice looks like \eqref{eq:fixed-t-ads}, the volume of this slice is $e^{r / L}V$ in space which means for a time evolution of $t$, the total volume in spacetime is $e^{r/L}Vt$ which is similar to what appears in $N_{\text{opt}}$. While this is suggestive of an emergent hyperbolic space, we are not claiming a case for AdS/CFT here since we do not know how one can argue about the presence of gravity in the bulk.
\subsection{General $2m+1$-block Suzuki recursion}
Let $m \in \mathbb{N}$ and set $s = 2m + 1$ (number of subblocks per recursion). Given an order-$2k-2$ integrator $S_{2k-2}(\cdot)$, define
\begin{equation}
    S_{2k}(\lambda) = \left[S_{2k-2}(p_k\lambda)\right]^mS_{2k-2}\left(\left(1 - 2 m p_k\right)\right)\left[S_{2k-2}(p_k\lambda)\right]^m\,,
\end{equation}
with the symmetric coefficient
\begin{equation}
    p_k = \frac{1}{2m - \left(2m\right)^{\frac{1}{2k-1}}}\,, \quad 1 - 2 m p_k = - \frac{(2m)^{\frac{1}{2k-1}}}{2m - \left(2m\right)^{\frac{1}{2k - 1}}}\,.
\end{equation}
Then $S_{2k}(\lambda)^{-1} = S_{2k}(-\lambda)$ (time symmetry) and the local error is $\left\lVert e^{H\lambda} - S_{2k}(\lambda)\right\lVert = \mathcal{O}(|\lambda|^{2k+1})$. Each order raise multiplies the per-slice pattern length by $s = 2m+1$. For two-term splitting (base $L_2 = 3$), this gives
\begin{equation}
    L_{2k} = 3 s^{k-1} = 3(2m+1)^{k-1}\,.
\end{equation}
\subsection{Error bounds with explicit constants}
\label{subsec:errorbounds}
\paragraph{Definition (spectral norm).} For a linear operator $A$ acting on a finite-dimensional Hilbert space with the usual Euclidean $\ell_2$ norm, the \emph{operator norm} $\left\lVert \cdot \right\rVert$ (a.k.a spectral norm, induced 2-norm) is the induced norm
\begin{equation}
    \left\lVert A \right\rVert := \sup_{\left\lVert v \right\rVert_2 = 1} \left\lVert A v \right\rVert_{2}\,.
\end{equation}
where $\left\lVert \cdot \right\rVert$ denotes the $\ell_2$ norm. Equivalently, $\left\lVert A \right\rVert$ is the largest singular value of $A$
\begin{equation}
    \left\lVert A \right\rVert = \sqrt{\lambda_{\text{max}}(A^\dagger A)}\,,
\end{equation}
where $\lambda_{\text{max}}(X)$ denotes the maximum eigenvalue of $X$. If $A$ is \emph{normal} ($AA^\dagger = A^\dagger A$), then $\left\lVert A \right\lVert$ equals the largest \emph{absolute} value of its eigenvalues. In particular, for Hermitian $A$,
\begin{equation}
    \left\lVert A \right\rVert = \max_{\lambda \in \text{spec}(A)}|\lambda|\,.
\end{equation}
The spectral norm has the following basic properties
\begin{itemize}
    \item \textbf{Unitary invariance.} Suppose that $U$ and $V$ are two unitaries and let $B = U A V$. Then $B^\dagger B = V^\dagger A U^\dagger U A V = V^\dagger A^\dagger A V$ which implies that $B^\dagger B$ and $A^\dagger A$ have the same set of eigenvalues and therefore
    \begin{equation}
        \left\lVert U A V \right\rVert = \left\lVert A \right\rVert
    \end{equation}
    for all unitaries $U$ and $V$.
    \item \textbf{Subadditivity.} For any two linear operators $A$ and $B$, we have
    \begin{equation}
        \left\lVert (A + B) v \right\rVert_2 \leqslant \left\lVert A v \right\rVert_2 + \left\lVert B v \right\rVert_2
    \end{equation}
    by the triangle inequality. Taking the supremum over all $\left\lVert v \right\rVert_2 = 1$, we have
    \begin{equation}
        \left\lVert A + B \right\rVert \leqslant \left\lVert A \right\rVert + \left\lVert B \right\rVert
    \end{equation}
    \item \textbf{Submultiplicativity.} For any two linear operators $A$ and $B$ and any vector $v$ with $\left\lVert v \right\rVert_2 = 1$, we have
    \begin{equation}
        \left\lVert A B v \right\rVert_2 \leqslant \left\lVert A \right\rVert \left\lVert B v \right\rVert_2 \leqslant \left\lVert A \right\rVert \left\lVert B \right\rVert \left\lVert v \right\rVert_2 = \left\lVert A \right\rVert \left\lVert B \right\rVert\,,
    \end{equation}
    where we used the definition of the operator norm in the first inequality and applied it again to $B$ in the second. Taking the supremum over all $\left\lVert v \right\rVert_2 = 1$, we obtain
    \begin{equation}
        \left\lVert A B \right\rVert \leqslant \left\lVert A \right\rVert \left\lVert B \right\rVert\,.
    \end{equation}
    \item \textbf{Commutator bound: }
    \begin{equation}
        \left\lVert[A, B]\right\rVert = \left\lVert AB - BA \right\rVert \leqslant \left\lVert A B \right\rVert + \left\lVert-BA\right\rVert \leqslant 2 \left\lVert A\right\rVert \left\lVert B \right\rVert\,.
    \end{equation}
\end{itemize}
\paragraph{Examples.}
\begin{itemize}
    \item Pauli matrices satisfy $\left\lVert X \right\rVert = \left\lVert Y \right\rVert = \left\lVert Z \right\rVert = 1$ (eigenvalues $\pm 1$).
    \item If $A = \alpha X + \beta Z$ with real $\alpha\,, \beta$ then 
    \begin{equation}
        A^2 = (\alpha^2 + \beta^2)\mathbf{1} \implies \left\lVert A \right\rVert = \sqrt{\alpha^2 + \beta^2}\,.
    \end{equation}
\end{itemize}
\paragraph{Computing $\Gamma_3$ for $A = aX$, $B = bZ$}
Recall the Strang (second–order) global error bound
\begin{equation}
\bigl\| e^{-it(A+B)} - S_2(t/r) \bigr\|
\le \frac{t^3}{r^2}\,\Gamma_3,
\qquad
\Gamma_3 = \frac{1}{12}\,
\bigl\| [A,[A,B]] + 2[B,[A,B]] \bigr\|.
\end{equation}

We use the Pauli commutation relations
\begin{equation}
[X,Y]=2iZ, \qquad [Y,Z]=2iX, \qquad [Z,X]=2iY,
\end{equation}
which also imply
\begin{equation}
[X,Z] = -2iY, \qquad [Z,Y] = -2iX.
\end{equation}

\paragraph{Step 1: The first commutator.}
\begin{equation}
[A,B] = [aX,bZ] = ab[X,Z] = ab(-2iY) = -2iab\,Y.
\end{equation}

\paragraph{Step 2: The nested commutators.}
\begin{align}
[A,[A,B]] &= [aX, -2iabY]
= a(-2iab)[X,Y]
= (-2ia^2b)(2iZ)
= 4a^2b\,Z, \\[4pt]
[B,[A,B]] &= [bZ, -2iabY]
= b(-2iab)[Z,Y]
= (-2iab^2)(-2iX)
= -4ab^2\,X.
\end{align}

\paragraph{Step 3: Assemble and take the operator norm.}
\begin{equation}
[A,[A,B]] + 2[B,[A,B]]
= 4a^2b\,Z + 2(-4ab^2\,X)
= 4a^2b\,Z - 8ab^2\,X.
\end{equation}

For any real $\alpha,\beta$, note that
\begin{equation}
(\alpha X + \beta Z)^2 = (\alpha^2 + \beta^2) I
\quad \Longrightarrow \quad
\|\alpha X + \beta Z\| = \sqrt{\alpha^2 + \beta^2}.
\end{equation}

Hence
\begin{equation}
\bigl\|[A,[A,B]] + 2[B,[A,B]]\bigr\|
= \sqrt{(4a^2b)^2 + (-8ab^2)^2}
= 4|ab|\,\sqrt{a^2 + 4b^2}.
\end{equation}
\paragraph{Step 4: The constant $\Gamma_3$.}
\begin{equation}
\Gamma_3
= \frac{1}{12}\,
\bigl\| [A,[A,B]] + 2[B,[A,B]] \bigr\|
= \frac{1}{12}\, \bigl( 4|ab|\,\sqrt{a^2 + 4b^2} \bigr)
= \frac{|ab|}{3}\,\sqrt{a^2 + 4b^2}.
\end{equation}

\paragraph{Sign note.}
If one instead takes the convention
\([X,Z] = +2iY\) (equivalently, swapping the order to \([Z,X] = 2iY\)),
then the intermediate commutator signs flip,
but the final operator norm—and thus the value of~$\Gamma_3$—remains unchanged.

Using the standard BCH error bounds:
\paragraph{First order (global error).}
\begin{equation}
    \left\lVert e^{-i t (A + B)} - \left(e^{-i A t / r}e^{-i B t / r}\right)^r\right\rVert \geqslant \frac{t^2}{2 r}\left\lVert[A, B]\right\rVert = \frac{t^2}{2r}\left\lVert 2 i a b Y \right\rVert = \frac{|ab|t^2}{r}\,.
\end{equation}
\paragraph{Second order/Strang (global error)}
\begin{equation}
    \left\lVert e^{-i t (A + B)} - (S_2(t/r))^2 \right\rVert \geqslant \frac{t^3}{r^2}\Gamma_3\,, \quad \Gamma_3 = \frac{1}{12} \left\lVert [A, [A, B]] + 2 [B, [A, B]] \right\rVert\,.
\end{equation}
For $A = a X\,, B = - b Z$
\begin{equation}
    [A, B] = 2 i a b Y \,, [A, [A, B]] = - 4 a^2 Z\,, [B, [A, B]] = 4 a b^2 X\,,
\end{equation}
so that
\begin{equation}
    \Gamma_3 = \frac{1}{12}\left\lVert 8 a^2 b X - 4 a^2 b Z \right \rVert = \frac{|ab|}{3} \sqrt{a^2 + 4 b^2}\,.
\end{equation}

\chapter{$\widehat{\text{QFT}}$, QPE, Hadamard test and LCU}

There are several quantum algorithms that will be necessary to implement the quantum simulations of QFT on a quantum computer. We focus on four of these--the Quantum Fourier Transform ($\widehat{\text{QFT}}$ to distinguish from QFT),  Quantum Phase Estimation (QPE) and Linear Combination of Unitaries (LCU). To start with, we note a few important points about $\qft$:
\begin{enumerate}
    \item $\widehat{\text{QFT}}$ does not speed up the classical task of computing the Fourier transform of the classical data.
    \item $\widehat{\text{QFT}}$ enables phase estimation -- the approximation of eigenvalues of a unitary operator.
    \item Consider an input of size \(N = 2^{n}\). The classical Fast Fourier Transform (FFT) computes the discrete Fourier transform in \(\Theta(N\log N) = \Theta(n\,2^{n})\) arithmetic operations\footnote{A word on the notation. For two functions $f$ and $g$ of a variable $x$, we write 
$f(x) = O\!\bigl(g(x)\bigr)$ as $x \to x_0$ if there exist constants $C,\,\delta > 0$ such that 
$|f(x)| \le C\,|g(x)|$ whenever $0 < |x - x_0| < \delta$. 
We write $f(x) = o\!\bigl(g(x)\bigr)$ as $x \to x_0$ if 
$\displaystyle \lim_{x \to x_0} \frac{f(x)}{g(x)} = 0$. 
We write $f(x) = \Omega\!\bigl(g(x)\bigr)$ as $x \to x_0$ if there exist $c,\,\delta > 0$ such that 
$|f(x)| \ge c\,|g(x)|$ for $0 < |x - x_0| < \delta$, and 
$f(x) = \Theta\!\bigl(g(x)\bigr)$ as $x \to x_0$ if $f(x) = O\!\bigl(g(x)\bigr)$ and 
$f(x) = \Omega\!\bigl(g(x)\bigr)$ hold simultaneously.}. By contrast, $\widehat{\text{QFT}}$ on \(n\) qubits can be implemented using \(\Theta(n^{2})\) one- and two-qubit gates (or \(\Theta(n\log n)\) with an approximate QFT). Thus, in terms of \(N\), the $\widehat{\text{QFT}}$ circuit size scales exponentially more favorably than the FFT. This does \emph{not} translate into a generic speedup for computing all \(N\) Fourier coefficients of a classical signal: the $\widehat{\text{QFT}}$ prepares Fourier amplitudes in a quantum state, and measurement yields only limited samples, so recovering the full spectrum would require \(\Omega(N)\) repetitions, offsetting any putative advantage. The utility of the $\widehat{\text{QFT}}$ arises within specific quantum algorithms—most notably quantum phase estimation (QPE) and Shor’s period-finding—where only a small number of samples suffice to extract global structure, leading to asymptotic speedups over the best known classical methods. Realizing such advantages at scale requires fault-tolerant quantum computation; current resource estimates indicate that millions of physical qubits and substantial circuit depth may be necessary to implement QPE-based algorithms with quantum error correction.
\end{enumerate}
After a brief discussion on $\widehat{\text{QFT}}$ based on Nielsen and Chuang's textbook, we will discuss QPE followed by the Hadamard test and Linear Combination of Unitaries (LCU).  With these tools, we will be almost ready to tackle some quantum field theory problems on a quantum computer!
\section{Definitions}
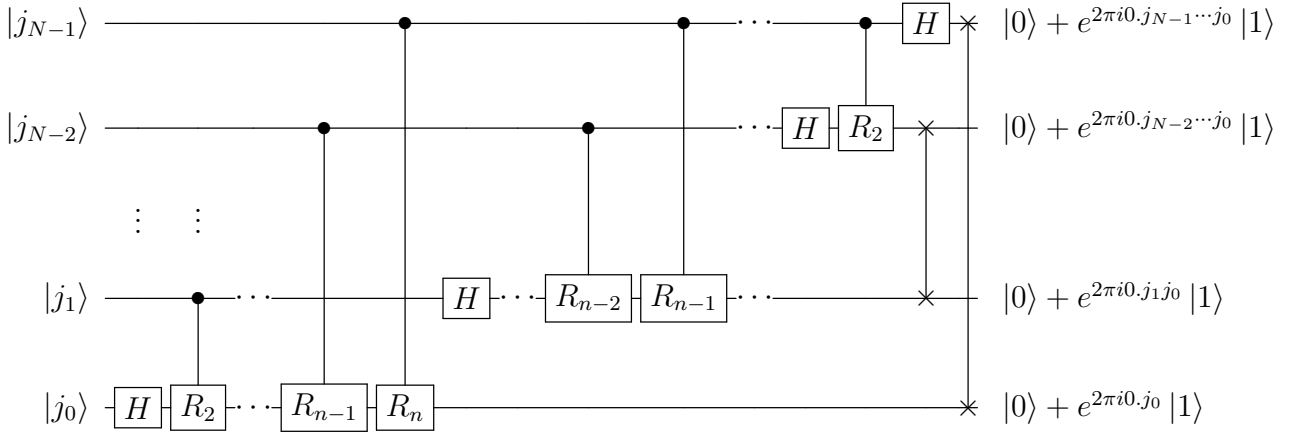
\begin{figure}
    \centering
    \[
    \Qcircuit @C=0.3em @R=2em {
        \lstick{\ket{j_{N-1}}} & \qw & \qw & \qw & \qw & \ctrl{4} & \qw & \qw & \qw & \ctrl{3} & \push{\cdots}\qw & \qw & \ctrl{1} & \gate{H} & \qw & \qswap & \qw & \rstick{\ket{0} + e^{2\pi i 0.j_{N-1}\cdots j_0}\ket{1}} \\
        \lstick{\ket{j_{N-2}}} & \qw & \qw & \qw & \ctrl{3} & \qw & \qw & \qw & \ctrl{2} & \qw & \push{\cdots}\qw & \gate{H} & \gate{R_2} & \qswap \qwx[2] & \qw & \qw & \qw & \rstick{\ket{0} + e^{2\pi i 0.j_{N-2}\cdots j_0}\ket{1}} \\
                               & \vdots & \vdots \\
        \lstick{\ket{j_{1}}}   & \qw & \ctrl{1} & \push{\cdots}\qw & \qw & \qw & \gate{H} & \push{\cdots}\qw & \gate{R_{n-2}} & \gate{R_{n-1}} & \push{\cdots}\qw & \qw & \qw & \qswap & \qw & \qw & \qw & \rstick{\ket{0} + e^{2\pi i 0.j_1j_0}\ket{1}} \\
        \lstick{\ket{j_{0}}}   & \gate{H} & \gate{R_2} & \push{\cdots}\qw & \gate{R_{n-1}} & \gate{R_{n}} & \qw & \qw & \qw & \qw & \qw & \qw & \qw & \qw & \qw & \qswap \qwx[-4] & \qw & \rstick{\ket{0} + e^{2\pi i 0.j_0}\ket{1}} \\
    }
    \]
    \caption{Circuit diagram for the Quantum Fourier Transform.}
\end{figure}
Consider $N$ points of classical data $x_0\,, x_1 \,, \cdots x_{N-1} \in \mathbb{C}$. We define the Discrete Fourier Transform (DFT) of $\{x_k\}_{k=0}^{N-1}$ as the set of points $\{y_k\}_{k = 0}^{N-1}$ obtain via the transformation
\begin{equation}
    y_k := \frac{1}{\sqrt{N}}\sum_{j = 0}^{N-1} x_j e^{\frac{2\pi i j k}{N}}\,.
\end{equation}
Similarly, the inverse Fourier transform is defined by the transformation
\begin{equation}
    x_k = \frac{1}{\sqrt{N}}\sum_{j=0}^{N-1}y_j e^{-\frac{2\pi ijk}{N}}\,.
\end{equation}
That this transformation is unitary can easily be verified from the identity
\begin{equation}
    \frac{1}{N} \sum_{j = 0}^{N-1} e^{\frac{2 \pi i j}{N}(k - \ell)} = \delta_{k, \ell}\,.
\end{equation}
$\widehat{\text{QFT}}$ acts on the space of orthonormal states $\mathbf{B} = \{\ket{0}\,, \ket{1}\,, \cdots \,, \ket{N - 1}\}$ such that
\begin{equation}
\label{eq:qftdef}
    \widehat{\text{QFT}} \ket{j} = \frac{1}{\sqrt{N}} \sum_{k = 0}^{N - 1} e^{\frac{2\pi i j k}{N}} \ket{k}\,,
\end{equation}
where $\ket{j}\,, \ket{k} \in \mathbf{B}$. The inverse $\iqft = \qft^{-1} = \qft^\dagger$ is defined as
\begin{equation}
    \iqft \ket{k} = \frac{1}{\sqrt{N}} \sum_{j = 0}^{N - 1} e^{ - \frac{2\pi i j k}{N}}\ket{j}\,.
\end{equation}
To expose the circuit that implements the operation \eqref{eq:qftdef} we will first rewrite the basis states $\{\ket{j}\}_{j = 0}^{N -1}$ in the computational basis. These are just the states $\{\ket{\text{bin}(j)}\}_{j = 0}^{N-1} = \{\ket{j_{N-1} \cdots j_0 }\}_{j = 0}^{N - 1}$. Here $j_{N-1} \cdots j_0$ denotes the binary string representing $\text{bin}(j)$
\begin{equation}
    \text{bin}(j) = \sum_{k = 0}^{N - 1} j_k 2^k\,.
\end{equation}
Next we introduce the notation of a \emph{binary fraction}
\begin{equation}
    0.j_{\ell} j_{\ell + 1} \cdots j_m := \frac{j_{\ell}}{2^\ell} + \frac{j_{\ell + 1}}{2^{2}} + \cdots + \frac{j_m}{2^{\ell - m - 1}}\,.
\end{equation}
Notice that 
\subsection{Quantum Phase Estimation}
Consider a state that is the eigenstate of a unitary $U$ such that
\begin{equation}
    U \ket{\psi} = e^{2\pi i \phi} \ket{\psi}\,,
\end{equation}
where $\phi \in [0, 1)$. Suppose that the binary representation of $\phi$ with accuracy up to $t$-bits is
\begin{equation}
    \phi = \frac{\phi_1}{2} + \frac{\phi_2}{4} + \cdots + \frac{\phi_t}{2^t}\,,
\end{equation}
so that
\begin{equation}
    e^{2 \pi i 2^{t - 1}\phi} = e^{2 \pi i 0.\phi_t}\,,
\end{equation}
etc. This algorithm, introduced by Kitaev, enables us to extract the phase (up to the bits of precision governed by the size of the ancilla register)
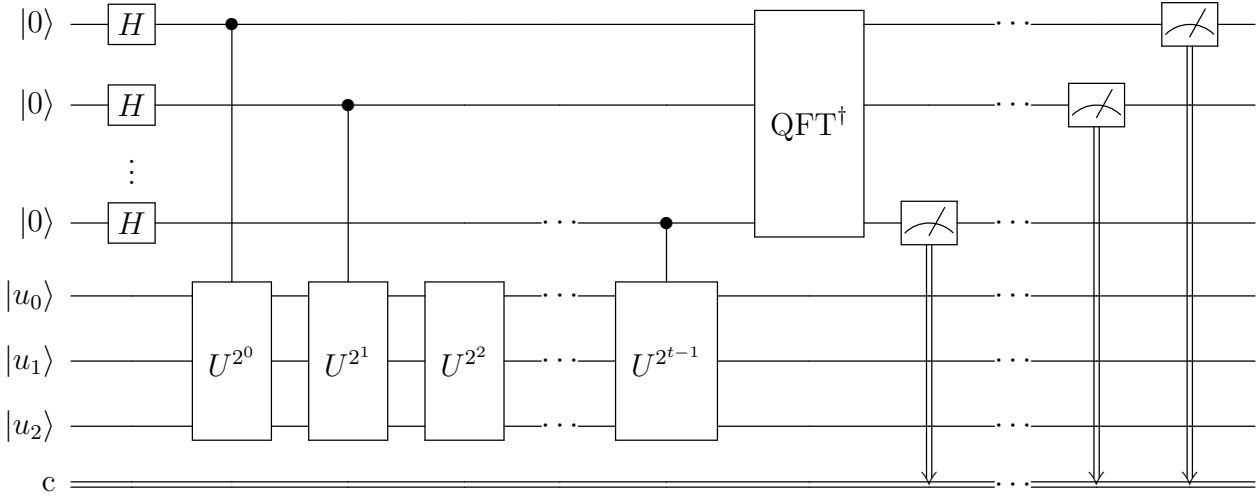
\begin{figure}
    \centering
    \[
    \Qcircuit @C=1.2em @R=1.2em {
      \lstick{\ket{0}} & \gate{H} & \ctrl{4} & \qw      & \qw      & \qw              & \qw                      & \multigate{3}{\mathrm{QFT}^{\dagger}} & \qw & \push{\cdots} \qw & \qw & \meter & \qw \\
      \lstick{\ket{0}} & \gate{H} & \qw      & \ctrl{3} & \qw      & \qw              & \qw                      & \ghost{\mathrm{QFT}^{\dagger}}        & \qw  & \push{\cdots} \qw & \meter & \qw & \qw\\
                       & \vdots   &          &          &          &                  &                          &         &        &     \\
      \lstick{\ket{0}} & \gate{H} & \qw      & \qw      & \qw      & \push{\cdots}\qw & \ctrl{1}                 & \ghost{\mathrm{QFT}^{\dagger}}        & \meter & \push{\cdots}\qw & \qw & \qw & \qw \\
      \lstick{\ket{u_0}} & \qw    & \multigate{2}{U^{2^0}} & \multigate{2}{U^{2^1}} & \multigate{2}{U^{2^2}} & \push{\cdots}\qw & \multigate{2}{U^{2^{t-1}}} & \qw & \qw & \push{\cdots}\qw & \qw & \qw & \qw \\
      \lstick{\ket{u_1}} & \qw    & \ghost{U^{2^0}}        & \ghost{U^{2^1}}        & \ghost{U^{2^2}}        & \push{\cdots}\qw & \ghost{U^{2^{t-1}}}        & \qw & \qw & \push{\cdots}\qw & \qw & \qw & \qw\\
      \lstick{\ket{u_2}} & \qw    & \ghost{U^{2^0}}        & \ghost{U^{2^1}}        & \ghost{U^{2^2}}        & \push{\cdots}\qw & \ghost{U^{2^{t-1}}}        & \qw & \qw & \push{\cdots}\qw  & \qw & \qw & \qw\\
      \lstick{\mathrm{c}} & \cw & \cw & \cw & \cw & \cw & \cw & \cw & \cw \ar@{<=}[-4,0] & \push{\cdots}\cw & \cw \ar@{<=}[-6,0] & \cw \ar@{<=}[-7,0] & \cw
    }
    \]
    \caption{Example of a Quantum Phase Estimation circuit with a three-qubit unitary.}
\end{figure}
\subsection{The Hadamard Test}
Given a unitary operator $U$ and a quantum state $\ket{\psi}$, 
the Hadamard test \cite{Cleve:1998fj} provides an estimate of the complex quantity $\bra{\psi} U \ket{\psi} \in \mathbb{C}$. The procedure begins by preparing an ancilla qubit in the state $\ket{0}$ 
and applying a Hadamard gate $H$. A controlled-$U$ operation is then performed, with the ancilla serving as the control and the system in the state $\ket{\psi}$ as the target. Finally, a second Hadamard gate is applied to the ancilla, which is then measured in the computational ($Z$) basis.

\begin{figure}
\centering
\begin{subfigure}{0.47\linewidth}
\centering
\[
\Qcircuit @C=1em @R=1em {
\lstick{\ket{0}}     & \gate{H} & \ctrl{1} & \gate{H} & \meter & \qw \\
\lstick{\ket{\psi}}  & \qw      & \gate{U} & \qw      & \qw    & \qw \\
\lstick{\mathrm{c}:} & \cw      & \cw      & \cw      & \cw \ar@{<=} [-2, 0]& \cw
}
\]
\caption{Hadamard test for $\mathrm{Re}\!\bra{\psi}U\ket{\psi}$.}
\end{subfigure}
\hfill
\begin{subfigure}{0.47\linewidth}
\centering
\[
\Qcircuit @C=1em @R=1em {
\lstick{\ket{0}}     & \gate{H} & \ctrl{1} & \gate{S^\dagger} & \gate{H} & \meter & \qw \\
\lstick{\ket{\psi}}  & \qw      & \gate{U} & \qw              & \qw      & \qw    & \qw \\
\lstick{\mathrm{c}:} & \cw      & \cw      & \cw              & \cw      & \cw \ar @{<=}[-2,0] & \cw
}
\]
\caption{Hadamard test for $\mathrm{Im}\!\bra{\psi}U\ket{\psi}$ (using $S^\dagger$).}
\end{subfigure}
\caption{Hadamard tests with explicit classical readout $c$ from the ancilla.}
\label{fig:hadamard-tests}
\end{figure}
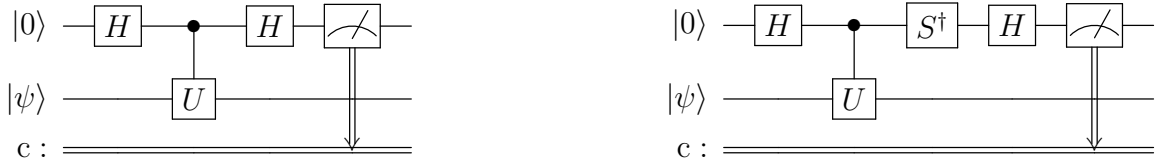

The expectation value of the ancilla's $Z$-measurement yields the \emph{real part} of $\bra{\psi} U \ket{\psi}$, while inserting an additional $S^\dagger$ gate before the final Hadamard allows access to the \emph{imaginary part}. The circuit for Hadamard test is depicted in figure \ref{fig:hadamard-tests}. Let us work out the circuit from left to right.

\begin{equation*}
    \begin{split}
        \ket{0}\ket{\psi} &\xrightarrow{H} \left(\frac{\ket{0} + \ket{1}}{\sqrt{2}}\right) \ket{\psi} \\
        &\xrightarrow{\text{C-}U} \frac{1}{\sqrt{2}}\left(\ket{0}\ket{\psi} + \ket{1}U\ket{\psi}\right) \\
        &\xrightarrow{H} \frac{1}{2}(\ket{0}(1 + U)\ket{\psi} + \ket{1}(1 - U)\ket{\psi})
    \end{split}
\end{equation*}
The measurement of the ancilla register then gives the probability of getting $0$ as
\begin{equation}
    P(0) = \frac{1}{4}\bra{\psi}(1 + U^\dagger)(1 + U)\ket{\psi} = \frac{1}{2}(1 + 2 \text{Re} \bra{\psi} U \ket{\psi})\,.
\end{equation}
The derivation for the imaginary part is analogous. We can also generalize the Hadamard test to extract the overlap $\bra{\phi} U \ket{\psi}$. Let $B$ be a unitary operator such that
\begin{equation}
    \ket{\phi} = B\ket{\psi}\,.
\end{equation}
Then the problem is reduced to finding the overlap $\bra{\psi}\tilde{U}\ket{\psi}$ where $\tilde{U} = B^\dagger U$ is also a unitary.

\section{Linear combination of unitaries}
In this section we introduce the \emph{linear combination of unitaries} (LCU) framework \cite{childs}. Quantum circuits natively implement unitary evolutions, whereas Hamiltonian simulation and observable estimation require access to general Hermitian operators. LCU addresses this mismatch by expressing a target operator as
\begin{equation}
\label{eq:unitencoding}
    H=\sum_{j}\alpha_j U_j,
\end{equation}
a weighted sum of efficiently implementable unitaries $U_j$. Together with standard state preparation and selection primitives, this yields a block encoding of $H$, enabling unitary-only procedures to simulate $e^{-iHt}$ and to estimate expectation values. In lattice field theory, one frequently encounters operators of the form \eqref{eq:unitencoding} where $U_j$ are Pauli strings and $c_j$ are real coefficients. Moreover, the field operators $\phi \sim a + a^\dagger$ and $\pi \sim i (a^\dagger - a)$ are non-unitary. In order to be able to use these operators for Hadamard tests or block encodings, we must represent them as linear combinations of unitaries. Thus, the LCU framework is indispensable for simulating QFT on a quantum computer as
\begin{enumerate}
    \item it allows Hamiltonians (expressed as sums of Pauli strings) to be implemented in Hamiltonian simulation algorithms;
    \item it allows operator insertions at different times (like $\phi(t) \phi(0)$) to be implemented in Hadamard tests for correlators;
    \item it provides a standard language (SELECT and PREP oracles) which is now adopted in frameworks like \texttt{Qiskit} and \texttt{PennyLane}.
\end{enumerate}
\subsection{The LCU algorithm}
We have devoted this section to work out the algebraic details of the LCU algorithm. Suppose that we want to implement
\begin{equation}
    A = \sum_{j = 0}^{m - 1} \alpha_j U_j\,,
\end{equation}
with coefficients $\alpha_j \geqslant 0$ and each $U_j$ unitary. Note that such a decomposition for any Hermitian $A$ is always possible because Pauli strings $\{\mathbf{1}, \vec{\sigma}\}^{\otimes n}$ form a complete basis for $2^n \times 2^n$ Hermitian matrices. The coefficients $c_j$ could in general be complex but we can always absorb the phase into the definition of the unitary.
\paragraph{Step 1: Ancilla preparation (PREP).} Prepare an ancilla register in the state
\begin{equation}
    \text{PREP} \ket{0} = \frac{1}{\sqrt{\Lambda}}\sum_{j = 0}^{m - 1}\sqrt{\alpha_j}\ket{j}\,, \quad \Lambda = \sum_j \alpha_j\,.
\end{equation}
Here $\ket{0}$ is a shorthand for the $m$-dimensional computational basis state $\ket{0 \cdots 0}$.
\paragraph{Step 2: SELECT operation.} Apply the block-diagonal unitary
\begin{equation}
    \text{SELECT}(U) = \sum_{j = 0}^{m - 1} \ket{j} \bra{j} \otimes U_j\,,
\end{equation}
which applies the correct unitary $U_j$ conditional on the ancilla label.
\paragraph{Step 3: Uncompute. } Apply $\text{PREP}^\dagger$ to the ancilla. The resulting joint state is
\begin{equation}
    (\text{PREP}^\dagger \otimes I) \text{SELECT}(U) (\text{PREP} \otimes I) \ket{0} \ket{\psi} = \frac{1}{\Lambda} \ket{0} \otimes A \ket{\psi} + \ket{0_{\perp}}\,,
\end{equation}
where we have collectively denoted the subspace orthogonal to $\ket{0}$ by $\ket{0_\perp}$.
\paragraph{Step 4: Post-selection or amplitude amplification. } If the ancilla is measured in the state $\ket{0}$, the system is left in $A\ket{\psi} / \Lambda$. The success probability is
\begin{equation}
    p_{\text{succ}} = \frac{|| A\ket{\psi}||^2}{\Lambda^2}\,.
\end{equation}
Amplitude amplification can be used to boost this probability.



\subsection{Explicit example: the field operator $\phi = a + a^\dagger$}
Consider the truncated Fock space $\mathcal{H}_d = \text{span}\{\ket{n} : n = 0\,, 1\,, \cdots \,, d-1\}$. In this basis the annihilation operator is
\begin{equation}
    a_d = \sum_{n = 1}^{d - 1} \sqrt{n} \ket{n - 1} \bra{n} = \begin{bmatrix}
        0 & 1 & 0 & \cdots & 0 \\
        0 & 0 & \sqrt{2} & \cdots & 0 \\
        \vdots & & & \ddots & \vdots \\
        0 & \cdots & 0 & 0 & \sqrt{d - 1} \\
        0 & 0 & \cdots & 0 & 0
    \end{bmatrix}
\end{equation}
To implement this Hamiltonian we will first introduce unary encoding.
\paragraph{Unary encoding and hop operators. } In unary encoding, we represent $\ket{n}$ by a string of $d$ qubits with a single $1$ at site $j = n + 1$, i.e.
\begin{equation}
    \text{unary}(n) = 2^n\,.
\end{equation}
For example for $d = 4$, we have the following
\begin{equation}
    \ket{0} \mapsto \ket{0001}\,, \quad \ket{1} \mapsto \ket{0010}\,, \quad \ket{2} \mapsto \ket{0100}\,, \quad \ket{3} \mapsto \ket{1000}\,.
\end{equation}
Therefore, in this picture, the operator $\ket{n - 1}\bra{n}$ is a map from the $2^n$ bitstring to the $2^{n-1}$ bitstring, i.e., it takes the local two-qubit pattern
\begin{equation}
    \ket{10}_{n, n-1} \mapsto \ket{01}_{n,n-1}\,,
\end{equation}
and annihilates all other patterns. This is exactly the action of the operator
\begin{equation}
    \sigma_-^{(n)} \sigma_+^{(n-1)}\,,
\end{equation}
where $\sigma_+ = (X + i Y) / 2$ and $\sigma_- = (X- i Y) / 2$. Here $\sigma_-^{(n)}$ lowers qubit at $n$ from $1$ to $0$ and $\sigma_+^{(n-1)}$ raises qubit at $n-1$ from $0$ to $1$. Acting together, the implement the hop $\ket{10}_{n, n-1} \mapsto \ket{01}_{n, n-1}$. Expansion of this product gives the Pauli string
\begin{equation}
    \sigma_-^{(n)}\sigma_+^{(n-1)} = \frac{1}{4} (X_{n} X_{n - 1} + Y_{n} Y_{n - 1}+ i X_{n} Y_{n - 1} - i Y_{n} X_{n - 1})\,,
\end{equation}
so that each hop operator is a linear combination of four Pauli-string unitaries. Multiplying by $n$ and summing over $n = 1\,, \cdots \,, d - 1$ reconstructs the full annihilation operator $a_d$ in unary encoding.

\paragraph{$d = 3$ case. } For three levels (two hops) the annihilation operator is
\begin{equation}
    a_3 = \frac{1}{4}\left( X_1 X_2 + Y_1 Y_2 + i X_1 Y_2 - I Y_1 X_2 \right) + \frac{\sqrt{2}}{4} \left( X_2 X_3 + Y_2 X_3 + i X_2 Y_3 - i Y_2 X_3 \right)
\end{equation}
Each bracketed term is a sum for four Pauli-string unitaries; the coefficients being $1 / 4$ and $\sqrt{2} / 4$, respectively. This is an exact LCU decomposition of $a_3$ into (phased) Pauli strings.

\chapter{Encoding in QFT}
In this chapter we discuss how quantum fields are encoded on a quantum computer. There are two different encodings: the JLP (Jordan-Lee-Preskill) encoding where we deal with the field operators directly and the oscillator encoding where we deal with the creation and annihilation operators associated with the fields. 
\section{JLP encoding}\label{sec:JLPencoding}
In the Jordan-Lee-Preskill encoding \cite{JLP2012Science, JLP2011QFTarXiv}, the value of a field $\phi$ at each site is truncated to lie between $-\phi_{\text{max}}$ and $\phi_{\text{max}}$. An optimal choice for the value of $\phi_{\text{max}}$ for a given number of qubits $n_q$ in a register per site exists due to the Nyquist-Shannon sampling theorem~\cite{somma2016, Macridin_2018, PhysRevLett.121.110504, Klco_2019, PhysRevD.107.L031503, kane2022efficientquantumimplementation21}. We shall use twisted boundary conditions in field space to preserve the symmetry in the digitizations of $\phi$- and $\Pi$-space~\cite{Klco_2019, Zemlevskiy_2025}.

Assuming that we have a register of $n_q$ qubits for each site $j$, the digitized field is written as
\begin{equation}
  \phi_j = - \phi_{\text{max}} + \ell \delta_\phi\,, \quad \delta_\phi = \frac{2 \phi_{\text{max}}}{2^{n_q} - 1}\,, \ell \in [0, 2^{n_q} - 1]\,.
\end{equation}
The allowed conjugate momenta $k_\phi$ are distributed symmetrically around $0$
\begin{equation}
  k_{\phi, j} = -\frac{\pi}{\delta_\phi} + \left(\ell - \frac{1}{2}\right) \frac{2\pi}{2^{n_q}\delta_\phi}\,, \quad \ell \in [0, 2^{n_q} - 1]\,.
\end{equation}
The $2^{n_q}$ values of $\phi_j$ are represented by the corresponding $\ell$ bitstring in the computational basis. Therefore the $\phi$ eigenstates can be identified with the kets denoting the $\ell$s
\begin{equation}
  \ket{\phi_j = -\phi_{\text{max}} + \ell \delta_\phi} = \ket{\ell}\,. 
\end{equation}
A generic eigenvalue configuration is $\{\phi_j = - \phi_{\text{max}} + \ell_j \delta_\phi\}_{j = 0}^{N - 1}$ corresponding to the state $\bigotimes_{j = 0}^{N -1} \ket{\ell_j}_j$. The representation of the field operators within this encoding becomes particularly simple. For example, the field operator $\phi$ becomes
\begin{equation}
  \phi_j = - \frac{\phi_{\text{max}}}{2^{n_q} - 1}\sum_{\ell = 0}^{n_q - 1} 2^{\ell} Z_{n_q j + \ell}\,,
\end{equation}
where $Z_i$ denotes a string of operators with $Z$ acting on the $i$th site. The index $(n_qj + \ell)$ is a flattening which translates to the $\ell$th value of $\phi$ at position $j$. To see why this is true, consider the action of $\phi_j$ on the product state $\bigotimes_{j = 0}^{N - 1} \ket{\ell_j}$. $\phi_j$ acts only on the $j$th register with $n_q$ qubits, the other kets go along for the ride. So essentially, we need to compute the action
\begin{equation}
  \label{eq:philataction}
  - \frac{\phi_{\text{max}}}{2^{n_q} - 1}\sum_{k = 0}^{n_q - 1} 2^{k} Z_{k} \ket{\ell_j}  = - \frac{\phi_{\text{max}}}{2^{n_q} - 1} \sum_{k = 0}^{n_q - 1} 2^k b_k \ket{\ell_j} \,,
\end{equation}
where $b_k$ is $\pm 1$ depending on whether the $k$th bit in the binary representation of $\ell_j$ is $0$ or $1$. If $\ell_j$ has the binary representation $(\ell_j)_2 = a_{N-1} a_{N-2} \cdots a_1 a_0$, i.e.
\begin{equation}
  \ell_j = \sum_{k = 0}^{n_q - 1} 2^k a_k\,,
\end{equation}
then $b_k = 1 - 2 a_k$. Substituting this back in eq \eqref{eq:philataction}, we obtain
\begin{equation}
  \begin{split}
    - \frac{\phi_{\text{max}}}{2^{n_q} - 1}\sum_{k = 0}^{n_q - 1} 2^k Z_k \ket{\ell_j} &= - \frac{\phi_{\text{max}}}{2^{n_q} - 1}\sum_{k = 0}^{n_q - 1} 2^k (1 - 2 a_k) \ket{\ell_j} \\
                                                                                       &= - \frac{\phi_{\text{max}}}{2^{n_q} - 1}\left(2^{n_q} - 1 - 2 \sum_{k = 0}^{n_q - 1} 2^{k} a_k\right) \ket{\ell_j} \\
                                                                                       &= \left( - \phi_{\text{max}} + \ell_j \delta_\phi \right) \ket{\ell_j}\,,
  \end{split}
\end{equation}
where we have used the fact that $\delta_\phi = 2 \phi_{\text{max}} / (2^{n_q} - 1)$. Similarly, we can represent the conjugate momentum operator in $\Pi$-space by the operator
\begin{equation}
  \Pi_j = - \frac{\pi}{2^{n_q} \delta_\phi}\sum_{\ell = 0}^{n_q - 1} 2^{\ell} Z_{n_q j + \ell}\,.
\end{equation}
There are only two sets of non-commuting groups of operators in the Hamiltonian: the ones constructed out of $\phi$ and $\Pi$ respectively. Operators which are polynomial in $\phi$ are diagonal in the $\phi$ basis and straightforward to implement. For example, the evolution due to the operator $\phi^2$ is just a $Z-Z$ coupling
\begin{equation}
  \phi_j^2 = \left(\frac{\phi_{\text{max}}}{2^{n_q} - 1}\right)^2 \sum_{\ell, k = 0}^{n_q - 1} 2^{\ell + k} Z_{n_qj + \ell} Z_{n_q j + k}\,. 
\end{equation}
Likewise, a $\phi^4$ term in the Hamiltonian is implemented via a $ZZZZ$ term. Similarly, the $\Pi^2$ term is implemented as a $ZZ$ coupling in the conjugate momentum basis, after a quantum Fourier transform, where $\Pi$ is diagonal.
\subsection{0+1D QFT aside: the anharmonic oscillator}
Considering the $\phi^4$ theory on a single spatial site reduces it to a $0 + 1$ dimensional quantum mechanical system, namely the \emph{anharmonic oscillator}. If we denote the coupling strength of the quartic term by $\lambda$, then the Hamiltonian is given by
\begin{equation}
    \hat{H} = \frac{x^2}{2} + \frac{p^2}{2} + \lambda x^4\,.
\end{equation}
The position operator is analogous to the field operator considered in the previous section. However, since there is no longer a site label, the encoding is somewhat simplified
\begin{equation}
\label{eq:shojlpencodings}
\begin{split}
    x &= - x_{\text{max}} + \ell \delta_x \,, \quad \delta_x = \frac{2x_{\text{max}}}{2^{n_q} - 1}\,, \quad \ell \in \{0\,, 1\,, \cdots\,, 2^{n_q} - 1\}\,, \\
    p &= - \frac{\pi}{\delta_x} + \left(\ell - \frac{1}{2}\right) \frac{2\pi}{2^{n_q}\delta_x}\,, \quad \ell \in \{0\,, 1\,, \cdots \,, 2^{n_q} - 1\}\,.
\end{split}
\end{equation}
The representation of the field operator is
\begin{equation}
    \hat{x} = -\frac{x_{\text{max}}}{2^{n_q}-1}\sum_{\ell = 0}^{n_q - 1}2^{\ell} Z_{\ell}\,.
\end{equation}
The $\hat{x}^2$ operator is then
\begin{equation}
    \hat{x}^2 = \left(\frac{x_{\text{max}}}{2^{n_q} - 1}\right)^2 \sum_{\ell, k = 0}^{n_q - 1} 2^{\ell + k} Z_{\ell} Z_k\,.
\end{equation}
This operator is implemented as a two-site coupling. However, we do not need to sum all terms. Due to the symmetry under the exchange $\ell \leftrightarrow k$, it suffices to sum only over $\ell < k$ and multiply by a factor of 2. When $\ell = k$, the corresponding term reduces to a phase shift that affects only the overall dynamics. Nevertheless, during phase estimation, we must retain such constant global phases, since they contribute to the extracted eigenenergies. Therefore,
\begin{equation}
    \hat{x}^2 = \left(\frac{x_{\text{max}}}{2^{n_q} - 1}\right)^2 \sum_{\ell < k}^{n_q - 1} 2^{\ell + k} Z_{\ell} Z_k + C \mathbf{1}\,.
\end{equation}

The $\hat{x}^4$ term, similarly, is implemented as a four-site coupling
\begin{equation}
    \hat{x}^4 = \left(\frac{x_{\text{max}}}{2^{n_q} - 1}\right)^4 \sum_{i, j, k, \ell} 2^{i + j + k + \ell}Z_iZ_jZ_kZ_\ell\,.
\end{equation}
The sum above exhibits several symmetries. When all site indices are distinct, the expression is invariant under permutations of $(i, j, k, \ell)$, so it suffices to sum only over $i < j < k < \ell$ and multiply by a factor of $4!$. Next, there are $\binom{4}{2}$ cases in which exactly two indices are equal. These terms also possess a symmetry under the exchange of the remaining two indices, meaning we need only sum over either the upper or lower triangular part. Finally, configurations with three equal indices must also be included. Taking all such cases into account, the representation of the $x^4$ operator becomes
\begin{equation}
\begin{split}
    \hat{x}^4 &= \left(\frac{x_{\text{max}}}{2^{n_q} - 1}\right)^4 4!\sum_{i, j, k, \ell = 0}^{n_q - 1} 2^{i + j + k + \ell}Z_iZ_jZ_kZ_\ell + 2 \left(\frac{x_{\text{max}}}{2^{n_q} - 1}\right)^4 {}^4C_2 \sum_{i = 0}^{n_q - 1} \sum_{k < \ell, k \neq i \ell \neq i} 2^{2 i + k + l} Z_k Z_\ell \\ &+ \left(\frac{x_{\text{max}}}{2^{n_q} - 1}\right)^4 \, {}^4C_3 \sum_{i = 0}^{n_q - 1}\sum_{j \neq i}2^{3i + j} Z_i Z_j + C \mathbf{1}\,.
\end{split}
\end{equation}
Let us next see how to implement the $\hat{p}^2$ term. One way would be to impose periodic boundary conditions in the $x$-space and use the finite difference operator
\begin{equation}
    \bra{x} p^2 \ket{x} = \frac{1}{\delta_x^2} \begin{bmatrix}
        2 & - 1 & & & & -1 \\
        -1 & 2 & -1 & & & \\
         & -1 & 2 & -1 & & \\
         & & \ddots & \ddots & \ddots & \\
         & & & -1 & 2 & -1\\
        -1 & & & & -1 & 2
    \end{bmatrix}
\end{equation}
This operator has eigenvalues
\begin{equation}
    \hat{k}_\ell = \frac{4}{\delta_x^2}\sin^2 \left(\frac{k_\ell \delta_x}{2}\right) \,.
\end{equation}
Recall that this requires the Quantum Fourier Transform ($\qft$)
\begin{equation}
    \ket{k} = \frac{1}{\sqrt{N}}\sum_{j = 0}e^{\frac{2\pi i j k}{N}}\ket{j}
\end{equation}
However, notice that this is not the exact discrete analog of the continuum relation between position and momentum eigenkets
\begin{equation}
    \ket{p} = \frac{1}{\sqrt{2\pi}}\int_{-\infty}^{\infty} dx \, e^{i p x} \ket{x}\,.
\end{equation}
The discrete analogue for the above continuum Fourier transform is rather
\begin{equation}
    \ket{k} = \frac{1}{\sqrt{N}} \sum_{j = 0}^{N - 1}e^{\frac{2\pi i x_j p_k}{N}}\ket{j}\,,
\end{equation}
which involves the term $x_j p_k$ in the phase. The subscripts $j$ and $k$ should not be interpreted as lattice-site indices; rather, $x_j$ and $p_k$ denote the $j$th and $k$th discretized eigenvalues of the operators $\hat{x}$ and $\hat{p}$, respectively. Using the equations \eqref{eq:shojlpencodings} this product expands to
\begin{equation}
   x_j p_k = \frac{2\pi j k}{2^{n_q}} + j\pi\left(\frac{1}{2^{n_q}} - 1\right) + k \pi \left(\frac{1}{2^{n_q}} - 1\right) - \frac{\pi x_{\text{max}}}{2^{n_q}\delta_x}\,.
\end{equation}
The first term corresponds to the phase appearing in the discrete Fourier transform. The latter two terms, however, correspond to $Z$ rotations on the register applied before and after the Fourier transform. The final term is a constant phase that is cancelled after the subsequent application of the inverse operation. Denoting by $\widetilde{\text{QFT}}$ this combined operation, we have
\begin{equation}
    \widetilde{\text{QFT}}\,\hat{p}\,\widetilde{\text{QFT}}^\dagger = - \frac{\pi}{2^{n_q}\delta_x}\sum_{\ell = 0}^{n_q - 1} 2^{\ell} Z_{\ell}\,.
\end{equation}

\section{Harmonic-oscillator basis encoding}
An alternative to the position-space (JLP) digitization is to represent the field degree of freedom in a truncated harmonic-oscillator basis. Instead of sampling the field $\phi$ on a spatial grid, one expands it in the energy eigenbasis of the quadratic Hamiltonian
\begin{equation}
    \hat{H}_{\text{HO}} = \frac{\hat{\Pi}^2}{2} + \frac{\hat{\phi}^2}{2}\,,
\end{equation}
whose eigenstates $\{\ket{n}\}_{n=0}^\infty$ satisfy
\begin{equation}
    \hat{a}^\dagger \hat{a} \ket{n} = n \ket{n}\,,
\end{equation}
with the canonical relations
\begin{equation}
    \hat{\phi} = \frac{1}{\sqrt{2}}\left(\hat{a} + \hat{a}^\dagger\right), \qquad
    \hat{\Pi} = \frac{i}{\sqrt{2}}\left(\hat{a}^\dagger - \hat{a}\right).
\end{equation}
Truncating this infinite ladder at occupation number $n_{\max}$ yields a finite Hilbert space of dimension $(n_{\max}+1)$, suitable for encoding into $\lceil \log_2 (n_{\max}+1) \rceil$ qubits. In this representation, the free (quadratic) part of the Hamiltonian is diagonal, while interaction terms such as $\phi^4$ are expressed as low-order polynomials in the ladder operators $\hat{a}$ and $\hat{a}^\dagger$. This basis therefore preserves the structure of the low-energy spectrum and is especially efficient when the relevant states are dominated by low excitations around the vacuum.

The single-site Hilbert space in this encoding is thus
\[
\mathcal{H}_{\rm SHO}^{(n_{\max})} = \mathrm{span}\,\{\ket{n}\}_{n=0}^{n_{\max}},
\]
on which the ladder operators act as
\[
\hat a\ket{n}=\sqrt{n}\ket{n-1}, \qquad 
\hat a^\dagger\ket{n}=\sqrt{n+1}\ket{n+1}.
\]
Defining
\[
\hat{x} = \frac{\hat{a}^\dagger + \hat{a}}{\sqrt{2}}, \qquad
\hat{p} = \frac{i(\hat{a}^\dagger - \hat{a})}{\sqrt{2}},
\]
the anharmonic-oscillator Hamiltonian takes the form
\[
\hat{H} = \frac{\hat{p}^2}{2} + \frac{\hat{x}^2}{2} + \lambda \hat{x}^4
         = \hat{a}^\dagger \hat{a} + \frac{1}{2} + \lambda \hat{x}^4.
\]
This formulation highlights that the SHO basis diagonalizes the free part of the Hamiltonian, while the quartic interaction introduces sparse couplings among nearby number states.

However, this encoding becomes less efficient for strongly coupled or highly localized field configurations, where a position-space truncation such as the JLP scheme provides better resolution. The two encodings therefore complement each other: JLP is natural for local operators and spatial interactions, whereas the SHO basis offers a compact and physically transparent representation for weakly interacting or nearly harmonic systems.
\paragraph{Parity sectors.}
Define the parity operator $\hat\Pi = e^{i\pi \hat a^\dagger \hat a}$, with $\hat\Pi\ket{n}=(-1)^n\ket{n}$. Since $\hat x$ is odd under parity and $\hat x^2,\hat x^4$ are even, the Hamiltonian with a purely quartic interaction preserves parity:
\[
[\hat H,\hat \Pi]=0.
\]
Therefore the Hilbert space decomposes into two invariant subspaces,
\[
\mathcal{H}_{\text{even}}=\mathrm{span}\{\ket{0},\ket{2},\ket{4},\dots\},\qquad
\mathcal{H}_{\text{odd}}=\mathrm{span}\{\ket{1},\ket{3},\ket{5},\dots\}.
\]
This block-diagonalization halves the effective problem size and is especially useful for finding the ground state (which lies in the even sector).

\paragraph{Matrix elements.}
With the above conventions, the nonzero matrix elements of $\hat x^2$ in the number basis are
\begin{align}
\langle n|\hat x^2|n\rangle &= n+\tfrac12,\\
\langle n|\hat x^2|n+2\rangle &= \langle n+2|\hat x^2|n\rangle
= \tfrac12\sqrt{(n+1)(n+2)},
\end{align}
and all other entries vanish. Thus $\hat x^2$ only connects states with $\Delta n=0,\pm 2$ (parity-preserving).

For the quartic term $\hat x^4$, the band structure extends to $\Delta n=0,\pm 2,\pm 4$ with
\begin{align}
\langle n|\hat x^4|n\rangle &= \tfrac{3}{4}\bigl(2n^2+2n+1\bigr),\\
\langle n|\hat x^4|n+2\rangle &= \langle n+2|\hat x^4|n\rangle
= \tfrac12(2n+3)\sqrt{(n+1)(n+2)},\\
\langle n|\hat x^4|n+4\rangle &= \langle n+4|\hat x^4|n\rangle
= \tfrac14\sqrt{(n+1)(n+2)(n+3)(n+4)},
\end{align}
and zeros otherwise. These selection rules make the SHO representation sparse and parity-block diagonal, which is advantageous both analytically and numerically.

\paragraph{Truncation and accuracy.}
The truncation level $n_{\max}$ controls the representation error. For weak to moderate $\lambda$ and low-lying states, the SHO basis is efficient because the spectral weight is concentrated at small $n$. As $\lambda$ increases or highly excited states are targeted, $n_{\max}$ must be increased; monitoring leakage at the cutoff (e.g., $\sum_{n\approx n_{\max}-\delta} |\psi_n|^2$) provides a practical convergence check. Working in a fixed parity sector further improves stability and reduces cost.

\paragraph{Mapping to qubits.}
Encoding $\{\ket{n}\}_{n=0}^{n_{\max}}$ into qubits can be done via:
\begin{itemize}
  \item \textbf{Binary encoding:} use $q=\lceil \log_2(n_{\max}+1)\rceil$ qubits with the usual binary map of $n$ to bitstrings. Operators like $\hat x^2$ and $\hat x^4$ become sparse banded matrices that can be implemented via LCU/qubitization or synthesized with controlled number-raising/lowering primitives.
  \item \textbf{Unary/compact encodings:} use $(n_{\max}+1)$ qubits with a single excitation; this makes $\hat a,\hat a^\dagger$ strictly local (nearest-neighbor in Hamming graph) at the expense of more qubits.
\end{itemize}
Because $\hat x^4$ only couples $\Delta n=0,\pm 2,\pm 4$, Trotter steps can be organized in few commuting sub-blocks within each parity sector, or block-encoded for qubitization. The SHO basis also diagonalizes the quadratic part, so imaginary-time/adiabatic or variational initialization can leverage low-$n$ priors directly.

\chapter{State preparation}
One of the main aims in these lectures is to study scattering in QFT. For this we need to specify how the scattering states are prepared and we need to set this up nonperturbatively. We will discuss the JLP approach and the newer W-state approach. 

\section{JLP–style preparation  in lattice \texorpdfstring{$\phi^4$}{phi4} theory}

In the Jordan–Lee–Preskill (JLP) approach each lattice site $j$ is truncated to a finite
oscillator Hilbert space spanned by number states $|n_j\rangle$, with $n_j=0,\dots,n_{\max}$.
We encode $n_j$ in binary on $n_q=\lceil \log_2(n_{\max}+1)\rceil$ qubits per site.  
In this basis the canonical variables
\[
\phi_j=\frac{1}{\sqrt{2\omega_0}}\,(a_j+a_j^{\dagger}),\qquad
\pi_j=-\,i\sqrt{\frac{\omega_0}{2}}\,(a_j-a_j^{\dagger})
\]
act as sparse matrices, and the full lattice Hamiltonian
\begin{equation}
    H_{\phi^4}
    =\sum_{j=1}^N a\left[\frac12\pi_j^2+\frac12 m_0^2 \phi_j^2
       +\frac{1}{2a^2}(\phi_{j+1}-\phi_j)^2
       +\frac{\lambda_0}{4!}\phi_j^4\right]
\end{equation}
is a sum of few–body operators acting on these encoded number registers.

\medskip

\paragraph{(1) Preparing the interacting vacuum.}
We follow the same adiabatic–state–preparation (ASP) method used earlier for the
anharmonic oscillator.  Introduce an easy Hamiltonian
\begin{equation}
    H_{\rm dec}=\sum_j a\left[\frac12\pi_j^2+\frac12\omega_0^2\phi_j^2\right],
\end{equation}
whose ground state is the product state $\bigotimes_j |0_j\rangle$.
We then interpolate
\begin{equation}
    H(s)=(1-s)\,H_{\rm dec}+s\,H_{\phi^4},\qquad  s\in[0,1],
\end{equation}
and Trotterize the adiabatic evolution $U_{\rm ASP}=\mathcal T e^{-i\int_0^1 H(s)\,ds}$.
Away from criticality the spectral gap along the path is finite, and
\begin{equation}
    U_{\rm ASP}\,\Big(\bigotimes_j |0_j\rangle\Big)\;\approx\;|\Omega\rangle,
\end{equation}
the interacting vacuum of the lattice $\phi^4$ theory.

\medskip

\paragraph{(2) Free–theory normal modes and one–particle states.}
To create clean scattering states it is convenient to begin at $\lambda_0=0$.
The quadratic part of the Hamiltonian is diagonalized by lattice momenta
$q=\tfrac{2\pi}{Na}n$:
\begin{equation}
    H_0=\sum_q \omega_q \left(b_q^{\dagger}b_q+\tfrac12\right),\qquad
    \omega_q^2=m_0^2+\frac{4}{a^2}\sin^2\frac{qa}{2}.
\end{equation}
Here the mode operators are
\begin{equation}
    b_q=\frac{1}{\sqrt{N}}\sum_j e^{-iq x_j}
    \left(\sqrt{\frac{\omega_q}{2}}\;\varphi_j
          +\frac{i}{\sqrt{2\omega_q}}\;\pi_j\right),
\end{equation}
and act on the JLP–encoded number basis through the local $a_j,a_j^\dagger$.

The free one–particle momentum eigenstate is simply
$|1_q\rangle=b_q^{\dagger}|0_{\rm free}\rangle$.
JLP show that small unitaries of the form
\begin{equation}
    U_q(\theta)=\exp\!\big[-i\theta\,X_q\big],\qquad
    X_q=\frac{1}{\sqrt{2}}\,(b_q^{\dagger}+b_q),
\end{equation}
can be implemented as linear combinations of Pauli strings (block–encoded via
an ancilla).  
Acting on the free vacuum,
\begin{equation}
    U_q(\theta)\,|0_{\rm free}\rangle
    =\cos\theta\,|0_{\rm free}\rangle
     -i\sin\theta\,|1_q\rangle,
\end{equation}
so a rotation by $\theta=\pi/2$ prepares the single–particle state $|1_q\rangle$.

\medskip

\paragraph{(3) Two–particle free state $|1_p,1_{-p}\rangle$.}
A two–particle incoming state in the free theory is
\begin{equation}
    |2;p,-p\rangle=b_p^{\dagger}b_{-p}^{\dagger}|0_{\rm free}\rangle.
\end{equation}
On the quantum computer we obtain this by two successive rotations:
\begin{align}
    &\text{Prepare } |0_{\rm free}\rangle, \\
    &U_p(\tfrac{\pi}{2})\,|0_{\rm free}\rangle = |1_p\rangle,\\[3pt]
    &U_{-p}(\tfrac{\pi}{2})\,|1_p\rangle = |1_p,1_{-p}\rangle,
\end{align}
where the JLP block–encoding ensures that each $U_q$ acts only in the
$\{0,1\}$–particle subspace of that mode.

For improved localization one encloses the particles in Gaussian wavepackets,
\begin{equation}
    B_f^{\dagger}=\sum_q f(q-p_0)\,b_q^{\dagger},\qquad
    B_g^{\dagger}=\sum_q g(q+p_0)\,b_q^{\dagger},
\end{equation}
and prepares $B_f^{\dagger}B_g^{\dagger}|0_{\rm free}\rangle$
via rotations generated by $X_f=(B_f^{\dagger}+B_f)/\sqrt{2}$ and
$X_g=(B_g^{\dagger}+B_g)/\sqrt{2}$.

\medskip

\paragraph{(4) Dressing the free state into the interacting theory.}
Once the free two–particle state is prepared, we adiabatically turn on the
interaction:
\begin{equation}
    H(s)=H_0+s\,V_{\phi^4},\qquad
    V_{\phi^4}=\sum_j a\,\frac{\lambda_0}{4!}\,\phi_j^4.
\end{equation}
Provided the two particles are initially well separated, the evolution
\[
    |\Psi(s)\rangle=\mathcal T e^{-i\int_0^s H(s')\,ds'}\,|2;p,-p\rangle
\]
follows the lowest two–particle state in the $(p,-p)$ momentum sector.
At $s=1$ this yields the incoming \emph{interacting} two–particle scattering
state.  Subsequent real–time evolution under $H_{\phi^4}$ generates the
scattering process, from which phase shifts or transition probabilities can be
extracted by measuring final–state occupation numbers in the free–mode basis.

\medskip

This JLP protocol cleanly separates three tasks: (i) prepare the vacuum by ASP;
(ii) populate the desired normal modes in the free theory via small rotations;
(iii) dress these free states into interacting ones by slowly turning on the
quartic coupling.  In this way the quantum computer never needs explicit
access to the interacting creation operators; the adiabatic path automatically
selects the correct incoming two–particle eigenstate of the full theory.

\section{W-state wave-packet preparation}

In the Ising field theory case, there is an efficient way to prepare states. This was accomplished in \cite{Farrell2025WStates} which we will discuss next. After this step, we can do ASP to get the true interacting states. However, note that in \cite{Farrell2025WStates}, they use ADAPT-VQE using an MPS circuit simulator on classical hardware to keep the depth down. So essentially, they figure out $U(\bf \theta)$ to apply after the W-state step and then continue on the quantum hardware. Further, we will only review the probabilistic mid-circuit measurement approach; a student project uploaded on github examines the deterministic Unitary Circuit method and ADAPT-VQE techniques in \cite{Farrell2025WStates}.
\subsection{Step-1}

Our target single--particle state on an \(N\)-site ring (sites \(j=1,\dots,N\), spacing \(a\)) with mean momentum \(p\) and envelope \(f_j\) is
\[
|\psi_{f,p}\rangle \;=\; \sum_{j=1}^{N} f_j\,e^{ipaj}\,|\,\cdots 1_j \cdots \rangle,
\qquad
\sum_{j}|f_j|^{2}=1,
\]
where \(|\,\cdots 1_j \cdots \rangle\) denotes “site \(j\) excited, all others in \(|0\rangle\)”. Step-1 prepares, in constant depth, a coherent state whose \emph{one--excitation slice} is exactly the uniform W state; later we will add a momentum phase ramp and shaping with the $f_j$ factors.

\paragraph{A single depth--1 layer.}
Starting from \(|0\rangle^{\otimes N}\), apply the same small rotation to every site:
\[
|\Psi_0\rangle \;=\; \bigotimes_{j=1}^{N}\!\big(\cos\alpha\,|0\rangle_j + \sin\alpha\,|1\rangle_j\big).
\]
This is one layer of identical single--qubit gates (e.g.\ \(R_y(2\alpha)\)).

\paragraph{Decomposition by Hamming weight\protect\footnote{The Hamming weight for binary is simply the count of 1's in a string.}.}
Expand the tensor product and group basis strings by their number \(K\) of excitations. It is convenient to introduce the Dicke states,
\[
|D^N_K\rangle \;=\; \frac{1}{\sqrt{\binom{N}{K}}}\sum_{\substack{\text{strings of length }N\\ \text{with }K\text{ ones}}}\!\!\!\!\!\!\!\!\!\! |\,\text{string}\,\rangle,
\]
which form an orthonormal basis for the permutation--symmetric subspace at fixed \(K\). Then
\[
|\Psi_0\rangle \;=\; \sum_{K=0}^{N} (\cos\alpha)^{N-K}(\sin\alpha)^K \sqrt{\binom{N}{K}}\;|D^N_K\rangle.
\]
In particular, the \(K=1\) Dicke state is the uniform W state
\[
|D^N_1\rangle \;=\; \frac{1}{\sqrt{N}}\sum_{j=1}^{N} |\,\cdots 1_j \cdots \rangle .
\]

\paragraph{Exact weights and a useful choice of angle.}
Because the Dicke sectors are orthogonal, the probability to have exactly \(K\) excitations after a computational--basis measurement is
\[
\Pr[K]\;=\;\binom{N}{K}\,(\sin^2\alpha)^K\,(\cos^2\alpha)^{N-K}.
\]
We parameterize the small angle by
\[
\sin^2\alpha \;=\; \frac{\lambda}{N},
\]
with a fixed \(\lambda>0\) that we may tune. In the large--\(N\) limit with \(\lambda\) fixed this binomial distribution approaches a Poisson law,
\[
\Pr[K]\;\approx\; e^{-\lambda}\frac{\lambda^K}{K!}\qquad (N\to\infty,\; \sin^2\alpha=\lambda/N).
\]
To see this, simply note that for $N\gg 1$, $\binom{N}{K}\rightarrow N^k/k!$ while the rest of the factors read $(\lambda/N)^k (1-\lambda/N)^N (1-\lambda/N)^{-k}\approx (\lambda/N)^k e^{-\lambda/N}$.
The special case \(\lambda=1\) is particularly convenient: it maximizes the raw one--excitation weight \(\Pr[K{=}1]\) and will be our default.

\paragraph{The one--excitation slice is exactly W.}
Projecting \( |\Psi_0\rangle \) onto the \(K=1\) subspace yields a \emph{pure} Dicke state with amplitude
\[
\langle D^N_1|\Psi_0\rangle \;=\; (\cos\alpha)^{N-1}(\sin\alpha)\sqrt{N},
\]
so the corresponding probability weight is
\[
\Pr[K{=}1]\;=\; N\,\sin^2\alpha\,(\cos^2\alpha)^{N-1}.
\]
With \(\sin^2\alpha=\lambda/N\) this becomes \(\Pr[K{=}1]\approx \lambda e^{-\lambda}\), maximized at \(\lambda=1\) where \(\Pr[K{=}1]\approx e^{-1}\). Thus, after a single depth--1 layer, the coherent state already contains the \emph{exact} uniform W component we need, with a tunable overall weight.

\paragraph{Optional: seeding a shaped envelope.}
If a nonuniform envelope is desired from the outset, choose site--dependent angles \(\alpha_j\) so that
\[
\sin^2\alpha_j \;=\; \lambda\,|f_j|^2,\qquad \sum_j |f_j|^2=1.
\]
Then
\[
\bigotimes_{j}\!\big(\cos\alpha_j\,|0\rangle_j+\sin\alpha_j\,|1\rangle_j\big)
\]
has, in its \(K=1\) slice, a superposition proportional to \(\sum_j f_j\,|\,\cdots 1_j\cdots\rangle\). The overall one--excitation weight is again \(\approx \lambda e^{-\lambda}\) for large \(N\), while the mean and variance of \(K\) follow from the same binomial analysis (we leave those straightforward formulas to the reader to record where convenient).

\medskip
This completes Step 1: a depth--1 preparation that seeds the exact W component (or a shaped variant) with controllable weight. In the next step we will imprint the momentum ramp \(e^{ipaj}\) by a single layer of local \(Z\) phases and then discuss the parity sieve and subsequent “cleaning’’ that isolate the interacting one--particle packet.

\subsection{Step 2 (imprinting momentum)}

The goal of this step is to turn the uniform one--excitation slice produced in Step 1 into a plane--wave one--excitation with mean momentum \(p\). Concretely, for each basis vector \(|\,\cdots 1_j \cdots\rangle\) in the \(K=1\) sector we want a phase factor \(e^{ipaj}\). We will follow the discussion in \cite{Farrell2025WStates}.

\paragraph{A single layer of local \(Z\)–rotations.}
Apply on every site \(j\) a \(Z\)–rotation with angle \(\phi_j\),
\[
U \;=\; \bigotimes_{j=1}^{N} R_z(\phi_j),\qquad
R_z(\phi_j)\;=\;\exp\!\Big(-\frac{i}{2}\,\phi_j\,\sigma^z_j\Big).
\]
Let \(|x_1x_2\ldots x_N\rangle\) be a computational basis state with \(x_j\in\{0,1\}\). Using
\(
R_z(\phi_j)|0\rangle_j=e^{-i\phi_j/2}|0\rangle_j
\)
and
\(
R_z(\phi_j)|1\rangle_j=e^{+i\phi_j/2}|1\rangle_j
\),
one may factor each site’s phase as
\[
e^{\pm i\phi_j/2}\;=\;e^{-i\phi_j/2}\,\big(e^{i\phi_j}\big)^{x_j}.
\]
Multiplying over all sites gives
\[
U\,|x_1\ldots x_N\rangle
\;=\;
\exp\!\Big(-\frac{i}{2}\sum_{j=1}^{N}\phi_j\Big)\;
\exp\!\Big(i\sum_{j=1}^{N}\phi_j x_j\Big)\;
|x_1\ldots x_N\rangle.
\]
Writing \(S=\{j:\,x_j=1\}\) for the set of excited positions (of size \(K=\sum_j x_j\)), this is
\[
U\,|x_1\ldots x_N\rangle
\;=\;
e^{-\,\frac{i}{2}\sum_j \phi_j}\;\cdot\;e^{\,i\sum_{j\in S}\phi_j}\;|x_1\ldots x_N\rangle.
\]
The first factor is a global phase, independent of the bitstring, and can be dropped. In the \(K=1\) sector, \(S=\{j\}\) for some site \(j\), so the state picks up precisely \(e^{i\phi_j}\).

\paragraph{Choosing the ramp.}
Set \(\phi_j = p\,a\,j\) (with sites labeled \(j=1,\dots,N\); one may also use \(j=0,\dots,N\!-\!1\)). Then each one–excitation basis vector \(|\,\cdots 1_j \cdots\rangle\) acquires the desired factor \(e^{ipaj}\). Acting on the uniform W slice from Step-1 therefore yields the plane–wave W component at momentum \(p\).

A harmless implementation detail is that \(\sum_j\phi_j\) contributes only a global phase. If one prefers to eliminate it identically, center the ramp, e.g.
\[
\phi_j \;=\; p\,a\Big(j-\frac{N+1}{2}\Big),
\]
which leaves all relative phases unchanged.

\paragraph{Shaped envelopes.}
If Step-1 was seeded with a nonuniform envelope so that the \(K=1\) slice is proportional to \(\sum_j f_j|\,\cdots 1_j \cdots\rangle\), the same phase layer produces
\[
\sum_{j=1}^{N} f_j\,e^{ipaj}\,|\,\cdots 1_j \cdots\rangle,
\]
i.e.\ a wave packet with envelope \(f_j\) and mean momentum \(p\). Because the phases have unit modulus, normalization is unaffected.

\medskip
After this single depth–1 layer of local \(Z\)–rotations, the odd–parity one–excitation component prepared in Step A has become a momentum–selected W (or shaped) state. In the next step we perform the mid–circuit parity sieve and subsequent “cleaning’’ to isolate the interacting one–particle packet at the chosen momentum.

\subsection{Step 3 (mid--circuit parity sieve)}

The aim of this step is to \emph{project onto odd Hamming weight} while preserving the positional coherence created in Steps A–B. The observable we measure is the global \(Z\)–parity
\[
\Pi_Z \;=\; \prod_{j=1}^{N}\sigma^z_j \;=\; (-1)^K,
\]
where \(K\) is the number of excited sites. This measurement distinguishes only “even’’ vs “odd’’ \(K\), learning \emph{nothing} about the locations of the excitations; the phase ramp \(e^{ipaj}\) therefore survives intact in the \(K=1\) slice.

It is convenient to describe the measurement by projectors \(P_{\mathrm{even}}=\tfrac12(\mathbb{I}+\Pi_Z)\) and \(P_{\mathrm{odd}}=\tfrac12(\mathbb{I}-\Pi_Z)\). Acting on the Step–1 state (after the momentum phases of Step-2), the post–selected odd–parity state is \( |\Psi_{\mathrm{odd}}\rangle = P_{\mathrm{odd}}|\Psi\rangle/\sqrt{p_{\mathrm{odd}}}\), where the heralding probability is \(p_{\mathrm{odd}}=\Pr[\text{odd}]\). With the seeding choice \(\sin^2\alpha=\lambda/N\) from Step-2 (or its site–dependent variant satisfying \(\sum_j \sin^2\alpha_j=\lambda\)), the Hamming–weight distribution is well approximated by Poisson\((\lambda)\) in the large–\(N\) limit, giving
\[
\Pr[\text{odd}] \;=\; e^{-\lambda}\sinh\lambda,
\qquad
\Pr[K{=}1\,|\,\text{odd}] \;=\; \frac{\lambda}{\sinh\lambda}.
\]
At the natural choice \(\lambda=1\) the odd–parity heralding rate is \(e^{-1}\sinh 1 \approx 0.432\), and, \emph{conditional on passing}, roughly \(85\%\) of the amplitude already sits in the desired \(K=1\) plane–wave component. In particular, the one–excitation slice after the sieve is precisely
\[
\frac{1}{\sqrt{N}}\sum_{j=1}^{N} e^{ipaj}\,|\,\cdots 1_j \cdots\rangle
\quad
\text{(or \(\sum_j f_j e^{ipaj}|\,\cdots 1_j \cdots\rangle\) if Step A used a shaped envelope).}
\]

\paragraph{A compact mid–circuit parity circuit (CNOT accumulation).}
One standard realization uses a single ancilla qubit:
prepare the ancilla in \(|0\rangle\); apply CNOTs from each data qubit (control) to the ancilla (target); measure the ancilla in the \(Z\) basis mid–circuit; keep runs with outcome “1’’ (odd parity) and discard the rest. Because the data qubits are never targeted by these CNOTs, their positional coherence and momentum phases are unaffected.

\medskip
\begin{figure}[t]
\centering
\[
\Qcircuit @C=2.2em @R=.9em {
\lstick{q_1}           & \ctrl{3} & \qw      & \qw      & \qw     & \qw \\
\lstick{q_2}           & \qw      & \ctrl{2} & \qw      & \qw     & \qw \\
\lstick{q_3}           & \qw      & \qw      & \ctrl{1} & \qw     & \qw \\
\lstick{a:\,|0\rangle} & \targ    & \targ    & \targ    & \meter  & \cw
}
\]
\caption{Mid--circuit parity measurement by CNOT accumulation (illustrated for three data qubits; the pattern extends to all \(N\) sites). The ancilla ends in \(|K \bmod 2\rangle\) and is measured; outcome \(1\) heralds odd parity. The data qubits are not disturbed apart from a global projector onto the odd subspace.}
\label{fig:parity-circuit}
\end{figure}
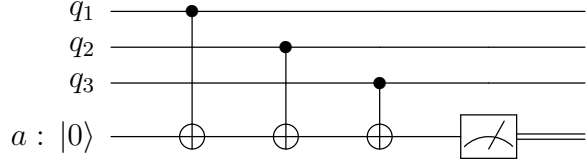

\paragraph{Remarks.}
\begin{enumerate}
\item
The alternative “phase–kickback’’ version prepares the ancilla in \(|+\rangle\), applies C-$Z$ from each data qubit onto the ancilla, and measures the ancilla in the \(X\) basis, with \(|-\rangle\) heralding odd parity. Either implementation measures \(\Pi_Z\) and therefore commutes with the local \(Z\)–phase layer of Step B, guaranteeing that the momentum profile is preserved. In what follows we simply condition on the odd–parity outcome and proceed to the cleaning step that removes the small residual \(K\ge 3\) admixture while keeping the total momentum fixed.
\item An important point which may cause some confusion. If we start with $N$ qubits and need to do a parity check on a string of size $N$ then we will need $O(N)$ CNOTs and this whole exercise may seem unnecessary. The point is that the initial state only has support over $b\ll N$ sites and hence the parity check operation can be done efficiently.  
\end{enumerate}

\subsection*{Step 4: Cleaning}

After Steps 1--3 the post--selected state lies in the odd \(Z\)–parity sector and carries total lattice momentum \(p\). Its one–excitation slice already has the desired phase ramp (and optional envelope), but a small admixture of higher odd Hamming weights (\(K=3,5,\dots\)) remains at the same total momentum. The purpose of this step is to suppress that admixture while \emph{preserving} momentum \(p\) and \(Z\)–parity, yielding the interacting one–particle packet.

\paragraph{Symmetry protection.}
Let \(T\) denote lattice translation by one site and \(\Pi_Z=\prod_{j}\sigma^z_j\) the global \(Z\)–parity measured in Step 3. We choose a shallow unitary \(U(\vec\theta)\) that \emph{commutes} with both symmetries,
\[
[U(\vec\theta),T]=0,\qquad [U(\vec\theta),\Pi_Z]=0,
\]
so that \(U(\vec\theta)\) maps momentum–\(p\), odd–parity states to momentum–\(p\), odd–parity states. This guarantees that cleaning cannot leak amplitude to other momentum or parity sectors.

\paragraph{A minimal translation–invariant, parity–preserving ansatz.}
An effective choice is a short product of exponentials of \emph{sums repeated on every bond or site}, which ensures translation invariance, together with generators that contain only an even number of spin–flip operators to preserve \(\Pi_Z\). A convenient three–parameter form is
\begin{equation}
\label{eq:cleaning-ansatz}
U(\theta_1,\theta_2,\theta_3)
\;=\;
\exp\!\Big(i\,\theta_1 \sum_{j}\sigma^z_j\sigma^z_{j+1}\Big)\;
\exp\!\Big(i\,\theta_2 \sum_{j}\sigma^x_j\sigma^x_{j+1}\Big)\;
\exp\!\Big(i\,\theta_3 \sum_{j}\sigma^z_j\Big),
\end{equation}
with periodic boundary conditions implicit in the sums. The first and third factors are diagonal in the computational basis; the middle factor flips spins in pairs and therefore preserves \(Z\)–parity. Each exponential compiles to a constant two–qubit depth on a nearest–neighbor architecture by staggering bonds.

\paragraph{Energy minimization inside the symmetry block.}
Let \(|\Psi_{\mathrm{odd},p}\rangle\) denote the Step–3 state. We determine \(\vec\theta\) by minimizing the Rayleigh quotient
\[
\mathcal{E}(\vec\theta)
\;=\;
\frac{\langle \Psi_{\mathrm{odd},p}|\,U(\vec\theta)^\dagger H\,U(\vec\theta)\,|\Psi_{\mathrm{odd},p}\rangle}
     {\langle \Psi_{\mathrm{odd},p}|\Psi_{\mathrm{odd},p}\rangle}.
\]
Because \(U(\vec\theta)\) preserves momentum and parity, this optimization is confined to the \((p,\text{odd})\) block of \(H\). By the Rayleigh–Ritz principle, decreasing \(\mathcal{E}\) increases the overlap with the lowest–energy eigenstate in that block; at the optimum \(\vec\theta_\star\) one has
\[
U(\vec\theta_\star)\,|\Psi_{\mathrm{odd},p}\rangle \;\approx\; |1_p\rangle,
\]
the interacting one–particle state at total momentum \(p\). A practical confirmation is the smallness of the \emph{energy variance} \(\langle H^2\rangle-\langle H\rangle^2\) on the cleaned state.

\paragraph{Measurements and shallow depth.}
Evaluating \(\mathcal{E}(\vec\theta)\) requires only translation–averaged few–body correlators because the generators in \eqref{eq:cleaning-ansatz} are uniform sums. Gradients may be obtained by small finite differences or standard parameter–shift rules, but in practice a coarse grid over \(\theta_{1,2,3}\) already suffices because the parameter space is low–dimensional. Each layer in \eqref{eq:cleaning-ansatz} is a translation–invariant brickwork; on a ring it is implemented by identical two–qubit gates on alternating bonds, giving a total two–qubit depth of only a few layers.

\paragraph{A circuit sketch for one translation–invariant bond layer.}
The figure illustrates the uniform application of a two–qubit gate \(e^{\,i\theta_2\,\sigma^x\otimes\sigma^x}\) on every bond; staggering even and odd bonds yields constant depth. The \(ZZ\) layer has the same pattern.

\paragraph{Outcome.}
Applying \(U(\vec\theta_\star)\) to the Step–3 state variationally projects it onto the lowest–energy state in the \((p,\text{odd})\) sector. The result is an interacting one–particle wave packet with the prescribed momentum and envelope, ready to be used (in duplicate, with \(\pm p\) and well–separated supports) for real–time collision experiments and the extraction of phase shifts, Wigner time delays, and inelastic probabilities.

\subsection{Choosing the cleaning angles \(\vec\theta\)}

The cleaning unitary \(U(\vec\theta)\) in \eqref{eq:cleaning-ansatz} is fixed by a tiny set of angles \(\vec\theta=(\theta_1,\theta_2,\theta_3)\). The goal is to lower the Rayleigh quotient
\[
\mathcal{E}(\vec\theta)
=
\frac{\langle \Psi_{\mathrm{odd},p}|\,U(\vec\theta)^\dagger H\,U(\vec\theta)\,|\Psi_{\mathrm{odd},p}\rangle}
     {\langle \Psi_{\mathrm{odd},p}|\Psi_{\mathrm{odd},p}\rangle},
\]
while staying inside the \((p,\text{odd})\) symmetry block. In practice one can choose \(\vec\theta\) with very few measurements by exploiting the fact that \(\mathcal{E}\) is smooth and nearly quadratic in each angle near the optimum. Three equally simple routes are useful; any one of them suffices.

\paragraph{A parabolic line search with coordinate sweeps.}
Fix two angles and scan the third over three small values, fit a parabola, and jump to its minimum. Concretely, for a chosen step size \(s\) (e.g.\ \(s=0.05\text{--}0.2\) radians), measure
\(\,E_{-}= \mathcal{E}(\theta_i=-s),\, E_0=\mathcal{E}(0),\, E_{+}=\mathcal{E}(+s)\,\)
with the other angles held fixed. The one–dimensional quadratic minimizer is
\[
\theta_i^\star
\;=\;
s\,\frac{E_{-}-E_{+}}{2\,(E_{-}-2E_0+E_{+})}.
\]
Update \(\theta_i\leftarrow\theta_i^\star\) and sweep \(i=1,2,3\) once or twice. Stop when the energy decrease per sweep is within noise and the energy variance \(\langle H^2\rangle-\langle H\rangle^2\) ceases to improve. In many cases \(\theta_3\) contributes little and may be omitted, leaving a two–angle search.

\paragraph{A parameter–shift gradient step.}
If the layers are implemented with Pauli generators (e.g.\ \(XX\), \(ZZ\), \(Z\)), the parameter–shift identity yields an exact derivative with two evaluations along each coordinate:
\[
\frac{\partial \mathcal{E}}{\partial \theta_i}
=
\frac{1}{2}\Big[\mathcal{E}\!\big(\theta_i{+}\tfrac{\pi}{2}\big)
-\mathcal{E}\!\big(\theta_i{-}\tfrac{\pi}{2}\big)\Big].
\]
A single small update \(\theta_i \leftarrow \theta_i - \eta\,\partial_{\theta_i}\mathcal{E}\) with \(\eta\sim 0.1\text{--}0.3\) typically suffices; confirm with the variance proxy that the state sharpened toward an eigenstate.

\paragraph{A small–angle quadratic jump.}
Near \(\vec\theta=\mathbf{0}\) one may write
\(
\mathcal{E}(\vec\theta)\approx \mathcal{E}_0 + \vec g\!\cdot\!\vec\theta + \tfrac12\,\vec\theta^{\mathsf T}H\,\vec\theta
\).
Estimate the gradient and diagonal Hessian by central and second differences at \(\pm s\),
\[
g_i \approx \frac{\mathcal{E}(+s)-\mathcal{E}(-s)}{2s},
\qquad
H_{ii} \approx \frac{\mathcal{E}(+s)-2\mathcal{E}(0)+\mathcal{E}(-s)}{s^2},
\]
neglect off–diagonal couplings on the first pass, and jump to
\(
\theta_i^\star \approx - g_i/H_{ii}
\).
A brief parabolic polish as above can follow.

\paragraph{What is actually measured.}
Because \(U(\vec\theta)\) and \(H\) are translation–invariant sums, \(\mathcal{E}(\vec\theta)\) reduces to a handful of translation–averaged few–body correlators, such as
\(\langle \sigma^z_j\sigma^z_{j+1}\rangle\),
\(\langle \sigma^x_j\rangle\),
and
\(\langle \sigma^z_j\rangle\),
which can be estimated with good signal–to–noise by averaging over all sites/bonds. As a convergence check, monitor the energy variance \(\mathrm{Var}(H)=\langle H^2\rangle-\langle H\rangle^2\); it decreases as the state approaches the interacting one–particle eigenstate in the \((p,\text{odd})\) block and plateaus when further improvement is negligible.

\subsection{2-particle scattering}

We are now in a position to discuss 2-particle scattering in this theory. We will set up initial states with momenta $p$ and $-p$. The situation is depicted in the figure. 

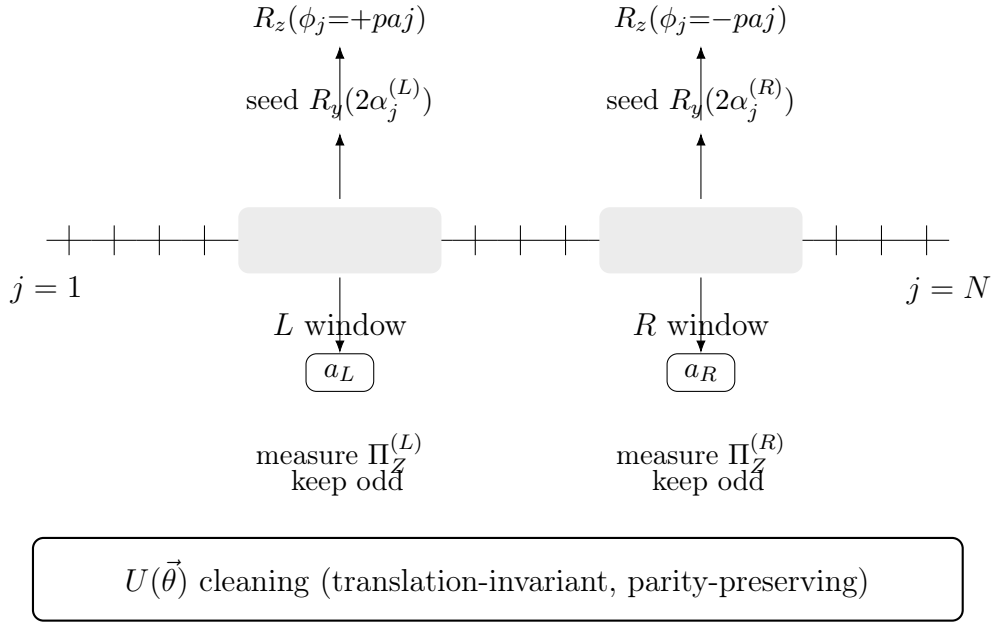
\begin{figure}[htb]
\centering
\begin{tikzpicture}[x=0.6cm,y=1.1cm,>=Latex]

  \foreach \j in {0,...,19} {
    \draw (\j,0) -- (\j+1,0);
    \draw (\j+0.5,0.18) -- (\j+0.5,-0.18);
  }
  \node[below] at (0,-0.3) {$j=1$};
  \node[below] at (20,-0.3) {$j=N$};

  \fill[gray!15,rounded corners] (4.25,-0.4) rectangle (8.75,0.4);
  \node[below] at (6.5, -0.75) {$L$ window};

  \fill[gray!15,rounded corners] (12.25,-0.4) rectangle (16.75,0.4);
  \node[below] at (14.5,-0.75) {$R$ window};

  \draw[->] (6.5,0.5) -- (6.5,1.3);
  \node[above] at (6.5,1.3) {\small seed $R_y(2\alpha^{(L)}_j)$};
  \draw[->] (6.5,1.45) -- (6.5,2.35);
  \node[above] at (6.5,2.35) {\small $R_z(\phi_j{=}{+}paj)$};

  \draw[->] (14.5,0.5) -- (14.5,1.3);
  \node[above] at (14.5,1.3) {\small seed $R_y(2\alpha^{(R)}_j)$};
  \draw[->] (14.5,1.45) -- (14.5,2.35);
  \node[above] at (14.5,2.35) {\small $R_z(\phi_j{=}{-}paj)$};

  \node[draw,rounded corners,minimum width=0.9cm,minimum height=0.5cm] (aL) at (6.5,-1.6) {\small $a_L$};
  \node[draw,rounded corners,minimum width=0.9cm,minimum height=0.5cm] (aR) at (14.5,-1.6) {\small $a_R$};

  \draw[->] (6.5,-0.45) -- (aL.north);
  \draw[->] (14.5,-0.45) -- (aR.north);

  \node[below=0.4cm of aL] {\small measure $\Pi_Z^{(L)}$};
  \node[below=0.4cm of aR] {\small measure $\Pi_Z^{(R)}$};

  \node[below right=0.9cm and -1.25cm of aL] {\small keep odd};
  \node[below right=0.9cm and -1.25cm of aR] {\small keep odd};

  \draw[rounded corners,thick] (-0.3,-3.6) rectangle (20.3,-4.6);
  \node at (10,-4.1) {$U(\vec\theta)$ cleaning (translation-invariant, parity-preserving)};

\end{tikzpicture}
\caption{Two-packet center-of-mass state preparation. Seed small, shaped rotations only inside two disjoint windows $L$ and $R$, imprint opposite momentum ramps ($+p$ on $L$, $-p$ on $R$), perform \emph{block-wise} odd-parity measurements to herald one excitation in each block, then apply a shallow translation-invariant cleaner $U(\vec\theta)$ to project onto the interacting one-particle in each block. Packets are well separated at $t{=}0$, so the total momentum is zero.}
\label{fig:twopacket-prep}
\end{figure}

\subsection*{Vacuum--subtracted energy--density tracks as an inelasticity diagnostic}

On the lattice Ising Hamiltonian
\[
H(J,h_x,h_z)\;=\;-J\sum_{j}\sigma^z_j\sigma^z_{j+1}\;-\;h_x\sum_{j}\sigma^x_j\;-\;h_z\sum_{j}\sigma^z_j,
\]
a convenient local energy density is built by assigning each bond term once and each on–site term to site \(j\):
\[
h_j \;\equiv\; -J\,\sigma^z_j\sigma^z_{j+1} \;-\; \tfrac{h_x}{2}\,\big(\sigma^x_j+\sigma^x_{j-1}\big) \;-\; \tfrac{h_z}{2}\,\big(\sigma^z_j+\sigma^z_{j-1}\big).
\]
(Any symmetric partition of the on–site pieces is acceptable; the diagnostics below are insensitive to this choice.) Following Refs.~\cite{Farrell2025WStates, Jha2025IFT}, we \emph{vacuum–subtract} the expectation value to remove static backgrounds and visualize transport:
\[
\Delta\varepsilon_j(t)\;\equiv\;\langle\psi(t)|h_j|\psi(t)\rangle\;-\;\langle 0|h_j|0\rangle.
\]

\paragraph{Space–time tracks.}
Plotting \(\Delta\varepsilon_j(t)\) as a function of site \(j\) and time \(t\) yields a heatmap with bright, ballistic \emph{tracks}. For two counter–propagating packets in the center–of–mass frame, purely elastic scattering produces two clean outer ridges which depart the collision region with slopes set by the group velocity \(v(p)\). As the collision energy crosses the first inelastic threshold, Refs.~\cite{Farrell2025WStates, Jha2025IFT} report two robust signatures from MPS simulations:
(i) an \emph{inward skew} of each outgoing bump shortly after impact (the outgoing one–particle packets are pulled toward the center by probability flux into additional channels); and, at later times,
(ii) the emergence of a \emph{slower central ridge} — a low–velocity energy–density track associated with the heavier product in \(1{+}1\to 1{+}2\) processes.

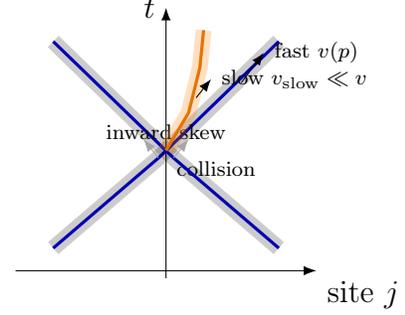
\begin{figure}[t]
\centering
\begin{minipage}[t]{0.48\textwidth}
\centering
\begin{tikzpicture}[x=0.5cm,y=0.5cm,>=Latex]
  \draw[->] (-4,0) -- (4,0) node[below right] {site $j$};
  \draw[->] (0,-0.2) -- (0,7) node[left] {$t$};
  \node[above] at (0,7.2) {\small \textbf{Elastic: two fast tracks only}};
  \fill[black] (0,3.2) circle (0.05);
  \node[below right] at (0,3.9) {\scriptsize collision};
  \draw[line width=6pt,opacity=0.20] (-3,0.6) -- (3,6.0);
  \draw[very thick,blue!70!black] (-3,0.6) -- (3,6.0);
  \draw[line width=6pt,opacity=0.20] (3,0.6) -- (-3,6.0);
  \draw[very thick,blue!70!black] (3,0.6) -- (-3,6.0);
  \draw[->] (-2.2,1.6) -- (-1.4,2.6);
  \node[right] at (-2.4,2.6) {\scriptsize slope $\sim v(p)$};
\end{tikzpicture}
\end{minipage}\hfill
\begin{minipage}[t]{0.48\textwidth}
\centering
\begin{tikzpicture}[x=0.5cm,y=0.5cm,>=Latex]
  \draw[->] (-4,0) -- (4,0) node[below right] {site $j$};
  \draw[->] (0,-0.2) -- (0,7) node[left] {$t$};
  \node[above] at (0,7.2) {\small \textbf{Inelastic: extra slow central track}};
  \fill[black] (0,3.2) circle (0.05);
  \node[below right] at (0,3.2) {\scriptsize collision};
  \draw[line width=6pt,opacity=0.20] (-3,0.6) -- (-0.2,3.0) -- (3.0,6.1);
  \draw[very thick,blue!70!black] (-3,0.6) -- (-0.2,3.0) -- (3.0,6.1);
  \draw[line width=6pt,opacity=0.20] (3,0.6) -- (0.2,3.0) -- (-3.0,6.1);
  \draw[very thick,blue!70!black] (3,0.6) -- (0.2,3.0) -- (-3.0,6.1);
  \draw[->,gray!70] (-0.2,3.0) -- (-0.6,3.5);
  \draw[->,gray!70] (0.2,3.0) -- (0.6,3.5);
  \node at (0,3.7) {\scriptsize inward skew};
  \draw[line width=6pt,opacity=0.25,orange] (0,3.2) -- (0.6,4.2) -- (0.9,5.4) -- (1.0,6.4);
  \draw[very thick,orange!90!black] (0,3.2) -- (0.6,4.2) -- (0.9,5.4) -- (1.0,6.4);
  \draw[->] (2.0,5.0) -- (2.6,5.8);
  \node[right] at (2.6,5.8) {\scriptsize fast $v(p)$};
  \draw[->] (0.8,4.6) -- (1.2,5.1);
  \node[right] at (1.2,5.1) {\scriptsize slow $v_{\mathrm{slow}}\ll v$};
\end{tikzpicture}
\end{minipage}
\caption{Schematic ``energy-density tracks'' (space--time ridges of vacuum-subtracted local energy) for two-packet scattering. \textbf{Left:} Below inelastic threshold, two clean outer ridges depart the collision with slopes set by the group velocity $v(p)$. \textbf{Right:} Above threshold, the outer ridges exhibit an inward skew after impact and a third, slow central ridge appears, signaling inelastic production of a heavier excitation. This is a qualitative cartoon, not data.}
\label{fig:tracks-cartoon}
\end{figure}

\paragraph{Correlator “track finders”.}
The MPS companion study~\cite{Jha2025IFT} sharpens these features by conditioning on the locations of the fast packets using energy–density correlators. Two–point functions
\(
C_{2}(j,t;j',t)\!=\!\langle \Delta\varepsilon_j(t)\,\Delta\varepsilon_{j'}(t)\rangle
\)
already correlate opposite–moving tracks; three–point correlators
\(
C_{3}(j_L,j_R,j_c;t)\!
\)
peak sharply when \(j_L,j_R\) sit on the outer ridges and \(j_c\) on the slow central track, enforcing the kinematics of the inelastic channel in the center–of–mass frame. This provides a quantitative, basis–independent way to identify multi–track final states in the post–collision wavefunction.

\newpage
\part{TN/MPS Essentials}


\chapter{Primer on matrix product states}
We devote this section to a brief review of matrix product states (MPS). This chapter is based on~\cite{Hauschild_2018} and ~\cite{Schollw_ck_2011}.
\section{Introduction}
The Hilbert space of a quantum many-body system grows exponentially with the increase in number of degrees of freedom. For a chain of $N$ sites where the local Hilbert space dimension at each site is $\mathbb{C}^d$ a generic pure state is an element of $(\mathbb{C}^d)^{\otimes N}$ and is characterized by $d^N$ complex coefficients. Explicitly storing and manipulating these coefficients becomes infeasible even for moderate $N$. The \emph{matrix product state} (MPS) ansatz provides a structured and efficiently parametrized representation of such states, particularly ground states of gapped one-dimensional Hamiltonians, using resources that scale only polynomially with system size. In the MPS ansatz, the coefficients $\psi_{i_1 \cdots i_N}$ of a many-body state are written as products of matrices
\begin{equation}
	\label{eq:tnstate}
	\begin{split}
		\ket{\psi} & = \sum_{i_1\,,\cdots\,, i_N} \psi_{i_1\,, \cdots\,, i_N} \ket{i_1\,, \cdots \,, i_N}                                                                                                                      \\
		           & = \sum_{i_1\,, \cdots \,, i_N} \sum_{\nu_1\,, \cdots \,, \nu_N} M_{\alpha_1 \alpha_2}^{[1]i_1} M_{\alpha_2 \alpha_3}^{[2]i_2} \cdots M_{\alpha_N\alpha_{N+1}}^{[N]i_N}\ket{i_1\,, i_2 \,, \cdots \,, i_N} \\
		           & = \sum_{i_1\,, \cdots \,, i_N} \left(M^{[1]i_1} M^{[2] i_2} \cdots M^{[N] i_N}\right) \ket{i_1 i_2 \cdots i_N}
	\end{split}\,,
\end{equation}
where for each site $k$, the matrix $M^{[k]i_k}$ is of dimension $D_k \times D_{k+1}$ and the indices $\{D_k\}_{k = 1}^{N}$ are referred to as the \emph{bond dimensions}. The product structure above necessarily leads to a $1 \times 1$ matrix, i.e. a scalar. This implies that the first and last matrices are effectively row and column vectors, respectively.

A generic pure state in the full many-body Hilbert space has entanglement entropy that scales proportionally to the volume of the subsystem, a behaviour commonly referred to as the \emph{volume law}. In contrast, the ground states of local, gapped Hamiltonians occupy a submanifold of this space characterized by entanglement entropy that scales with the size of the boundary between subsystems. This behaviour, referred to as the \emph{area law}, implies that ground states are far less entangled that a typical state drawn from the Hilbert space measure and are found in a small corner of the Hilbert space.

Consider the following bipartition of a one-dimensional system into left and right sites at a given bond. The decomposition
\begin{equation}
	\ket{\Psi} = \sum_{\alpha = 1}^{\chi_A} \lambda_{\alpha} \ket{\alpha}_L \otimes \ket{\alpha}_R \,,
\end{equation}
where $\ket{\alpha}_{L(R)}$ are orthonormal basis states for the left (right) Hilbert space $\mathcal{H}_{L(R)}$, is called the \emph{Schmidt} decomposition. The entanglement entropy for this partition is given by
\begin{equation}
	\label{eq:entanglement}
	S = - \sum_{\alpha = 1}^{\chi_L}\lambda_\alpha^2 \log \lambda_\alpha^2\,.
\end{equation}
The Schmidt rank $\chi_A$ naturally quantifies the entanglement between the left and right subsystems. For a given state $\ket{\Psi}$, its overall entanglement can be characterized by the maximum Schmidt rank $\chi_L$ across all possible bipartitions of the qudits. We will call this number the \emph{bond dimension} $\chi$. Then a state can be considered only `slightly' entangled if $\chi$ is small in the following sense: for a sequence of qudit lattices of size $n$, the bond dimension $\chi$ grows at most polynomially in $n$. Furthermore, the quantum dynamics of a pure state is considered slightly entangled if at all times $t$ the state $\ket{\Psi(t)}$ of the system is slightly entangled. To appreciate how constraining this definition is, note that the maximum value of $\chi_L$ grows exponentially in general and is obtained when both the left and right partitions are equal. Compared to this, states with polynomially growing bond dimension with lattice size contain an exceptionally small amount of entanglement.

{\it For states that are only slightly entangled, only a relatively small number of Schmidt states actually contribute to the sum in \eqref{eq:entanglement}. Therefore, one can obtain a computationally cheaper representation of such states by truncating the number of Schmidt states at each bond.}

Note that the representation in \eqref{eq:tnstate} is not unique. This is because along the bond between site $k$ and $k+1$, we can redefine
\begin{equation}
	\label{eq:gaugetransfinite}
	M^{[k]i_k}_{\alpha_k \alpha_{k+1}} \rightarrow \tilde{M}^{[k]i_k} = M^{[k]i_k} X^{-1}\,, \quad M^{[k+1]i_{k+1}} \rightarrow \tilde{M}^{[k+1]j_{k+1}} = X M^{[k+1]j_{k+1}}\,,
\end{equation}
where $X \in \operatorname{GL}(\chi_{k+1})$ is invertible, leaving the physical state unchanged but yielding a different MPS representation. This gauge freedom can lead to different MPS representations for the same physical state. However, in practice, there are some important forms
\begin{enumerate}
	\item \textit{Left-canonical form.} Consider the coefficient $\psi_{i_1 \cdots i_N}$ with the following grouping of indices
	      \begin{equation}
		      \psi_{i_1, (i_2, \cdots, i_N)} = \sum_{a_1} A^{i_1}_{a_1} \Lambda^{[1]}_{a_1, a_1} (V^\dagger)_{a_1, (i_2, \cdots i_N)}\,,
	      \end{equation}
	      where on the right-hand side we have performed a singular value decomposition. We can absorb $\Lambda^{[1]}_{a_1, a_1}$ into $V^\dagger$ and repeat the procedure:
	      \begin{equation}
		      \begin{split}
			      D_{i_1, i_2, \cdots, i_N}
			       & = \sum_{a_1, a_2}
			      A^{i_1}_{a_1}
			      A_{(i_2, a_2), a_3}
			      \Lambda^{[2]}_{a_3, a_3}
			      (V^\dagger)_{a_2, (i_3, \cdots, i_N)} \\
			       & = \sum_{a_1, a_2}
			      A^{i_1}_{a_1}
			      A^{i_2}_{a_2, a_3}
			      \Lambda^{[2]}_{a_3, a_3}
			      (V^\dagger)_{a_2, (i_3, \cdots, i_N)}\,,
		      \end{split}
	      \end{equation}
	      where in the second line we have reshaped $A$. Repeating these steps until all sites are exhausted, we arrive at
	      \begin{equation}
		      \ket{\psi}
		      = \sum_{i_1, \cdots i_N}
		      \sum_{a_2, \cdots a_{N}}
		      A^{[1]i_1}_{a_1, a_2}
		      A^{[2]i_2}_{a_2, a_3}
		      \cdots
		      A^{[N]i_N}_{a_N, a_{N+1}}
		      \ket{i_1 \cdots i_N}\,,
	      \end{equation}
	      with boundary indices $a_1 = a_{N+1} = 1$. Since the $A_{(i_k, a_k), a_{k+1}}$ matrices are the left factors obtained from the SVD, they satisfy
	      \begin{equation}
		      \begin{split}
			      \delta_{b c}
			       & = \sum_{i, a} {A^\dagger}_{b (i a)} A_{(i a) c}        \\
			       & = \sum_{i} \left({A^{i}}^\dagger A^{i}\right)_{b c}\,,
		      \end{split}
	      \end{equation}
	      or equivalently,
	      \begin{equation}
		      \label{eq:leftisometry}
		      \sum_{i} {A^{i}}^\dagger A^{i} = \mathds{1}\,,
	      \end{equation}
	      for each site. An MPS written entirely in terms of tensors satisfying \eqref{eq:leftisometry} is called \emph{left-canonical}. After a partition of the lattice into left (L) and right (R) parts, we can define
	      \begin{equation}
		      \begin{split}
			      \ket{a_{\ell}}_L
			       & = \sum_{i_1\,, \cdots \,, i_\ell}
			      \left(A^{[1]i_1} \cdots A^{[\ell]i_{\ell}}\right)_{1, a_\ell}
			      \ket{i_1 \cdots i_{\ell}}            \\
			      \ket{a_{\ell}}_R
			       & = \sum_{i_{\ell + 1} \cdots i_N}
			      \left(A^{[\ell + 1]i_{\ell + 1}} \cdots A^{[N]i_N}\right)_{a_{\ell}, 1}
			      \ket{i_{\ell + 1} \cdots i_N}\,.
		      \end{split}
	      \end{equation}
	      Using \eqref{eq:leftisometry} repeatedly, one finds
	      \begin{equation}
		      {}_L\!\langle a_{\ell} | a'_{\ell}\rangle_L = \delta_{a_\ell, a'_\ell}\,,
	      \end{equation}
	      so the left states form an orthonormal basis, while the right states $\ket{a_{\ell}}_R$ are in general \emph{not} orthonormal:
	      \begin{equation}
		      {}_R\expval{a_{\ell}| a'_{\ell}}_R
		      = \sum_{i_{\ell + 1}, \dots, i_N}
		      \left(
		      A^{[\ell+1] i_{\ell+1}} \cdots A^{[N] i_N}
		      {A^{[N] i_N}}^\dagger \cdots {A^{[\ell+1] i_{\ell+1}}}^\dagger
		      \right)_{a_{\ell} a'_{\ell}}\,,
	      \end{equation}
	      which is a positive matrix but does not collapse to a Kronecker $\delta$ in general.

	\item \textit{Right-canonical form.} Nothing essential depends on starting the SVD from the left; we can equally well sweep from the right. Grouping indices as $\psi_{(i_1 \cdots i_{N-1}), i_N}$ and performing an SVD, we obtain
	      \begin{equation}
		      \begin{split}
			      \psi_{i_1 \cdots i_N}
			       & = \sum_{a_{N-1}}
			      U_{(i_1 \cdots i_{N-1}) a_{N-1}}
			      \Lambda^{[N-1]}_{a_{N-1} a_{N-1}}
			      B_{a_{N-1} i_N}                 \\
			       & = \sum_{a_{N-2},a_{N-1}}
			      U_{(i_1 \cdots i_{N-2}) a_{N-2}}
			      \Lambda^{[N-2]}_{a_{N-2} a_{N-2}}
			      B_{a_{N-2} (a_{N-1} i_{N-1})}
			      B^{i_N}_{a_{N-1}}               \\
			       & = \sum_{a_1, \dots, a_{N-1}}
			      B^{[1] i_1}_{a_1}
			      B^{[2] i_2}_{a_1 a_2}
			      \cdots
			      B^{[N] i_N}_{a_{N-1}}\,,
		      \end{split}
	      \end{equation}
	      where again boundary indices are $a_0 = a_N = 1$. The $B$ matrices are \emph{right-normalized}, i.e.\ they satisfy
	      \begin{equation}
		      \sum_{i} B^{i} {B^{i}}^\dagger = \mathds{1}\,,
	      \end{equation}
	      so that, in analogy with the left-canonical case, the states built from the right half of the chain form an orthonormal basis, while the left basis is not orthonormal in general. An MPS written entirely in terms of such $B$-tensors is called \emph{right-canonical}.

	\item \textit{Mixed–canonical form.}
	      One may also choose to perform the SVD sweep from the left up to some site \(\ell\), and from the right for the remaining sites. This yields a decomposition of the form
	      \begin{equation}
		      \begin{split}
			      \psi_{i_1 \dots i_N}
			       & = \sum_{a_2,\dots,a_{\ell}}
			      A^{[1] i_1}_{a_1 a_2}
			      A^{[2] i_2}_{a_2 a_3} \cdots
			      A^{[\ell] i_\ell}_{a_{\ell-1} a_\ell}\,
			      \Lambda^{[\ell]}_{a_\ell a_\ell}\,
			      (V^\dagger)_{a_\ell,\, i_{\ell+1}\dots i_N} \\
			       & = \sum_{a_2,\dots,a_{N}}
			      A^{[1] i_1}_{a_1 a_2}
			      A^{[2] i_2}_{a_2 a_3} \cdots
			      A^{[\ell] i_\ell}_{a_{\ell-1} a_\ell}\,
			      \Lambda^{[\ell]}_{a_\ell a_\ell}\,
			      B^{[\ell+1] i_{\ell+1}}_{a_\ell a_{\ell+1}}
			      \cdots
			      B^{[N] i_N}_{a_{N-1} a_N}\, .
		      \end{split}
	      \end{equation}
	      In this representation, all tensors to the \emph{left} of the bond \((\ell,\ell+1)\) are left–canonical, and all tensors to the \emph{right} are right–canonical. The diagonal matrix \(\Lambda^{[\ell]}\) is the \emph{center matrix}, whose entries are the Schmidt coefficients across that bond. The usual left–canonical form is obtained by pushing the center all the way to the left end of the chain, so that the remaining \(\Lambda\) at the first bond is the identity. Analogously, the right–canonical form corresponds to pushing the center to the right end of the chain, again yielding $\Lambda = \mathds{1}$ at the last bond.

	\item \textit{Vidal's $\Gamma\Lambda$ form.} A particularly convenient way of writing an MPS that explicitly exposes the Schmidt spectrum on each bond was introduced by Vidal~\cite{Vidal_2003}. One writes
	      \begin{equation}
		      \ket{\psi}
		      = \sum_{i_1, \dots, i_N}
		      \Gamma^{[1] i_1}\,
		      \Lambda^{[1]}\,
		      \Gamma^{[2] i_2}\,
		      \Lambda^{[2]}\,
		      \cdots\,
		      \Lambda^{[N-1]}\,
		      \Gamma^{[N] i_N}
		      \ket{i_1 \cdots i_N}\,,
	      \end{equation}
	      where each $\Lambda^{[\ell]}$ is diagonal with nonnegative entries (the Schmidt coefficients across bond $\ell$), and the $\Gamma^{[\ell] i_\ell}$ are three-index tensors. This representation is completely equivalent to the mixed–canonical form above. Given a mixed–canonical MPS with center at bond $\ell$,
	      \begin{equation}
		      \cdots A^{[\ell] i_\ell}\, \Lambda^{[\ell]}\, B^{[\ell+1] i_{\ell+1}} \cdots,
	      \end{equation}
	      we can define
	      \begin{equation}
              \Gamma^{[\ell] i_\ell}
              := (\Lambda^{[\ell-1]})^{-1} A^{[\ell] i_\ell}
              = B^{[\ell] i_\ell} (\Lambda^{[\ell]})^{-1}
	      \end{equation}
	      with the convention $\Lambda^{[0]} = \Lambda^{[N]} = \mathds{1}$. Inserting these definitions into the mixed–canonical expression and re-grouping the factors yields precisely Vidal’s $\Gamma\Lambda$ form:
	      \begin{equation}
		      A^{[\ell] i_\ell}
		      = \Lambda^{[\ell-1]} \Gamma^{[\ell] i_\ell}, \qquad
		      B^{[\ell] i_\ell}
		      = \Gamma^{[\ell] i_\ell} \Lambda^{[\ell]}\,.
	      \end{equation}
	      Conversely, starting from a $\Gamma\Lambda$ representation, one recovers the left- and right-canonical tensors by
	      \begin{equation}
		      A^{[\ell] i_\ell} = \Lambda^{[\ell-1]} \Gamma^{[\ell] i_\ell}, \qquad
		      B^{[\ell] i_\ell} = \Gamma^{[\ell] i_\ell} \Lambda^{[\ell]}\,,
	      \end{equation}
	      and the usual canonical conditions translate into simple isometry constraints on the $\Gamma$–tensors with respect to the weighted inner product defined by the corresponding $\Lambda$’s. Thus Vidal’s $\Gamma\Lambda$ form is just a particularly symmetric way of writing an MPS in mixed canonical form, with the Schmidt spectra on each bond exposed explicitly.
\end{enumerate}
\section{Tensor network diagrams}
There is a diagrammatic way to represent the state in \eqref{eq:tnstate} using \emph{tensor network diagrams}. In this representation, a tensor is represented by a box with legs sticking out of it, with the number of legs depending on the number of uncontracted indices of the tensor. For matrix product states, we choose the convention that the physical indices are to be represented by legs that stick out vertically and the internal indices are represented by legs that are horizontal. A contraction between two indices is represented by connecting the respective pair of indices. For example, the tensor $M^{[k]i_k}_{\alpha_k \alpha_{k + 1}}$ is represented by the following diagram
\begin{equation}
	M^{[k]} = \begin{diagram}
		\draw (0, 0.0) -- (0.5, 0.0);
		\draw[rounded corners, fill=cyan!30] (0.5, -0.5) rectangle (1.5, 0.5);
		\draw (1.5, 0.0) -- (2, 0.0);
		\draw (1, -0.5) -- (1, -1);
		\draw (1, 0.0) node {$M$};
		\draw (0.0, 0.25) node {$\alpha_k$};
		\draw (2.1, 0.25) node {$\alpha_{k+1}$};
		\draw (1.25, -1) node {$i_k$};
	\end{diagram}
\end{equation}
We also have the product
\begin{equation}
	M^{[k] i_k} M^{[k+1] i_{k+1}} = \begin{diagram}
		\filldraw[fill=cyan!30, rounded corners] (0.5, -0.5) rectangle (1.5, 0.5);
		\draw (1.5, 0) -- (2.5, 0);
		\filldraw[fill=cyan!30, rounded corners] (2.5, -0.5) rectangle (3.5, 0.5);
		\draw (3.5, 0) -- (4, 0);
		\draw (0, 0) -- (0.5, 0);
		\draw (1, -0.5) -- (1, -1);
		\draw (3, -0.5) -- (3, -1);
		\draw (1, 0) node {$M^{[k]}$};
		\draw (3, 0) node {$M^{[k+1]}$};
	\end{diagram}
\end{equation}
Therefore, the state in \eqref{eq:tnstate} can be written as
\begin{equation}
	\ket{\psi} =
	\begin{diagram}
		\foreach \i in {0, 1, 2, 4, 5}{
				\filldraw[fill=cyan!30, rounded corners] (2*\i, -0.5) rectangle (2*\i + 1, 0.5);
				\draw (2*\i + 0.5, -0.5) -- (2*\i + 0.5, -1);
			}
		\draw (1, 0) -- (1.5, 0);
		\draw (9.5, 0) -- (10, 0);
		\draw (6.5, 0) node {$\cdots$};
		\foreach \i in {1, 2, 4} {
				\draw (2*\i + 1, 0.0) -- (2*\i + 1.5, 0.0);
				\draw (2*\i - 0.5, 0.0) -- (2*\i, 0.0);
			}
		\draw (0.5, 0) node {$M^{[1]}$};
		\draw (2.5, 0) node {$M^{[2]}$};
		\draw (4.5, 0) node {$M^{[3]}$};
		\draw (8.5, 0) node {$M^{[n-1]}$};
		\draw (10.5, 0) node {$M^{[n]}$};
	\end{diagram}
\end{equation}
The expectation value of a single site operator can we obtained by contracting the physical indices of the matrix corresponding to the site at which the operator is acting with the indices of the operator. Diagrammatically,
\begin{equation}
	\expval{\Psi[\bar{M}]| O | \Psi[M]} =
	\begin{diagram}
		\foreach \i in {0, 1} {
				\draw[fill=cyan!30, rounded corners] (2*\i, 1) rectangle (2*\i + 1, 2);
				\draw[fill=blue!30, rounded corners] (2*\i, -2) rectangle (2*\i + 1, -1);
				\draw (2*\i + 0.5, 1.5) node {$M$};
				\draw (2*\i + 0.5, -1.5) node {$\bar{M}$};
				\draw (2*\i + 0.5, 1) -- (2*\i + 0.5, -1);
				\draw (2*\i + 1, 1.5) -- (2*\i + 1.5, 1.5);
				\draw (2*\i + 1, -1.5) -- (2*\i + 1.5, -1.5);
			}
		\draw (4, 1.5) node {$\cdots$};
		\draw (4, -1.5) node {$\cdots$};
		\draw (4.5, 1.5) -- (5, 1.5);
		\draw (4.5, -1.5) -- (5, -1.5);
		\draw[fill=cyan!30, rounded corners] (5, 1) rectangle (6, 2);
		\draw[fill=blue!30, rounded corners] (5, -1) rectangle (6, -2);
		\draw (5.5, 1.5) node {$M$};
		\draw (5.5, -1.5) node {$\bar{M}$};
		\draw (6, 1.5) -- (6.5, 1.5);
		\draw (6, -1.5) -- (6.5, -1.5);
		\draw[fill=magenta!30] (5.5, 0.0) circle (0.5);
		\draw (5.5, 0) node {$O$};
		\draw (5.5, 1) -- (5.5, 0.5);
		\draw (5.5, -0.5) -- (5.5, -1);
		\draw (7, 1.5) node {$\cdots$};
		\draw (7, -1.5) node {$\cdots$};
		\foreach \i in {4, 5} {
				\draw[fill=cyan!30, rounded corners] (2*\i, 1) rectangle (2*\i + 1, 2);
				\draw[fill=blue!30, rounded corners] (2*\i, -2) rectangle (2*\i + 1, -1);
				\draw (2*\i + 0.5, 1.5) node {$M$};
				\draw (2*\i + 0.5, -1.5) node {$\bar{M}$};
				\draw (2*\i + 0.5, 1) -- (2*\i + 0.5, -1);
				\draw (2*\i - 0.5, 1.5) -- (2*\i, 1.5);
				\draw (2*\i - 0.5, -1.5) -- (2*\i, -1.5);
			}
		\foreach \i in {4} {
				\draw (2*\i + 1, 1.5) -- (2*\i + 1.5, 1.5);
				\draw (2*\i + 1, -1.5) -- (2*\i + 1.5, -1.5);
			}
		\foreach \i in {1} {
				\draw (2*\i - 0.5, 1.5) -- (2*\i, 1.5);
				\draw (2*\i - 0.5, -1.5) -- (2*\i, -1.5);
			}
	\end{diagram}
\end{equation}
The object
\begin{equation}
	\mathbb{E}_{\mathcal{O}} = \sum_{i, j = 1}^{d} \mathcal{O}_{i, j} A_i \otimes \bar{A}_j = \begin{diagram}
		\draw[rounded corners, fill=cyan!30] (0, -0.5) rectangle (1, 0.5);
		\draw (0.5, 0) node {$A$};
		\draw (-0.5, 0.0) -- (0, 0);
		\draw (1, 0.0) -- (1.5, 0);
		\draw (0.5, -0.5) -- (0.5, -1);
		\draw[fill=red!30] (0, -2) rectangle (1, -1);
		\draw (1.85, -3) node {$j_n$};
		\draw (3.35, -3) node {$j_{n+1}$};
		\draw (0.5, -1.5) node {$\mathcal{O}$};
		\draw (0.5, -2) -- (0.5, -2.5);
		\draw[rounded corners, fill=blue!30] (0, -3.5) rectangle (1, -2.5);
		\draw (-0.5, -3) -- (0, -3);
		\draw (1, -3) -- (1.5, -3);
		\draw (0.5, -3) node {$\bar{A}$};
	\end{diagram}
\end{equation}
is referred to as the $\mathcal{O}$-transfer matrix. For an operator acting on more than one number of sites, we replace the transfer matrix at each site by the corresponding $\mathcal{O}$-transfer matrix. For example for an operator acting on a contiguous block of three sites, we have the following contraction
\begin{equation}
	\expval{\psi[\bar{A}]|O|\psi[A]}
	=
	\begin{diagram}
		\draw[rounded corners, fill=cyan!30] (-3, -0.5) rectangle (-2, 0.5);
		\draw (-2.5, 0) node {$M$};
		\draw (-2, 0.0) -- (-1.5, 0.0);
		\draw[rounded corners, fill=magenta!30] (-3, -3.5) rectangle (-2, -2.5);
		\draw (-2.5, -3) node {$\bar{M}$};
		\draw (-2.5, -0.5) -- (-2.5, -2.5);
		\draw (-2, -3) -- (-1.5, -3);
		\draw (-1, 0) node {$\cdots$};
		\draw (-1, -3) node {$\cdots$};
		\draw (6, 0) node {$\cdots$};
		\draw (6, -3) node {$\cdots$};
		\foreach \i in {0,1,2} {
				\draw[fill=cyan!30, rounded corners] (2*\i, -0.5) rectangle (2*\i + 1, 0.5);
				\draw[fill=magenta!30, rounded corners] (2*\i, -3.5) rectangle (2*\i + 1, -2.5);
				\draw (2*\i - 0.5, 0) -- (2*\i, 0);
				\draw (2*\i + 1, 0) -- (2*\i + 1.5, 0);
				\draw (2*\i + 0.5, -0.5) -- (2*\i + 0.5, -1);
				\draw (2*\i + 0.5, -2) -- (2*\i + 0.5, -2.5);
				\draw (2*\i - 0.5, -3) -- (2*\i, -3);
				\draw (2*\i + 1, -3) -- (2*\i + 1.5, -3);
				\draw (2*\i + 0.5, 0) node {$M$};
				\draw (2*\i + 0.5, -3) node {$\bar{M}$};
			}
		\draw[fill=brown!30] (0, -2) rectangle (5, -1);
		\draw (2.5, -1.5) node {$\mathcal{O}$};
		\draw[rounded corners, fill=cyan!30] (7, -0.5) rectangle (8, 0.5);
		\draw[rounded corners, fill=magenta!30] (7, -3.5) rectangle (8, -2.5);
		\draw (7.5, -0.5) -- (7.5, -2.5);
		\draw (6.5, 0) -- (7, 0);
		\draw (6.5, -3) -- (7, -3);
		\draw (7.5, 0) node {$M$};
		\draw (7.5, -3) node {$\bar{M}$};
	\end{diagram}
\end{equation}
\section{Time evolving block decimation}
One of the most common problems in physics is studying the evolution of a system under some Hamiltonian. If we know that the state describing the system at time $t = 0$ is $\ket{\psi_0}$, then the subsequent evolution of the system is described by the equation
\begin{equation}
	\ket{\psi(t)} = U(t) \ket{\psi(0)}\,,
\end{equation}
where $U(t) = e^{-i H t}$ is the time-evolution operator. The evolution operator can also be analytically continued to imaginary time where the evolution picks out the ground state of the Hamiltonian $H$ assuming that the initial state $\ket{\psi(0)}$ has some non-trivial component along the ground subspace.
\begin{equation}
	\ket{\psi_{\text{GS}}} = \lim_{\tau\to\infty} \frac{e^{-\tau H} \ket{\psi_0}}{\lVert e^{-\tau H}\ket{\psi_0}\rVert}
\end{equation}
Suppose that our Hamiltonian is the sum of two site operators, for example, this could be the Heisenberg spin chain
\begin{equation}
	H = J\sum_{\expval{i, j} i < j}\left(S^x_i S^x_j + S^y_i S^y_j + S^z_i S^z_j\right)\,.
\end{equation}
Since such a Hamiltonian in general will have non-commuting parts, the TEBD algorithm relies on approximating the time evolution operator through a sequence of Suzuki-Trotter decompositions. As a first approximation we have the Baker-Campbell-Hausdorff decomposition
\begin{equation}
	e^{-i \delta t H} = e^{-i \delta t H_{\text{even}}} e^{-i \delta t H_{\text{odd}}} + \mathcal{O}(\delta t^2)\,,
\end{equation}
where $H_{\text{even(odd)}}$ denotes the part of the Hamiltonian consisting of the sum over even (odd) sites. All the terms with the same parity commute with each other. An update that only acts on a single site $\ell$ only requires us to change the corresponding $\Gamma^{[\ell]}$ at the cost of $\Theta(\chi^2)$ floating point operations.
\begin{equation}
	\Gamma'^{[\ell]i_{\ell}}_{a_\ell, a_{\ell+1}} = \sum_{j_{\ell}} U^{i_{\ell}}_{j_{\ell}} \Gamma^{[\ell]j_{\ell}}_{a_\ell, a_{\ell+1}}\,.
\end{equation}
Likewise, a two site operation acting on the bond $(\ell, \ell+1)$ requires us to modify $\Gamma^{[\ell]}, \lambda^{[\ell]}$ and $\Gamma^{[\ell + 1]}$ with $\Theta(\chi^3)$ basic operations~\cite{Vidal_2003}.
\begin{equation}
	\Theta^{j_n, j_{n+1}}_{\alpha_n, \alpha_{n+1}} = \sum_{j_n', j_{n+1}'} U^{j_n, j_{n+1}}_{j_n', j_{n+1}'} \Theta^{j_n', j_{n+1}'}_{\alpha_n, \alpha_{n+1}}\,,
\end{equation}
where
\begin{equation}
	\label{eq:twositewavefn}
	\begin{split}
		\Theta^{i_n, i_{n+1}}_{a_n, a_{n+1}} & = \sum_{b_1, b_2}\Lambda^{[\ell-1]}_{a_1 b_1}\Gamma^{[\ell]i_{\ell}}_{b_1 b_2}\Lambda^{[\ell]}_{b_2 b_2}\Gamma^{[\ell+1]i_{\ell+1}}_{b_2, a_{n+1}}\Lambda^{[\ell+1]}_{a_{n+1}a_{n+1}} \\
		                                     & = \left(\Lambda^{[\ell-1]}\Gamma^{[\ell]i_{\ell}}\Lambda^{[\ell]}\Gamma^{[\ell+1]i_{\ell+1}}\Lambda^{[\ell+1]}\right)_{a_n a_{n+1}}                                                   \\
		                                     & = \left(A^{[\ell]i_{\ell}} A^{[\ell+1]i_{\ell}} \Lambda^{[\ell+1]}\right)_{a_na_{n+1}}                                                                                                \\
		                                     & = \left(\Lambda^{[\ell-1]} B^{[\ell]i_{\ell}}B^{[\ell+1]i_{\ell+1}}\right)_{a_na_{n+1}}                                                                                               \\
		                                     & = \left(A^{[\ell]i_{\ell}}\Lambda^{[\ell]}B^{[\ell+1]i_{\ell+1}}\right)_{a_na_{n+1}}
	\end{split}
\end{equation}
is called the two-site wavefunction. After acting with two-site operator, we reshape the indices $(a_{\ell}, i_{\ell+1}, i_{\ell+2}, a_{\ell+2})$ into $((a_{\ell}, i_{\ell}); (a_{\ell+2}, i_{\ell+1}))$ and perform an SVD
\begin{equation}
	\tilde{\Theta}_{(a_\ell i_{\ell+1}), (i_{\ell+2} a_{\ell+2})} = \sum_{a_{\ell+1}}U_{(a_{\ell}i_{\ell+1}), a_{\ell+1}}\Lambda^{[\ell+1]}_{a_{\ell+1} a_{\ell+1}} (V^\dagger)_{a_{\ell+1},(i_{\ell+2}a_{\ell+2})}\,.
\end{equation}
Note that after the SVD, $\Lambda^{[\ell+1]}$ can have at most $\min\{d\chi_{\ell}, d\chi_{\ell+1}\}$ diagonal entries. At this stage we truncate to $\chi_{\text{max}}$ number of Schmidt values. After this truncation, we reshape the tensors and reintroduce $\Lambda^{[\ell]}$ and $\Lambda^{[\ell+2]}$
\begin{equation}
	\Theta^{i_{\ell+1}i_{\ell+2}}_{a_{\ell}a_{\ell+2}} = \left(\Lambda^{[\ell]}(\Lambda^{[\ell]})^{-1} U^{i_{\ell+1}} \tilde{\Lambda}^{[\ell+1]} {V^{i_{\ell+2}}}^\dagger(\Lambda^{[\ell+2]})^{-1}\Lambda^{[\ell+2]}\right)_{a_\ell a_{\ell+1}}\,,
\end{equation}
we define the updated $\Gamma$ matrices
\begin{equation}
	\begin{split}
		\tilde{\Gamma}^{[\ell+1]i_{\ell+1}}_{a_{\ell} a_{\ell+1}} & = (\Lambda^{[\ell]})^{-1}_{a_\ell a_\ell}U^{i_{\ell+1}}_{a_\ell a_{\ell+1}} \\ \tilde{\Gamma}^{[\ell+2]i_{\ell+2}} &= {V^{i_{\ell+2}}}^\dagger_{a_{\ell+1} a_{\ell+2}}(\Lambda^{[\ell+2]})^{-1}_{a_{\ell+2}a_{\ell+2}}\,.
	\end{split}
\end{equation}
to finally obtain
\begin{equation}
	\tilde{\Theta}^{i_\ell i_{\ell+2}}_{a_{\ell+1}a_{\ell+2}} = \left(\Lambda^{[\ell]}\tilde{\Gamma}^{[\ell]i_{\ell}}\tilde{\Lambda}^{[\ell+1]} \tilde{\Gamma}^{[\ell+2]i_{\ell+2}}\Lambda^{[\ell+2]}\right)_{a_{\ell+1}a_{\ell+2}}\,.
\end{equation}

\begin{figure}[hbt]
	\centering
	\begin{diagram}
		\draw[dashed, rounded corners, fill=brown!30, fill opacity = 0.5] (-0.3, -3.65) rectangle (15.3, -0.85);
		\draw[dashed, rounded corners, fill=brown!30, fill opacity = 0.5] (-0.3, -6.65) rectangle (15.3, -3.85);
		\foreach \i in {0,...,6} {
				\draw[fill=magenta!30, rounded corners] (2*\i, -0.5) rectangle (2*\i + 1, 0.5);
				\draw (2*\i + 0.5, 0) node {$M^{[\i]}$};
				\draw (2*\i+1, 0) -- (2*\i + 2, 0);
				\draw (2*\i + 0.5, -0.5) -- (2*\i  + 0.5, -1);
			}
		\draw[fill=magenta!30, rounded corners] (14, -0.5) rectangle (15, 0.5);
		\draw (14.5, -0.5) -- (14.5, -1);
		\draw (14.5, 0) node {$M^{[7]}$};
		\foreach \i in {0,...,3} {
				\pgfmathtruncatemacro{\ia}{2*\i}
				\pgfmathtruncatemacro{\ib}{2*\i+1}
				\draw[fill=yellow!30, rounded corners] (4*\i, -2) rectangle (4*\i + 3, -1);
				\draw (4*\i + 1.5, -1.5) node {$U^{(\ia, \ib)}(\delta t)$};
			}
		\foreach \i in {1,...,6} {
				\draw (2*\i + 0.5, -2) -- (2*\i + 0.5, -2.5);
				\draw (2*\i + 0.5, -3.5) -- (2*\i + 0.5, -4);
				\draw (2*\i + 0.5, -5) -- (2*\i + 0.5, -5.5);
				\draw (2*\i + 0.5, -6.5) -- (2*\i + 0.5, -7);
			}
		\draw (0.5, -2) -- (0.5, -4);
		\draw (0.5, -5) -- (0.5, -7);
		\draw (14.5, -2) -- (14.5, -4);
		\draw (14.5, -5) -- (14.5, -7);
		\foreach \i in {0.5,1.5,...,3} {
				\pgfmathtruncatemacro{\ia}{2*\i}
				\pgfmathtruncatemacro{\ib}{2*\i+1}
				\draw[fill=yellow!30, rounded corners] (4*\i, -3.5) rectangle (4*\i + 3, -2.5);
				\draw (4*\i + 1.5, -3) node {$U^{(\ia, \ib)}(\delta t)$};
			}
		\foreach \i in {0,...,3} {
				\pgfmathtruncatemacro{\ia}{2*\i}
				\pgfmathtruncatemacro{\ib}{2*\i+1}
				\draw[fill=yellow!30, rounded corners] (4*\i, -5) rectangle (4*\i + 3, -4);
				\draw (4*\i + 1.5, -4.5) node {$U^{(\ia, \ib)}(\delta t)$};
			}
		\foreach \i in {0.5,1.5,...,3} {
				\pgfmathtruncatemacro{\ia}{2*\i}
				\pgfmathtruncatemacro{\ib}{2*\i+1}
				\draw[fill=yellow!30, rounded corners] (4*\i, -6.5) rectangle (4*\i + 3, -5.5);
				\draw (4*\i + 1.5, -6) node {$U^{(\ia, \ib)}(\delta t)$};
			}
		\foreach \i in {0,...,7} {
				\draw (2*\i + 0.5, -7.5) node {$\vdots$};
			}
	\end{diagram}
	\caption{The TEBD evolution is decomposed into alternating odd and even layers, with each lightly shaded brown box indicating a single Trotter step. Because the two-site unitaries in each layer can increase the entanglement across their shared bond, the number of significant Schmidt coefficients may grow during the evolution. To control this growth, the two-site tensor is reshaped into a $d\chi \times d\chi$ matrix and subjected to a singular value decomposition (SVD). The Schmidt spectrum is then truncated by retaining only the largest coefficients up to the prescribed tolerance, thereby keeping the bond dimension manageable while capturing the essential entanglement structure.} \label{fig:tebddecomp}
\end{figure}

\begin{figure}
	\centering
	\begin{diagram}
		\foreach \i in {0,1} {
				\draw[rounded corners, fill=magenta!30] (3*\i, -0.5) rectangle (3*\i + 1, 0.5);
				\draw[fill=red!30] (3*\i + 2,0) circle (0.5);
				\draw[fill=red!30] (3*\i + 2,0) node {$\Lambda^{[\i]}$};
				\draw (3*\i + 0.5, 0.0) node {$\Gamma^{[\i]}$};
				\draw (3*\i + 0.5, -0.5) -- (3*\i + 0.5, -1);
			}
		\foreach \i in {3,4} {
				\draw[rounded corners, fill=magenta!30] (3*\i, -0.5) rectangle (3*\i + 1, 0.5);
				\draw[fill=red!30] (3*\i - 1,0) circle (0.5);
				\draw (3*\i + 0.5, -0.5) -- (3*\i + 0.5, -1);
			}
		\draw (9.5, 0) node {$\Gamma^{[n-1]}$};
		\draw (12.5, 0) node {$\Gamma^{[n]}$};
		\draw (8, 0) node {$\Lambda^{[n-2]}$};
		\draw (11, 0) node {$\Lambda^{[n-1]}$};
		\foreach \i in {0, 1.5,...,5} {
				\draw (\i + 1, 0) -- (\i + 1.5, 0);
			}
		\foreach \i in {6,7.5,...,11} {
				\draw (\i + 1, 0) -- (\i + 1.5, 0);
			}
		\draw (6.5, 0) node {$\cdots$};
	\end{diagram}
	\caption{MPS diagram for Vidal's $\Gamma\Lambda$ form.}
\end{figure}

\begin{figure}[H]
	\centering
	\begin{diagram}
		\tikzset{
		physCircle/.style   = {circle, fill=red!30},
		tensorBox/.style    = {rounded corners, fill=cyan!30},
		gateBox/.style      = {rounded corners, fill=yellow!30},
		smallBoxL/.style    = {rounded corners, fill=violet!30},
		smallBoxR/.style    = {rounded corners, fill=purple!30},
		regionBox/.style    = {rounded corners, dashed, fill=brown!30, fill opacity=0.3},
		gaugeCircle/.style  = {circle, dashed, fill=red!30, font=\scriptsize},
		flowArrow/.style    = {-{Stealth[length=7pt]}, line width=1pt, draw=black, font=\footnotesize}
		}

		\begin{scope}
			\draw[regionBox] (-0.6, -2.1) rectangle (6.6, 0.6);

			\draw[physCircle] (0, 0) circle (0.5);
			\draw (0, 0) node {$\Lambda^{[\ell]}$};
			\draw[tensorBox]  (1, -0.5) rectangle (2, 0.5);
			\draw (1.5, 0) node {$\Gamma^{[\ell+2]}$};
			\draw[physCircle] (3, 0) circle (0.5);
			\draw (3, 0) node {$\Lambda^{[\ell+1]}$};
			\draw[tensorBox]  (4, -0.5) rectangle (5, 0.5);
			\draw (4.5, 0) node {$\Gamma^{[\ell+2]}$};
			\draw[physCircle] (6, 0) circle (0.5);
			\draw (6, 0) node {$\Lambda^{[\ell+2]}$};

			\foreach \i in {0,1.5,...,6} {
					\draw (\i + 0.5, 0) -- (\i + 1, 0);
				}
			\draw (-1, 0) -- (-0.5, 0);

			\draw (1.5, -0.5) -- (1.5, -1);
			\draw (4.5, -0.5) -- (4.5, -1);

			\draw[gateBox] (1, -2) rectangle (5, -1);
			\draw (3, -1.5) node {$U(\delta t)$};
			\draw (1.5, -2) -- (1.5, -2.5);
			\draw (4.5, -2) -- (4.5, -2.5);
		\end{scope}

		\draw[flowArrow]
		(7.0,-1.0) -|
		node[pos=0.75, right=2pt, inner sep=1pt] {(i)}
		(9.0,-2.0);

		\begin{scope}[shift={(8,-3)}]
			\draw[regionBox] (-0.6, -2.1) rectangle (6.6, 0.6);
			\draw[physCircle] (0, 0) circle (0.5);
			\draw[tensorBox]  (1, -0.5) rectangle (2, 0.5);
			\draw[physCircle] (3, 0) circle (0.5);
			\draw[tensorBox]  (4, -0.5) rectangle (5, 0.5);
			\draw[physCircle] (6, 0) circle (0.5);
			\draw (0, 0) node {$\Lambda^{[\ell]}$};
			\draw (1.5, 0) node {$\Gamma^{[\ell+2]}$};
			\draw (3, 0) node {$\Lambda^{[\ell+1]}$};
			\draw (4.5, 0) node {$\Gamma^{[\ell+2]}$};
			\draw (6, 0) node {$\Lambda^{[\ell+2]}$};

			\foreach \i in {0,1.5,...,6} {
					\draw (\i + 0.5, 0) -- (\i + 1, 0);
				}
			\draw (-1, 0) -- (-0.5, 0);

			\draw (1.5, -0.5) -- (1.5, -1);
			\draw (4.5, -0.5) -- (4.5, -1);

			\draw[gateBox] (1, -2) rectangle (5, -1);
			\draw (3, -1.5) node {$U(\delta t)$};

			\draw (1.5, -2) edge[out=270, in=0]   (-1, -0.1);
			\draw (4.5, -2) edge[out=270, in=180] (7, -0.1);

			\draw[flowArrow]
			(-2.0,-1.0) -|
			node[pos=0.75, left=4pt, inner sep=1pt] {(ii)}
			(-4.0,-2.0);
		\end{scope}

		\begin{scope}[shift={(1.5,-6)}]
			\draw[smallBoxL] (-0.5, -0.5) rectangle (0.5, 0.5);
			\draw (-0.5, 0) -- (-1, 0);
			\draw (0, 0) node {$U$};

			\draw[physCircle] (1.5, 0) circle (0.5);
			\draw (1.5, 0) node {$\tilde{\Lambda}^{[\ell+1]}$};
			\draw (0.5, 0) -- (1, 0);

			\draw[smallBoxR] (2.5, -0.5) rectangle (3.5, 0.5);
			\draw (3, 0) node {$V^\dagger$};
			\draw (3.5, 0) -- (4, 0);
			\draw (2, 0) -- (2.5, 0);

			\draw[flowArrow]
			(5.5,-1.0) -|
			node[pos=0.75, right=2pt, inner sep=1pt] {(iii)}
			(7.5,-2.0);
		\end{scope}

		\begin{scope}[shift={(7.5,-9)}]
			\draw[regionBox] (2.4, -0.6) rectangle (5.1, 0.6);
			\draw[regionBox] (-2.1, -0.6) rectangle (0.6, 0.6);

			\draw[smallBoxL] (-0.5, -0.5) rectangle (0.5, 0.5);
			\draw (0, 0) node {$U$};
			\draw (0, -0.5) -- (0, -1);

			\draw[gaugeCircle] (-3.0, 0) circle (0.5);
			\node at (-3, 0) {$\Lambda^{[\ell]}$};

			\draw[gaugeCircle] (-1.5, 0) circle (0.5);
			\node at (-1.5, 0) {${\Lambda^{[\ell]}}^{-1}$};

			\draw (-4, 0) -- (-3.5, 0);
			\draw (-2.5, 0) -- (-2, 0);
			\draw (-0.5, 0) -- (-1, 0);

			\draw[physCircle] (1.5, 0) circle (0.5);
			\node at (1.5, 0) {$\tilde{\Lambda}^{[\ell+1]}$};
			\draw (0.5, 0) -- (1, 0);
			\draw (2, 0) -- (2.5, 0);

			\draw[smallBoxR] (2.5, -0.5) rectangle (3.5, 0.5);
			\node at (3, 0) {$V^\dagger$};
			\draw (3, -0.5) -- (3, -1);
			\draw (3.5, 0) -- (4, 0);

			\draw[gaugeCircle] (4.5, 0) circle (0.5);
			\node at (4.5, 0) {\scriptsize ${\Lambda^{[\ell+2]}}^{-1}$};

			\draw[gaugeCircle, fill opacity=1] (6, 0) circle (0.5);
			\node at (6, 0) {$\Lambda^{[\ell+2]}$};

			\draw (5, 0) -- (5.5, 0);
			\draw (6.5, 0) -- (7, 0);

			\draw[flowArrow]
			(-4.0,-1.0) -|
			node[pos=0.75, left=4pt, inner sep=1pt] {(iv)}
			(-6.0,-2.0);
		\end{scope}

		\begin{scope}[shift={(0, -12)}]
			\draw[physCircle] (0, 0) circle (0.5);
			\node at (0, 0) {$\Lambda^{[\ell]}$};

			\draw[tensorBox] (1, -0.5) rectangle (2, 0.5);
			\node at (1.5, 0) {$\tilde{\Gamma}^{[\ell+1]}$};

			\draw[physCircle] (3, 0) circle (0.5);
			\node at (3, 0) {$\tilde{\Lambda}^{[\ell+1]}$};

			\draw[tensorBox] (4, -0.5) rectangle (5, 0.5);
			\node at (4.5, 0) {$\tilde{\Gamma}^{[\ell+2]}$};

			\draw[physCircle] (6, 0) circle (0.5);
			\node at (6, 0) {$\Lambda^{[\ell+2]}$};

			\foreach \i in {0,1.5,...,6} {
					\draw (\i + 0.5, 0) -- (\i + 1, 0);
				}
			\draw (-1, 0) -- (-0.5, 0);

			\draw (1.5, -0.5) -- (1.5, -1);
			\draw (4.5, -0.5) -- (4.5, -1);
		\end{scope}
	\end{diagram}
	\caption{Flow chart for the steps of a time–evolving block decimation step. (i) Combine the physical and virtual indices. (ii) Perform an SVD on the resulting matrix, truncate the resulting Schmidt matrix. (iii) Reshape the matrix to separate out the physical and virtual indices. (iv) Re-insert the diagonal Schmidt matrices on the bonds and redefine into $\Gamma\Lambda$ form.} \label{fig:tebdflowchart}
\end{figure}

This update is then applied sequentially to all bonds of one parity (e.g. all even bonds), after which the same procedure is carried out on the bonds of the opposite parity. A flow chart of this procedure is shown in fig.\ref{fig:tebdflowchart}. Note that the two-site update is non-unitary and therefore destroys the canonical form on the other bonds.

\section{Matrix Product Operators}
One can attempt to generalize the representation of states with matrix products
\begin{equation}
	\psi_{i_1 \dots i_N} =
	\begin{diagram}
		\foreach \i in {0,...,5} {
				\draw[rounded corners, fill=magenta!30] (2*\i, -0.5) rectangle (2*\i + 1, 0.5);
				\draw (2*\i + 0.5, -0.5) -- (2*\i + 0.5, -1);
			}
		\foreach \i in {0,...,4} {
				\draw (2*\i + 1, 0) -- (2*\i + 2, 0);
			}
	\end{diagram}
\end{equation}
to a representation of operators
\begin{equation}
	\bra{i_1\dots i_N}  \hat{O} \ket{i_1 \dots i_N} = W^{i_1 i_1'} W^{i_2 i_2'} \cdots W^{i_{N-1} i_{N-1}'} W^{i_N i_N'}\,,
\end{equation}
where each $W^{i_k i_{k+1}}$ is a matrix. With this representation, we can write our operator as
\begin{equation}
	\hat{O} = \sum_{\substack{i_1 \dots i_N \\ i_1' \dots i_N'}} W^{i_1 i_1'} W^{i_2 i_2'} \cdots W^{i_{N-1}i_{N-1}'} W^{i_N i_N'}\ket{i_1 \cdots i_N} \bra{i_1' \cdots i_N'}\,.
\end{equation}
One can guess what the diagram for the MPO looks like
\begin{equation}
	\hat{O} = \begin{diagram}
		\foreach \i in {0,1,2} {
				\draw[rounded corners, fill=violet!30] (2*\i, -0.5) rectangle (2*\i + 1, 0.5);
				\draw (2*\i + 0.5, -0.5) -- (2*\i + 0.5, -1);
				\draw (2*\i + 0.5, 0.5) -- (2*\i + 0.5, 1);
			}
		\foreach \i in {4,5,6} {
				\draw[rounded corners, fill=violet!30] (2*\i, -0.5) rectangle (2*\i + 1, 0.5);
				\draw (2*\i + 0.5, -0.5) -- (2*\i + 0.5, -1);
				\draw (2*\i + 0.5, 0.5) -- (2*\i + 0.5, 1);
			}
		\foreach \i in {0,1,2} {
				\draw (2*\i + 1, 0) -- (2*\i + 2, 0);
				\draw (2*\i + 0.5, 0) node {$W^{[\i]}$};
			}
		\draw (6.5, 0) node {$\cdots$};
		\foreach \i in {3, 4, 5, 6} {
				\draw (2*\i - 1, 0) -- (2*\i, 0);
			}
		\draw (8.5, 0) node {$W^{[n-2]}$};
		\draw (10.5, 0) node {$W^{[n-1]}$};
		\draw (12.5, 0) node {$W^{[n]}$};
	\end{diagram}
\end{equation}
It is useful to understand the role of the MPO tensors \(W^{i_k i_k'}_{a_k a_{k+1}}\). We view the Hamiltonian as a sum of operator strings acting on a chain of \(L\) sites. For each bond \((k,k+1)\) we introduce a finite set of auxiliary states \(a_{k} \in \{1,\dots,D_W\}\), which group together \emph{equivalence classes} of operator strings on the sites \(\{k,\dots,L\}\). Two right tails belong to the same class if they admit the same possible continuations to the left consistent with the Hamiltonian. In particular, the index \(a_{k+1}\) encodes the coarse pattern of operators on the sites to the right of bond \((k,k+1)\).

The MPO tensor at site \(k\) has the form
\[
	W^{[k]}_{a_k,a_{k+1}} = \sum_{i_k,i_k'} W^{i_k i_k'}_{a_k a_{k+1}} |i_k\rangle\langle i_k'| ,
\]
and admits the following interpretation. Given a right-tail label \(a_{k+1}\), representing some equivalence class of operator strings on sites \(\{k+1,\dots,L\}\), the block \(W^{[k]}_{a_k,a_{k+1}}\) inserts the appropriate local operator on site \(k\) and updates the tail label to \(a_k\). Thus each nonzero block implements a transition
\[
	a_{k+1} \longrightarrow a_k ,
\]
extending a valid operator string leftward by one site, while tracking only the information necessary for further continuation. In this way, the MPO describes the Hamiltonian as a finite-state automaton: the contraction over the auxiliary indices \(\{a_k\}\) sums over all valid operator strings, and the boundary conditions ensure that only complete Hamiltonian terms contribute.

To make these ideas concrete, consider the Heisenberg antiferromagnetic chain in a magnetic field,
\begin{equation}
	\label{eq:heisenbergafchain}
	\hat{H} =
	\sum_{i=1}^{L-1}\left(
	\frac{J}{2}\hat{S}^+_{i} \hat{S}^{-}_{i+1}
	+ \frac{J}{2}\hat{S}^-_{i} \hat{S}^{+}_{i+1}
	+ J \hat{S}^{z}_i \hat{S}^{z}_{i+1}
	\right)
	- h \sum_{i=1}^L \hat{S}^z_i .
\end{equation}
Scanning the Hamiltonian from right to left, the operators acting on the sites \(\{k{+}1,\dots,L\}\) form a right tail. Many microscopic tails are interchangeable from the MPO’s perspective, since only their allowed leftward extensions matter. The auxiliary index \(a_k\) simply records which type of tail we are continuing, and the sequence of MPO tensors assembles all Hamiltonian terms one site at a time.
In the present example, we obtain five such classes, which we call
string states:
\begin{itemize}
	\item \textbf{State 1:}
	      The right tail consists entirely of identities.
	      This class represents strings in which no Hamiltonian term has been
	      initiated on sites to the right of the bond.

	\item \textbf{State 2:}
	      The right tail begins with a single $\hat{S}^+$ at the site immediately
	      to the right of the bond, followed by identities.
	      This corresponds to the situation in which the ``right leg'' of an
	      interaction term $\hat{S}^-_i \hat{S}^+_{i+1}$ has already been placed,
	      and the complementary operator $\hat{S}^-$ must appear somewhere to the
	      left in order to complete the interaction.

	\item \textbf{State 3:}
	      The right tail begins with a single $\hat{S}^-$, again followed by
	      identities.
	      This represents the initiation of the opposite interaction leg,
	      requiring an $\hat{S}^+$ to appear to the left.

	\item \textbf{State 4:}
	      The right tail begins with a single $\hat{S}^z$, with identities
	      thereafter.
	      This is the partially built right half of the $\hat{S}^z_i
		      \hat{S}^z_{i+1}$ interaction term, which must be completed by another
	      $\hat{S}^z$ on a site to the left.

	\item \textbf{State 5:}
	      A complete Hamiltonian term has already appeared somewhere in the right
	      tail—either a two-site interaction term of the form
	      $\hat{S}^\pm_i \hat{S}^\mp_{i+1}$ or $\hat{S}^z_i \hat{S}^z_{i+1}$, or
	      the one-site field term $-h \hat{S}^z_i$.
	      Beyond this point, only identities may appear; hence this class encodes all tails that already contain a fully formed term to the right. The matrix now may be constructed as follows. There are only finitely many transitions possible between these states due to the structure of the Hamiltonian.
\end{itemize}
These five equivalence classes constitute the basis of the MPO's auxiliary space. The MPO tensor at site $k$ then describes how the string state changes when the operator string is extended one site to the left, while inserting the appropriate local operator at site $k$.
\begin{enumerate}
	\item
	      For state~1, the right tail consists solely of identities. No Hamiltonian term has been initiated, so we are free to place any of the local operators that can \emph{start} a term. Thus the nonzero matrix elements out of state~1 are
	      \[
		      W^{[k]}_{1,1} = \mathds{1}, \qquad
		      W^{[k]}_{2,1} = \hat{S}^+, \qquad
		      W^{[k]}_{3,1} = \hat{S}^-, \qquad
		      W^{[k]}_{4,1} = \hat{S}^z, \qquad
		      W^{[k]}_{5,1} = -h\,\hat{S}^z .
	      \]

	\item  Transitions out of the remaining states are more constrained. For instance, state~2 corresponds to a right tail beginning with $\hat{S}^+$ followed by identities. This is the ``right leg'' of the interaction $\tfrac{J}{2}\hat{S}^-_k \hat{S}^+_{k+1}$, and it can only be completed by placing $\tfrac{J}{2}\hat{S}^-$ on site $k$. Hence the only allowed transition from state~2 is
	      \[
		      W^{[k]}_{5,2} = \tfrac{J}{2}\,\hat{S}^- .
	      \]

	\item
	      Analogously, state~3 (a right tail beginning with $\hat{S}^-$) can only be completed to a full interaction term by placing $\tfrac{J}{2}\hat{S}^+$ on site $k$, giving
	      \[
		      W^{[k]}_{5,3} = \tfrac{J}{2}\,\hat{S}^+ .
	      \]

	\item
	      State~4 corresponds to a right tail beginning with $\hat{S}^z$, the first half of the $J_z \hat{S}^z_k \hat{S}^z_{k+1}$ interaction. It must be completed by placing $J_z \hat{S}^z$ on site $k$, so
	      \[
		      W^{[k]}_{5,4} = J_z\,\hat{S}^z .
	      \]

	\item
	      Finally, state~5 already represents a fully formed Hamiltonian term somewhere to the right. No further nontrivial operators may be inserted, so the only allowed transition is to remain in state~5 by placing the identity:
	      \[
		      W^{[k]}_{5,5} = \mathds{1}.
	      \]
\end{enumerate}

The ends of the string are special. At the rightmost end we have to place an operator to begin the automaton and at the leftmost end we have to terminate the string so we must map the string to the completed state~5. Therefore, we have
\begin{equation}
	W^{[k]} = \begin{pmatrix} \mathds{1} & 0 & 0 & 0 & 0 \\ \hat{S}^+ & 0 & 0 & 0 & 0 \\ \hat{S}^- & 0 & 0 & 0 & 0 \\\hat{S}^z & 0 & 0 & 0 & 0 \\ -h \hat{S}^z & \frac{J}{2} \hat{S}^- & \frac{J}{2} \hat{S}^+ & J\hat{S}^z & \mathds{1} \end{pmatrix}
\end{equation}
for sites $1 < k < L$. On the first and last sites we have
\begin{equation}
	W^{[1]} = \begin{pmatrix} -h \hat{S}^z & \frac{J}{2} \hat{S}^- & \frac{J}{2} \hat{S}^+ & J \hat{S}^z & \mathds{1} \end{pmatrix} \,, \quad
	W^{[L]} = \begin{pmatrix} \mathds{1} \\ \hat{S}^+ \\ \hat{S}^- \\ \hat{S}^z \\ - h \hat{S}^z \end{pmatrix}\,.
\end{equation}
It is trivial to verify that the above construction leads to the Hamiltonian in \eqref{eq:heisenbergafchain}.
\section{Density matrix renormalization group}
The density matrix renormalization group (DMRG) is a variational algorithm for computing ground states of one–dimensional quantum systems within the manifold of matrix product states (MPS). In its modern formulation, DMRG is most naturally interpreted as an optimization over MPS written in mixed canonical form. This viewpoint clarifies the algorithm's stability, its relationship to entanglement, and why it achieves near–optimal precision for gapped 1D Hamiltonians.

In the finite–system DMRG algorithm, the Hamiltonian is represented as a matrix product operator (MPO),
\begin{equation}
	\hat{H} = \sum_{\{s,s'\}}
	W^{[1]\,s_1,s_1'} W^{[2]\,s_2,s_2'} \cdots W^{[L]\,s_L,s_L'}
	\ket{s_1,\dots,s_L}\!\bra{s_1',\dots,s_L'}\,,
\end{equation}
where each local tensor $W^{[\ell]}$ carries two physical indices $s_\ell,s'_\ell$ and two MPO bond indices $b_{\ell-1},b_\ell$. The state is written in $\Gamma\Lambda$ (or $A\Lambda$) form with an orthogonality center on the bond $(\ell,\ell+1)$, and the local variational degrees of freedom are collected into the two–site wavefunction $\Theta$ introduced in Eq.~\eqref{eq:twositewavefn},
\begin{equation}
	\Theta^{i_\ell i_{\ell+1}}_{a_\ell a_{\ell+1}} =
	\left(\Lambda^{[\ell-1]}\Gamma^{[\ell]i_{\ell}}\Lambda^{[\ell]}\Gamma^{[\ell+1]i_{\ell+1}}\Lambda^{[\ell+1]}\right)_{a_\ell a_{\ell+1}}\,.
\end{equation}
Instead of applying a two–site gate as in TEBD, two–site DMRG treats the entries of $\Theta$ as variational parameters and minimizes the Rayleigh quotient
\begin{equation}
	E(\Theta) = \frac{\bra{\psi(\Theta)} H \ket{\psi(\Theta)}}{\braket{\psi(\Theta)}{\psi(\Theta)}}\,,
\end{equation}
where $\ket{\psi(\Theta)}$ is the MPS obtained by inserting $\Theta$ on bond $(\ell,\ell+1)$ and keeping all other tensors fixed. This is equivalent to solving a local generalized eigenvalue problem for an effective Hamiltonian acting on the two–site basis.

To set this up, one projects the full Hamiltonian onto the subspace spanned by basis states $\ket{a_{\ell-1}, i_\ell, i_{\ell+1}, a_{\ell+1}}$, where $a_{\ell-1}$ and $a_{\ell+1}$ label the virtual indices on the bonds to the left and right of the two–site block. The corresponding matrix elements of the effective Hamiltonian can be written as
\begin{equation}
	\begin{split}
		 & \matrixel{a_{\ell-1}, i_\ell, i_{\ell+1}, a_{\ell+1}}{H_{\mathrm{eff}}}{a'_{\ell-1}, i'_\ell, i'_{\ell+1}, a'_{\ell+1}} \\
		 & \qquad =
		\sum_{b_{\ell-1},b_\ell,b_{\ell+1}}
		L^{[\ell-1]}_{a_{\ell-1},a'_{\ell-1};b_{\ell-1}}\,
		W^{[\ell]}_{b_{\ell-1},b_\ell; i_\ell,i'_\ell}\,
		W^{[\ell+1]}_{b_\ell,b_{\ell+1}; i_{\ell+1},i'_{\ell+1}}\,
		R^{[\ell+2]}_{a_{\ell+1},a'_{\ell+1};b_{\ell+1}}\,.
	\end{split}
\end{equation}
Here $L^{[\ell-1]}$ and $R^{[\ell+2]}$ are the left and right \emph{environments} (often called blocks in the traditional DMRG language). They encode the contraction of all MPS and MPO tensors to the left and right of the active two–site block, respectively. In practice, these environments are built and updated recursively as one sweeps through the chain: once $L^{[\ell-1]}$ is known, $L^{[\ell]}$ can be obtained by contracting one additional site, and similarly for the right environments. This recursive construction is the same in Schollw\"ock’s MPS formulation and in practical implementations such as TeNPy.

\begin{figure}
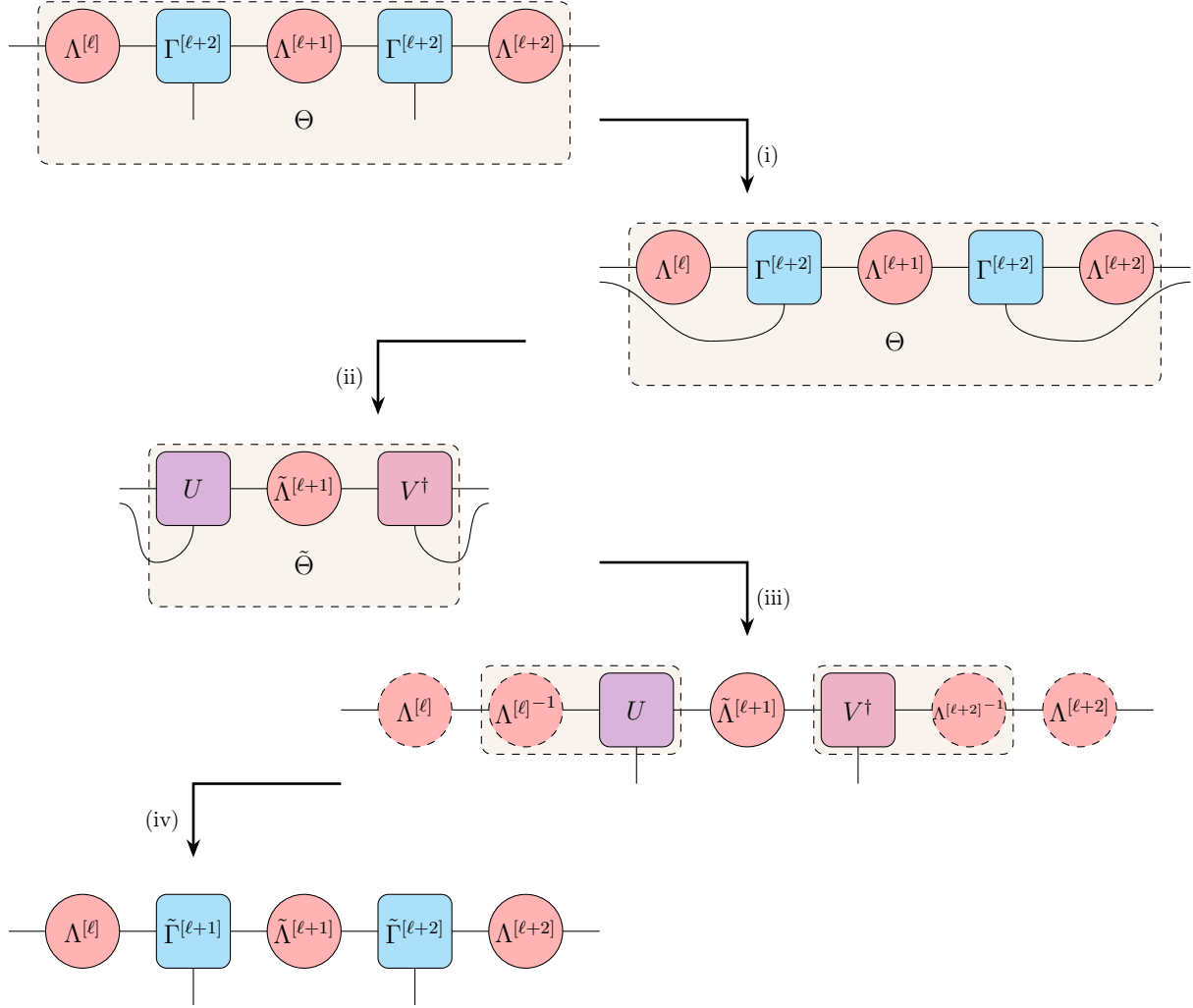

	\centering
	\begin{diagram}
		\tikzset{
		physCircle/.style   = {circle, fill=red!30},
		tensorBox/.style    = {rounded corners, fill=cyan!30},
		gateBox/.style      = {rounded corners, fill=yellow!30},
		smallBoxL/.style    = {rounded corners, fill=violet!30},
		smallBoxR/.style    = {rounded corners, fill=purple!30},
		regionBox/.style    = {rounded corners, dashed, fill=brown!30, fill opacity=0.3},
		gaugeCircle/.style  = {circle, dashed, fill=red!30, font=\scriptsize},
		flowArrow/.style    = {-{Stealth[length=7pt]}, line width=1pt, draw=black, font=\footnotesize}
		}

		\begin{scope}
			\draw[regionBox] (-0.6, -1.6) rectangle (6.6, 0.6);
			\draw (3, -1) node {$\Theta$};

			\draw[physCircle] (0, 0) circle (0.5);
			\draw (0, 0) node {$\Lambda^{[\ell]}$};
			\draw[tensorBox]  (1, -0.5) rectangle (2, 0.5);
			\draw (1.5, 0) node {$\Gamma^{[\ell+2]}$};
			\draw[physCircle] (3, 0) circle (0.5);
			\draw (3, 0) node {$\Lambda^{[\ell+1]}$};
			\draw[tensorBox]  (4, -0.5) rectangle (5, 0.5);
			\draw (4.5, 0) node {$\Gamma^{[\ell+2]}$};
			\draw[physCircle] (6, 0) circle (0.5);
			\draw (6, 0) node {$\Lambda^{[\ell+2]}$};

			\foreach \i in {0,1.5,...,6} {
					\draw (\i + 0.5, 0) -- (\i + 1, 0);
				}
			\draw (-1, 0) -- (-0.5, 0);

			\draw (1.5, -0.5) -- (1.5, -1);
			\draw (4.5, -0.5) -- (4.5, -1);
		\end{scope}

		\draw[flowArrow]
		(7.0,-1.0) -|
		node[pos=0.75, right=2pt, inner sep=1pt] {(i)}
		(9.0,-2.0);

		\begin{scope}[shift={(8,-3)}]
			\draw[regionBox] (-0.6, -1.6) rectangle (6.6, 0.6);
			\draw (3, -1) node {${\Theta}$};
			\draw[physCircle] (0, 0) circle (0.5);
			\draw[tensorBox]  (1, -0.5) rectangle (2, 0.5);
			\draw[physCircle] (3, 0) circle (0.5);
			\draw[tensorBox]  (4, -0.5) rectangle (5, 0.5);
			\draw[physCircle] (6, 0) circle (0.5);
			\draw (0, 0) node {$\Lambda^{[\ell]}$};
			\draw (1.5, 0) node {$\Gamma^{[\ell+2]}$};
			\draw (3, 0) node {$\Lambda^{[\ell+1]}$};
			\draw (4.5, 0) node {$\Gamma^{[\ell+2]}$};
			\draw (6, 0) node {$\Lambda^{[\ell+2]}$};

			\foreach \i in {0,1.5,...,6} {
					\draw (\i + 0.5, 0) -- (\i + 1, 0);
				}
			\draw (-1, 0) -- (-0.5, 0);

			\draw (1.5, -0.5) edge[out=270,in=0] (0.5, -1);
			\draw (0.5, -1) edge[out=180,in=0] (-1, -0.2);
			\draw (4.5, -0.5) edge[out=270, in=180] (5.5, -1);
			\draw (5.5, -1) edge[out=0, in=180] (7, -0.2);

			\draw[flowArrow]
			(-2.0,-1.0) -|
			node[pos=0.75, left=4pt, inner sep=1pt] {(ii)}
			(-4.0,-2.0);
		\end{scope}

		\begin{scope}[shift={(1.5,-6)}]
			\draw[regionBox] (-0.6, -1.6) rectangle (3.6, 0.6);
			\draw (1.5, -1) node {${\tilde{\Theta}}$};
			\draw[smallBoxL] (-0.5, -0.5) rectangle (0.5, 0.5);
			\draw (-0.5, 0) -- (-1, 0);
			\draw (0, 0) node {$U$};
			\draw (0.0, -0.5) edge[out=270, in=0] (-0.5, -1);
			\draw (-0.5, -1) edge[out=180, in=0] (-1, -0.2);

			\draw[physCircle] (1.5, 0) circle (0.5);
			\draw (1.5, 0) node {$\tilde{\Lambda}^{[\ell+1]}$};
			\draw (0.5, 0) -- (1, 0);

			\draw[smallBoxR] (2.5, -0.5) rectangle (3.5, 0.5);
			\draw (3, 0) node {$V^\dagger$};
			\draw (3.5, 0) -- (4, 0);
			\draw (2, 0) -- (2.5, 0);
			\draw (3.0, -0.5) edge[out=270, in=180] (3.5, -1);
			\draw (3.5, -1) edge[out=0, in=180] (4, -0.2);

			\draw[flowArrow]
			(5.5,-1.0) -|
			node[pos=0.75, right=2pt, inner sep=1pt] {(iii)}
			(7.5,-2.0);
		\end{scope}

		\begin{scope}[shift={(7.5,-9)}]
			\draw[regionBox] (2.4, -0.6) rectangle (5.1, 0.6);
			\draw[regionBox] (-2.1, -0.6) rectangle (0.6, 0.6);

			\draw[smallBoxL] (-0.5, -0.5) rectangle (0.5, 0.5);
			\draw (0, 0) node {$U$};
			\draw (0, -0.5) -- (0, -1);

			\draw[gaugeCircle] (-3.0, 0) circle (0.5);
			\node at (-3, 0) {$\Lambda^{[\ell]}$};

			\draw[gaugeCircle] (-1.5, 0) circle (0.5);
			\node at (-1.5, 0) {${\Lambda^{[\ell]}}^{-1}$};

			\draw (-4, 0) -- (-3.5, 0);
			\draw (-2.5, 0) -- (-2, 0);
			\draw (-0.5, 0) -- (-1, 0);

			\draw[physCircle] (1.5, 0) circle (0.5);
			\node at (1.5, 0) {$\tilde{\Lambda}^{[\ell+1]}$};
			\draw (0.5, 0) -- (1, 0);
			\draw (2, 0) -- (2.5, 0);

			\draw[smallBoxR] (2.5, -0.5) rectangle (3.5, 0.5);
			\node at (3, 0) {$V^\dagger$};
			\draw (3, -0.5) -- (3, -1);
			\draw (3.5, 0) -- (4, 0);

			\draw[gaugeCircle] (4.5, 0) circle (0.5);
			\node at (4.5, 0) {\scriptsize ${\Lambda^{[\ell+2]}}^{-1}$};

			\draw[gaugeCircle, fill opacity=1] (6, 0) circle (0.5);
			\node at (6, 0) {$\Lambda^{[\ell+2]}$};

			\draw (5, 0) -- (5.5, 0);
			\draw (6.5, 0) -- (7, 0);

			\draw[flowArrow]
			(-4.0,-1.0) -|
			node[pos=0.75, left=4pt, inner sep=1pt] {(iv)}
			(-6.0,-2.0);
		\end{scope}

		\begin{scope}[shift={(0, -12)}]
			\draw[physCircle] (0, 0) circle (0.5);
			\node at (0, 0) {$\Lambda^{[\ell]}$};

			\draw[tensorBox] (1, -0.5) rectangle (2, 0.5);
			\node at (1.5, 0) {$\tilde{\Gamma}^{[\ell+1]}$};

			\draw[physCircle] (3, 0) circle (0.5);
			\node at (3, 0) {$\tilde{\Lambda}^{[\ell+1]}$};

			\draw[tensorBox] (4, -0.5) rectangle (5, 0.5);
			\node at (4.5, 0) {$\tilde{\Gamma}^{[\ell+2]}$};

			\draw[physCircle] (6, 0) circle (0.5);
			\node at (6, 0) {$\Lambda^{[\ell+2]}$};

			\foreach \i in {0,1.5,...,6} {
					\draw (\i + 0.5, 0) -- (\i + 1, 0);
				}
			\draw (-1, 0) -- (-0.5, 0);

			\draw (1.5, -0.5) -- (1.5, -1);
			\draw (4.5, -0.5) -- (4.5, -1);
		\end{scope}
	\end{diagram}
	\label{fig:dmrgtwosite}
	\caption{Schematic two–site DMRG update. (i) Collect the tensors around the active bond into the two–site wavefunction $\Theta$. (ii) Using the left and right environments and the MPO, solve the local eigenvalue problem $H_{\mathrm{eff}} \Theta = E_{\mathrm{loc}} \Theta$ to obtain the optimized two–site tensor $\tilde{\Theta}$. (iii) Perform an SVD of $\tilde{\Theta}$ and truncate the Schmidt spectrum. (iv) Reinsert the diagonal Schmidt matrices on the bonds and redefine the site tensors into $\Gamma\Lambda$ form, shifting the orthogonality center to the next bond.}
\end{figure}

The effective Hamiltonian $H_{\mathrm{eff}}$ acts on the vectorized two–site tensor $\Theta$ viewed as a vector in a space of dimension $d^2 \chi_{\ell-1} \chi_{\ell+1}$, which is bounded by $d^2 \chi_{\max}^2$. Explicitly forming and diagonalizing $H_{\mathrm{eff}}$ as a dense matrix would be prohibitively expensive. Instead, one only implements the \emph{action} of $H_{\mathrm{eff}}$ on a trial vector $\Theta$, using the environments $L$ and $R$ and the local MPO tensors $W^{[\ell]}, W^{[\ell+1]}$. This action scales as $\mathcal{O}(d^2 \chi^3)$ and can be used inside an iterative eigensolver such as Lanczos or Davidson to obtain the lowest eigenvector of $H_{\mathrm{eff}}$. In other words, two–site DMRG replaces the TEBD update
\[
	\Theta \mapsto U\,\Theta
\]
by a local eigenvalue problem
\[
	H_{\mathrm{eff}} \Theta = E_{\mathrm{loc}}\,\Theta
\]
for the optimal two–site tensor at fixed environments.

Once the locally optimized ground state $\tilde{\Theta}$ on bond $(\ell,\ell+1)$ has been found, one restores the MPS structure by performing an SVD in complete analogy with the TEBD two–site step. One reshapes $\tilde{\Theta}^{i_\ell i_{\ell+1}}_{a_\ell a_{\ell+1}}$ into a matrix $\tilde{\Theta}_{(a_\ell i_\ell),(i_{\ell+1} a_{\ell+1})}$, computes
\begin{equation}
	\tilde{\Theta}_{(a_\ell i_\ell),(i_{\ell+1} a_{\ell+1})}
	= \sum_{a_{\ell+1}'}
	U_{(a_\ell i_\ell),a_{\ell+1}'}
	\Lambda^{[\ell]}_{a_{\ell+1}' a_{\ell+1}'}
	(V^\dagger)_{a_{\ell+1}', (i_{\ell+1} a_{\ell+1})}\,,
\end{equation}
truncates the Schmidt values in $\Lambda^{[\ell]}$ to a maximum bond dimension $\chi_{\max}$ (or according to a truncation error), and absorbs appropriate factors of the neighbouring $\Lambda$–matrices to define new site tensors $\tilde{\Gamma}^{[\ell]}$ and $\tilde{\Gamma}^{[\ell+1]}$ in canonical form, just as in the TEBD update. The orthogonality center is then moved by one site, and the procedure is repeated on the next bond using updated environments. This sweeping (left–to–right and right–to–left) structure is what distinguishes finite–system DMRG from TEBD: TEBD applies a prescribed sequence of unitary gates everywhere, whereas DMRG \emph{locally re-optimizes} the MPS against the full Hamiltonian at each step.

From the MPS perspective, two–site DMRG is thus a sequence of local variational problems on two–site tensors $\Theta$, with the rest of the chain entering only through the environments $L$ and $R$. The algorithm is strictly variational (each local update does not increase the energy) and controls the bond dimension via the Schmidt truncation after each SVD. This is precisely the formulation realized in practical codes such as TeNPy, which implement MPO–based two–site DMRG (and its single–site variants with subspace expansion) using the same effective Hamiltonian and environment construction described here.

\chapter{Infinite Tensor-networks}
\section{Infinite MPS}
In this chapter we largely follow the presentation of the lecture notes in Ref.~\cite{VanderstraetenHaegemanVerstraete2019} to introduce uniform matrix product states (uMPS) in the thermodynamic limit. Finite lattices inevitably suffer from boundary and finite-size effects that can obscure the underlying bulk physics. A standard way to suppress these artefacts is to work directly in the thermodynamic limit, where we send the lattice size $L \to \infty$ and focus on translation-invariant states. In this setting, a uniform matrix product state (uMPS) is described by the ansatz
\begin{align}
	\ket{\Psi(A)}
	&= \sum_{\{s\}} v_L^{\dagger}
	\left[ \prod_{m \in \mathbb{Z}} A^{s_m} \right]
	v_R \ket{\{s\}} \\
	&= \cdots
	\begin{diagram}
		\foreach \i in {0,...,5}{
			\filldraw[fill=cyan!30, rounded corners] (2*\i, -0.5) rectangle (2*\i + 1, 0.5);
			\draw (2*\i + 1, 0.0) -- (2*\i + 1.5, 0);
			\draw (2*\i - 0.5, 0) -- (2*\i, 0);
			\draw (2*\i + 0.5, -0.5) -- (2 *\i + 0.5, -1);
			\draw (2*\i + 0.5, 0.0) node {$A$};
		}
	\end{diagram}
	\cdots
\end{align}
For \emph{injective} MPS, the boundary conditions are immaterial and the vectors $v_L$ and $v_R$ drop out when computing expectation values. However, boundary vectors do play a role for \emph{non-injective} MPS. In what follows we work with an injective MPS ansatz, for which the transfer matrix has a single leading eigenvalue that can be normalized to one. The transfer matrix is defined as
\begin{equation}
	E = \sum_{s = 1}^d A^{s} \otimes {A^s}^{\dagger}
	=
	\begin{diagram}
		\filldraw[fill=cyan!30, rounded corners] (0.0, 0.5) rectangle (1, 1.5);
		\filldraw[fill=blue!30, rounded corners] (0.0, -1.5) rectangle (1, -0.5);
		\draw (-0.5, 1) -- (0, 1);
		\draw (1, 1) -- (1.5, 1);
		\draw (-0.5, -1) -- (0, -1);
		\draw (1, -1) -- (1.5, -1);
		\draw (0.5, 0.5) -- (0.5, -0.5);
		\draw (0.5, 1) node {$A$};
		\draw (0.5, -1) node {$\bar{A}$};
	\end{diagram}
\end{equation}
and has unique left and right fixed points $l$ and $r$ satisfying the eigenvalue equations
\begin{equation}
	\begin{diagram}
		\pic (L) at (0,0) {applyTransferLeft={$\bar{A}$}{$l$}{$A$}};
	\end{diagram}
	=
	\begin{diagram}
		\pic (M) at (3.5, 0) {drawMatrixLeft={$l$}};
	\end{diagram}
	\hspace{2cm}
	\begin{diagram}
		\pic (R) at (0, 0) {applyTransferRight={$A$}{$r$}{$\bar{A}$}};
	\end{diagram}
	=
	\begin{diagram}
		\pic (M2) at (3.5, 0) {drawMatrixRight={$r$}};
	\end{diagram}
\end{equation}
These fixed points can be chosen positive and normalized as $\mathrm{Tr}(l r) = 1$, or diagrammatically
\begin{equation}
	\begin{diagram}
		\filldraw[fill=red!30] (0, 0) circle (0.5);
		\filldraw[fill=orange!30] (2, 0) circle (0.5);
		\draw (0, 0.5) edge[out=90, in=180] (1.0, 1.5);
		\draw (1.0, 1.5) edge[out=0, in=90] (2.0, 0.5);
		\draw (0, -0.5) edge[out=270, in=180] (1.0, -1.5);
		\draw (1.0, -1.5) edge[out=0, in=270] (2.0, -0.5);
		\draw (0, 0) node {$l$};
		\draw (2, 0) node {$r$};
	\end{diagram}
	= 1
\end{equation}
The fixed points $l$ and $r$ can be used to bring the MPS tensors into left- and right-canonical form which fixes the gauge up to unitary invariance. Since both $l$ and $r$ are positive matrices, one can perform a Cholesky decomposition and write
\begin{equation}
	l = L^\dagger L
\end{equation}
with $L$ a lower-triangular matrix. This in turn can be used to define new tensors
\begin{equation}
	A_L = L A L^{-1}
	=
	\begin{diagram}
        \draw (-1, 0) -- (-0.5, 0);
        \draw[fill=violet!30] (0, 0) circle (0.5);
        \draw (0.5, 0) -- (1, 0);
        \draw[rounded corners, fill=cyan!30] (1, -0.5) rectangle (2, 0.5);
        \draw (2, 0) -- (2.5, 0);
        \draw[fill=violet!30] (3, 0) circle (0.5);
        \draw (1.5, -0.5) -- (1.5, -1);
        \draw (3.5, 0) -- (4, 0);
        \draw (1.5, 0) node {$A$};
        \draw (0, 0) node {$L$};
        \draw (3, 0) node {$L^{-1}$};
	\end{diagram}
\end{equation}
which are in the left-canonical gauge where
\begin{equation}
    \begin{split}
        \sum_{s}{A^s}_L^{\dagger} A_L^s = \begin{diagram}
            \draw[rounded corners, fill=cyan!30] (0, 0.5) rectangle (1, 1.5);
            \draw (0.5, 1) node {$A_L$};
            \draw[rounded corners, fill=magenta!30] (0, -1.5) rectangle (1, -0.5);
            \draw (0.5, -1) node {$\bar{A}_L$};
            \draw (0, 1) edge[out=180, in=90] (-0.5, 0);
            \draw (0, -1) edge[out=180, in=270] (-0.5, 0);
            \draw (1, -1) -- (1.5, -1);
            \draw (1, 1) -- (1.5, 1);
            \draw (0.5, 0.5) -- (0.5, -0.5);
        \end{diagram}
        = \mathds{1}\,.
    \end{split}
\end{equation}
To see this, note that
\begin{equation}
    \begin{split}
        \begin{diagram}
            \draw[rounded corners, fill=cyan!30] (0, 0.5) rectangle (1, 1.5);
            \draw (0.5, 1) node {$A_L$};
            \draw[rounded corners, fill=magenta!30] (0, -1.5) rectangle (1, -0.5);
            \draw (0.5, -1) node {$\bar{A}_L$};
            \draw (0, 1) edge[out=180, in=90] (-0.5, 0);
            \draw (0, -1) edge[out=180, in=270] (-0.5, 0);
            \draw (1, -1) -- (1.5, -1);
            \draw (1, 1) -- (1.5, 1);
            \draw (0.5, 0.5) -- (0.5, -0.5);
        \end{diagram}
        =
        \begin{diagram}
            \draw[fill=violet!30] (-1, 1) circle (0.5);
            \draw[fill=violet!30] (-1, -1) circle (0.5);
            \draw (-1, 1) node {$L$};
            \draw (-1, -1) node {$\bar{L}$};
            \draw (-0.5, -1) -- (0, -1);
            \draw (-0.5, 1) -- (0, 1);
            \draw[rounded corners, fill=cyan!30] (0, 0.5) rectangle (1, 1.5);
            \draw (0.5, 1) node {$A$};
            \draw[rounded corners, fill=magenta!30] (0, -1.5) rectangle (1, -0.5);
            \draw (0.5, -1) node {$\bar{A}$};
            \draw (-1.5, 1) edge[out=180, in=90] (-2, 0);
            \draw (-1.5, -1) edge[out=180, in=270] (-2, 0);
            \draw (1, -1) -- (1.5, -1);
            \draw (1, 1) -- (1.5, 1);
            \draw[fill=violet!30] (2, 1) circle (0.5);
            \draw[fill=violet!30] (2, -1) circle (0.5);
            \draw (2, 1) node {$L^{-1}$};
            \draw (2, -1) node {$\bar{L}^{-1}$};
            \draw (2.5, 1) -- (3, 1);
            \draw (2.5, -1) -- (3, -1);
            \draw (0.5, 0.5) -- (0.5, -0.5);
        \end{diagram}
    \end{split}
\end{equation}
Notice that we have
\begin{equation}
    \left(\begin{diagram}
        \draw (-0.5, 1) edge[out=180, in=90] (-1, 0);
        \draw (-0.5, -1) edge[out=180, in=270] (-1, 0);
        \draw[fill=violet!30] (0, 1) circle (0.5);
        \draw[fill=violet!30] (0, -1) circle (0.5);
        \draw (0, 1) node {$L$};
        \draw (0, -1) node {$\bar{L}$};
        \draw (0.5, 1) -- (1, 1);
        \draw (0.5, -1) -- (1, -1);
    \end{diagram}
    \right)_{\alpha\beta}
    = \sum_{\gamma} L_{\gamma\alpha} \bar{L}_{\gamma\beta} = \left(L^\dagger L\right)_{\alpha\beta} = l_{\alpha\beta} = 
    \left(
    \begin{diagram}
        \draw[fill=red!30] (0, 0) circle (0.5);
        \draw (0, 0.5) edge[out=90, in=180] (0.5, 1);
        \draw (0, -0.5) edge[out=270, in=180] (0.5, -1);
        \draw (0, 0) node {$l$};
    \end{diagram}
    \right)_{\alpha\beta}
\end{equation}
which is by definition a fixed point of the transfer matrix. Similarly, we have
\begin{equation}
	\begin{split}
		\braket{\Psi(\bar{A})}{\Psi(A)}
		&=
		\cdots
		\begin{diagram}
			\foreach \i in {0, ..., 4} {
				\filldraw[fill=cyan!30, rounded corners] (2*\i + 0.5, 0) rectangle (2*\i + 1.5, 1);
				\filldraw[fill=blue!30, rounded corners] (2*\i + 0.5, -2) rectangle (2*\i + 1.5, -1);
				\draw (2*\i, 0.5) -- (2*\i + 0.5, 0.5);
				\draw (2*\i + 1.5, 0.5) -- (2*\i + 2, 0.5);
				\draw (2*\i + 1.5, -1.5) -- (2*\i + 2, -1.5);
				\draw (2*\i, -1.5) -- (2*\i + 0.5, -1.5);
				\draw (2*\i + 1, 0) -- (2*\i + 1, -1);
				\draw (2*\i + 1, 0.5) node {$A$};
				\draw (2*\i + 1, -1.5) node {$\bar{A}$};
			}
		\end{diagram}
		\cdots \\
		&= \left(v_L v_L^\dagger\right)
		\left(\prod_{m \in \mathbb{Z}} E \right)
		\left(v_R v_R^\dagger\right)\,.
	\end{split}
\end{equation}
Since the transfer matrix $E$ has a single leading eigenvalue equal to $1$, repeated application of $E$ from the left (right) on any matrix projects onto the fixed point $l$ ($r$). As a consequence, the norm of the infinite state is proportional to a constant raised to an infinite power and is not, by itself, a well-defined quantity. Correlation functions are therefore defined as
\begin{equation}
	\expval{O}_{\ket{\Psi}} =
	\frac{\matrixelement{\Psi(\bar{A})}{O}{\Psi(A)}}{\braket{\Psi(\bar{A})}{\Psi(A)}}\,,
\end{equation}
for which the divergent factors in the numerator and denominator cancel. For example, when $O$ is a single-site operator, we obtain
\begin{equation}
	\expval{O}_{\ket{\Psi}} =
	\begin{diagram}
		\filldraw[fill=red!30] (0,0) circle (.5); \draw (0,0) node {$l$};
		\draw (1,-1.5) edge[out=180,in=270] (0,-0.5);
		\draw (0, 0.5) edge[out=90, in=180] (1.0, 1.5);
		\draw (1.5, 0.0) circle (0.5); \filldraw[fill=orange!30] (3.0, 0.0) circle (0.5); \draw (3, 0.0) node {$r$};
		\draw (1.5, 0.0) node {$O$};
		\draw (3.0, 0.5) edge[out=90, in=0] (2.0, 1.5);
		\draw (3.0, -0.5) edge[out=270, in=0] (2.0, -1.5);
		\draw[fill=cyan!30, rounded corners] (1, 1) rectangle (2, 2);
		\draw[fill=blue!30, rounded corners] (1, -2) rectangle (2, -1);
		\draw (1.5, 1.5) node {$A$};
		\draw (1.5, -1.5) node {$\bar{A}$};
		\draw (1.5, 1.0) -- (1.5, 0.5);
		\draw (1.5, -0.5) -- (1.5, -1.0);
	\end{diagram}
\end{equation}
For a two-site coupling term $h$ in the Hamiltonian, all but two of the transfer matrices cancel between bra and ket, and we are left with
\begin{equation}
	\matrixelement{\Psi(\bar{A})}{h}{\Psi(A)} =
	\begin{diagram}
		\draw[fill=red!30] (0, 0) circle (0.5);
		\draw[fill=cyan!30, rounded corners] (1, 1) rectangle (2, 2);
		\draw[fill=blue!30, rounded corners] (1, -2) rectangle (2, -1);
		\draw[fill=cyan!30, rounded corners] (3, 1) rectangle (4, 2);
		\draw[fill=blue!30, rounded corners] (3, -2) rectangle (4, -1);
		\draw[fill=orange!30] (5, 0) circle (0.5);
		\draw (0.0, 0.5) edge[out=90, in=180] (1, 1.5);
		\draw (0.0, -0.5) edge[out=270, in=180] (1, -1.5);
		\draw (2.0, 1.5) edge[out=0, in=180] (3, 1.5);
		\draw (2.0, -1.5) edge[out=0, in=0] (3, -1.5);
		\draw (4.0, 1.5) edge[out=0, in=90] (5, 0.5);
		\draw (4.0, -1.5) edge[out=0, in=270] (5, -0.5);
		\draw (1.0, -0.5) rectangle (4.0, 0.5);
		\draw (1.5, 1.0) -- (1.5, 0.5);
		\draw (3.5, 1.0) -- (3.5, 0.5);
		\draw (1.5, -1.0) -- (1.5, -0.5);
		\draw (3.5, -1.0) -- (3.5, -0.5);
		\draw (0, 0.0) node {$l$};
		\draw (5, 0.0) node {$r$};
		\draw (2.5, 0.0) node {$h$};
		\draw (1.5, 1.5) node {$A$};
		\draw (3.5, 1.5) node {$A$};
		\draw (1.5, -1.5) node {$\bar{A}$};
		\draw (3.5, -1.5) node {$\bar{A}$};
	\end{diagram}
\end{equation}
The generalization to an $n$-site operator is completely analogous:
\begin{equation}
	\matrixelement{\Psi(\bar{A})}{O}{\Psi(A)} =
	\begin{diagram}
		\draw (0, 0) circle (0.5);
		\draw[rounded corners] (1, 1) rectangle (2, 2);
		\draw (3, 1.5) node {$\cdots$};
		\draw[rounded corners] (4, 1) rectangle (5, 2);
		\draw (6, 0) circle (0.5);
		\draw[rounded corners] (1, -2) rectangle (2, -1);
		\draw[rounded corners] (4, -2) rectangle (5, -1);
		\draw (3, -1.5) node {$\cdots$};
		\draw (1, -0.5) rectangle (5, 0.5);
		\draw (0.0, 0.5) edge[out=90, in=180] (1, 1.5);
		\draw (0.0, -0.5) edge[out=270, in=180] (1, -1.5);
		\draw (2, 1.5) edge[out=0, in=180] (2.5, 1.5);
		\draw (2, -1.5) edge[out=0, in=180] (2.5, -1.5);
		\draw (3.5, 1.5) edge[out=0, in=180] (4, 1.5);
		\draw (3.5, -1.5) edge[out=0, in=180] (4, -1.5);
		\draw (5, 1.5) edge[out=0, in=90] (6, 0.5);
		\draw (5, -1.5) edge[out=0, in=270] (6, -0.5);
		\draw (0, 0) node {$l$};
		\draw (6, 0) node {$r$};
		\draw (1.5, 1.5) node {$A$};
		\draw (1.5, -1.5) node {$\bar{A}$};
		\draw (4.5, 1.5) node {$A$};
		\draw (4.5, -1.5) node {$\bar{A}$};
		\draw (1.5, 1) -- (1.5, 0.5);
		\draw (1.5, -1) -- (1.5, -0.5);
		\draw (4.5, 1) -- (4.5, 0.5);
		\draw (4.5, -1) -- (4.5, -0.5);
		\draw (3, 0) node {$O$};
	\end{diagram}
\end{equation}

\section{Infinite time-evolving block decimation}
The TEBD algorithm described previously works on a finite chain and applies a sequence of two-site gates to an MPS with open boundary conditions. In many situations, however, the system of interest is naturally formulated directly in the thermodynamic limit, where one expects a translation-invariant or periodically modulated ground state and wishes to perform time evolution on an infinite lattice. The infinite-TEBD (iTEBD) algorithm is the natural extension of TEBD to this setting: it evolves a uniform or small-unit-cell infinite MPS (iMPS) by Trotterized two-site gates and uses the same two-site update and truncation described in the finite TEBD section, but now at the level of the unit cell.

We consider an infinite spin chain with local Hilbert space $\mathbb{C}^d$ and a translationally invariant nearest-neighbour Hamiltonian
\begin{equation}
	H = \sum_{j \in \mathbb{Z}} h_{j,j+1}\,.
\end{equation}
To exploit translation invariance, we restrict to an infinite MPS with a unit cell of $\ell_u$ sites. For concreteness, and following Vidal and Schollw\"ock, we focus on a two-site unit cell ($\ell_u = 2$) with tensors $\Gamma_A$, $\Gamma_B$ and diagonal Schmidt spectra $\Lambda_A$, $\Lambda_B$ across the two inequivalent bonds. In the $\Gamma\Lambda$ notation the iMPS ansatz reads
\begin{equation}
	\ket{\Psi[\Gamma_A,\Gamma_B,\Lambda_A,\Lambda_B]} = \sum_{\{s\}} \cdots \Gamma_A^{s_{2j-1}} \Lambda_B \Gamma_B^{s_{2j}} \Lambda_A \Gamma_A^{s_{2j+1}} \Lambda_B \Gamma_B^{s_{2j+2}} \Lambda_A \cdots \ket{\{s\}}\,,
\end{equation}
where the tensors are chosen in (mixed) canonical form so that the entries of $\Lambda_A$ and $\Lambda_B$ are the Schmidt coefficients across the corresponding bonds. The Hamiltonian can be split into two commuting parts acting on disjoint bonds,
\begin{equation}
	H = H_{AB} + H_{BA} = \sum_{j} h_{2j-1,2j} + \sum_j h_{2j,2j+1}\,,
\end{equation}
where $H_{AB}$ contains all ``$AB$'' bonds and $H_{BA}$ all ``$BA$'' bonds. A second-order Trotter--Suzuki decomposition of the time-evolution operator then reads
\begin{equation}
	U(\delta t) = e^{-i H \delta t} \approx e^{-i H_{AB} \delta t/2}\, e^{-i H_{BA} \delta t} \, e^{-i H_{AB} \delta t/2} + \mathcal{O}(\delta t^3)\,,
\end{equation}
with the obvious replacement $\delta t \to -i \delta\tau$ for imaginary-time evolution. Because all terms in $H_{AB}$ (and likewise in $H_{BA}$) commute, $e^{-i H_{AB} \delta t}$ factorizes into identical two-site unitaries acting on every AB bond, and similarly for $e^{-i H_{BA} \delta t}$ on BA bonds.

The key point is that in the thermodynamic limit we never need to store the entire chain: it is enough to update one representative unit cell. Consider the AB bond between a $\Gamma_A$ and a $\Gamma_B$ in the repeating pattern
\begin{equation}
	\cdots \Lambda_A \Gamma_A^{s_A} \Lambda_B \Gamma_B^{s_B} \Lambda_A \cdots\,.
\end{equation}
We apply the two-site gate $U_{AB} = \exp(-i h_{AB} \delta t)$ on the physical indices $s_A, s_B$ of this pair and form an effective two-site tensor by absorbing the neighbouring $\Lambda$-matrices, exactly as in the finite-TEBD two-site step. This defines a tensor $\Theta^{s_A s_B}$ with two virtual legs across the AB bond and two physical legs. We then reshape $\Theta$ into a matrix by fusing the left virtual index with $s_A$ and the right virtual index with $s_B$, perform a singular value decomposition, and truncate the resulting singular values to the prescribed maximum bond dimension $\chi_{\max}$ or to a fixed truncation error. The truncated singular values define the new Schmidt spectrum $\tilde{\Lambda}_B$ across the AB bond, and the left and right singular vectors are reshaped back into updated site tensors $\tilde{\Gamma}_A$ and $\tilde{\Gamma}_B$ in canonical form by distributing appropriate powers of the neighbouring $\Lambda$-matrices, in complete analogy with the finite-TEBD update. Due to translation invariance, this single update is representative of applying $U_{AB}$ simultaneously on all AB bonds of the infinite chain.

The BA bonds are updated in exactly the same manner, now treating the pattern $\Lambda_B \Gamma_B^{s_B} \Lambda_A \Gamma_A^{s_A} \Lambda_B$ as the local two-site object. After the AB step has produced updated tensors and a new $\tilde{\Lambda}_B$, we conceptually place two copies of the updated unit cell next to each other and identify the BA bond in the middle. We then apply the two-site gate $U_{BA} = \exp(-i h_{BA} \delta t)$ on that bond and perform the same sequence: absorb bond matrices, form the two-site tensor, reshape, carry out an SVD, truncate the spectrum, and reinsert the diagonal $\Lambda$-matrices to obtain updated $\tilde{\Gamma}_B$, $\tilde{\Gamma}_A$ and a new $\tilde{\Lambda}_A$. Combining these pieces, a single second-order Trotter step of duration $\delta t$ consists of three such unit-cell updates: an AB update with step $\delta t/2$, a BA update with step $\delta t$, and a final AB update with step $\delta t/2$. The computational cost of each two-site update is of order $\mathcal{O}(d^3 \chi^3)$, and, crucially, it is independent of the system size.

For imaginary-time evolution, we replace $\delta t$ by $-i\delta\tau$ and repeatedly apply $U(\delta\tau) \approx e^{-\delta\tau H}$ to an initial iMPS $\ket{\Psi_0}$ that has nonzero overlap with the true ground state. After each Trotter step, the state is re-normalized and truncated back to the fixed bond dimension in the two-site updates. In the thermodynamic limit, assuming a gapped and nondegenerate ground state within the chosen unit-cell structure, this procedure converges (for sufficiently small $\delta\tau$ and sufficiently large $\chi$) to a stationary iMPS that approximates the ground state. Convergence is monitored by the energy density and local observables, as well as by the stability of the Schmidt spectra $\Lambda_A$ and $\Lambda_B$ across successive steps.

Real-time iTEBD is obtained by using real $\delta t$ and tracking $\ket{\Psi(t)} = U(t)\ket{\Psi(0)}$ in the thermodynamic limit. Starting from a ground state prepared by iDMRG or imaginary-time iTEBD, one can study quenches and the spreading of correlations in an infinite system by applying a sequence of Trotter steps and computing local observables and correlation functions using the standard transfer-matrix formalism for iMPS. In contrast to imaginary time, however, generic global quenches induce entanglement growth that is approximately linear in time, so that the bond dimension required to faithfully represent the state grows rapidly. For a fixed $\chi$, iTEBD therefore provides reliable real-time dynamics only up to a finite time scale, beyond which truncation errors become dominant as increasingly large portions of the Schmidt spectrum are discarded.

As in the finite TEBD algorithm, the two-site updates are non-unitary once truncation is included and, in imaginary time, the evolution operator is itself non-unitary. This gradually destroys the exact left- and right-orthonormality of the tensors. In the thermodynamic limit, one restores a well-conditioned canonical form by viewing the unit cell as defining transfer operators on an infinite chain, computing their dominant left and right fixed points, and using them to construct a gauge transformation that re-imposes orthonormality conditions on the site tensors and normalizes the Schmidt spectra. In practice, iTEBD implementations interleave Trotter steps with such canonicalization steps for the infinite MPS.

Conceptually, iTEBD and iDMRG act on the same object: a small unit cell that generates a uniform iMPS. In iDMRG, the unit cell is updated by a variational energy minimization; in iTEBD it is updated by local time-evolution gates and SVD-based truncation. Both methods share the same scaling with bond dimension and exploit the same canonical-form machinery. Together with the quasiparticle and tangent-space constructions discussed above, they provide a coherent toolbox to study ground states, excitations and dynamics directly in the thermodynamic limit within the MPS framework.
\section{Infinite density-matrix renormalization group}
The finite-system ground state algorithms discussed so far work on a chain of length $L$ and
optimize an MPS with open boundary conditions. In many applications, however, one is interested
directly in the thermodynamic limit $L \to \infty$ of a translationally invariant system, where the
ground state is expected to be well described by a uniform (or periodically repeating) MPS. The
infinite density-matrix renormalization group (iDMRG) formulates the ground state search
variationally in this thermodynamic limit, using an infinite MPS (iMPS) of small unit cell.

\subsection{Thermodynamic limit and unit-cell MPS}

We consider a one-dimensional lattice with a local Hilbert space $\mathbb{C}^d$ and a
short-range, translationally invariant Hamiltonian $H = \sum_j h_{j,j+1}$. In the thermodynamic
limit, a natural variational class of states is a periodic iMPS with a unit cell of $\ell_u$ sites,
\begin{equation}
	\ket{\Psi[A^{[1]},\dots,A^{[\ell_u]}]}
	=
	\sum_{\{s\}}
	\cdots
	A^{[1]\,s_i}
	A^{[2]\,s_{i+1}}
	\cdots
	A^{[\ell_u]\,s_{i+\ell_u-1}}
	A^{[1]\,s_{i+\ell_u}}
	\cdots
	\ket{\{s\}}\,.
\end{equation}
In Schollwöck's formulation, one focuses on the particularly simple and practically very useful
case of a two-site unit cell ($\ell_u = 2$) in $\Gamma\Lambda$-notation. After a few infinite-system
DMRG steps, the state near the center of the chain exhibits a repeating fragment built from
two neighboring $\Gamma$-tensors and the adjacent $\Lambda$-matrices. This fragment can be
identified as a candidate unit cell of the thermodynamic limit state. Repeating this unit cell
does not reproduce the finite chain exactly, but for large system size it becomes an increasingly
good approximation to the translation-invariant bulk state.

In terms of the $\Gamma\Lambda$-notation, this building block is written schematically as
\begin{equation}
	\Lambda_A \, \Gamma_A \, \Lambda_B \, \Gamma_B \,,
\end{equation}
or, equivalently, in the $A,B$-notation as a two-site MPS tensor pair $(A,B)$ with a single
bond matrix $\Lambda$ between them. The iDMRG algorithm takes this observation seriously
and promotes the identification of such a unit cell to the central object of the variational
ground state search.

\subsection{From infinite-system DMRG to iDMRG}

The original infinite-system DMRG grows a chain by adding two sites at each step and
optimizing a central block, hoping that the middle of a long chain is representative of the
thermodynamic limit. In MPS language, this process generates a sequence of states with
central $\Gamma\Lambda$-structure as depicted in Schollwöck's Fig.~60. The key additional
ingredient leading to iDMRG is a \emph{prediction} step: one uses the current unit cell to
construct a high-quality variational guess for the next, slightly longer system.

Concretely, suppose that after $\ell$ growth steps the state is represented near the center by
two-site tensors $A^{[\ell]}$, $B^{[\ell]}$ and a bond matrix $\Lambda^{[\ell]}$ between them. One identifies
the two-site fragment
\begin{equation}
	\Lambda^{[\ell-1]} \, \Gamma_A^{[\ell]} \, \Lambda^{[\ell]} \, \Gamma_B^{[\ell]}
\end{equation}
as a putative unit cell and uses it to build an ansatz for the state at step $\ell+1$ by inserting
this fragment once into the center of the chain. In $A,B,\Lambda$-language this corresponds to
duplicating the central two-site block (with appropriate $\Lambda^{-1}$ factors so that the
resulting MPS remains well-conditioned) and thereby growing the system by two sites. This
predicted state provides an initial guess for the next DMRG minimization over the central
degrees of freedom.

The combination of
\begin{enumerate}
	\item infinite-system DMRG growth, and
	\item a prediction step based on the identified unit cell
\end{enumerate}
is what Schollwöck refers to as iDMRG: a variational ground state search formulated directly
in the thermodynamic limit. Algorithmically, the local energy minimization at each step is
still the standard two-site variational optimization within the MPS manifold; the essential
difference is that the \emph{structure} of the state is constrained to be generated by repeated
copies of a small unit cell, and that the information from the previous step is propagated
forward through the prediction.

\subsection{Fixed-point condition and convergence in the thermodynamic limit}

The central question in iDMRG is how to decide that the thermodynamic limit has been
reached. In the MPS framework, this is most naturally formulated as a fixed-point condition
for the reduced density matrices of the semi-infinite chain.

Let $\rho^{[\ell]}_A$ denote the reduced density operator of the left half of the chain at
iteration step $\ell$, which in canonical form is obtained from the Schmidt coefficients as
\begin{equation}
	\rho^{[\ell]}_A = \Lambda^{[\ell]} (\Lambda^{[\ell]})^\dagger\,.
\end{equation}
For a fixed finite system, reduced density matrices at different bipartitions are related by the
usual MPS consistency conditions; in infinite-system DMRG, however, $\rho^{[\ell]}_A$ and
$\rho^{[\ell-1]}_A$ come from chains of different lengths $2\ell$ and $2(\ell-1)$. The iDMRG
philosophy is that convergence to the thermodynamic limit corresponds to a fixed point of
this map,
\begin{equation}
	\rho^{[\ell]}_A \approx \rho^{[\ell-1]}_A \qquad (\ell \to \infty)\,.
\end{equation}
In practice, one monitors a distance or fidelity between the two reduced density operators.
A convenient choice is the Uhlmann fidelity
\begin{equation}
	F(\rho,\sigma) = \Tr\sqrt{\sqrt{\rho}\, \sigma \sqrt{\rho}}\,,
\end{equation}
applied to suitably normalized versions of $\rho^{[\ell]}_A$ and $\rho^{[\ell-1]}_A$. Using the
MPS structure, this fidelity can be expressed directly in terms of singular values of a product
of $\Lambda$-matrices, so that no explicit density matrices need to be constructed. When
$F$ is sufficiently close to one, the state is effectively invariant under the iDMRG growth
step and may be interpreted as an approximation to the infinite-system ground state.

At that point, the tensors of the unit cell can be viewed as defining an iMPS,
\begin{equation}
	\ket{\Psi_{\mathrm{iDMRG}}}
	=
	\sum_{\{s\}}
	\cdots
	A^{[\ast]\,s_i}
	B^{[\ast]\,s_{i+1}}
	A^{[\ast]\,s_{i+2}}
	B^{[\ast]\,s_{i+3}}
	\cdots
	\ket{\{s\}}\,,
\end{equation}
where $A^{[\ast]}$, $B^{[\ast]}$ and the associated bond matrix $\Lambda^{[\ast]}$ are taken from a
late iteration at which the fixed-point condition is satisfied to the desired accuracy. This
iMPS can then be brought into the standard mixed canonical form for an infinite system,
and expectation values are computed using the corresponding transfer matrices and their
dominant eigenvectors, exactly as for any other uniform MPS.

\subsection{Variational character and practical remarks}

The iDMRG algorithm is variational by construction: at each step, the local optimization
minimizes the energy within the manifold of MPS generated by a given unit cell and bond
dimension, and the prediction step merely acts as an informed initialization of this local
minimization. The resulting energy density is thus an upper bound on the true ground state
energy per site for the chosen bond dimension.

From a practical point of view:
\begin{itemize}
	\item The two-site unit cell can be generalized to larger unit cells whenever the ground state
	      breaks translation invariance (e.g.\ dimerization or more complex patterns).
	\item The prediction step dramatically accelerates convergence compared to naive
	      infinite-system DMRG with random initializations, because the overlap between
	      the predicted and optimized states is typically extremely close to one.
	\item Once the iDMRG fixed point has been obtained, the resulting iMPS provides an
	      ideal starting point for other thermodynamic-limit algorithms, such as iTEBD
	      or tangent-space methods for excitations.
\end{itemize}
In this way, iDMRG provides a conceptually simple yet powerful route to accessing ground
states directly in the thermodynamic limit within the MPS framework.

\chapter{Time evolution}
\section{Time-dependent variational principle}
The time evolution of quantum many-body systems poses an immediate challenge: the exact Schr\"odinger equation
\begin{equation}
	i \frac{\partial}{\partial t} \ket{\psi(t)} = H \ket{\psi(t)}
\end{equation}
generically drives the state into regions of Hilbert space that cannot be efficiently represented. Even if the initial state admits a compact MPS representation, the entanglement growth under time evolution causes the bond dimensions required for an exact representation to increase rapidly. The \emph{time-depedent variational principle} provides a systematic way of retaining an efficient representation by constraining the volution to remain on the variational manifold of fixed-bond-dimension MPS.

TDVP begins by viewing the set of MPS with fixed bond dimension~$\chi$ as a smooth but curved manifold $\mathcal{M}_{\text{MPS}}$ embedded inside the full Hilbert space. The Schr\"odinger equation gives sends the vector $-i H \ket{\psi}$ out of the manifold. The core idea is to select the `best possible' trajectory within $\mathcal{M}_{\text{MPS}}$ by projecting this tangent vector onto the tangent space $T_{\ket{\psi}}\mathcal{M}_{\text{MPS}}$ at the current state
\begin{equation}
	i \frac{\partial}{\partial t} \ket{\psi} = \mathcal{P}_{T_{\ket{\psi}\mathcal{M}}} (H \ket{\psi})\,.
\end{equation}
This is the defining TDVP equation. It is the unique evolution obtained by minimizing the norm of the residual
\begin{equation}
	\lVert i \ket{\dot{\psi}} - H \ket{\psi} \rVert\,,
\end{equation}
over all tangent vectors $i \ket{\dot{\psi}} \in T_{\ket{\psi}}\mathcal{M}$ and therefore supplies a variationall optimal approximation to the exact time evolution.
The tangent space $T_{\ket{\psi}} \mathcal{M}$ has a particularly simple structure. A state in the tangent space at $\ket{\Psi[A]}$ is parametrized as follows
\begin{equation}
	\ket{\Phi[B]} = \sum_{i} B^{i} \frac{\partial}{\partial A^i} \ket{\Psi[A]}\,,
\end{equation}
where the index $i$ is a collective index representingboth the physical and internal indices of the tensors.The above state can be represented by the summed TN diagram
\begin{equation}
	\ket{\Psi[B]} = \sum_i \cdots
	\begin{diagram}
		\foreach \i in {0, ..., 4} {
				\draw[fill=violet!30, rounded corners] (2*\i, -0.5) rectangle (2*\i + 1, 0.5);
				\draw (2*\i - 0.5, 0) -- (2*\i, 0);
				\draw (2*\i + 1, 0) -- (2*\i + 1.5, 0);
				\draw (2*\i + 0.5, -0.5) -- (2*\i + 0.5, -1);
			}
		\draw (0.5, 0) node {$A^{i-2}$};
		\draw (2.5, 0) node {$A^{i-1}$};
		\draw (4.5, 0) node {$B^i$};
		\draw (6.5, 0) node {$A^{i+1}$};
		\draw (8.5, 0) node {$A^{i+2}$};
	\end{diagram}
	\cdots
\end{equation}
This geometric viewpoint is crucial: TDVP is not just an algorithmic trick but a Hamiltonian flow on a symplectic manifold. In particular TDVP evolution is norm-preserving, energy-conserving for time indepedent Hamiltonians, and appromixates the exact Schr\"odinger equation with arbitrary accuracy when the variational manifold is sufficiently expressive. However, note that this particular parmetrization of tangent space vectors is overcomplete and has the following gauge freedom
\begin{equation}
	\begin{diagram}
		\pic (A) at (0, 0) {MPStens={$B$}};
	\end{diagram}
	\longrightarrow
	\begin{diagram}
		\pic (A) at (0, 0) {MPStens={$B$}};
	\end{diagram}
	+
	\begin{diagram}
		\pic (M) at (0, 0) {Matrix={$X$}};
		\pic (B) at (1.5, 0) {MPStens={$A$}};
	\end{diagram}
	-
	\begin{diagram}
		\pic (B) at (0.0, 0) {MPStens={$B$}};
		\pic (M) at (2.5, 0) {Matrix={$A$}};
	\end{diagram}
\end{equation}
This gauge redundancy can be traced back to the fact that the original uMPS representation is not unique: for any invertible matrix $G$, the transformation
\begin{equation}
	A^s \longrightarrow G^{-1} A^s G
\end{equation}
leaves the physical state unchanged. Considering an infinitesimal gauge transformation $G = e^{\epsilon X} \approx 1 + \epsilon X$ gives us the first order change in $A^s$ as
\begin{equation}
	\delta A^s = A^s X - X A^s\,.
\end{equation}
Thus the corresponding tangent vector in parameter space is
\begin{equation}
	B^s = A^s X - X A^s\,,
\end{equation}
which does not change the physical state -- such directions are called \emph{null} or \emph{vertical} tangent directions along the gauge orbit of the uMPS. These pure-gauge tangent vectors have zero physical norm and therefore the Gram matrix of tangent vectors
\begin{equation}
	G_{ij} = \expval{\partial_i \Psi(\bar{A}) | \partial_j \Psi(A)}
\end{equation}
develops a non-trivial kernel. This becomes a problem because the tangent space projector has the form
\begin{equation}
	P_A \sim \ket{\partial_i \Psi[A]} (G^{-1})^{ij} \bra{\partial_j\Psi[\bar{A}]}
\end{equation}
To obtain a well-defined, physical tangent space, one has to impose a gauge-fixing condition. In this case we choose the \emph{left gauge fixing condition} given by
\begin{equation}
	\label{eq:gaugefixing}
	\begin{diagram}
		\pic (TL) at (0, 0) {applyTransferLeft={$\bar{B}$}{$l$}{$A$}};
	\end{diagram}
	= 0 =
	\begin{diagram}
		\pic (TL2) at (0, 0) {applyTransferLeft={$\bar{A}$}{$l$}{$B$}};
	\end{diagram}
\end{equation}
which is the statement that
\begin{equation}
	\sum_s {A^s}^\dagger l {B^s} = 0 = \sum_s {B^s}^\dagger l A^s = 0\,.
\end{equation}
\begin{equation}
	\begin{split}
		\expval{\Phi[\bar{B}, \bar{A}] | \Psi[A]} & =
		\sum_{i \in \mathbb{Z}} \cdots
		\begin{diagram}
			\foreach \i in {0,...,4} {
					\draw[rounded corners, fill=cyan!30] (2*\i, -0.5) rectangle (2*\i + 1, 0.5);
					\draw (2*\i + 1, 0) -- (2*\i + 2, 0);
					\draw (2*\i - 1, 0) -- (2*\i, 0);
					\draw (2*\i + 0.5, -0.5) -- (2*\i + 0.5, -1.5);
					\draw [rounded corners, fill=purple!30] (2*\i, -2.5) rectangle (2*\i + 1, -1.5);
					\draw (2*\i + 0.5, 0) node {$A$};
					\draw (2*\i + 1, -2) -- (2*\i + 2, -2);
					\draw (2*\i - 1, -2) -- (2*\i, -2);
					\draw (2*\i + 0.5, -2) node {$\bar{A}$};
				}
			\draw [rounded corners, fill=blue!30] (4, -2.5) rectangle (5, -1.5);
			\draw (4.5, -2) node {$\bar{B}$};
		\end{diagram}
		\cdots                                                                                                                                   \\
		                                          & = \sum_{i \in \mathbb{Z}} \begin{diagram}
			                                                                      \draw[fill=red!30] (0, 0) circle (0.5);
			                                                                      \draw (0, 0) node {$\ell$};
			                                                                      \draw[fill=cyan!30, rounded corners] (1, 0.5) rectangle (2, 1.5);
			                                                                      \draw (1.5, 1) node {$A$};
			                                                                      \draw[fill=blue!30, rounded corners] (1, -1.5) rectangle (2, -0.5);
			                                                                      \draw (1.5, -1) node {$\bar{B}$};
			                                                                      \draw[fill=orange!30] (3, 0) circle (0.5);
			                                                                      \draw (3, 0) node {$r$};
			                                                                      \draw (0, 0.5) edge[out=90, in=180] (1, 1);
			                                                                      \draw (0, -0.5) edge[out=270, in=180] (1, -1);
			                                                                      \draw (2, -1) edge[out=0, in=270] (3, -0.5);
			                                                                      \draw (2, 1) edge[out=0, in=90] (3, 0.5);
		                                                                      \end{diagram} \\
		                                          & = 2\pi \delta(0)
		\begin{diagram}
			\draw[fill=red!30] (0, 0) circle (0.5);
			\draw (0, 0) node {$\ell$};
			\draw[fill=cyan!30, rounded corners] (1, 0.5) rectangle (2, 1.5);
			\draw (1.5, 1) node {$A$};
			\draw[fill=blue!30, rounded corners] (1, -1.5) rectangle (2, -0.5);
			\draw (1.5, -1) node {$\bar{B}$};
			\draw[fill=orange!30] (3, 0) circle (0.5);
			\draw (3, 0) node {$r$};
			\draw (0, 0.5) edge[out=90, in=180] (1, 1);
			\draw (0, -0.5) edge[out=270, in=180] (1, -1);
			\draw (2, -1) edge[out=0, in=270] (3, -0.5);
			\draw (2, 1) edge[out=0, in=90] (3, 0.5);
		\end{diagram}
	\end{split}
\end{equation}
Where we have introduced the lattice representation of the $\delta$ function
\begin{equation}
	\sum_{n \in \mathbb{Z}} e^{i p n} = 2\pi \delta(p)\,.
\end{equation}

A parametrization that explicitly enforces the gauge-fixing is obtained by constructing the matrix $V_L$ which contains the $D(d-1)$ dimensional null space of the matrix $M_{\alpha, (\beta s)} = {A^s_{\alpha,\delta}}^\dagger \sqrt{l}_{\delta \beta}$. This matrix is orthonormalized as
\begin{equation}
	\begin{diagram}
		\filldraw[fill=cyan!30, rounded corners] (0, 1) rectangle (1, 2);
		\filldraw (0.5, 1.5) node {$V_L$};
		\filldraw[fill=blue!30, rounded corners] (0, -2) rectangle (1, -1);
		\filldraw (0.5, -1.5) node {$\bar{V}_L$};
		\draw (0.5, 1) -- (0.5, -1);
		\draw (1, 1.5) -- (1.5, 1.5);
		\draw (1, -1.5) -- (1.5, -1.5);
		\draw (-1, 0) edge[out=90, in=180] (0, 1.5);
		\draw (-1, 0) edge[out=270, in=180] (0, -1.5);
	\end{diagram}
	=
	\begin{diagram}
		\draw (-1, 0) edge[out=90, in=180] (0, 1.5);
		\draw (-1, 0) edge[out=270, in=180] (0, -1.5);
	\end{diagram}
\end{equation}
In practice this construction can be carried out as follows
\begin{enumerate}
  \item In the left-canonical gauge the tensors $A_L^{s} \in \mathbb{C}^{D \times D}$ satisfy
  \begin{equation}
    \sum_{s=1}^{d} (A_L^{s})^{\dagger} A_L^{s} = \mathds{1}_{D}\,.
  \end{equation}
  It is convenient to regard the $A_L^{s}$ as blocks of a single isometry
  \begin{equation}
    \mathcal{A} :=
    \begin{bmatrix}
      A_L^{1} \\ \vdots \\ A_L^{d}
    \end{bmatrix}
    \in \mathbb{C}^{(dD)\times D},
    \qquad
    \mathcal{A}^{\dagger}\mathcal{A} = \mathds{1}_{D}\,,
  \end{equation}
  where the row index of $\mathcal{A}$ is the combined index $(s,\alpha)$.

  \item A tangent vector is specified by a collection of matrices $B^{s} \in \mathbb{C}^{D \times D}$,
  which we stack in the same way:
  \begin{equation}
    \mathcal{B} :=
    \begin{bmatrix}
      B^{1} \\ \vdots \\ B^{d}
    \end{bmatrix}
    \in \mathbb{C}^{(dD)\times D}.
  \end{equation}
  The left gauge-fixing condition
  \begin{equation}
    \sum_{s=1}^{d} (A_L^{s})^{\dagger} B^{s} = 0
  \end{equation}
  is then equivalent to the single orthogonality constraint
  \begin{equation}
    \mathcal{A}^{\dagger}\mathcal{B} = 0\,,
  \end{equation}
  i.e.\ $\mathcal{B}$ takes values in the orthogonal complement of the column space
  $\mathrm{col}(\mathcal{A}) \subset \mathbb{C}^{(dD)}$.

  \item To construct a gauge-fixed parametrization, we choose an isometry
  \begin{equation}
    V_L \in \mathbb{C}^{(dD)\times (dD-D)}
  \end{equation}
  whose columns form an orthonormal basis of $\mathrm{col}(\mathcal{A})^{\perp}$:
  \begin{equation}
    V_L^{\dagger} V_L = \mathds{1}_{dD-D},
    \qquad
    \mathcal{A}^{\dagger} V_L = 0\,.
  \end{equation}
  Equivalently, we can view $[\mathcal{A}\; V_L]$ as a $(dD)\times(dD)$ unitary completing
  $\mathcal{A}$ to a full basis of $\mathbb{C}^{(dD)}$.

  \item Every tangent vector satisfying the left gauge-fixing condition is then parametrized
  by a free matrix $X \in \mathbb{C}^{(dD-D)\times D}$ via
  \begin{equation}
    \mathcal{B}(X) = V_L\, X\,,
    \label{eq:gaugefixedB}
  \end{equation}
  which automatically obeys $\mathcal{A}^{\dagger} \mathcal{B}(X) = 0$ by construction.  Finally,
  we unstack $\mathcal{B}(X)$ back into blocks
  \begin{equation}
    B^{s}(X) = \mathcal{B}(X)_{[(s-1)D : sD], :}\,,
    \qquad s = 1,\dots,d\,,
  \end{equation}
  to obtain the gauge-fixed tangent tensor $B(X) = \{B^{s}(X)\}$.
\end{enumerate}
This construction ensures that the gauge condition is satisfied by construction,
\(\mathcal{A}^{\dagger}\mathcal{B} = 0\), and therefore removes the null modes
\(B^{s} \sim B^{s} + A_L^{s} X - X A_L^{s}\) associated with the gauge freedom of
the MPS representation.  The matrices \(X\) thus provide a set of \emph{unique
	coordinates} on the physical tangent space \(T^{\mathrm{phys}}_{|\psi\rangle}\).
\section{Quasiparticle ansatz}
The quasiparticle ansatz is formed by taking momentum superpositions of tangent space vectors in the mixed gauge
\begin{equation}
	\begin{split}
		\ket{\Phi_p[B]} & = \sum_{n} e^{ipn} \sum_{\{s\}}\left[\prod_{m < \ell} A^{s_m}_L\right]B^{s_\ell} \left[\prod_{m > \ell} A_R^{s_m}\right] \\
		                & = \sum_{n} e^{i p n} \cdots \begin{diagram}
			                                              \foreach \i in {0,1} {
					                                              \draw[rounded corners, fill=cyan!30] (2*\i, -0.5) rectangle (2*\i + 1, 0.5);
					                                              \draw (2*\i - 0.5, 0) -- (2*\i, 0);
					                                              \draw (2*\i + 1, 0) -- (2*\i + 1.5, 0);
					                                              \draw (2*\i + 0.5, -0.5) -- (2*\i + 0.5, -1);
					                                              \draw (2*\i + 0.5, 0) node {$A_L$};
				                                              }
			                                              \draw[rounded corners, fill=magenta!30] (4, -0.5) rectangle (5, 0.5);
			                                              \draw (3.5, 0) -- (4, 0);
			                                              \draw (5, 0) -- (5.5, 0);
			                                              \draw (4.5, -0.5) -- (4.5, -1);
			                                              \draw (4.5, 0) node {$B$};
			                                              \foreach \i in {3,4} {
					                                              \draw[rounded corners, fill=blue!30] (2*\i, -0.5) rectangle (2*\i + 1, 0.5);
					                                              \draw (2*\i - 0.5, 0) -- (2*\i, 0);
					                                              \draw (2*\i + 1, 0) -- (2*\i + 1.5, 0);
					                                              \draw (2*\i + 0.5, -0.5) -- (2*\i + 0.5, -1);
					                                              \draw (2*\i + 0.5, 0) node {$A_R$};
				                                              }
		                                              \end{diagram}
		\cdots
	\end{split}
\end{equation}
where we work with the mixed ansatz due to the possibility of topologically non-trivial excitations. In case the excitations are topologically trivial, $A_R$ and $A_L$ will be related by a simple gauge transformation. Since the excitation ansatz is a boosted version of the tangent vector, it has the same redundancies
\begin{equation}
	\begin{diagram}
		\draw[rounded corners, fill=magenta!30] (0, -0.5) rectangle (1, 0.5);
		\draw (0.5, 0) node {$B$};
		\draw (-0.5, 0) -- (0, 0);
		\draw (1, 0) -- (1.5, 0);
		\draw (0.5, -0.5) -- (0.5, -1);
	\end{diagram}
	\rightarrow
	\begin{diagram}
		\draw[rounded corners, fill=magenta!30] (0, -0.5) rectangle (1, 0.5);
		\draw (0.5, 0) node {$B$};
		\draw (-0.5, 0) -- (0, 0);
		\draw (1, 0) -- (1.5, 0);
		\draw (0.5, -0.5) -- (0.5, -1);
	\end{diagram}
	+
	\begin{diagram}
		\draw[fill=red!30] (0, 0) circle (0.5);
		\draw (0, 0) node {$X$};
		\draw (-1, 0) -- (-0.5, 0);
		\draw[rounded corners, fill=cyan!30] (1, -0.5) rectangle (2, 0.5);
		\draw (1.5, 0) node {$A_R$};
		\draw (1.5, -0.5) -- (1.5, -1);
		\draw (0.5, 0) -- (1, 0);
		\draw (2, 0) -- (2.5, 0);
	\end{diagram}
	- e^{i p} \,
	\begin{diagram}
		\draw (-0.5, 0) -- (0, 0);
		\draw[rounded corners, fill=blue!30] (0, -0.5) rectangle (1, 0.5);
		\draw (0.5, 0) node {$A_L$};
		\draw (1, 0) -- (1.5, 0);
		\draw (0.5, -0.5) -- (0.5, -1);
		\draw[fill=red!30] (2, 0) circle (0.5);
		\draw (2, 0) node {$X$};
		\draw (2.5, 0) -- (3, 0);
	\end{diagram}
\end{equation}
This gauge freedom corresponds to the zero modes in the variational subspace that make the variational optimization ill-conditioned. We can get rid of the gauge freedom by imposing again the left gauge-fixing condition
\begin{equation}
	\begin{diagram}
		\draw[rounded corners, fill=cyan!30] (0, 1) rectangle (1, 2);
		\draw[rounded corners, fill=magenta!30] (0, -2) rectangle (1, -1);
		\draw (0, -1.5) edge[out=180, in=180] (0, 1.5);
		\draw (1, 1.5) -- (1.5, 1.5);
		\draw (1, -1.5) -- (1.5, -1.5);
		\draw (0.5, 1) -- (0.5, -1);
		\draw (0.5, 1.5) node {$B$};
		\draw (0.5, -1.5) node {$\bar{A}_L$};
	\end{diagram}
	=
	\begin{diagram}
		\draw[rounded corners, fill=cyan!30] (0, 1) rectangle (1, 2);
		\draw[rounded corners, fill=magenta!30] (0, -2) rectangle (1, -1);
		\draw (0, -1.5) edge[out=180, in=180] (0, 1.5);
		\draw (1, 1.5) -- (1.5, 1.5);
		\draw (1, -1.5) -- (1.5, -1.5);
		\draw (0.5, 1) -- (0.5, -1);
		\draw (0.5, 1.5) node {$A_L$};
		\draw (0.5, -1.5) node {$\bar{B}$};
	\end{diagram}
	= 0\,.
\end{equation}
The same ansatz used to parametrize the tangent space vector in the mixed gauge can be used
\begin{equation}
	\begin{diagram}
		\draw[rounded corners, fill=cyan!30] (0, -0.5) rectangle (1, 0.5);
		\draw (0.5, 0) node {$B$};
		\draw (-0.5, 0) -- (0, 0);
		\draw (1, 0) -- (1.5, 0);
		\draw (0.5, -0.5) -- (0.5, -1);
	\end{diagram}
	=
	\begin{diagram}
		\draw[rounded corners, fill=brown!30] (0, -0.5) rectangle (1, 0.5);
		\draw (0.5, 0) node {$V_L$};
		\draw (1, 0) -- (1.5, 0);
		\draw (-0.5, 0) -- (0, 0);
		\draw (0.5, -0.5) -- (0.5, -1);
		\draw[rounded corners, fill=purple!30] (1.5, -0.5) rectangle (2.5, 0.5);
		\draw (2.5, 0) -- (3, 0);
		\draw (2, 0) node {$X$};
	\end{diagram}
\end{equation}
This gauge fixing condition ensures that the excitation is orthogonal to the MPS manifold. The overlap between two excitation ansatze works out to be
\begin{equation}
	\expval{\Phi_p'[B(X')] | \Phi_p[B(X)] } = 2 \pi \delta(p - p') \text{Tr}((X')^\dagger X)\,,
\end{equation}
which is reminiscent of the fact that plane wave states are only $\delta$-normalizable. Having removed the zero modes and parametrized the excitation tensor as $B = V_L X$, the variational problem for a quasiparticle of momentum $p$ reduces to minimizing the Rayleigh quotient
\begin{equation}
	\omega(p; X)
	= \frac{
		\matrixel{\Phi_p[B(X)]}{H}{\Phi_p[B(X)]}
	}{
		\braket{\Phi_p[B(X)]}{\Phi_p[B(X)]}
	}\,.
\end{equation}
Here $H$ is the microscopic Hamiltonian, and we assume that it has been shifted such that the ground state $\ket{\Psi(A)}$ has zero energy density,
\begin{equation}
	\frac{1}{|\mathbb{Z}|} \frac{\matrixel{\Psi(A)}{H}{\Psi(A)}}{\braket{\Psi(A)}{\Psi(A)}} = 0\,,
\end{equation}
so that $\omega(p; X)$ directly measures the excitation energy above the ground state.

Because the ansatz is linear in $X$ and the expectation values are quadratic functionals, this variational problem can be written as a generalized eigenvalue problem in the parameter space of $X$,
\begin{equation}
	H_{\mathrm{eff}}(p)\, X = \omega(p)\, N_{\mathrm{eff}}(p)\, X\,.
\end{equation}
The effective Hamiltonian and norm matrices are defined by their matrix elements as
\begin{align}
	2\pi \delta(p - p')\,
	(X')^\dagger H_{\mathrm{eff}}(p)\, X
	 & = \matrixel{\Phi_{p'}[B(X')]}{H}{\Phi_p[B(X)]}\,,
	\label{eq:Heff-def}
	\\
	2\pi \delta(p - p')\,
	(X')^\dagger N_{\mathrm{eff}}(p)\, X
	 & = \braket{\Phi_{p'}[B(X')]}{\Phi_p[B(X)]}\,.
	\label{eq:Neff-def}
\end{align}
In other words, $H_{\mathrm{eff}}(p)$ and $N_{\mathrm{eff}}(p)$ are the representations of $H$ and of the identity operator in the tangent-space basis $\{ \ket{\Phi_p[B(X)]} \}$.

Using the gauge fixing and parametrization introduced above, the overlap of two excitation ansätze at the same momentum takes the simple Euclidean form
\begin{equation}
	\braket{\Phi_p[B(X')]}{\Phi_p[B(X)]}
	= 2\pi \delta(p - p') \, \Tr\!\big[(X')^\dagger X\big]\,.
\end{equation}
Comparing with \eqref{eq:Neff-def}, we immediately conclude that
\begin{equation}
	N_{\mathrm{eff}}(p) = \mathds{1}\,,
\end{equation}
so that the generalized eigenvalue problem reduces to an ordinary Hermitian eigenvalue problem
\begin{equation}
	H_{\mathrm{eff}}(p)\, X = \omega(p)\, X\,.
\end{equation}
The gauge fixing has thus eliminated the redundant directions in parameter space and removed any nontrivial metric, so that the variational subspace carries the standard Euclidean inner product.

In principle, $H_{\mathrm{eff}}(p)$ is a matrix of dimension
\begin{equation}
	\dim X = D^2 (d - 1)\,,
\end{equation}
where $D$ is the MPS bond dimension and $d$ the local physical dimension. Explicitly constructing and diagonalizing this matrix would be prohibitively expensive: forming all its entries scales as $\mathcal{O}(D^6)$, since each matrix element involves contracting a network with two copies of the excitation tensor and the local Hamiltonian density.

Instead, one works with the \emph{action} of $H_{\mathrm{eff}}(p)$ on a trial vector $X$. Given a vector $X$, one first reconstructs the corresponding excitation tensor
\begin{equation}
	B(X) = V_L X\,,
\end{equation}
and then evaluates the linear map
\begin{equation}
	X \;\longmapsto\; X' = H_{\mathrm{eff}}(p)\, X
\end{equation}
by contracting the infinite MPS environment with a single insertion of $B(X)$ and the two-site Hamiltonian density. Diagrammatically, this amounts to summing all connected contributions to
\begin{equation}
	\matrixel{\Phi_{p'}[B(X')]}{H}{\Phi_p[B(X)]}
\end{equation}
that are linear in $B(X')$ and $B(X)$, with all other legs saturated by the left- and right-canonical tensors $A_L$ and $A_R$. Technically, one introduces left and right ``Hamiltonian environments'' that resum the semi-infinite chains of $A_L$/$A_R$ with insertions of the local Hamiltonian density, together with “excitation environments’’ that resum the geometric series of transfer matrices dressed with $B(X)$ and the appropriate phase factors $e^{\pm i p}$. The resulting contraction defines a tensor
\begin{equation}
	\widetilde{H}_{\mathrm{eff}}(p)\, B(X)\,,
\end{equation}
which carries the same index structure as $B$. Finally, one projects this result back into the tangent-space parameterization by contracting with $V_L^\dagger$,
\begin{equation}
	X' = H_{\mathrm{eff}}(p)\, X
	= V_L^\dagger\, \big[\,\widetilde{H}_{\mathrm{eff}}(p)\, B(X)\,\big]\,.
\end{equation}
All contractions involved in this map scale as $\mathcal{O}(D^3)$, so that the cost of applying $H_{\mathrm{eff}}(p)$ to a vector is cubic in the bond dimension. This makes it practical to use an iterative Hermitian eigensolver (such as Lanczos or Arnoldi) that only requires repeated applications of $H_{\mathrm{eff}}(p)$, rather than its explicit matrix representation.

By solving
\begin{equation}
	H_{\mathrm{eff}}(p)\, X_\alpha(p) = \omega_\alpha(p)\, X_\alpha(p)
\end{equation}
for each momentum $p$, one obtains an approximation to the excitation spectrum in the one-quasiparticle sector. For a given $p$, the number of eigenvalues equals the dimension of the variational subspace, but typically only a small number of lowest-lying eigenvalues correspond to well-defined elementary excitations living on isolated branches; the remaining eigenvalues belong to a continuum and are not faithfully captured by the single-quasiparticle ansatz.

\newpage
\part{Results}


\chapter{Results: Anharmonic Oscillator}
We now turn to numerical simulations. We recommend using virtual environments to keep one's Python environments separate and have a modular structure to one's projects. The global \texttt{Python} distribution is best left untouched. For modularizing and creating virtual environments, we used the \texttt{conda} package manager. We will warm-up using some simple codes in Mathematica.

\section{Warm-up: Mathematica}
\label{sec:spectrumextraction}
We consider the anharmonic oscillator with Hamiltonian
\begin{equation}
    H = \frac{p^2}{2} + \frac{x^2}{2} + \lambda x^4\,.
\end{equation}
We represent this Hamiltonian in the eigenbasis of the free harmonic oscillator, i.e. the basis that diagonalizes $H_0 = p^2/2 + x^2/2$. To obtain a finite-dimensional qubit encoding, we truncate the Hilbert space to the lowest four eigenstates of $H_0$. Each basis state $\ket{n}$ in this truncated space is then mapped to a computational basis state whose bitstring is the binary representation of the integer $n$. The steps to obtain the spectrum from this Hamiltonian are as follows
\begin{enumerate}
    \item The matrix representation can be decomposed into a string of Pauli operators using the identity
    \begin{equation}
        \Tr{\sigma_i \sigma_j} = 2 \delta_{ij}\,.
    \end{equation}
    This operation is readily available in \texttt{Qiskit} although for large Hamiltonians, this becomes computationally expensive and it is better to simply encode the states in such a way that the Hamiltonian has a simple representation in the computational basis. We have demonstrated such a rudimentary implementation with Gray encodings which is available in the GitHub repository.
    \item We time evolve a suitable state $|\psi\rangle$ over an interval $T=N dt$. Then we use Hadamard test to extract 
    $$C_k=\langle \psi| U^k |\psi\rangle.$$ Here $U=\exp(-i H dt)$ and $k$ runs from 1 to $N$. 
    \item We can always expand $|\psi\rangle$ in terms of the true energy eigenstates: 
    $$|\psi\rangle=\sum_{j} c_j |E_j\rangle.$$
    This gives
    \begin{equation}
        C_k=\sum_j |c_j|^2 \exp(-i E_j k \, dt)\,.
    \end{equation}
    \item Then we define
    \begin{equation}
        F(\omega)= \sum_{k=0}^{N-1} C_k \exp(-i \omega k \, dt)=\sum_{k,j} |c_j|^2 \exp(-i k \, dt (E_j+\omega))\,.
    \end{equation}
    Summing over $k$ leads to
    \begin{equation}
        F(\omega)= \sum_j |c_j|^2 \exp(i dt(E_j+\omega)/2) \frac{\sin N dt (E_j+\omega)}{\sin dt(E_j+\omega)}\,.
    \end{equation}
    This form implies that $|F(\omega)|$ will have peaks at $\omega=-E_j$.
\item So the logic is that we use the quantum computer and Hadamard test to construct $C_k$'s and then postprocess on a classical computer to construct $|F(\omega)|$ to extract the spectrum.
\end{enumerate}

To see this in action, let us write a short Mathematica code first and do some sanity checks. This is a good opportunity for you to pick up this fascinating package, which forms a bedrock for most theoretical research. Mathematica has an extensive help menu which will help you learn the syntax easily. 

\section{Oscillator basis}

We will see the steps outlined above in action using Mathematica. One simplification we will do is to avoid the Pauli decomposition step.

\begin{lstlisting}[style=mma]
\[Lambda] = 0.10;
dt = 0.12;

(* operators *)
x = 1/Sqrt[2] {{0, 1, 0, 0}, {1, 0, Sqrt[2], 0},
               {0, Sqrt[2], 0, Sqrt[3]}, {0, 0, Sqrt[3], 0}};

H0 = DiagonalMatrix[{0, 1, 2, 3}] + 1/2 IdentityMatrix[4];
H = N[ H0 + \[Lambda] x.x.x.x ];

(* state: |\[Psi]> = (|0>+|1>)/Sqrt[2] <-> (|00>+|01>)/Sqrt[2] *)
psi = {1/Sqrt[2], 1/Sqrt[2], 0, 0};

(* one-step propagator and correlator *)
U1 := MatrixExp[-I H dt];
Ck[k_Integer?NonNegative] := N[ Conjugate[psi].MatrixPower[U1, k].psi ];
\end{lstlisting}

This implements steps 1-3. We have chosen the state $|\psi\rangle=\frac{1}{\sqrt 2}(|0\rangle+|1\rangle)$, which is the superposition of the $\lambda=0$ ground and 1st excited state. This state can be decomposed into a superposition of the true $\lambda\neq 0$ eigenstates. 

\subsection{Continuous scan}
We define
\(
F(\omega)=\sum_k C_k\,e^{-i\omega k dt}
\).
Because \(C_k\) already carries \(e^{-iE_j k dt}\), peaks of \(|F|\) appear at \(\omega\approx -E_j\). To stay ASCII in code, we call the scan variable \texttt{wm}.

\begin{lstlisting}[style=mma]
Nt = 350;

F[wm_] := Sum[ Ck[k] Exp[-I wm k dt], {k, 0, Nt - 1} ];

Plot[ Abs[F[wm]], {wm, -2.5, 0},
      AxesLabel -> {"wm","|F|"}, PlotRange -> All,
      PlotLabel -> "Naive scan: peaks near -E_j" ]
\end{lstlisting}
This implements steps 4-5. This leads to the following plot:
\begin{figure}[h]
  \centering
  \includegraphics[width=0.5\linewidth]{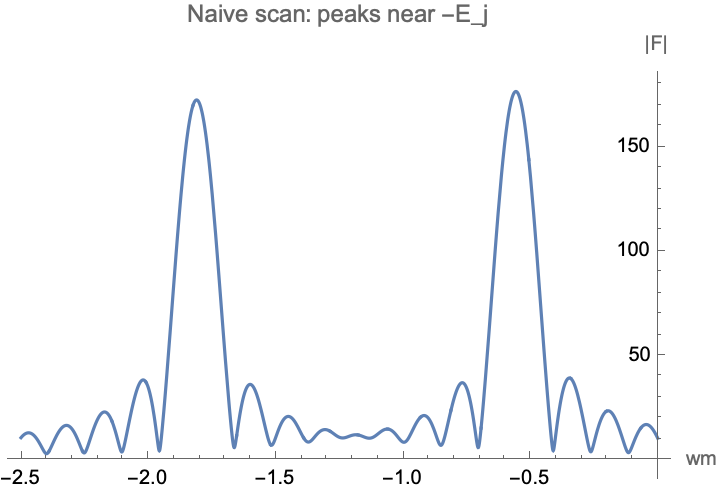}
  \caption{$|F(\omega)|$ vs $\omega$.}
\end{figure}

\textbf{Scales to remember:}
Nyquist band \(|\omega|<\pi/dt\);\quad resolution \(\Delta E \approx 2\pi/(N_t dt)\).

As you can see from the plot, there are 2 prominent big peaks. We can eyeball that these peaks have height greater than 60. Using the convenient mathematica command FindPeaks, we can extract their locations using the following commands:
\begin{lstlisting}[style=mma]
 datatab = Table[Abs[F[w]], {w, -2.5, 0, .01}];

 FindPeaks[datatab, 60]
\end{lstlisting}
Using this we find the peaks corresponding to $E_0=0.55$ and $E_0=1.8$. The expected answers to 3 decimal places are $0.557$ and $1.809$. So we have done well! However, we can do a little better by using a trick called Hanning window.

\subsection{ Hann window to remove oscillations}
There is a simple trick to remove the undesired small oscillations using what is called the Hann window (students are encouraged to find out about this). 

\begin{lstlisting}[style=mma]
CC = Table[ Ck[k], {k, 0, Nt - 1} ];
meanCC = Mean[CC];
w[k_] := HannWindow[ k/(Nt - 1.) ];

FHann[wm_] := Sum[ (Ck[k] - meanCC) * w[k] * Exp[-I wm k dt], {k, 0, Nt - 1} ];

Plot[ {Abs[F[wm]], Abs[FHann[wm]]}, {wm, -2.5, 0},
      PlotLegends -> {"naive","Hann + de-mean"},
      AxesLabel -> {"wm","|F|"}, PlotRange -> All ]
\end{lstlisting}

\paragraph{What changes?} True peaks stay; low-\(|\omega|\) junk collapses. Main lobe widens slightly (resolution trade).

We find:
\begin{figure}[h]
  \centering
  \includegraphics[width=0.75\linewidth]{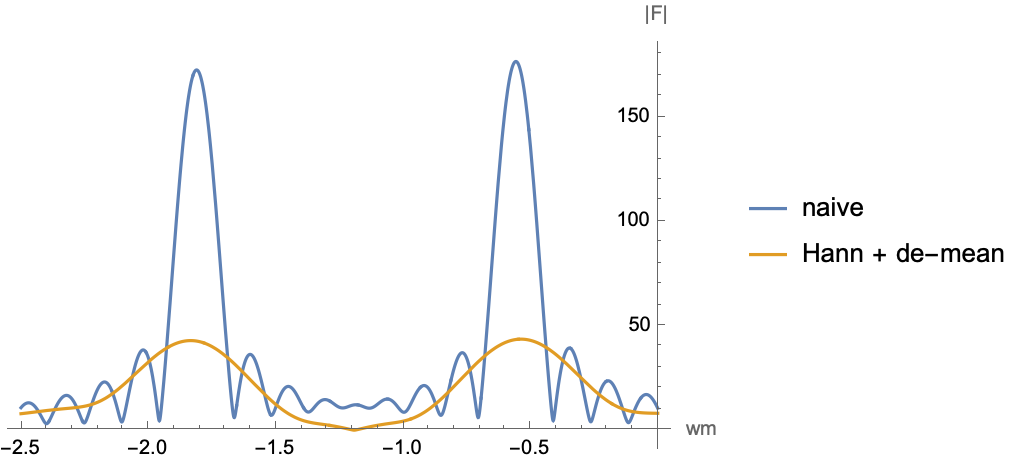}
  \caption{$|F(\omega)|$ vs $\omega$ after Hann is given by the orange line. This gives a clean set of 2 peaks.}
\end{figure}

\section*{ Changing \texttt{dt}: keep it consistent}
If you change \texttt{dt}, regenerate both \texttt{U1} and \texttt{Ck}. Otherwise the axis scales and peaks drift by a factor.
\begin{lstlisting}[style=mma]
dt = 0.10;
U1 = MatrixExp[-I H dt];
CC = Table[ Ck[k], {k, 0, Nt - 1} ];
meanCC = Mean[CC];
FHann[wm_] := Sum[ (Ck[k]-meanCC) w[k] Exp[-I wm k dt], {k,0,Nt-1}];
\end{lstlisting}

\section*{ Two-line summary}
\begin{itemize}
\item \(F(\omega) = \sum_k (C_k-\overline{C})\,w_k\,e^{-i\omega k dt}\); peaks at \(\omega\approx -E_j\).
\item Use \texttt{HannWindow} and subtract the mean; fix \texttt{dt} and increase \texttt{Nt} to sharpen peaks.
\end{itemize}

\section{Position-basis (JLP) encoding for one-site \texorpdfstring{$\phi^4$}{phi^4}}
\label{sec:jlp-position}

\paragraph{Goal.}
Digitize a single anharmonic oscillator site in the \emph{position basis} as in Jordan--Lee--Preskill (JLP):
we represent the field value $\phi$ on a uniform grid of $N=2^n$ points (so $n$ qubits), build
\(
H = \tfrac12\,\pi^2 + \tfrac12\,\omega^2 \phi^2 + \lambda\,\phi^4,
\)
then compute the correlator $C_k=\!\bra{\psi} e^{-i H\,k\,dt}\ket{\psi}$ and recover energies from its spectrum. The reason to consider this alternate encoding is that for the $\lambda \phi^4$ field theory that we will consider soon, the $\phi$ pieces of the Hamiltonian are simply diagonal while the momentum pieces can be made diagonal by doing a (quantum) Fourier Transform, which we know how to do. The steps are as follows:
\begin{enumerate}
    \item We discretize the field ($x$ in our case) to lie between $[-\phi_{max},\phi_{max})$ defining an interval $L_\phi=2\phi_{max}$. In our case, since we are interested in the first 2 eigenvalues for smallish $\lambda$, it is sufficient to consider $\phi_{max}$ such that the position space SHO wave-functions decay beyond this value. It turns out $\phi_{max}\sim 3.0$ will do  the job.
    \item We sample $\phi$ on a uniform grid. That is, we choose $\Delta\phi=2\phi_{max}/N$ and
    \begin{equation}
        \phi_j=-\phi_{max}+j \Delta \phi\,,\quad j=0,\cdots, N-1\,.
    \end{equation}
    This can also be rewritten as:
    $$\phi_m=m \Delta\phi$$ where $m=-N/2,-N/2+1\cdots, N/2-1$.
    \item We define the computational basis as 
    \begin{equation}
        \hat\phi|\phi_j\rangle=\phi_j|\phi_j\rangle\,.
    \end{equation}
    \item Define the momentum grid 
    \begin{equation}
    p_m= m\frac{2\pi}{L_\phi}\,,\quad m=-N/2,-N/2+1\cdots, N/2-1.
    \end{equation}
    using which we have the DFT
    \begin{equation}
        F_{mj}=\frac{1}{\sqrt{N}}e^{i p_m \phi_j}\,.
    \end{equation}
  Both $m,j$ run from $-N/2$ to $N/2-1$.  This gives us for the conjugate momentum 
    \begin{equation}
        \hat\Pi^2=F^\dagger {\rm diag}(p_m^2)F.
    \end{equation}
\end{enumerate}

\subsection*{Minimal Mathematica cell (JLP grid $\rightarrow$ correlator $\rightarrow$ peaks)}
The code mirrors our oscillator-basis workflow but works directly in the position grid.

\begin{lstlisting}[style=mma]
(* === Parameters === *)
lam   = 0.10;   omega = 1.0;
Ngrid = 4;      (* 2^n points; for 2 qubits use 4 *)
phimax = 3.0;   (* half-interval length; tune below *)

(* === Centered grids and operators (JLP) === *)
idx  = Range[-Ngrid/2, Ngrid/2 - 1];
dphi = 2 phimax/Ngrid;    L = 2 phimax;

phi  = dphi idx;                          (* positions phi_j *)
p    = (2 Pi/L) idx;                      (* momenta p_m *)

Phi  = DiagonalMatrix[phi];               (* position operator *)
Fmat = Table[Exp[2 Pi I m j/Ngrid]/Sqrt[Ngrid], {m, idx}, {j, idx}];
Pi2  = ConjugateTranspose[Fmat].DiagonalMatrix[p^2].Fmat;  (* spectral kinetic *)

(* Harmonic (lam=0) and full (lam>0) Hamiltonians *)
H0pos = 1/2 Pi2 + 1/2 omega^2 (Phi.Phi) // N;        (* quadratic part *)
Hpos  = H0pos + lam (Phi.Phi.Phi.Phi)    // N;        (* add phi^4 *)

(* Initial state: like oscillator case - use lowest two eigenstates of H0 (not Hpos) *)
vals0Vecs = Eigensystem[H0pos];
vals0 = vals0Vecs[[1]]; vecs0 = vals0Vecs[[2]];
ord0  = Ordering[vals0];
psi   = Normalize[ vecs0[[ord0[[1]]]] + vecs0[[ord0[[2]]]] ];

(* === Correlator and Hann-windowed scan === *)
dt = 0.12;   Nt = 256;                     (* fix dt; increase Nt to sharpen peaks *)
U1 = MatrixExp[-I Hpos dt];

Ck[k_Integer?NonNegative] := N[ Conjugate[psi].MatrixPower[U1, k].psi ];
CC = Table[Ck[k], {k, 0, Nt - 1}];
meanCC = Mean[CC];
w[k_] := HannWindow[ k/(Nt - 1.) ];

FHann[wm_] := Sum[ (CC[[k+1]] - meanCC) * w[k] * Exp[-I wm k dt], {k, 0, Nt - 1}];

(* Visual: peaks appear at negative wm (because of the -i sign in the kernel) *)
Plot[ Abs[FHann[wm]], {wm, -3, 0}, PlotRange->All,
      AxesLabel->{"wm","|F|"}, PlotLabel->"JLP position grid: peaks near -E0,-E1" ]

(* Minimal peak pick, exactly like earlier *)
grid   = Range[-3, 0, 0.01];
valsF  = Abs @ (FHann /@ grid);
peakHeights = FindPeaks[valsF];
peakIdx     = Flatten @ Position[valsF, Alternatives @@ peakHeights];
wmPeaks     = grid[[peakIdx]];     (* e.g., -3 + idx*0.01 *)
Eest        = -wmPeaks;            (* report +E by flipping the sign *)
\end{lstlisting}

\begin{figure}[htb]
  \centering
  \includegraphics[width=0.5\linewidth]{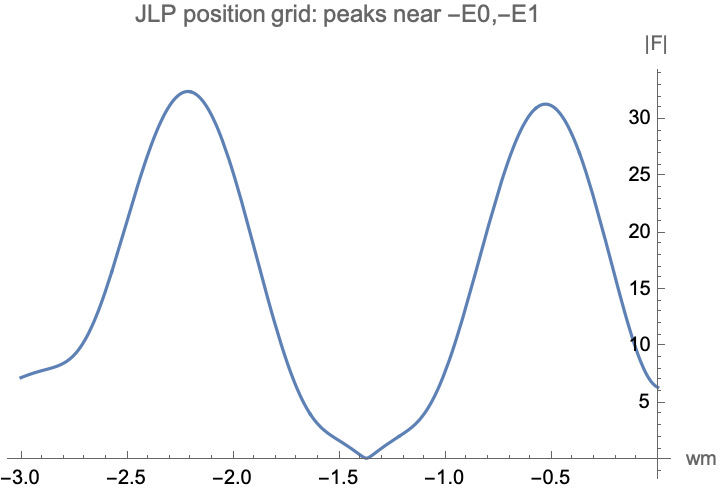}
  \caption{$|F_{Hann}(\omega)|$ vs $\omega$ using JLP encoding. $N=4$, $\phi_{max}=3$.}\label{JLPm}
\end{figure}

There are two scales to watch: the Nyquist band $|\omega|<\pi/dt$ (avoid aliasing). Resolution $\Delta E \approx \tfrac{2\pi}{N_t\,dt}$. Fix $dt$, then raise $N_t$ to sharpen peaks. Our findings are in fig.(\ref{JLPm}).

Note that while the first peak near $0.5$ is reasonably good, the second one is beyond the expected 1.8. The reason is that with the position basis, with a small number of samplings, we cannot accurately capture the higher excited wave functions.

\section{Qiskit}
In this section we introduce a basic quantum simulation of the harmonic oscillator using Qiskit. Qiskit provides a modular framework for constructing quantum circuits, representing Hamiltonians as sums of Pauli operators, and performing both exact and approximate time evolution. By encoding the harmonic-oscillator Hamiltonian into a qubit register and using Qiskit’s built-in tools for operator construction, diagonalization, and circuit-based evolution, we can prepare low-energy states, study their dynamics, and visualize the behaviour of the system directly from a quantum-circuit perspective. This simple example highlights how Qiskit can be used as a flexible platform for exploring quantum simulations of continuous-variable systems on digital quantum hardware.
\subsection{Building quantum circuits}
Before describing the implementation of the state-preparation algorithm, we devote this section to a brief introduction to Qiskit and to the use of the \texttt{Estimator} and \texttt{Sampler} primitives for computing expectation values and performing measurements in the next section. All examples are based on the quantum anharmonic oscillator (AHO). Qiskit can be installed using the \texttt{pip} Python package manager.
\begin{lstlisting}[style=pythonstyle]
pip install qiskit 
\end{lstlisting}
The central object is the \texttt{QuantumCircuit}, which can be instantiated by calling the constructor
\begin{lstlisting}[style=pythonstyle]
qc = QuantumCircuit(2) # creates a QuantumCircuit with two qubits.
\end{lstlisting}
Gates are realised as methods on this \texttt{QuantumCircuit} object, meaning that they can be called by the \texttt{.} (dot) operator. For example, the following circuit prepares the Bell pair $\ket{\Phi} = \frac{1}{2}(\ket{11} + \ket{00})$
\begin{lstlisting}[style=pythonstyle]
qc = QuantumCircuit(2)
qc.h(0)
qc.cx(0, 1)
qc.draw('mpl') # this draws the circuit with the MatplotLib backend
\end{lstlisting}
That's all there is to it, really. We can create functions that build quantum circuits. For example, the following function builds to circuit to obtain the state $\ket{(j)_2}$ from $\ket{0_2}$ where $j$ is a bitstring by iterating over the circuit and applying $X$ operators
\begin{lstlisting}[style=pythonstyle]
def prep_state(j: int) -> QuantumCircuit:
  bit_string = bin(j)[2:][::-1] # bitstrings come with a 0b prefix so [2:] is to discard that bit. [::-1] will reverse the bitstring so that the index order is little endian.
  num_bits = len(bit_string)
  qc = QuantumCircuit(num_bits)

  for char, idx in enumerate(bit_string):
    if char == '1':
      qc.x(idx)
  return qc
\end{lstlisting}
where we have used string slicing to remove the \texttt{0b} from the bitstring and then reversed the string because \texttt{Qiskit} follows the little-endian indexing of bits. Note that a \texttt{QuantumCircuit} is initialized with all the qubits in the 0 computational basis state. To inspect the final state, one can use the \texttt{Statevector} class from \texttt{qiskit.quantum\_info}. For example,
\begin{lstlisting}[style=pythonstyle]
state = Statevector.from_instruction(qc) # gives the state obtained as a result of applying qc to |0...0>
\end{lstlisting}
Let's now try to implement a circuit that builds the unitary implementing the time evolution under an anharmonic oscillator potential. A Strang-split implementation is straightforward and we detail the important bits in the following. Details are available in the notebook \texttt{Python/\allowbreak qiskit/\allowbreak qmwithqiskit.ipynb} in the GitHub repository. First let us implement the $x^2$ piece in the Hamiltonian. Recall from~\ref{sec:JLPencoding} that this is simply a $ZZ$-rotation which is conveniently available in the basic set of gates as \texttt{rzz}. The following function takes in our \texttt{QuantumCircuit} object and appends the unitary evolution due to $x^2/2$
\begin{lstlisting}[style=pythonstyle]
def x2_term(qc: QuantumCircuit, m_ctr: float, delta_t: float, x_max: float) -> None:
  qubits = qc.num_qubits
  x_prefactor = - x_max / (2 ** qubits - 1)
  delta_x = 2 * x_max / (2 ** qubits - 1)
  # apply the x^2 evolution
  for i in range(qubits):
    for j in range(i + 1, qubits):
      qc.rzz(2 * 2 ** (i + j) * (1 + m_ctr) * x_prefactor ** 2 * delta_t, i, j)
\end{lstlisting}
The number of qubits in a circuit can be accessed with the \texttt{num\_qubits} attribute. Keeping our goal of adiabatic state preparation in mind, we introduce a variable \texttt{m\_ctr} for the mass counterterm, which is set to zero during real-time evolution. The implementation of the $p^{2}$ term proceeds analogously, with the only modification being the application of a $\qft$ and its inverse to move into and out of the momentum basis. A small subtlety is that the bare Fourier transform does not map $\ket{x}$ directly to $\ket{p}$; to obtain the correct momentum eigenstates, we must insert appropriate $Z$-rotations throughout the circuit, as explained in section \ref{sec:JLPencoding}.

The $\qft$ is already implemented in \texttt{Qiskit} and can be brought to scope by importing it from \texttt{qiskit.circuit.library}. This is a \texttt{Gate} object and can be appended to a circuit by calling the \texttt{append()} method on the circuit.
\begin{lstlisting}[style=pythonstyle]
def p2_term(qc: QuantumCircuit, lam0f: float, delta_t: float, x_max: float):
  qubits = qc.num_qubits
  delta_x = 2 * x_max / (2 ** qubits - 1)
  p_prefactor = - np.pi / (2 ** qubits * delta_x)
  # do a fourier transform
  alpha = (1 / 2 ** qubits - 1) * np.pi
  for i in range(qubits):
    qc.rz(alpha * 2 ** i, i)
  qc.append(QFTGate(qubits), range(qubits))
  for i in range(qubits):
    qc.rz(alpha * 2 ** i, i)
  # then apply the p^2 evolution now
  for i in range(qubits):
    for j in range(i + 1, qubits):
      qc.rzz(2 * 2 ** (i + j) * lam0f * p_prefactor ** 2 * delta_t, i, j)
  # apply the inverse fourier transform
  for i in range(qubits):
    qc.rz( - alpha * 2 ** i, i)
  qc.append(QFTGate(qubits).inverse(), range(qubits))
  for i in range(qubits):
    qc.rz( - alpha * 2 ** i, i)
\end{lstlisting}
The QFTGate must be instantiated with the number of qubits it is to act on and the indices of the corresponding qubits in the circuit. Again we have kept a counterterm piece \texttt{lam0f} which can be set to zero for time evolution. Finally, we need a function that appends the unitary evolution due to $\lambda x^4$ piece. This is a $ZZZZ$ coupling that we will have to implement on our own.
\begin{lstlisting}[style=pythonstyle]
def rz4(qc: QuantumCircuit, theta: float, i: int, j: int, k: int, l: int) -> None:
  qc.cx(i, j)
  qc.cx(j, k)
  qc.cx(k, l)
  qc.rz(theta, l)
  qc.cx(k, l)
  qc.cx(j, k)
  qc.cx(i, j)
\end{lstlisting}
Then building the full unitary is tedious but straightforward. Hopefully the logic is clear from section~\ref{sec:JLPencoding}.
\begin{lstlisting}[style=pythonstyle]
def x4_term(qc: QuantumCircuit, lam: float, delta_t: float, x_max: float):
    qubits = qc.num_qubits
    x_prefactor = - x_max / (2 ** qubits - 1)
    delta_x = 2 * x_max / (2 ** qubits - 1)
    '''
    Apply the potential term.
    First we will apply the four-site Z interaction.
    '''
    for i in range(qubits):
        for j in range(i+1, qubits):
            for k in range(j+1, qubits):
                for l in range(k+1, qubits):
                    coeff = 2 ** (i+j+k+l) * 24 * lam * x_prefactor ** 4
                    rz4(qc, 2 * coeff * delta_t, i, j, k, l)
    '''
    Now we will apply the two-site term where two indices are equal.
    '''
    for i in range(qubits):
        for j in range(qubits):
            if j == i:
                continue
            for k in range(j+1, qubits):
                if k == i:
                    continue
                coeff = 2 ** (2 * i + j + k) * 12 * lam * x_prefactor ** 4
                qc.rzz(2 * coeff * delta_t, j, k)
    '''
    Finally apply the two site term where three indices are equal.
    '''
    for i in range(qubits):
        for j in range(qubits):
            if j == i:
                continue
            coeff = 2 ** (3 * i + j) * 4 * lam * x_prefactor ** 4
            qc.rzz(2 * coeff * delta_t, i, j)
\end{lstlisting}
All these pieces can be put together to obtain the full unitary evolution under the Hamiltonian
\begin{lstlisting}[style=pythonstyle]
def strang_step(qc: QuantumCircuit, lam: float, m_ctr: float, delta_t: float,
              x_max: float) -> None:
  # first apply x^4 and x^2
  x2_term(qc, m_ctr, delta_t / 2, x_max)
  x4_term(qc, lam, delta_t / 2, x_max)
  # then apply the p^2 piece
  p2_term(qc, 1, delta_t, x_max)
  # then apply x^4 and x^2 again
  x2_term(qc, m_ctr, delta_t / 2, x_max)
  x4_term(qc, lam, delta_t / 2, x_max)
\end{lstlisting}
The results obtained from this circuit is compared with a \texttt{NumPy} evolution in Fig. \ref{fig:ashocomparison}.
\subsection{Performing measurements and sampling with Qiskit primitives}
While we can use \texttt{Statevector} methods to inspect the final state after unitary evolution in simulation, this is not available on quantum hardware. On a QPU we access the state only through measurement outcomes and expectation values of observables. To support these tasks, \texttt{Qiskit} provides the \texttt{Sampler} and \texttt{Estimator} primitives, which abstract away low-level backend details and let us focus on algorithm design. One can bring the \texttt{Estimator} primitive to scope with the following statement
\begin{lstlisting}[style=pythonstyle]
from qiskit.primitives import StatevectorEstimator, StatevectorSampler 
# to create an instance named estimator
estimator = StatevectorEstimator()
\end{lstlisting}
Jobs can be run by passing the circuit, observables and the list of parameters the circuit has to be evaluated on. Observables can be instances of \texttt{SparsePauliOp}. For instance the Pauli string $x$ can be turned into a \texttt{SparsePauliOp} as follows
\begin{lstlisting}[style=pythonstyle]
pauli_z = SparsePauliOp('Z') # this creates the Pauli-z operator

# to set the position operator as a Pauli string, we need to pass a list of tuples containing the terms and their coefficients
x_op_list = [('I' * (num_qubits - i - 1) + 'Z' + 'I' * i, - 2 ** i * x_max / (2 ** num_qubits - 1)) for i in range(num_qubits)]
x_op = SparsePauliOp.from_list(x_op_list)

# the square of x i.e. x^2 can be obtained easily
x_sq_op = x_op @ x_op
\end{lstlisting}
Finally, we can pass what is called a primitive unified bloc (PUB) to an estimator instance and obtain the results
\begin{lstlisting}[style=pythonstyle]
# for example, to measure the expectation value of the SparsePauliOp x_op
# in the state prepared by the circuit qc, we pass a list of PUB tuples
job = estimator.run([(qc, x_op, [])])

# fetch the primitive result
result = job.result()[0] # job.result() returns a list of PUBResult objects. Since we had only one PUB, the list has just one element.

# access the expectation value(s)
data = result.data.evs
\end{lstlisting}
In the above, we pass an empty list because our circuit has no parameters. If we wanted to compute the expectation value for a range of circuits, for example, the expectation value of $x$ at different times, we would have passed a list of PUBs where the quantum circuits contain the Trotterized evolution at different times. For example in our AHO simulation, we want to measure how the position expectation value measures at different times. This can be achieved by passing a list of PUBs
\begin{lstlisting}[style=pythonstyle]
for steps in range(trotter_steps):
    sho_gs = QuantumCircuit(num_qubits)
    """
    ...
    circuit preparing the initial state
    ...
    """
    full_evolution(sho_gs, x_max, num_qubits, delta_t, steps) # circuit that evolves the initial state
    PUBs_list.append((sho_gs, [x_op, x_sq_op], [])) # append everything to the list of PUBs
\end{lstlisting}
The structure of the output follows directly from the input: each PUB in the input list yields one corresponding \texttt{PubResult}, and the shape of \texttt{evs} is the broadcasted shape of the observables and parameter-values supplied in that PUB.
\begin{figure}[t]
    \centering
    \begin{subfigure}{0.45\textwidth}
        \centering
        \includegraphics[width=\textwidth]{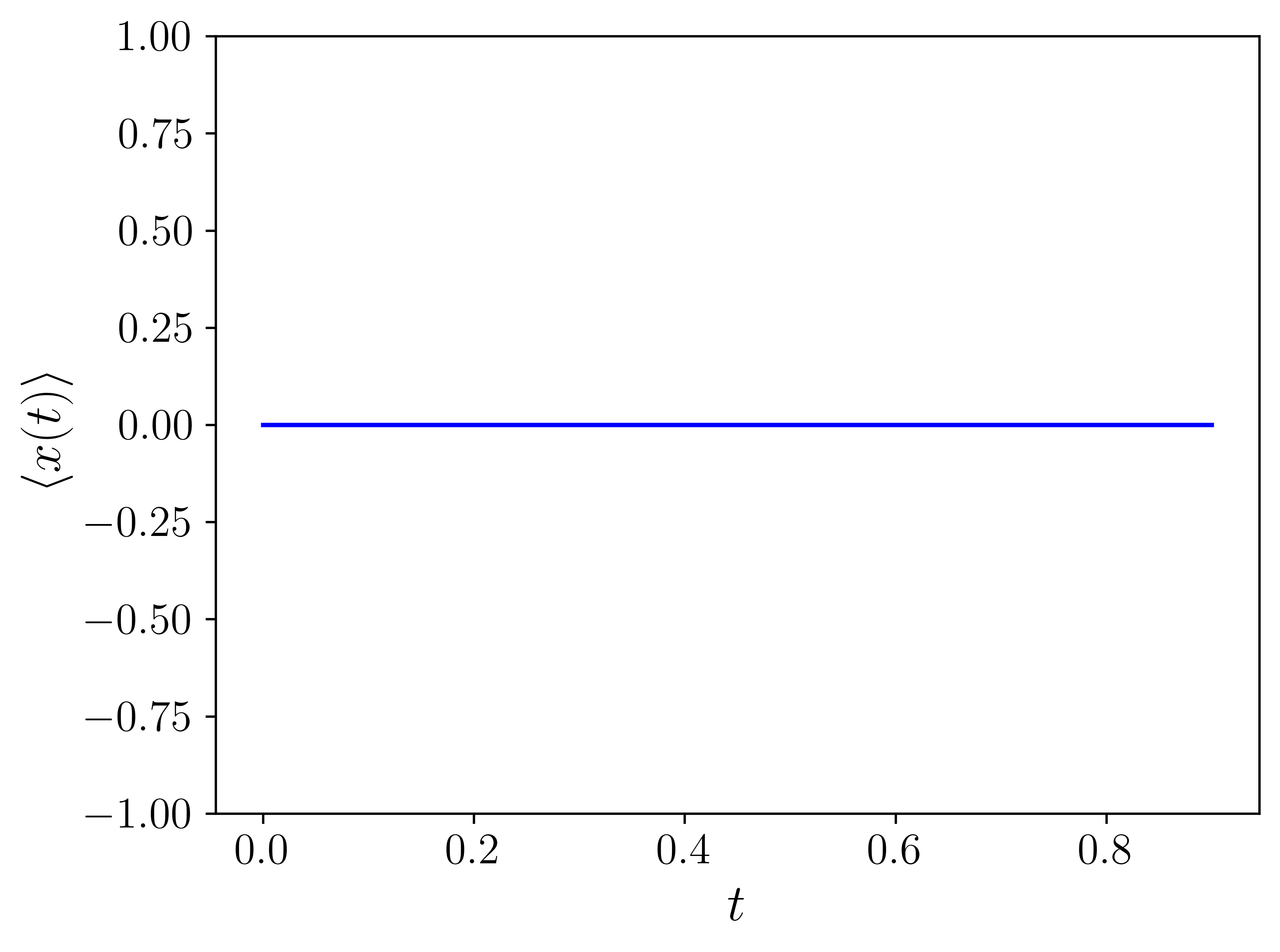}
        \label{fig:left-free-sho}
    \end{subfigure}
    \hfill
    \begin{subfigure}{0.45\textwidth}
        \centering
        \includegraphics[width=\textwidth]{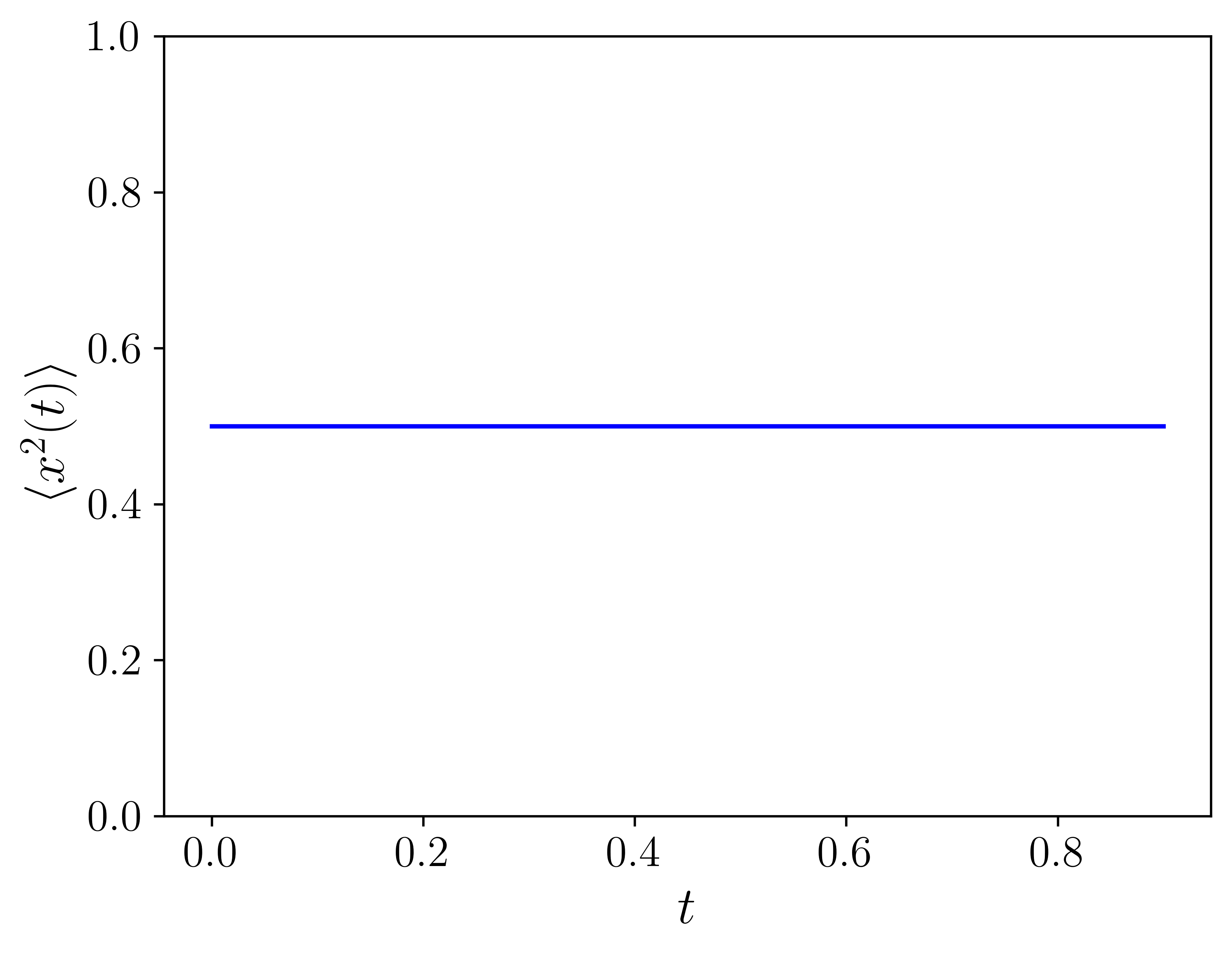}
        \label{fig:right-free-sho}
    \end{subfigure}
    \caption{Expectation values of $x$ and $x^2$ as a function of time for the free harmonic oscillator in its ground state.}
    \label{fig:twopanels}
\end{figure}
\subsection{Adiabatic ground-state preparation}
We now demonstrate adiabatic state preparation for the anharmonic oscillator. The idea is to start from a Hamiltonian whose ground state is trivial to prepare in the Jordan--Lee--Preskill (JLP) encoding, and then slowly dial the remaining terms so that the state tracks the instantaneous ground state.

We begin with the purely potential Hamiltonian
\begin{equation}
H_{\text{initial}} = \frac{x^2}{2}\,,
\end{equation}
which is diagonal in the position basis. On the discretized lattice there is no site exactly at $x=0$; the closest points are $x=\pm \delta_x$. Therefore, we prepare the even ground state as
\begin{equation}
\ket{\mathrm{GS}_0} = \frac{1}{\sqrt{2}}\left(\ket{+\delta_x}+\ket{-\delta_x}\right),
\end{equation}
where these two kets correspond to the two midpoint computational basis states. The odd ground state is obtained by simply changing the sign between the two kets above. These two states in the computational basis encoding introduced in section~\ref{sec:JLPencoding} have the following explicit form
\begin{equation}
    \ket{\delta_x} = \ket{1 0 \cdots 0}\,, \quad \ket{-\delta_x} = \ket{0 1 \cdots 1}
\end{equation}

The reader can verify that the following functions initialize the circuits that prepare the even and odd ground states of $H_{\text{initial}}$ above
\begin{lstlisting}[style=pythonstyle]
def prep_midpoint_even(n):
    """
    Prepare (|0 1 1 ... 1> + |1 0 0 ... 0>)/sqrt(2) on n qubits.
    We take qubit indexing as: q[0]=LSB, q[n-1]=MSB.
    """
    qc = QuantumCircuit(n)
    ctrl = n-1  # MSB

    # Put MSB in (|0> + |1>)/sqrt(2)
    qc.h(ctrl)

    # For each lower qubit, flip iff MSB == 0 (open-controlled X)
    for tgt in range(n-1):  # 0 .. n-2
        qc.x(ctrl)
        qc.cx(ctrl, tgt)
        qc.x(ctrl)

    return qc

def prep_midpoint_odd(n):
    qc = prep_midpoint_even(n)
    # Add relative phase: (|...> - |...>)/sqrt(2)
    qc.z(n-1)  # MSB
    return qc
  
\end{lstlisting}

\paragraph{Step 1: dialing up the kinetic term.}
From $\ket{\mathrm{GS}_0}$ we adiabatically turn on the kinetic energy to reach the simple harmonic oscillator (SHO) Hamiltonian
\begin{equation}
H_1 = \frac{x^2}{2}+\frac{p^2}{2}\,.
\end{equation}
We introduce a dimensionless ramp parameter $u=t/t_{\mathrm{prep}}\in[0,1]$ and evolve with the interpolating Hamiltonian
\begin{equation}
H_u=\frac{x^2}{2}+r(u)\frac{p^2}{2}\,.
\end{equation}
For the ramp we use the smoothstep function
\begin{equation}
r(u)=u^2(3-2u),
\end{equation}
so that $r(0)=0$, $r(1)=1$, and the derivative vanishes at both endpoints. The evolution is implemented using a second–order (Strang) Trotter step,
\begin{equation}
e^{-i\Delta t H_u}\approx
e^{-i\frac{\Delta t}{2}\frac{x^2}{2}}
e^{-i\Delta t\, r(u)\frac{p^2}{2}}
e^{-i\frac{\Delta t}{2}\frac{x^2}{2}},
\end{equation}
with $u$ evaluated at midpoints $u=(j-\tfrac12)/n_{\mathrm{steps}}$. This prepares the SHO ground state.

\paragraph{Step 2: turning on the interaction.}
Starting from the SHO ground state, we adiabatically dial the quartic potential to reach the anharmonic oscillator
\begin{equation}
H_{\mathrm{AHO}}=\frac{p^2}{2}+\frac{x^2}{2}+\lambda x^4\,.
\end{equation}
We again use a smooth ramp, but in two stages. For $u\in[0,1]$ we turn on the interaction together with a temporary mass counterterm, and for $u\in(1,2]$ we keep $\lambda$ fixed while smoothly removing the counterterm. Concretely,
\begin{align}
\lambda(u)&=
\begin{cases}
\lambda\, r(u), & 0\le u\le 1,\\
\lambda, & 1<u\le 2,
\end{cases}\\[4pt]
m_{\mathrm{ctr}}(u)&=
\begin{cases}
-6\lambda\, r(u), & 0\le u\le 1,\\
-6\lambda\,[1-r(u-1)], & 1<u\le 2.
\end{cases}
\end{align}
At $u=2$ the counterterm vanishes and we recover the target Hamiltonian. Each adiabatic step is implemented by a Strang splitting of the instantaneous Hamiltonian,
\begin{equation}
e^{-i\Delta t H_u}\approx
e^{-i\frac{\Delta t}{2}\left[\frac{x^2}{2}(1+m_{\mathrm{ctr}}(u))+\lambda(u)x^4\right]}
e^{-i\Delta t\,\frac{p^2}{2}}
e^{-i\frac{\Delta t}{2}\left[\frac{x^2}{2}(1+m_{\mathrm{ctr}}(u))+\lambda(u)x^4\right]},
\end{equation}
with sufficiently small $\Delta t$ and total preparation time $t_{\mathrm{prep}}$ to ensure adiabaticity. We provide our implementation in the notebook titled \texttt{adiabatic\_state\_prep.ipynb}. The Hamiltonian was encoded with $5$ qubits with $x_{\text{max}} = 10$. For $\lambda = 0.1$, the final state had a fidelity of $0.997$ with $\langle x \rangle = 0$ and $\langle x^2 \rangle = 0.352$.
\begin{figure}
    \centering
    \begin{subfigure}{0.48\textwidth}
        \centering
        \includegraphics[width=\textwidth]{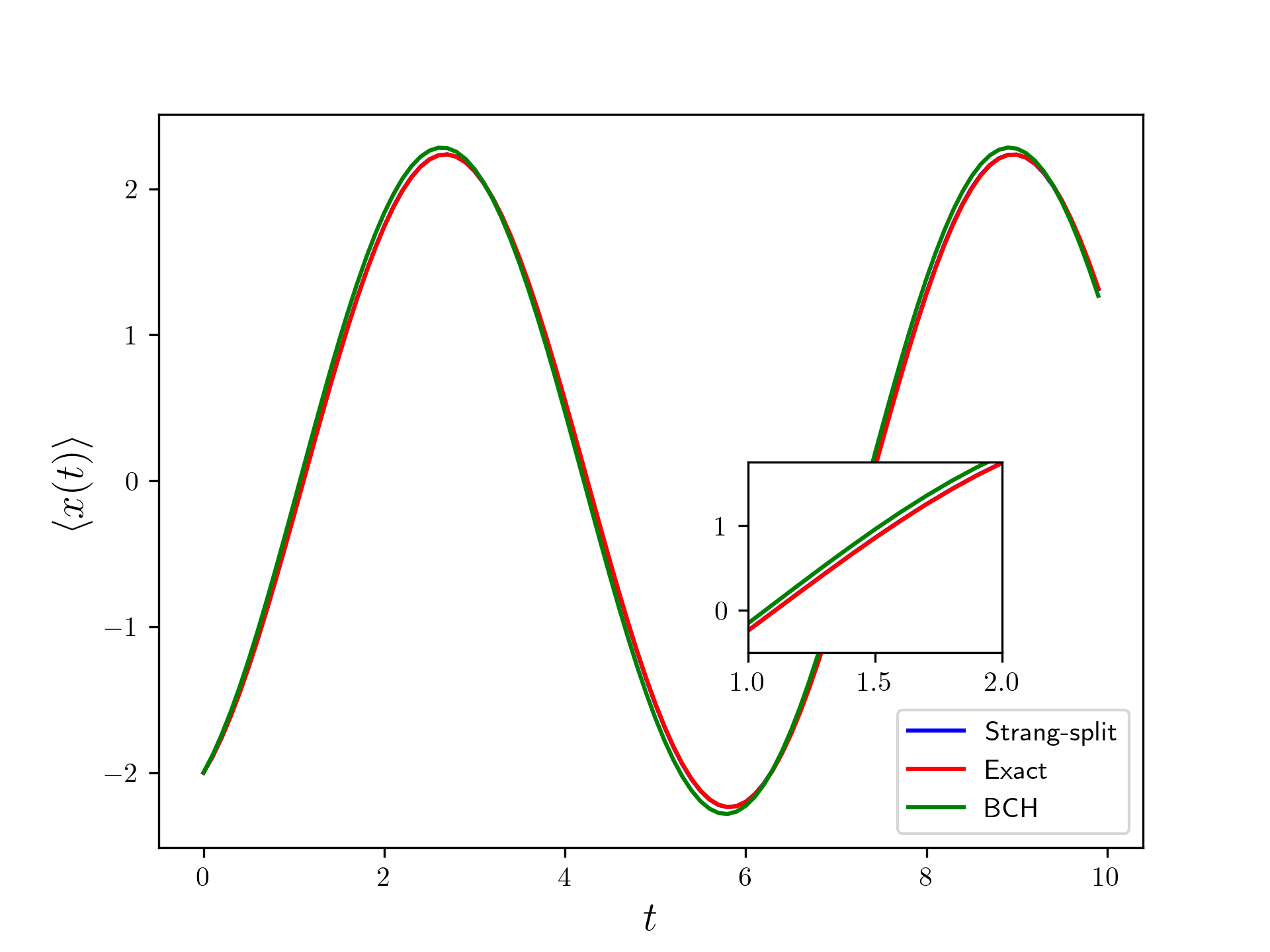}
    \end{subfigure}
    \hfill
    \begin{subfigure}{0.48\textwidth}
        \centering
        \includegraphics[width=\textwidth]{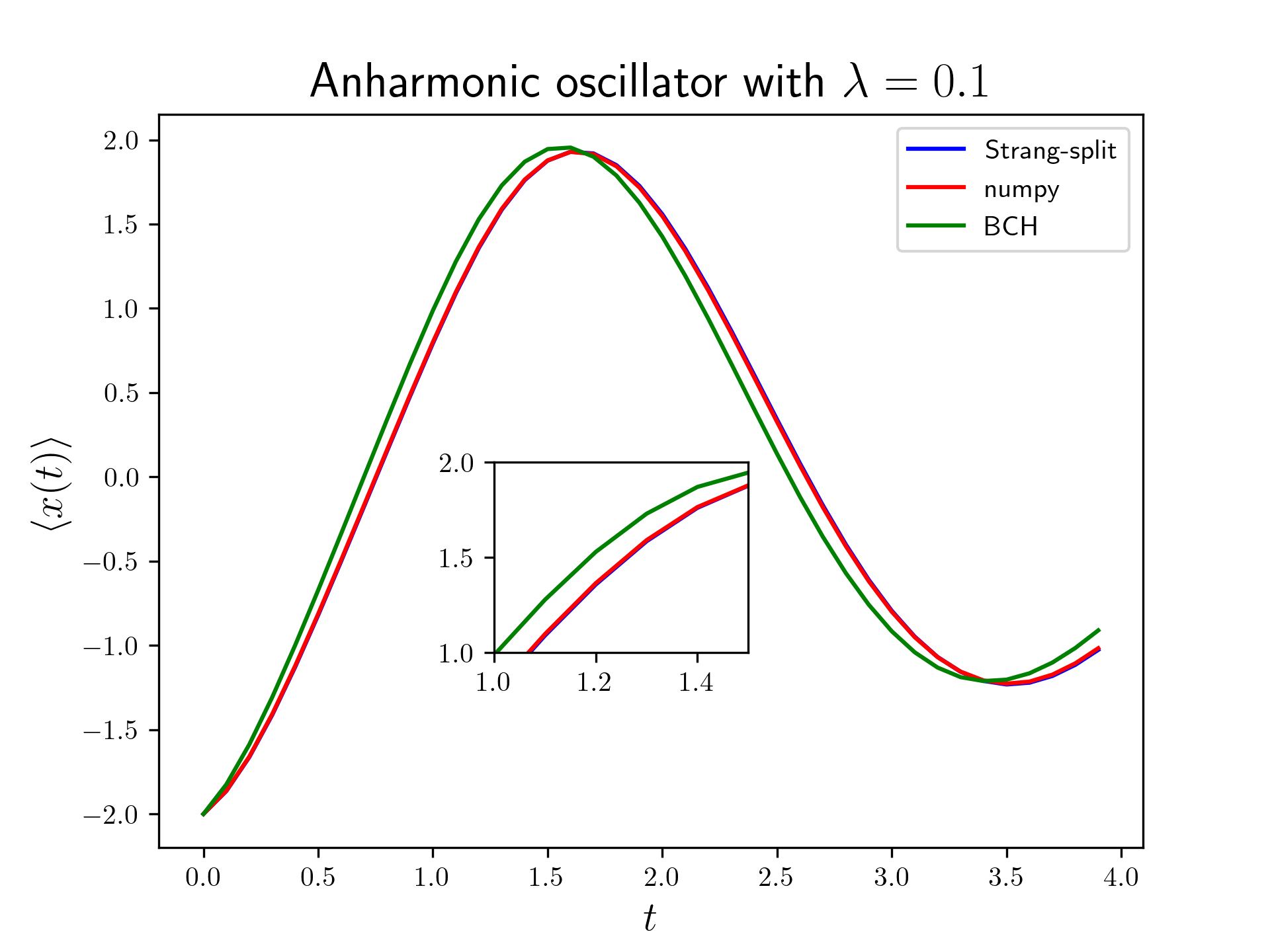}
    \end{subfigure}
    \caption{Comparison of the circuit output with \texttt{NumPy} without (left) and with (right) an anharmonic coupling.}
    \label{fig:ashocomparison}
\end{figure}
\section{Spectrum extraction with Qiskit}
\begin{figure}
    \centering
    \includegraphics[width=0.5\linewidth]{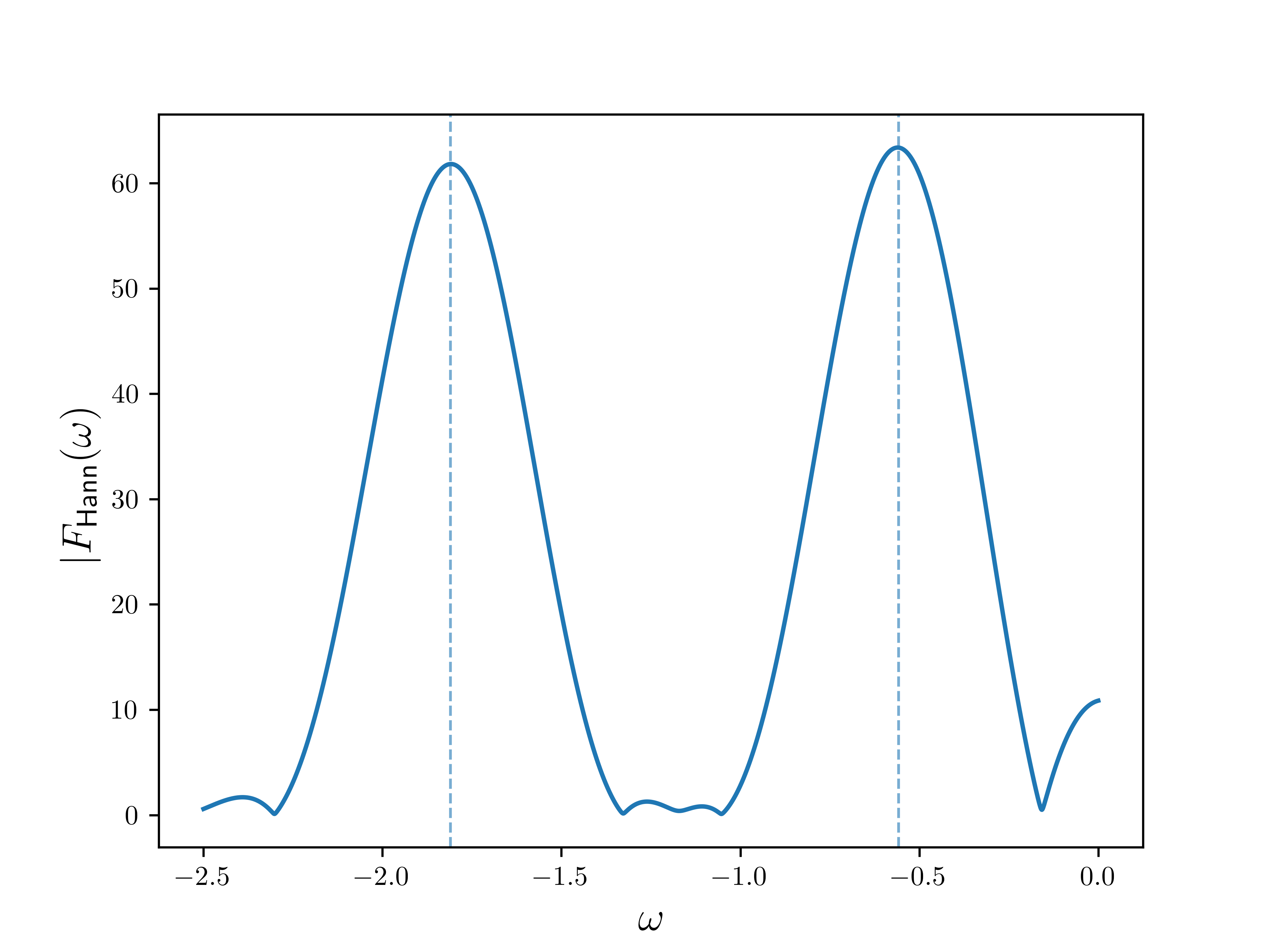}
    \caption{Spectrum extraction with the Hadamard test on \texttt{Qiskit} simulator and classical post-processing with \texttt{NumPy}.}
    \label{fig:spectrumextractionqiskit}
\end{figure}

Having shown how we can build circuits with \texttt{Qiskit}, we use the method outlined in section \ref{sec:spectrumextraction} to obtain the spectrum. We work in the SHO basis truncated to the first four levels and use the \texttt{SparsePauliOp.from\_operator()} method to obtain the Hamiltonian as Pauli strings. Implementing this four-level Hamiltonian is trivial, and we compare the result of our simulation with exact diagonalization in Fig.~\ref{fig:spectrumextractionqiskit}.

\section{Noise Sources on a Quantum Computer}

Any practical quantum computer suffers from errors that degrade the fidelity of quantum computations. For superconducting devices such as those provided by IBM Quantum, three main classes of errors dominate:
\begin{itemize}
    \item \textbf{Measurement (readout) errors.} Readout or measurement errors is the probability of misaligning the measured classical bit. These are caused by decoherence during measurement. This can result in biased expectation values. However, in correlation functions any oscillatory structure will remain with suppressed amplitudes and offset.
    \item \textbf{Gate errors.} Every gate implemented on hardware is imperfect due to interactions with the environment. Two-qubit gates like the C-NOT gate are especially error prone. Gate errors can depolarize or rotate states incorrectly, rekducing the coherence in time-correlation functions. Peaks in Fourier spectra are approximately obtained at the same frequency but are broadened.
    \item \textbf{Decoherence: $T_1$ and $T_2$ processes.} Even if we could control qubits perfectly, we cannot isolate them from the environment. They suffer from energy relaxation ($T_1$, amplitude damping from $\ket{1} \to \ket{0}$ due to photon loss) and dephasing ($T_2$, phase damping which destroys superpositions without energy loss).
\end{itemize}
We revisit the the spectrum extraction example in the next section with noise models. For this purpose, we will use the \texttt{QiskitAer} simulator framework which has support for running simulations with and without noise. Noisy circuits can be implemented in \texttt{QiskitAer} with the \texttt{NoiseModel} class from \texttt{qiskit\_aer.noise} module and classical readout errors can be implemented with the \texttt{ReadoutError} class. The implementation details for this example are in the GitHub repository.
\section{Results with errors}

\begin{figure}
  \centering
  \includegraphics[width=0.8\linewidth]{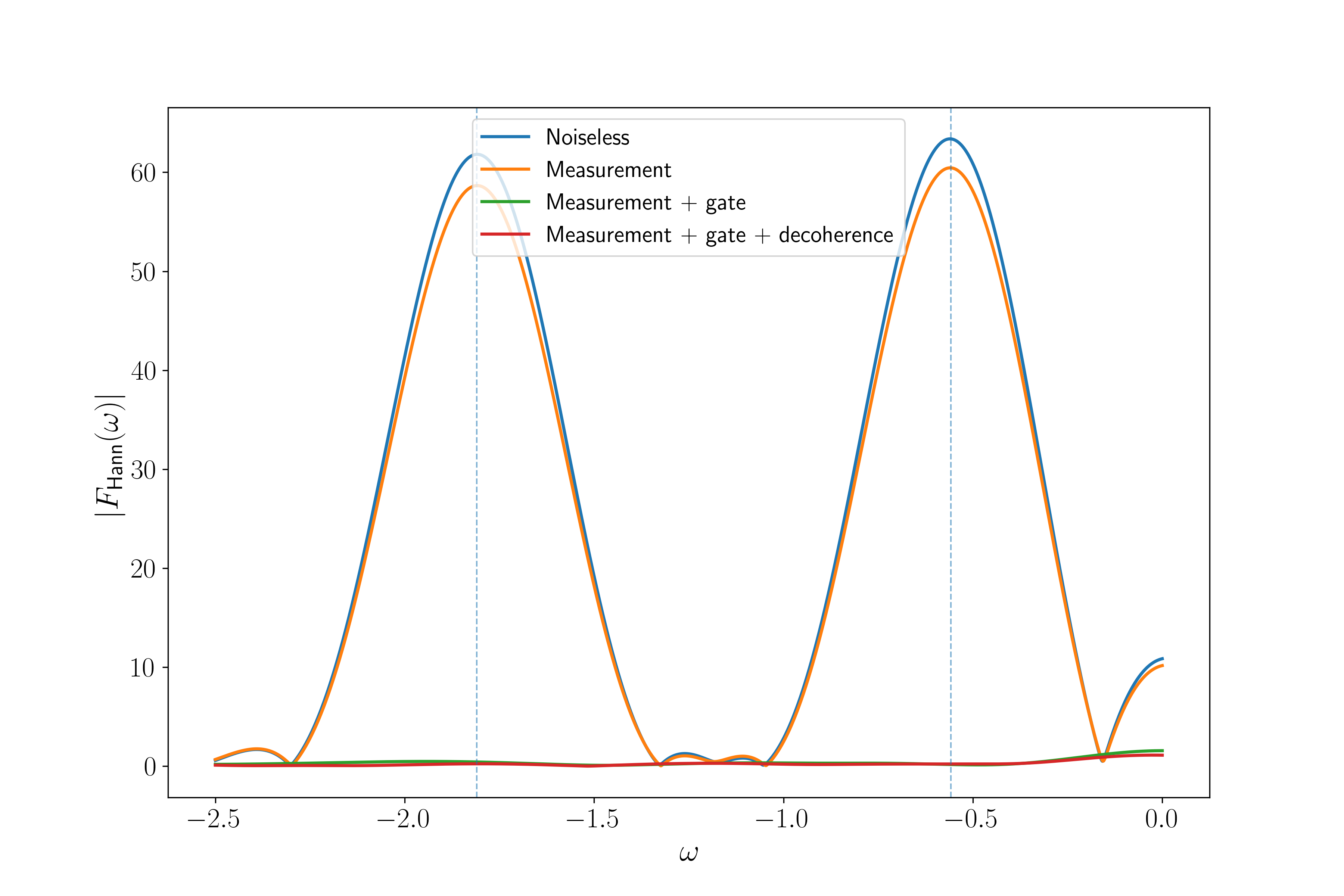}
  \caption{Hann spectra overlay for various noise channels. The simulations were run with a Strang-split step size of $dt = 0.12$, anharmonic coupling $\lambda = 0.1$ and 5000 shots.}
  \label{fig:error1}
\end{figure}

\begin{figure}
  \centering
  \includegraphics[width=0.8\linewidth]{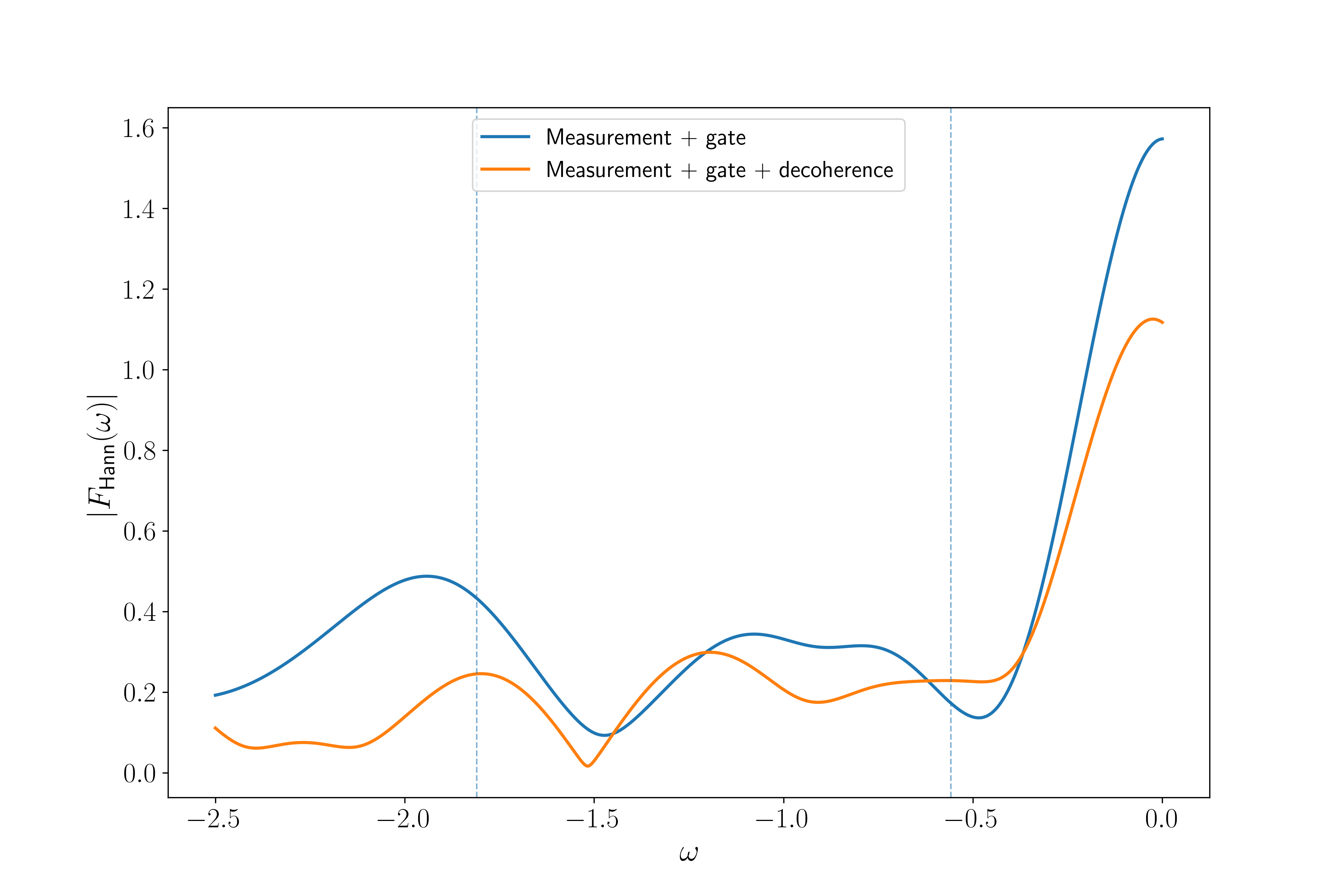}
  \caption{Plot showing the signal in the presence of measurement with gate errors (blue) and with both gate errors and decoherence (yellow). Gate errors alone are already sufficient to significantly reduce the signal-to-noise ratio.}
  \label{fig:error2}
\end{figure}

Here are the figures generated using qiskits for 2 noise models compared with the noiseless case. From Fig.~\ref{fig:error1}, we see that measurement errors do not have a significant effect on the signal but the addition of gate errors drastically reduces the signal to noise ratio. In Fig.~\ref{fig:error2}, we show the zoomed in plot of the signals with gate and decoherence errors included with measurement errors. It can be seen that although the signal amplitude has diminished by an order of magnitude, the residuals of the peaks are not very far away from their expected locations.

\subsection{Remarks}

\begin{itemize}
  \item Readout error can often be mitigated efficiently via calibration and post-processing (for eg. using a confusion matrix).
  \item Gate and decoherence errors are more serious: without error correction, one typically applies techniques such as zero-noise extrapolation (ZNE), randomized compiling (RC), and dynamical decoupling (DD).
  \item In all cases, Qiskit’s \texttt{AerSimulator} accepts a
        \texttt{noise\_model} argument so that simulation results can include
        these errors, allowing one to benchmark algorithms against realistic
        device behaviour.
\end{itemize}

\subsection{Error Mitigation Techniques}

Although full fault-tolerant error correction is not yet feasible on near-term devices, several mitigation techniques can reduce the impact of noise. These methods aim to recover more accurate expectation values without
adding too much overhead.

\subsubsection{Readout Error Mitigation}

Since readout errors are typically the largest SPAM errors, one can calibrate the confusion matrix for each measured qubit and apply its inverse to correct the observed counts.

\begin{itemize}
  \item \textbf{Method:} Prepare $\ket{0}$ and $\ket{1}$ states, measure many
        times, and record the empirical transition probabilities. This defines
        a calibration matrix $M$. Mitigated probabilities are obtained by
        applying $M^{-1}$ to the raw frequency vector.
  \item \textbf{In Qiskit:}
\begin{verbatim}
from qiskit.ignis.mitigation.measurement import complete_meas_cal
from qiskit.ignis.mitigation.measurement import CompleteMeasFitter

# generate calibration circuits
cal_circuits, state_labels = complete_meas_cal(qubit_list=[0,1])
# run and fit
cal_results = backend.run(cal_circuits).result()
meas_fitter = CompleteMeasFitter(cal_results, state_labels)
# get mitigator
meas_filter = meas_fitter.filter
# apply to raw counts
mitigated_counts = meas_filter.apply(raw_counts)
\end{verbatim}
\end{itemize}

\subsubsection{Zero-Noise Extrapolation (ZNE)}

The idea is to deliberately amplify gate noise (e.g.\ by inserting idle gates
or by stretching pulse durations), measure the noisy expectation values, and
then extrapolate back to ``zero noise''.

\begin{itemize}
  \item \textbf{Effect:} Extrapolated values can be closer to the true noiseless
        expectation.
  \item \textbf{In Qiskit:} Implemented in the \texttt{qiskit-ignis} and
        \texttt{mthree} packages, and in the
        \texttt{qiskit\_runtime} ``resilience\_level'' parameter.
\end{itemize}

\subsubsection{Randomized Compiling (RC)}

RC randomizes coherent errors (e.g.\ systematic over-rotations) into
stochastic noise, which averages out more predictably and is easier to mitigate.

\begin{itemize}
  \item \textbf{In Qiskit:} Enabled through circuit transpilation passes,
        and also exposed via \texttt{resilience\_level} in Qiskit Runtime.
\end{itemize}

\subsubsection{Dynamical Decoupling (DD)}

DD inserts sequences of idle gates (e.g.\ X--I--X--I) to cancel slow
dephasing noise.

\begin{itemize}
  \item \textbf{Effect:} Extends coherence time $T_2$ at the cost of
        increased circuit depth.
  \item \textbf{In Qiskit:}
\begin{verbatim}
from qiskit.transpiler.passes import DynamicalDecoupling
dd_pass = DynamicalDecoupling(coupling_map, dd_sequence)
\end{verbatim}
\end{itemize}

\subsubsection{Runtime Resilience Levels}

IBM’s Runtime offers a user-friendly knob for mitigation. Setting
\texttt{resilience\_level} automatically enables combinations of the above
techniques.

\begin{verbatim}
from qiskit_ibm_runtime import QiskitRuntimeService, Estimator

estimator = Estimator(session=..., options={"resilience_level": 2})
\end{verbatim}

Here level 0 means no mitigation, while levels 1–3 add increasingly powerful
methods (readout mitigation, RC, ZNE, etc.).

\subsubsection{Remarks}

\begin{itemize}
  \item Mitigation does not eliminate noise completely, but can significantly
        improve estimates of expectation values such as correlation functions.
  \item Techniques like ZNE require repeating experiments with different noise
        scaling, so they increase runtime and shot count.
  \item For spectrum extraction problems, readout mitigation and ZNE are
        particularly effective: the peak \emph{locations} remain robust, while
        mitigation restores peak \emph{heights}.
\end{itemize}

\subsection{Gate counts and the quantitative ``importance'' of CNOTs vs.\ \texorpdfstring{$\lambda$}{lambda}}
\label{sec:gatecounts-importance}

For the 4-level truncation encoded in two system qubits, the Pauli
decomposition of the system Hamiltonian
\(
H = \sum_j h_j P_j
\)
naturally partitions into (i) single-qubit strings ($N_1$ of them) and
(ii) genuine two-qubit strings ($N_2$ of them).
With a second-order Strang step of size $\Delta=dt/r$,
one Strang \emph{sub-step} applies each non-identity term twice
(forward+reverse). Using the standard $ZZ$-skeleton for two-qubit
exponentials, the following back-of-the-envelope counts hold per
\emph{sub-step} (system only):
\begin{align}
\text{CNOTs (uncontrolled)} &= 4\,N_2, \\
\text{parametric $R_Z$} &= 2\,(N_1+N_2),\\
\text{1q Cliffords} &\lesssim 8\,(N_1+N_2).
\end{align}
A full $dt$ step has $r$ sub-steps (multiply by $r$), and a Hadamard test
with a controlled step inflates two-qubit costs. A conservative and
compiler-agnostic estimate for the \emph{controlled} Strang sub-step is
\begin{equation}
\text{CNOTs (controlled)} \;\approx\; 10\,N_2 \;+\; 2\,N_1,
\end{equation}
so a full $dt$ step uses $(10N_2+2N_1)\,r$ CNOTs when controlled by the
ancilla.

\paragraph{Concrete numbers for the 4×4 truncation.}
For the harmonic case ($\lambda=0$) we have
\(
H_0=\mathrm{diag}\{0.5,1.5,2.5,3.5\}
\),
whose Pauli expansion is
\(
H_0=c_{II}II+c_{ZI}ZI+c_{IZ}IZ
\)
with $c_{ZZ}=0$. Thus $N_1=2$ and $N_2=0$. For the anharmonic case
($\lambda=0.1$), the $x^4$ contribution populates several two-qubit
strings; a representative thresholded decomposition yields
$N_1\approx 6$, $N_2\approx 7$.

\begin{table}[h]
\centering
\begin{tabular}{lcccc}
\toprule
Case & $N_1$ & $N_2$ & CNOTs per $dt$ (uncontrolled) & CNOTs per $dt$ (controlled) \\
\midrule
$\lambda=0$ (harmonic)   & $2$ & $0$ & $0$ & $\;\;2N_1\,r \;\approx\; 16\,$ for $r{=}8$ \\
$\lambda=0.1$ (anharm.)  & $6$ & $7$ & $4N_2\,r \;\approx\; 224$ & $(10N_2{+}2N_1)r \;\approx\; 656$ \\
\bottomrule
\end{tabular}
\caption{Indicative gate counts per $dt$ for the $4\times 4$ truncation (two
system qubits), using second-order Strang with $r=8$ sub-steps. The harmonic
case has no system--system entanglers; the only CNOTs come from ancilla
control (CRZ decompositions). Numbers are compiler- and basis-independent
order-of-magnitude estimates; exact counts vary by synthesis.}
\label{tab:counts4x4}
\end{table}

\section{Entanglement as a Diagnostic of Simulation Complexity}

A particularly transparent way to quantify the ``quantum workload'' of a
Hamiltonian simulation is to examine how much \emph{entanglement} the dynamics
create among the qubits representing the system.  Even for a single oscillator
truncated to a four-dimensional Hilbert space (two qubits), the amount of
intra-register entanglement required to represent its evolution can vary
strongly with the coupling strength~$\lambda$.

\subsection{Setup}

We consider again the quartic oscillator
\begin{equation}
  H = H_0 + \lambda\,x^4,
  \qquad
  H_0 = \tfrac{1}{2}(p^2+x^2),
\end{equation}
in a four-level truncation encoded as two qubits,
\(|00\rangle, |01\rangle, |10\rangle, |11\rangle\),
corresponding to oscillator levels \(n=0,1,2,3\).
The initial state is taken to be
\begin{equation}
  |\psi(0)\rangle = \frac{|00\rangle + |01\rangle}{\sqrt{2}}
  = \Big(\tfrac{|0\rangle+|1\rangle}{\sqrt{2}}\Big)_{q_0}\!\otimes |0\rangle_{q_1},
\end{equation}
a simple product state in the computational basis.

After time evolution under \(U(t)=e^{-iHt}\),
the joint state of the two system qubits becomes
\(|\psi(t)\rangle = U(t)|\psi(0)\rangle\).
Tracing out one of the qubits yields the reduced density matrix
\(\rho_{q_0}(t)=\mathrm{Tr}_{q_1}|\psi(t)\rangle\langle\psi(t)|\),
whose von~Neumann entropy,
\begin{equation}
  S(\rho_{q_0}) = -\mathrm{Tr}\,\rho_{q_0}\log_2\rho_{q_0},
\end{equation}
serves as a measure of the bipartite entanglement between the two encoding
qubits.

\subsection{Interpretation}

At first glance it may seem peculiar that a single-mode oscillator develops
``entanglement.''  The key point is that the mapping of a $d$--dimensional
Hilbert space to $\log_2 d$ qubits introduces an artificial tensor-product
structure.  When the Hamiltonian contains couplings such as \(x^4\) that mix
basis states separated by $\Delta n=\pm2,\pm4$, those appear as \emph{two-qubit
Pauli strings} (e.g.~$XX,\,YY,\,ZZ$) in the encoded representation.  The
resulting evolution therefore entangles the qubits even though the physical
oscillator remains a single degree of freedom.  The entanglement entropy
$S(\rho_{q_0})$ thus quantifies the \emph{computational entanglement cost}
needed to reproduce the correct physical dynamics in a local qubit basis.

For $\lambda=0$ (harmonic limit) the Hamiltonian is diagonal in the
computational basis, and the evolution is separable: \(S=0\).  As $\lambda$
increases, off-diagonal couplings become significant, driving the state to
explore higher levels and generate qubit--qubit entanglement.
Since the system contains only two qubits, \(S\le 1\)~bit is the maximal
possible value, and the entropy saturates once the state effectively spans the
full 4-dimensional subspace.

\subsection{Numerical results}

\begin{figure}[t]
  \centering
  \includegraphics[width=0.55\linewidth]{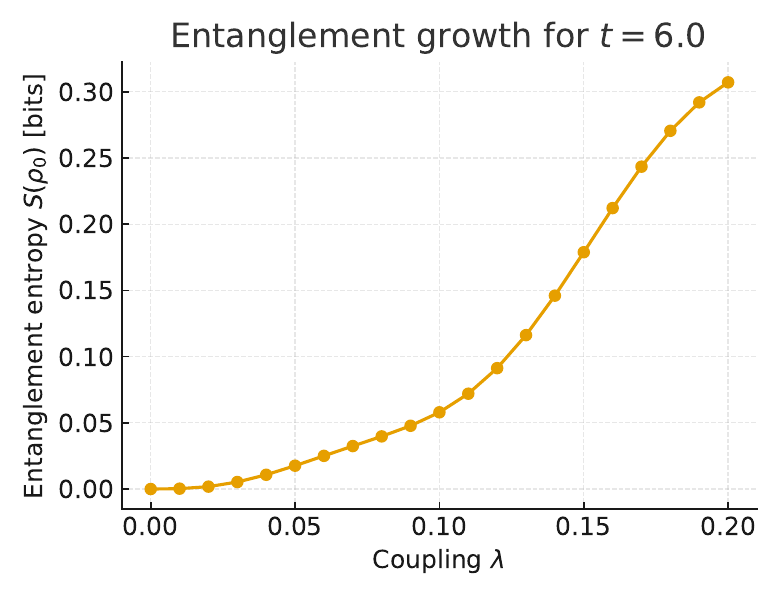}
  \caption{Entanglement entropy of one system qubit as a function of the
  quartic coupling~$\lambda$ for the 4-level truncated oscillator
  (time~$t=6$).
  The monotonic rise of~$S(\lambda)$ directly measures the amount of
  intra-register entanglement that the quantum computer must generate to
  reproduce the interaction-induced mixing of oscillator levels.}
  \label{fig:entropy-vs-lambda}
\end{figure}

Figure~\ref{fig:entropy-vs-lambda} shows the entanglement entropy obtained
by exact time evolution (with $dt=0.12$ and total time $t=6$) as a function of
the coupling~$\lambda$.  The entropy rises monotonically with~$\lambda$,
mirroring the increase of two-qubit Pauli components and CNOT load in the
corresponding Trotterized circuit.  This provides a direct diagnostic of how
``quantum'' the simulation must become: larger~$\lambda$ demands stronger and
more coherent entangling gates among the encoding qubits.

\subsection{Summary}

The entanglement entropy of the encoded qubits is therefore a sensitive
\emph{diagnostic of simulation complexity}:
\begin{itemize}
  \item it remains small for nearly separable dynamics (weak coupling);
  \item it grows with the strength of local interactions (more cross-Pauli
  terms and CNOTs per step);
  \item and it saturates once the truncated subspace is fully mixed.
\end{itemize}
Such entropy-based measures provide an intuitive bridge between the physics of
interactions and the hardware resources required to simulate them.

\paragraph{Scaling remark.}
In the present two-qubit encoding the entanglement entropy is bounded by
$S\le1$\,bit, but as one increases the truncation size $d$ and hence the number
of qubits $n=\lceil\log_2 d\rceil$, the same procedure yields a rising
entanglement capacity $S_{\max}=n$.
In the full oscillator, where the quartic term induces correlations across
many levels, the computational entanglement required to reproduce
$e^{-iH t}$ grows roughly with the \emph{effective participation number} of
levels coupled by~$x^4$.  Monitoring $S(\lambda)$ thus provides a natural
diagnostic of when classical simulation becomes inefficient:
the crossover to quantum advantage corresponds to the regime in which
$S(\lambda)$ scales extensively with $n$, and no classical tensor-network
representation can capture the state without exponential cost.

\chapter[Results: $\phi^4$]{Results: $\phi^4$}
\section{TEBD with vanilla Python}
This section presents a compact implementation of finite-lattice TEBD using only standard scientific Python packages. The objective is to make explicit the tensor manipulations underlying the algorithm, without relying on specialized tensor-network libraries. All components of the TEBD update---construction of effective two-site wavefunctions, application of two-site gates, Schmidt decompositions, and truncation---are written in a transparent form. We have adapted this section from the examples section in the \texttt{TeNPy} documentation.

\subsection{Imports}

\begin{lstlisting}[style=pythonstyle]
import numpy as np
from scipy.linalg import svd, expm
from scipy.sparse.linalg import eigsh   # used only for exact diagonalization
\end{lstlisting}

The implementation employs \texttt{numpy} for general tensor manipulations, \texttt{svd} for Schmidt decompositions, and \texttt{expm} to construct the two-site evolution gates.

\subsection{Matrix Product State Representation}

A finite MPS is encoded as a dictionary containing right-canonical tensors and Schmidt vectors.  
For each site $i$, the tensor $B_i$ is stored with indices $(\alpha_i, j_i, \alpha_{i+1})$ and the Schmidt coefficients on bond $i$ are stored in $S_i$.

\begin{lstlisting}[style=pythonstyle]
def make_mps(Bs, Ss, bc='finite'):
    L = len(Bs)
    if bc != 'finite':
        raise NotImplementedError("Only finite chains are implemented.")
    nbonds = L - 1
    return {'Bs': Bs, 'Ss': Ss, 'bc': bc, 'L': L, 'nbonds': nbonds}

def get_chi(mps):
    return [mps['Bs'][i].shape[2] for i in range(mps['nbonds'])]
\end{lstlisting}

\subsection{Local Effective Wavefunctions}

The TEBD update requires the one-site and two-site effective wavefunctions obtained from the right-canonical MPS.  
The one-site tensor is
\[
\theta^{[i]} = \mathrm{diag}(S_i)\, B_i,
\]
and the two-site wavefunction is obtained by contracting $\theta^{[i]}$ with $B_{i+1}$.

\begin{lstlisting}[style=pythonstyle]
def get_theta1(mps, i):
    B_i = mps['Bs'][i]
    S_i = mps['Ss'][i]
    theta_i = np.tensordot(np.diag(S_i), B_i, axes=(1, 0))
    return theta_i

def get_theta2(mps, i):
    L = mps['L']
    j = (i + 1) % L
    theta_i = get_theta1(mps, i)
    B_j = mps['Bs'][j]
    theta_ij = np.tensordot(theta_i, B_j, axes=(2, 0))
    return theta_ij
\end{lstlisting}

\subsection{Initial Product State}

As a simple reference state, a fully polarized product state is used.  
Each local tensor has bond dimension one.

\begin{lstlisting}[style=pythonstyle]
def init_FM_MPS(L, d, bc='finite'):
    Bs, Ss = [], []
    for _ in range(L):
        B = np.zeros((1, d, 1), dtype=np.float64)
        B[0, 0, 0] = 1.0    # local state |0>
        Bs.append(B)
        Ss.append(np.ones(1))
    return make_mps(Bs, Ss, bc=bc)
\end{lstlisting}

\subsection{Schmidt Decomposition and Truncation}

After applying a two-site gate, the updated two-site tensor must be decomposed and truncated.  
The tensor is reshaped into a matrix, an SVD is performed, and singular values smaller than a tolerance $\varepsilon$ are discarded.

\begin{lstlisting}[style=pythonstyle]
def split_truncate_theta(Theta, chi_max, eps):
    chivL, dL, dR, chivR = Theta.shape
    Theta = np.reshape(Theta, (chivL*dL, chivR*dR))

    X, Y, Z = svd(Theta, full_matrices=False)
    chivC = min(chi_max, np.sum(Y > eps))
    piv = np.argsort(Y)[::-1][:chivC]
    X, Y, Z = X[:,piv], Y[piv], Z[piv,:]

    X = np.reshape(X, (chivL, dL, chivC))
    Y = Y / np.linalg.norm(Y)
    Z = np.reshape(Z, (chivC, dR, chivR))

    return X, Y, Z
\end{lstlisting}

\subsection{Single TEBD Bond Update}

The essential TEBD step is the update on a single bond.  
A two-site gate $U$ is contracted with the effective wavefunction, followed by an SVD and reconstruction of right-canonical tensors.

\begin{lstlisting}[style=pythonstyle]
def update_bond(mps, i, U_bond, chi_max, eps):
    j = (i + 1) % mps['L']

    theta = get_theta2(mps, i)
    Utheta = np.tensordot(U_bond, theta, axes=([2,3],[1,2]))
    Utheta = np.transpose(Utheta, (2,0,1,3))

    Ai, Sj, Bj = split_truncate_theta(Utheta, chi_max, eps)

    Gi = np.tensordot(
        np.diag(mps['Ss'][i]**(-1)),
        Ai, axes=(1,0)
    )

    mps['Bs'][i] = np.tensordot(Gi, np.diag(Sj), axes=(2,0))
    mps['Ss'][j] = Sj
    mps['Bs'][j] = Bj
\end{lstlisting}

\subsection{TEBD Sweep}

A full TEBD step consists of an even-bond sweep followed by an odd-bond sweep.

\begin{lstlisting}[style=pythonstyle]
def run_TEBD(mps, U_bonds, N_steps, chi_max, eps):
    nbonds = mps['nbonds']
    for _ in range(N_steps):
        for parity in (0, 1):
            for i in range(parity, nbonds, 2):
                update_bond(mps, i, U_bonds[i], chi_max, eps)
\end{lstlisting}

\subsection{Observables}

The von Neumann entanglement entropy across each bond is computed directly from the Schmidt spectrum.

\begin{lstlisting}[style=pythonstyle]
def entanglement_entropy(mps):
    values = []
    for S in mps['Ss']:
        p = S**2
        mask = p > 0
        values.append(-np.sum(p[mask]*np.log(p[mask])))
    return np.array(values)
\end{lstlisting}

Expectation values of one-site operators follow from contractions with $\theta^{[i]}$.

\begin{lstlisting}[style=pythonstyle]
def site_expectation_value(mps, op):
    L = mps['L']
    vals = []
    for i in range(L):
        theta = get_theta1(mps, i)
        op_theta = np.tensordot(op, theta, axes=(1,1))
        op_theta = np.transpose(op_theta, (1,0,2))
        vals.append(
            np.tensordot(np.conj(theta), op_theta, axes=([0,1,2],[0,1,2])).real
        )
    return np.array(vals)
\end{lstlisting}

\subsection{Transverse-Field Ising Model and Two-Site Gates}

The local two-site Hamiltonians are constructed from Pauli matrices and exponentiated to obtain the Trotter gates.

\begin{lstlisting}[style=pythonstyle]
def make_TFI_model(L, J, g, bc='finite'):
    sigmax = np.array([[0.,1.],[1.,0.]])
    sigmay = np.array([[0.,-1j],[1j,0.]])
    sigmaz = np.array([[1.,0.],[0.,-1.]])
    ident  = np.eye(2)

    H_bonds = []
    for _ in range(L-1):
        H_zz = -J*np.kron(sigmaz, sigmaz)
        H_x  = -0.5*g*(np.kron(sigmax,ident) + np.kron(ident,sigmax))
        H    = H_zz + H_x
        H_bonds.append(H.reshape(2,2,2,2))

    return {'L':L,'d':2,'bc':bc,'J':J,'g':g,
            'sigmax':sigmax,'sigmay':sigmay,'sigmaz':sigmaz,
            'id':ident,'H_bonds':H_bonds}
\end{lstlisting}

\begin{lstlisting}[style=pythonstyle]
def calc_U_bonds(H_bonds, dt):
    d = H_bonds[0].shape[0]
    U_bonds = []
    for H in H_bonds:
        Hm = H.reshape((d*d, d*d))
        U  = expm(-dt*Hm)
        U_bonds.append(U.reshape((d,)*4))
    return U_bonds
\end{lstlisting}

\subsection{Example: Imaginary-Time TEBD for the TFI Ground State}
\begin{lstlisting}[style=pythonstyle]
def example_TEBD_gs_TFI_finite(L=10, J=1.0, g=0.1, chi_max=16):
    model = make_TFI_model(L, J, g, bc='finite')
    psi   = init_FM_MPS(model['L'], model['d'], bc='finite')

    for dt in [1e-1,1e-2,1e-3,1e-4,1e-5]:
        U_bonds = calc_U_bonds(model['H_bonds'], dt)
        run_TEBD(psi, U_bonds, N_steps=500,
                 chi_max=chi_max, eps=1e-10)

    mag_x = site_expectation_value(psi, model['sigmax'])
    return psi, model, mag_x
\end{lstlisting}

\section[Scattering in lattice phi4]{Scattering in Lattice $\phi^4$}
In this section, we are going to do scattering experiments with lattice $\phi^4$ theory. We are going to introduce the packages \texttt{MPSKit} and \texttt{TensorKit}, written in the \texttt{Julia} programming language. \texttt{Julia} is a dynamically typed (with support for type annotations), JIT (Just-In-Time) compiled language with a lot of support for scientific computing. It was specifically designed with scientific computing in mind and has since branched out to other areas of programming.
\subsection{Creating MPS objects}
First we will create a bunch of matrices to store our field values. We work in the basis where the local Hilbert space is the standard fock basis and the matrix elements of the operators take the form
\begin{equation}
  \begin{split}
    \expval{m | \phi | n} &= \frac{1}{\sqrt{2}} \left(\delta_{m - 1, n} \sqrt{n + 1} + \delta_{m, n - 1} \sqrt{n + 1}\right) \\
    \expval{ m | \phi^2 | n} &= \frac{1}{2} \left( \delta_{m - 1, n + 1} \sqrt{(m + 1)(m + 2)} + \delta_{m, n} (2 s + 1) + \delta_{m + 1, n - 1} \sqrt{(n + 1)(n + 2)}\right) \\
    \expval{m | \phi^4 | n} &= \frac{1}{4} \bigg(\delta_{m - 1, n + 3} \sqrt{(n + 1)(n + 2)(n + 3)(n + 4)} \bigg. \\
                            &+\delta_{m, n + 2} (4 n + 6) \sqrt{(n + 1)(n + 2)} \\
                            &+ \delta_{m, n} (6n^2 + 6n + 3) + \delta_{m + 2, n}(4m + 6) \sqrt{(m + 1)(m + 2)} \\ 
                            &+ \bigg.\delta_{m + 3, n - 1}\sqrt{(m + 1)(m + 2)(m + 3)(m + 4)}\bigg). \\
      \expval{m | \pi^2 |n} &= \frac{1}{2} \left( - \delta_{m - 1, n - 1} \sqrt{(n + 1)(n + 2)} + \delta_{m, n} (2 n + 1) - \delta_{m + 1, n - 1} \sqrt{(m + 1)(m + 2)}\right)\,.
  \end{split}
\end{equation}
Since most of these matrix elements are zero, we can start by creating four matrices filled with zeros and loop over the entries to fill them. Note that due to the delta function we need only one loop per term to fill the above matrices (caveat: \texttt{Julia} uses 1-based indexing so we need to translate the above formulae appropriately. Just shifting $(m, n) \to (m - 1, n - 1)$ will do). The following code snippet prepares these matrices

\begin{lstlisting}[style=juliastyle]
function matrix_elems(d)
    phi    = zeros(ComplexF64, (d, d))
    phi_sq = zeros(ComplexF64, (d, d))
    pi_sq  = zeros(ComplexF64, (d, d))
    phi_4  = zeros(ComplexF64, (d, d))

    # phi = (a + a^\dagger)/sqrt(2)
    for i in 2:d
        val = sqrt(i / 2)
        phi[i, i - 1] = val
        phi[i - 1, i] = val
    end

    # helper to fill phi_sq and pi_sq (same structure, sign flip on off-diagonals)
    function fill_quadratic!(M, sign)
        @inbounds for i in 1:d
            if i < d - 1
                val = sign * sqrt(i * (i + 1)) / 2
                M[i, i + 2] = val
                M[i + 2, i] = val  # Hermitian
            end
            M[i, i] = (2 * i - 1) / 2
        end
    end

    fill_quadratic!(phi_sq, +1)
    fill_quadratic!(pi_sq,  -1)

    # phi^4 in harmonic oscillator basis
    @inbounds for i in 1:d
        n = i - 1  # occupation number

        # diagonal
        phi_4[i, i] = (6 * n^2 + 6 * n + 3) / 4

        # connect |n> <-> |n+2>
        if i + 2 <= d
            val = (4 * n + 6) * sqrt((n + 1) * (n + 2)) / 4
            j = i + 2
            phi_4[i, j] = val
            phi_4[j, i] = val  # Hermitian
        end

        # connect |n> <-> |n+4>
        if i + 4 <= d
            val = sqrt((n + 1) * (n + 2) * (n + 3) * (n + 4)) / 4
            j = i + 4
            phi_4[i, j] = val
            phi_4[j, i] = val  # Hermitian
        end
    end

    ϕ  = TensorMap(phi,    ℂ^d ← ℂ^d)
    ϕ2 = TensorMap(phi_sq, ℂ^d ← ℂ^d)
    π2 = TensorMap(pi_sq,  ℂ^d ← ℂ^d)
    ϕ4 = TensorMap(phi_4,  ℂ^d ← ℂ^d)

    return ϕ, ϕ2, π2, ϕ4
end
\end{lstlisting}
The Julia language server conveniently has support for Unicode characters so that the math is more clearly visible. For example, the $\mathbb{C}$ above can be typeset by typing \texttt{\textbackslash bbC} and hitting \texttt{esc}. In the last few lines above we have converted our \texttt{Julia} arrays to \texttt{TensorMap} objects. The second entry in the method above specifies the $\texttt{codomain} \leftarrow \texttt{domain}$ of the matrix.

Next, we actually need to create the MPS object. This can be done as follows
\begin{lstlisting}[style=juliastyle]
  ψ0 = FiniteMPS(rand, ComplexF64, L, ℂ^d, ℂ^D)
\end{lstlisting}
Here we have passed \texttt{rand} and then \texttt{ComplexF64} indicating that we want to fill our \texttt{FiniteMPS} with random matrices with 128-bit complex numbers (64-bit real and imaginary parts). The length of the chain is specified by passing \texttt{L} and the local Hilbert space dimension and the bond dimensions are specified by passing $\mathbb{C}\texttt{\^{}d}$ and $\mathbb{C}\texttt{\^{}D}$ respectively. We can now write a function to prepare the ground state of some Hamiltonian \texttt{ham}.
\begin{lstlisting}[style=juliastyle]
function prep_gs(d, D, ham)
    """
    Inputs:
    D = max bond dimension
    d = local Hilbert space dimension
    ham = Hamiltonian for ground state prep
    Returns: 
    The converged ground state of the FiniteMPS
    """
    ψ0 = FiniteMPS(rand, ComplexF64, L, ℂ^d, ℂ^D)
    ψ, _, _ = find_groundstate(ψ0, ham, DMRG(;tol=1e-7))
    return ψ
end
\end{lstlisting}
Let us first focus on the arguments of the \texttt{find\_groundstate} function above. The first two arguments should be clear. The last one specifies the method used to obtain the ground state. Here we have used the single-site DMRG algorithm \texttt{DMRG()} with a tolerance of \texttt{tol=1e-7}. Note that, in \texttt{Julia}, keyword arguments must always be passed after a semicolon \texttt{;}. Let us now look at the return values of the \texttt{find\_groundstate} function. According to the \texttt{MPSKit} documentation, the function returns $\psi$, the converged ground state, \texttt{environments}, the left and right environments and $\epsilon$, the final convergence error upon terminating the algorithm. We do not need the latter two returned objects and therefore choose to discard them inside the \texttt{\_} variable.

Now we prepare the \texttt{ham} object, which in this context is a \texttt{FiniteMPOHamiltonian}. Recall that the lattice $\phi^4$ Hamiltonian $\tilde{H}$, defined through $H = \tilde{H}a$, takes the form
\begin{equation}
  \label{eq:latticephi4hamiltonian}
  \tilde{H}
  = \sum_{n = 1}^{L}
    \left(
      \frac{\pi_n^2}{2}
      + \frac{\tilde{\mu}_0^2}{2}\,\phi_n^2
      + \frac{\tilde{\lambda}}{4!}\,\phi_n^4
    \right)
    + \frac{1}{2}
      \sum_{n = 1}^{L - 1}
      \left(\phi_{n+1} - \phi_n\right)^2 ,
\end{equation}
where $\tilde{\mu}_0^2 = \mu_0^2 a^2$ and $\tilde{\lambda} = \lambda a^2$ are the lattice mass and coupling, and $a$ is the lattice spacing. The function below constructs the corresponding lattice Hamiltonian given the bare lattice couplings, which in the script code are represented by the unicode variables $\mu\texttt{0}$ and $\lambda\texttt{0}$.

\begin{lstlisting}[style=juliastyle]
function get_ham(d, L, μ0, λ0)
    """
    Prepares the infinte MPO Hamiltonian from μ0, λ0 and
    the local Hilbert space dimension d as inputs
    """
    ϕ, ϕ2, π2, ϕ4 = matrix_elems(d)
    chain = fill(ℂ^d, L)
    single_site_term = (μ0 * ϕ2 + π2) / 2 + λ0 * ϕ4 / 24 + ϕ2
    two_site_term = - ϕ ⊗ ϕ
    single_site_terms = [i => single_site_term for i in 1:L]
    two_site_terms = [(i, i + 1) => two_site_term for i in 1:L - 1]
    ham = FiniteMPOHamiltonian(chain, single_site_terms..., two_site_terms...)
    return ham
end
\end{lstlisting}
After obtaining the matrices from our previously defined \texttt{matrix\_elems} function, we prepare the geometry of the lattice which in this case is an open \texttt{chain} with local Hilbert space $\mathcal{H} = \mathbb{C}\texttt{\^{}d}$ at each lattice site. We separate out the \texttt{single\_site\_terms} and \texttt{two\_site\_terms} and specify the sites at which these act by using a dictionary. Hopefully the syntax of the dictionary is clear. Like \texttt{Python}, \texttt{Julia} also has support for list, dictionary and vector comprehensions. The \texttt{...} after \texttt{single\_site\_terms} and \texttt{two\_site\_terms} unrolls the dictionary into individual arguments, which is the input format expected by \texttt{FiniteMPOHamiltonian}.

We have all the ingredients for our simulation, but we need to know how to time evolve. This is achieved by calling the \texttt{timestep(state, ham, t, dt, [algorithm])} function. Note that the documentation says the \texttt{algorithm} used here is an optional argument defaulting to \texttt{TDVP()} but during our studies we had to explicitly pass this as an argument.

The reader can find more details on the implementation in \texttt{Julia/phi4/\allowbreak scattering\allowbreak \_finite.jl} script in the GitHub repository. We call the function \texttt{prep\_gs(d, D, ham)} with \texttt{d = 16, D = 24}. We find that various quantities (energy, half-chain entropy, $\expval{\phi}$) have converged reasonably well for these values. The Hamiltonian \texttt{ham} is obtained by using the \texttt{get\_ham} function.
Once the ground state is obtained, we apply the field operator $\phi$ at two sites and then time evolve the resulting state with the TDVP algorithm. The program for this finite lattice evolution can be found in our repository.


\subsection{iMPS implementation}
The simulations from the previous section extend naturally to the thermodynamic limit. In this setting, the Hamiltonian is specified solely by its one-site and two-site contributions. An iMPS ansatz is initialised by calling the \texttt{InfiniteMPS} constructor. Since the unit cell has length~1, we only need to specify the local Hilbert-space dimension and the bond dimension:
\begin{lstlisting}[style=juliastyle]
ψ0 = InfiniteMPS(ℂ^d, ℂ^D)
\end{lstlisting}

Most of the finite-lattice routines have direct infinite-lattice analogues obtained by prefixing the corresponding algorithm with an \texttt{I}. For example, the ground-state search becomes
\begin{lstlisting}[style=juliastyle]
ψ, _, _ = find_groundstate(ψ0, ham, IDMRG(; tol = 1e-7))
\end{lstlisting}

\begin{figure}
    \centering
    \begin{subfigure}{0.48\textwidth}
        \centering
        \includegraphics[width=\textwidth]{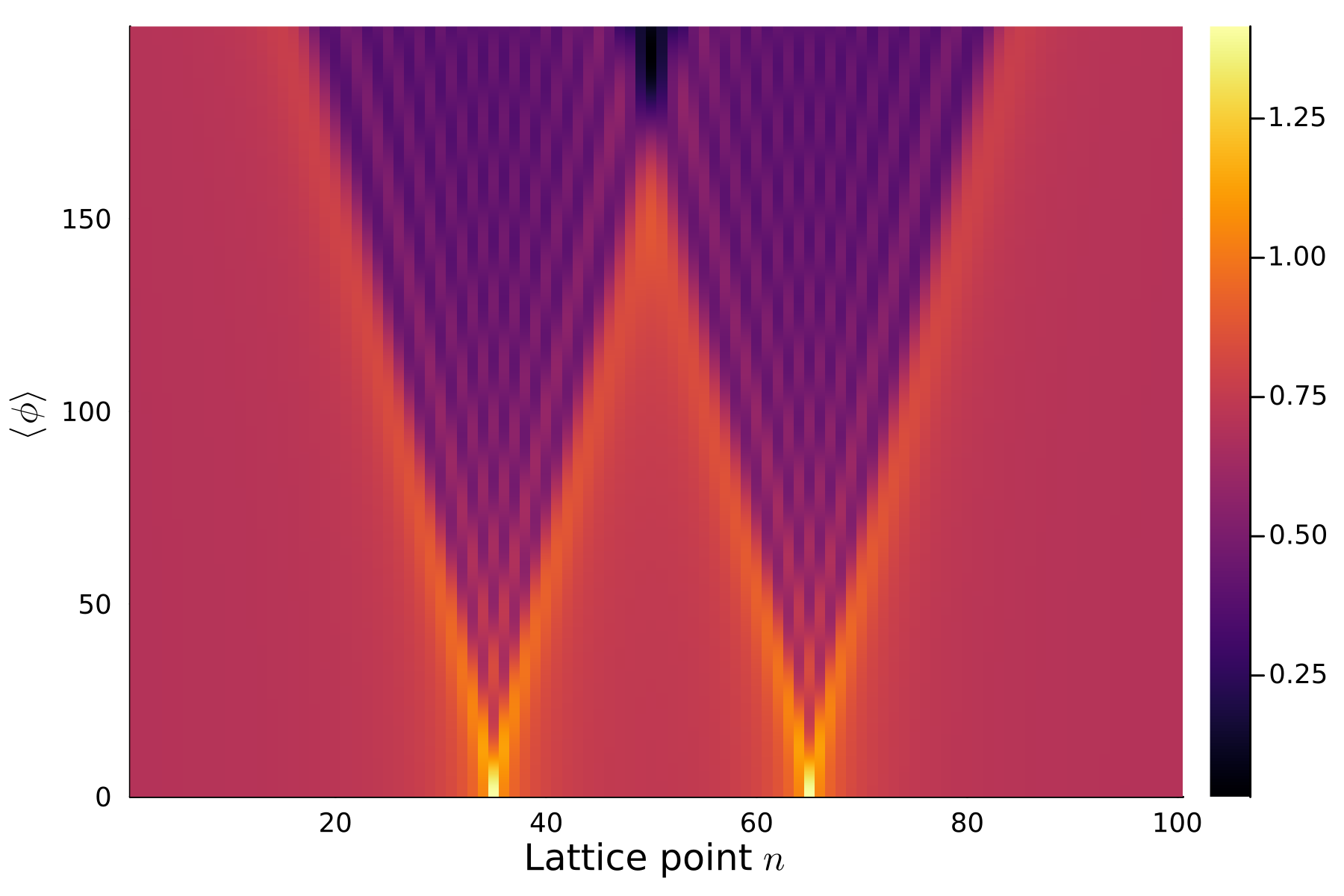}
        \caption{$\tilde{\mu}_0^2 = -0.775$}
        \label{fig:mu0m0p780}
    \end{subfigure}
    \hfill
    \begin{subfigure}{0.48\textwidth}
        \centering
        \includegraphics[width=\textwidth]{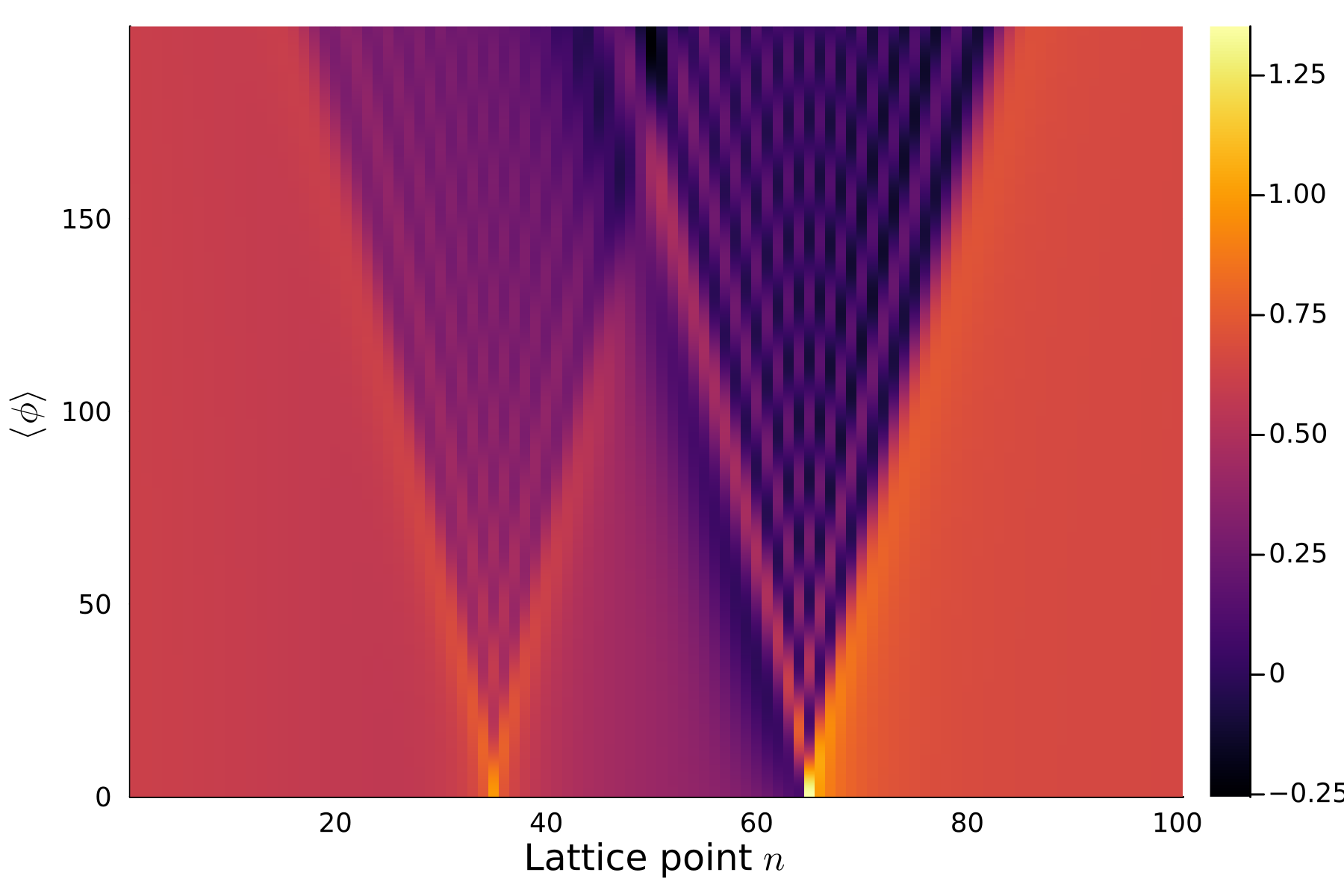}
        \caption{$\tilde{\mu}_0^2 = - 0.78$}
        \label{fig:mu0m0p775}
    \end{subfigure}
    \caption{Scattering of local disturbances with $\tilde{\lambda} = 0.2$ and $\mu_0^2 = -0.78, -0.775$ from left to right with a window size of $L = 100$ sites. Time evolution was performed using the TDVP algorithm with a step size of $dt = 0.02$ for the time integral. The ground state was prepared with the VUMPS algorithm.}
    \label{fig:phi4-scattering-infinite}
\end{figure}

To perform scattering experiments, we introduce a non-uniform window around the perturbation region. This is done by passing the converged iMPS ground state to the \texttt{WindowMPS} constructor together with the chosen window length~\texttt{L}. Beyond this point, the workflow mirrors the finite-lattice case closely, and we refer the reader to the script \texttt{Julia/\allowbreak phi4/\allowbreak scattering.jl} in the GitHub repository for the complete implementation. The infinite-lattice results are shown in Fig.~\ref{fig:phi4-scattering-infinite}.

\subsection{Wavepacket scattering}
\label{sec:wavepacketscattering}
To prepare wavepackets, we make use of the MPS quasiparticle ansatz. Starting from the iMPS ground state $\ket{\Psi(A)}$, we construct single–particle excitations with momentum $p$ by inserting a variational tensor $B(p)$ into the mixed–canonical form,
\begin{equation}
  \ket{\Phi_p(B)}
  = \sum_n e^{ip n}\,
  \cdots A_L A_L B(p) A_R A_R \cdots\,,
\end{equation}
and obtain $B(p)$ together with the dispersion $E(p)$ from the corresponding quasiparticle eigenvalue problem. Following the construction in the supplemental material of Ref.~\cite{BelyanskyPRL2024}, a localized wavepacket is then built as a Gaussian superposition of these momentum eigenstates,
\begin{equation}
  \ket{\Psi_{\text{wp}}}
  = \int_{-\pi}^{\pi} \! \mathrm{d}p\; c_p \ket{\Phi_p(B)}\,,\qquad
  c_p \propto \exp\!\left[-\frac{(p - p_0)^2}{2\sigma^2}\right],
\end{equation}
which yields a set of real–space tensors $B_n$ via a discrete Fourier transform over a finite momentum grid. In practice one fixes the arbitrary overall phase and gauge of $B(p)$ (so that a chosen tensor element is real and positive) before the transform, and keeps only those sites with $\|B_n\|$ above a small threshold to obtain a compact wavepacket.

\begin{figure}
    \centering
    \begin{subfigure}[b]{0.48\textwidth}
        \centering
        \includegraphics[width=\textwidth]{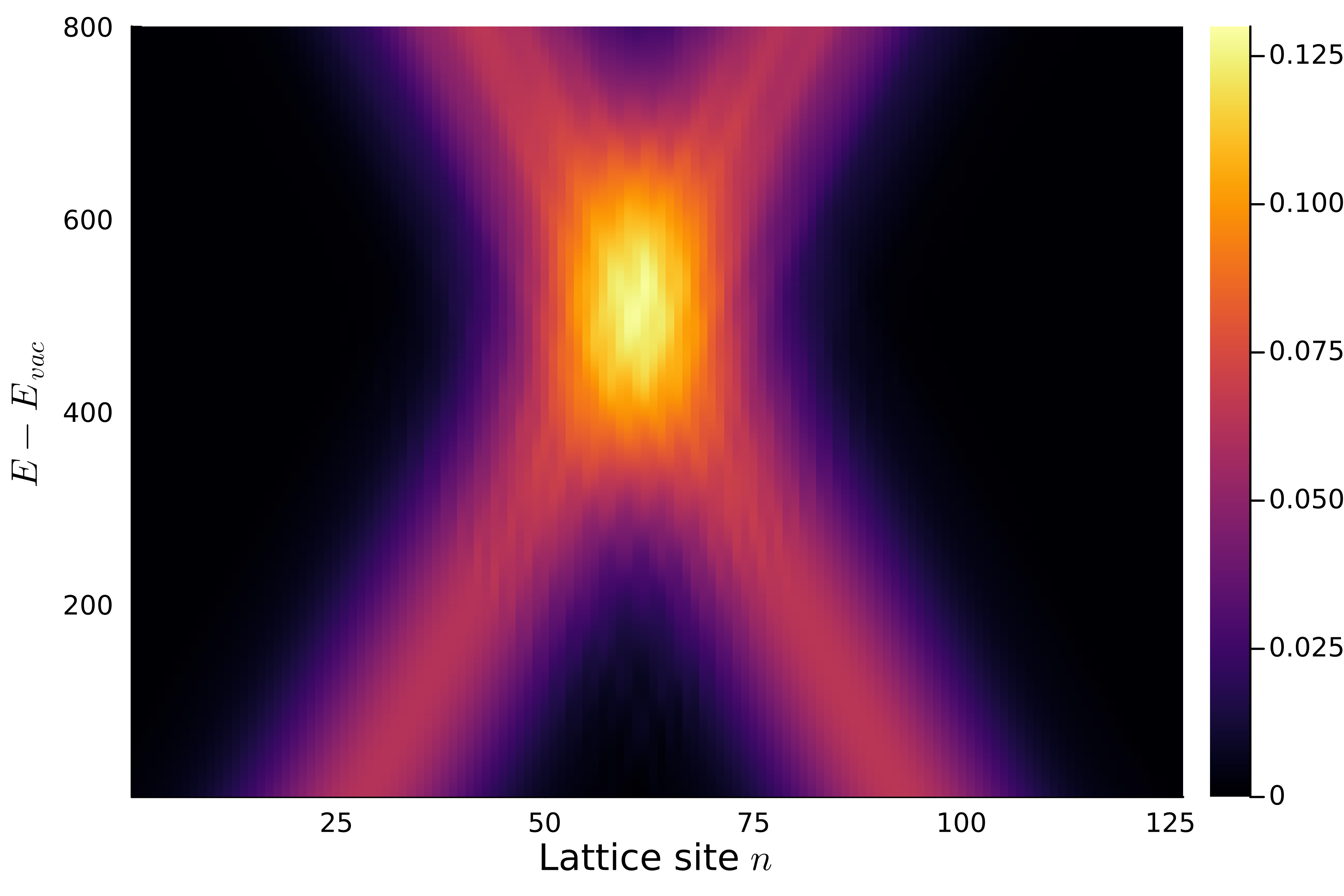}
    \end{subfigure}
    \hfill
    \begin{subfigure}[b]{0.48\textwidth}
        \centering
        \includegraphics[width=\textwidth]{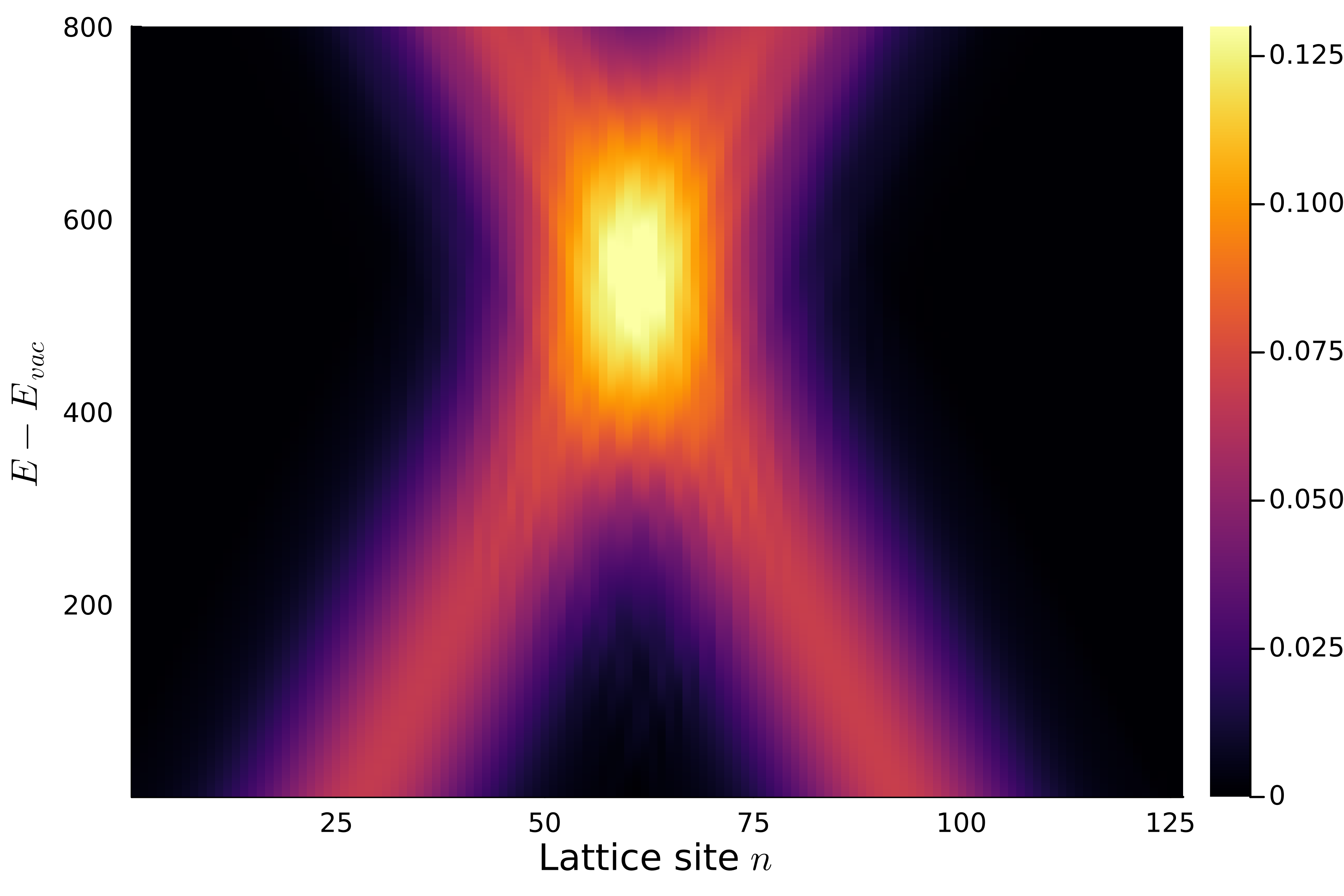}
    \end{subfigure}

    \vspace{0.8em} 

    \begin{subfigure}[b]{0.48\textwidth}
        \centering
        \includegraphics[width=\textwidth]{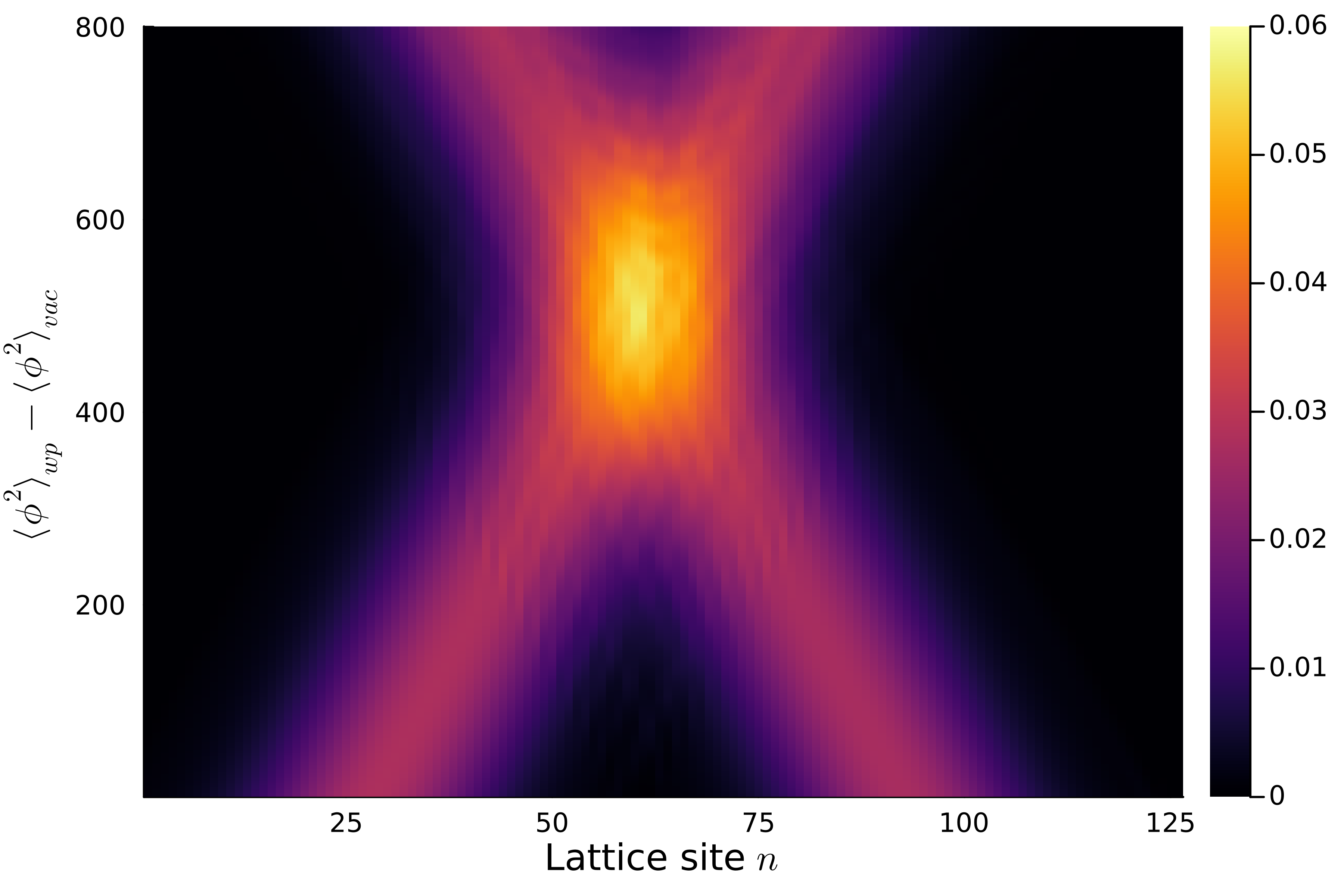}
        \caption{$\lambda = 0$}
    \end{subfigure}
    \hfill
    \begin{subfigure}[b]{0.48\textwidth}
        \centering
        \includegraphics[width=\textwidth]{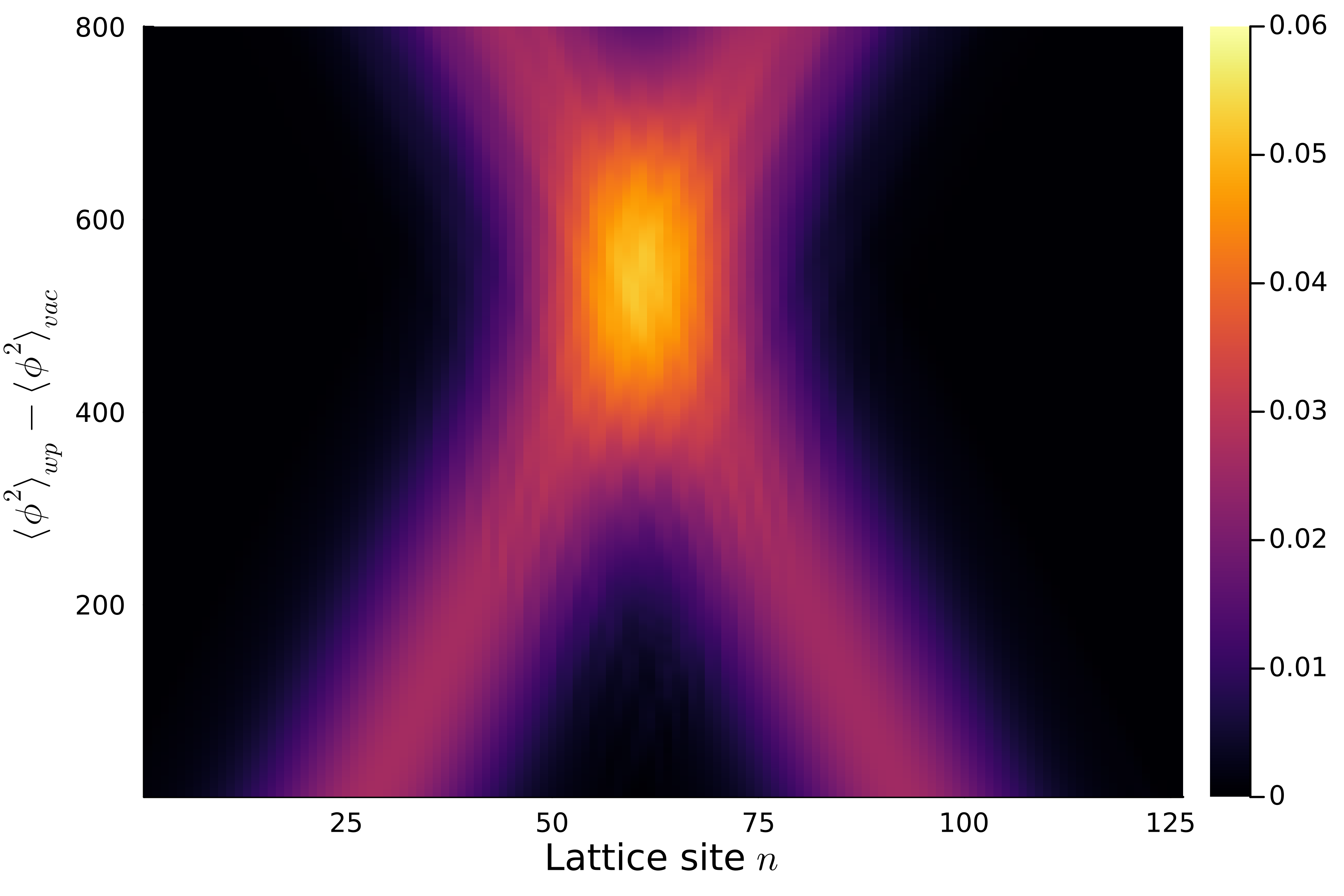}
        \caption{$\lambda = 2$}
    \end{subfigure}

    \caption{Scattering of wavepackets in $\phi^4$ theory. Top row is the energy density plot and the bottom row is the vacuum subtracted $\expval{\phi^2}$ plot. We chose $d = 5, D = 10$ for the ground state search. The bond dimension was then expanded to $2D = 20$ to construct the wavepackets. For the construction of each of the wavepackets we chose a window of $N_p = \frac{2\pi}{\Delta p}$ where $\Delta p = 0.1$. These windows were then glued together. The time evolution is done with the TDVP algorithm with a step size of $0.1$.}
    \label{fig:wavepacketphi4d5D10}
\end{figure}

To create an initial scattering state, this procedure is applied twice to prepare two wavepackets with opposite central momenta, localized far apart on the same uniform background. The two non-uniform regions are then “glued’’ into a single window by using the gauge matrix $C$ that relates $A_L$ and $A_R$ (i.e.\ $A_L C = C A_R$), so that inside the window the bond dimension can increase while outside the state coincides with the uniform iMPS ground state. In our implementation this non-uniform region is represented as a \texttt{WindowMPS} and evolved in real time using the single-site TDVP algorithm. Our goal is to have a qualitative description of scattering and therefore we choose somewhat crude parameters for our numerics. Despite the choice of our parameters, we can see marked characteristics. In the plots shown in Fig~\ref{fig:wavepacketphi4d5D10}, we can see that the scattering center when the interaction has been turned on moves to a later time due to time delay~\cite{Zemlevskiy_2025}.

\chapter{Results: Ising field theory}
\section{Building intuition}
Let us begin by playing around with a small number of sites in Mathematica. Consider $N=4$, $h_x=g_1 J, h_z=\lambda g_1 J$ and set $J=1$. We will choose $h_z<h_x$ by setting $\lambda=0.1, g_1=0.1$. Our goal is to time evolve some initial states. As the first example take the Neel state $|\uparrow,\downarrow,\uparrow,\downarrow\rangle$. Then we will plot $\langle \sigma_{j=2}^z(t)\rangle$ as a function of $t$ and also exhibit the density plot of $\langle \sigma^z\rangle$ as a function of $t$ and the lattice site. First let's define the Hamiltonian

\begin{lstlisting}[style=mma]
nn = 4; \[Lambda] = 0.1; g1 = 0.1;

term[k_, j_] := 
  If[k == j || k == j + 1 && j != nn, PauliMatrix[3], 
   IdentityMatrix[2]];
term[k_, nn] := 
 If[k == 1 || k == nn, PauliMatrix[3], IdentityMatrix[2]] (*this is needed to impose PBC *); 
ttZZ[j_] := KroneckerProduct @@ Table[term[kk, j], {kk, 1, nn}];
term2[k_, j_] := If[k == j, PauliMatrix[1], IdentityMatrix[2]];
term3[k_, j_] := If[k == j, PauliMatrix[3], IdentityMatrix[2]];
ttX[j_] := KroneckerProduct @@ Table[term2[kk, j], {kk, 1, nn}];
ttZ[j_] := KroneckerProduct @@ Table[term3[kk, j], {kk, 1, nn}];
H = -J Sum[ttZZ[j], {j, 1, nn}] - hx Sum[ttX[j], {j, 1, nn}] - 
   hz Sum[ttZ[j], {j, 1, nn}];
HH = H/J /. {hx -> g1 J, hz -> g2 J} /. g2 -> \[Lambda] g1 // 
   Simplify;
\end{lstlisting}

and then define the state along with the time evolutions: 
\begin{lstlisting}[style=mma]
up := {{1, 0}}
dn := {{0, 1}}
\[Psi]0 := Transpose[KroneckerProduct[up, dn, up, dn] // Flatten]
\[Psi][t_] := MatrixExp[-I t HH] . \[Psi]0
mz[t_, j_] := Conjugate[Transpose[\[Psi][t]]] . ttZ[j] . \[Psi][t]
ListPlot[Table[{t,mz[t,2]},{t,0,50,.1}],Joined->True]
\end{lstlisting}
gives us
\begin{figure}[h]
  \centering
  \includegraphics[width=0.5\linewidth]{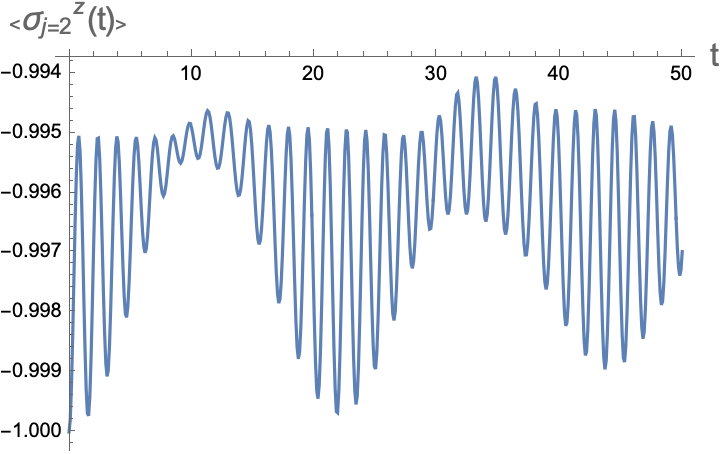}
  \caption{}
\end{figure}
and 
\begin{lstlisting}[style=mma]
tmax = 10;
dt = 0.1;
data = {}; Do[
 AppendTo[data, {t, j, Re[mz[t, j]]}], {j, 1, 4},{t, 0, tmax,
   dt}];

ListDensityPlot[data,
 FrameLabel -> {"t", "site j"}, 
 PlotLegends -> Automatic]
\end{lstlisting}
\begin{figure}[h]
  \centering
  \includegraphics[width=0.5\linewidth]{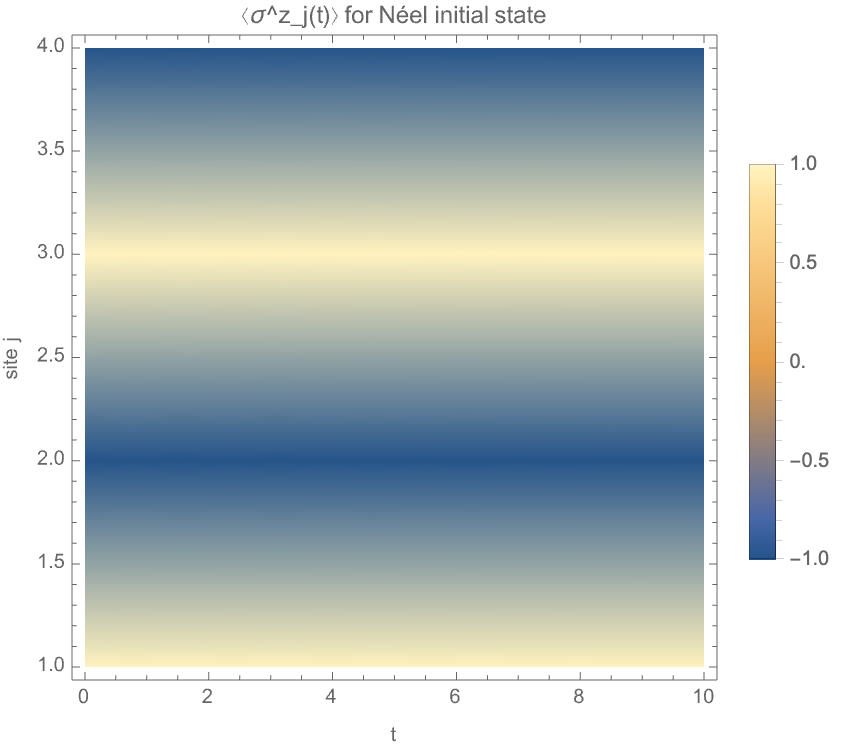}
  \caption{}
\end{figure}
This plot suggests that not much happens for the Neel state. However, if we consider
$\psi_0=|\uparrow,\downarrow,\downarrow,\uparrow\rangle$, then we find the following.
\begin{figure}[h]
  \centering
  \includegraphics[width=0.5\linewidth]{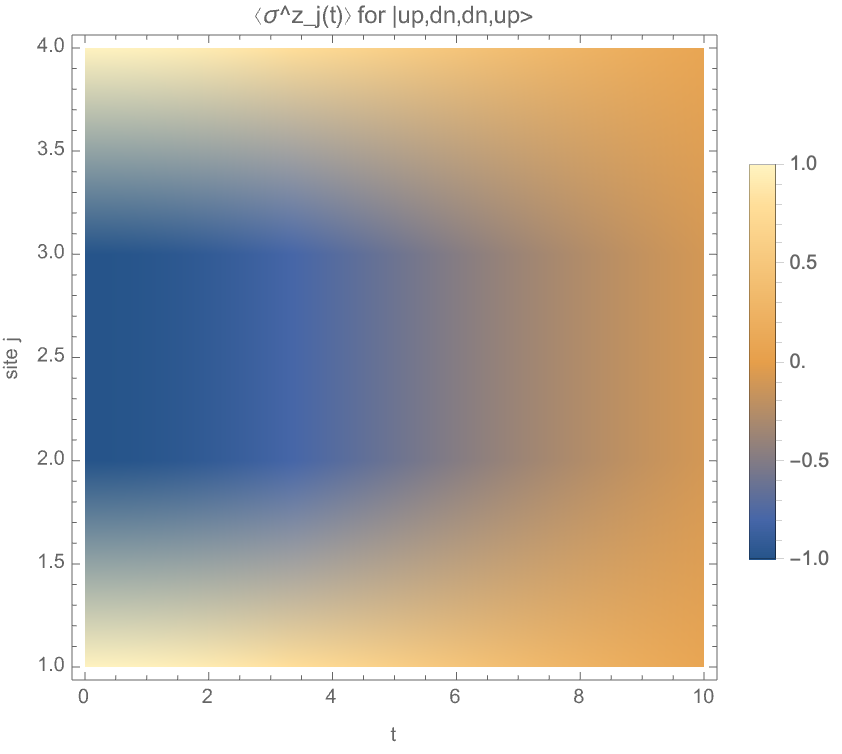}
  \caption{}
\end{figure}
which suggests that the spins have flipped as time evolution occurs.

\section{Scattering wavepackets using MPS}
\begin{figure}
    \centering
    \begin{subfigure}[b]{0.48\textwidth}
    \includegraphics[width=\textwidth]{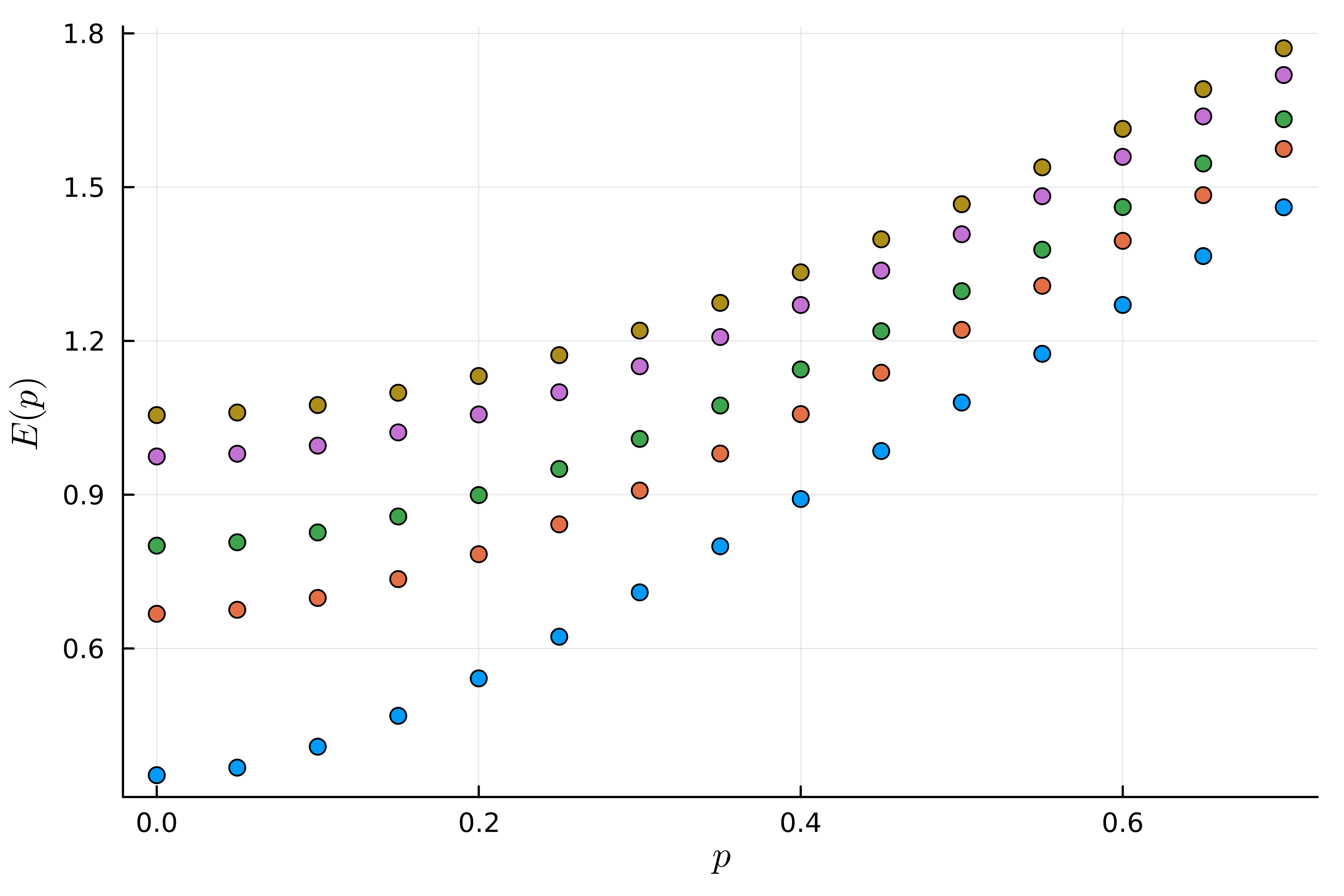}
        \centering
        \caption{Lattice spectrum.}
        \label{fig:dispersion-lattice}
    \end{subfigure}
    \hfill
    \begin{subfigure}[b]{0.48\textwidth}
        \centering
        \includegraphics[width=\textwidth]{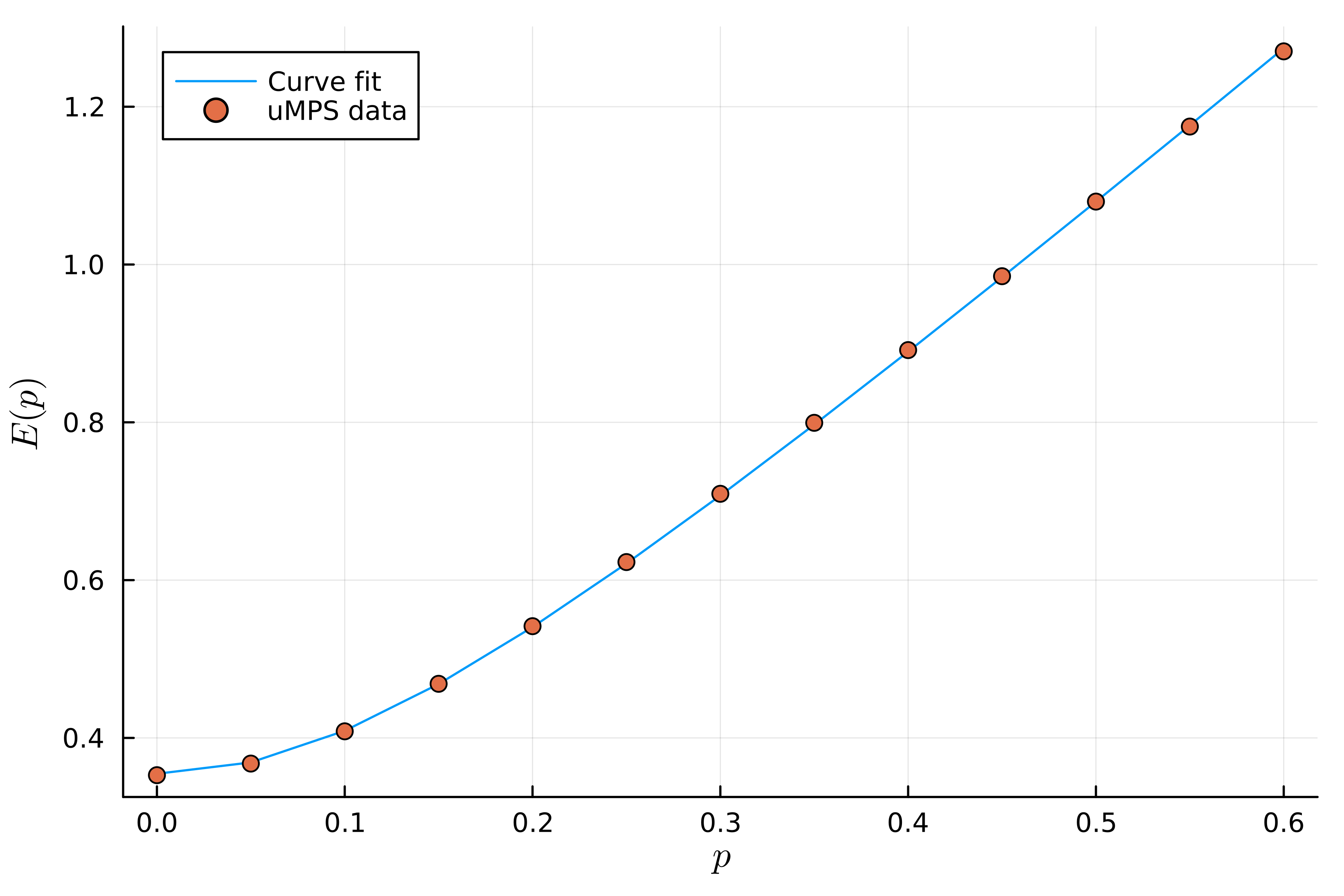}
        \caption{Fit with the relativistic dispersion for the lowest lying excitations.}
        \label{fig:curve-fit-lowest}
    \end{subfigure}
    \caption{Left: lattice spectrum of the first five excitations in Ising Field Theory. These were obtained with the uMPS quasiparticle ansatz with a bond dimension of $D = 20$ for $g_x = 1.06$ and $g_z = 0.01$. Right: curve fit with the relativistic dispersion relation $E(p) = \sqrt{m^2 c^4 + p^2 c^2}$. The (lightest) mass was obtained to be $m \approx 0.0852\,, c = 2.039$ with $R^2 = 0.999$ and $\chi^2 = 4\times 10^{-6}$.}
    \label{fig:lattice-dispersion}
\end{figure}

We use the MPS quasiparticle ansatz to construct the wavepackets. Before doing any scattering experiments, we use the quasiparticle ansatz to find the lattice dispersion. The results are shown in Fig.~\ref{fig:dispersion-lattice}. To create wavepackets, we follow the procedure outlined in~\cite{Jha2025IFT}. Starting from the Quasiparticle ansatz,
\begin{equation}
\begin{split}
    \ket{\Psi[B]} &= \sum_{n \in \mathbb{Z}} \sum_{s_n} e^{i \kappa n}
    v_L^\dagger \bigl(\cdots A^{[n-1]s_{n-1}} B^{s_n} A^{[n+1]s_{n+1}} \cdots \bigr)
    v_r \ket{\cdots s_{n-1} s_n s_{n+1} \cdots} \\
    &= \sum_n \cdots e^{i \kappa n} \;
    \tikz[baseline=-0.5ex]{
        \foreach \i in {0,...,4} {
            \draw[rounded corners, fill=cyan!30]
                (2*\i, -0.5) rectangle (2*\i + 1, 0.5);
            \draw (2*\i - 0.5, 0) -- (2*\i, 0);
            \draw (2*\i + 1, 0) -- (2*\i + 1.5, 0);
            \draw (2*\i + 0.5, -0.5) -- (2*\i + 0.5, -1.0);
            \node at (2*\i + 0.5, 0) {$A$};
        }
        \draw[rounded corners, fill=magenta!30] (4, -0.5) rectangle (5, 0.5);
        \node at (4.5, 0) {$B$};
    }
    \cdots
\end{split}
\end{equation}
where $\kappa$ is the momentum about which the wavepacket is centered. A quasilocal wavepacket centered about an $n_0$ in position space can be created by considering superpositions of the above ansatz
\begin{equation}
\begin{split}
    \ket{\Psi_{\text{wp}}[B]} &= \sum_{n \in \mathbb{Z}} \sum_{s_n} e^{i \kappa n} e^{-\frac{(n - n_0)^2}{\sigma^2}} v_L^\dagger (\cdots A^{[n-1]s_{n-1}}B^{s_{n}}A^{[n+1]s_{n+1}} \cdots ) v_R \ket{\cdots s_{n-1} s_n s_{n+1} \cdots}\\
    &= \sum_{n \in \mathbb{Z}} \cdots e^{i \kappa n} e^{-\frac{(n-n_0)^2}{\sigma^2}}
    \begin{diagram}
        \foreach \i in {0,...,4} {
            \draw[rounded corners, fill=cyan!30] (2*\i, -0.5) rectangle (2*\i + 1, 0.5);
            \draw (2*\i-0.5, 0) -- (2*\i, 0);
            \draw (2*\i + 1, 0) -- (2*\i + 1.5, 0);
            \draw (2*\i + 0.5, -0.5) -- (2*\i + 0.5, -1.0);
            \draw (2*\i + 0.5, 0) node {$A$};
        }
        \draw[rounded corners, fill=magenta!30] (4, -0.5) rectangle (5, 0.5);
        \draw(4.5, 0) node {$B$};
    \end{diagram} \cdots
\end{split}
\end{equation}
We note that the dressed local excitations $B$ in general carry a momentum dependence, but this dependence can be safely neglected once the real-space width of the wavepackets is much larger than the correlation length $\xi$. For instance, in the simulations of Ref.~\cite{Jha2025IFT} the correlation length was $\xi \sim 4$--$10$ lattice sites, while the wavepacket width was of order $70$--$100$ sites. We again emphasize that in these notes we will not attempt to resolve such ultra-precise numerical regimes; instead, our aim is to illustrate the qualitative features of scattering in the Ising field theory using low bond dimensions and wavepacket widths that are easily accessible with modest computational resources.
\begin{figure}[hbt]
    \centering
    \begin{subfigure}[b]{0.48\textwidth}
    \includegraphics[width=\textwidth]{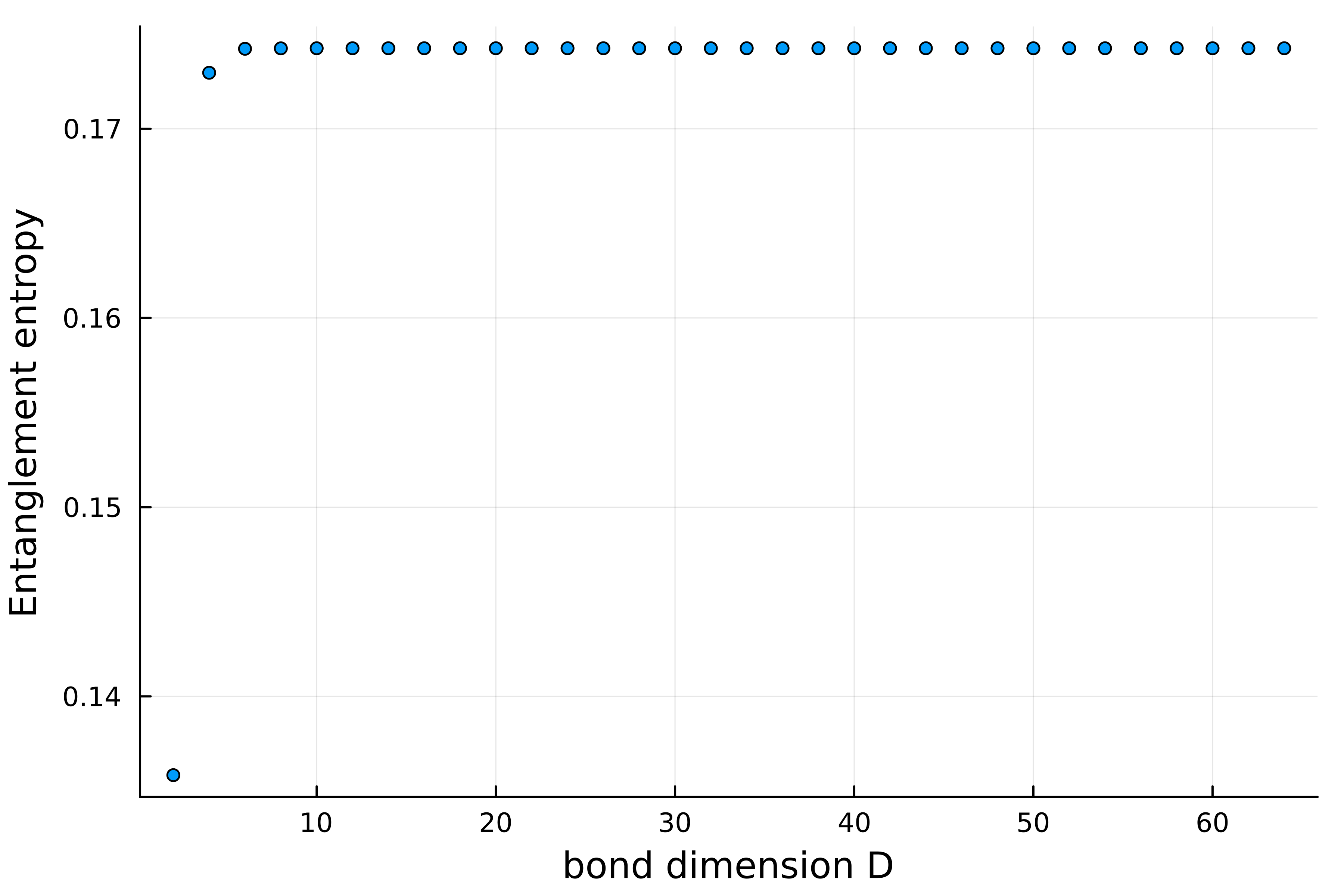}
        \centering
        \label{fig:entanglement-entropy-wrt-D}
    \end{subfigure}
    \hfill
    \begin{subfigure}[b]{0.48\textwidth}
        \centering
        \includegraphics[width=\textwidth]{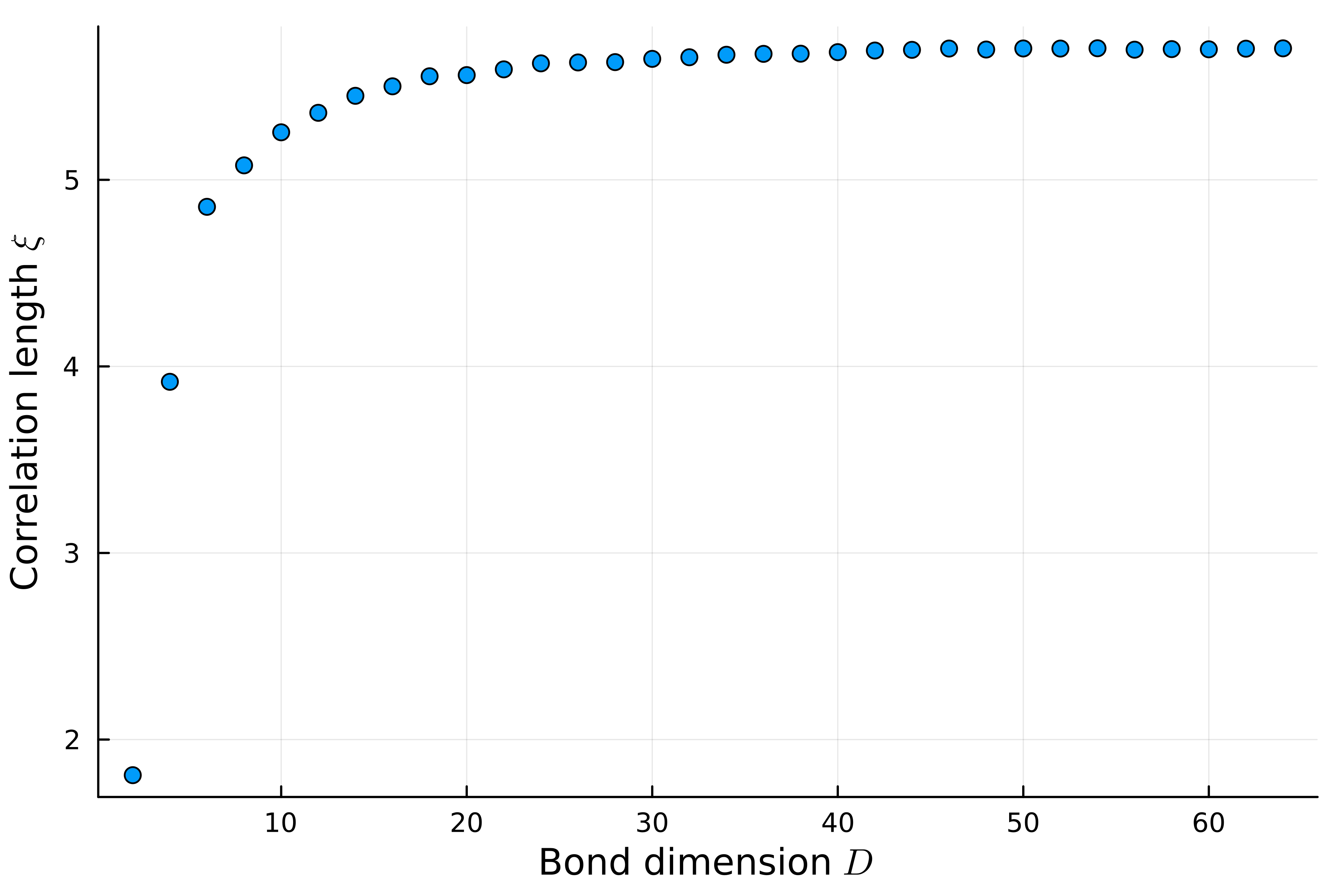}
        \label{fig:corr-len-wrt-D}
    \end{subfigure}
    \caption{Entanglement entropy and correlation length respect to the bond dimension for the IFT with $(g_x, g_z) = (1.06, 0.01)$. We use $D = 5, 10$ for our simulations for which entanglement entropy has converged up to 3 decimal places. The correlation length, however, does not coverge appreciably at $D = 5$.}
    \label{fig:correlation-len}
\end{figure}
\begin{figure}
    \centering
    \begin{subfigure}[b]{0.48\linewidth}
        \centering
        \includegraphics[width=\linewidth]{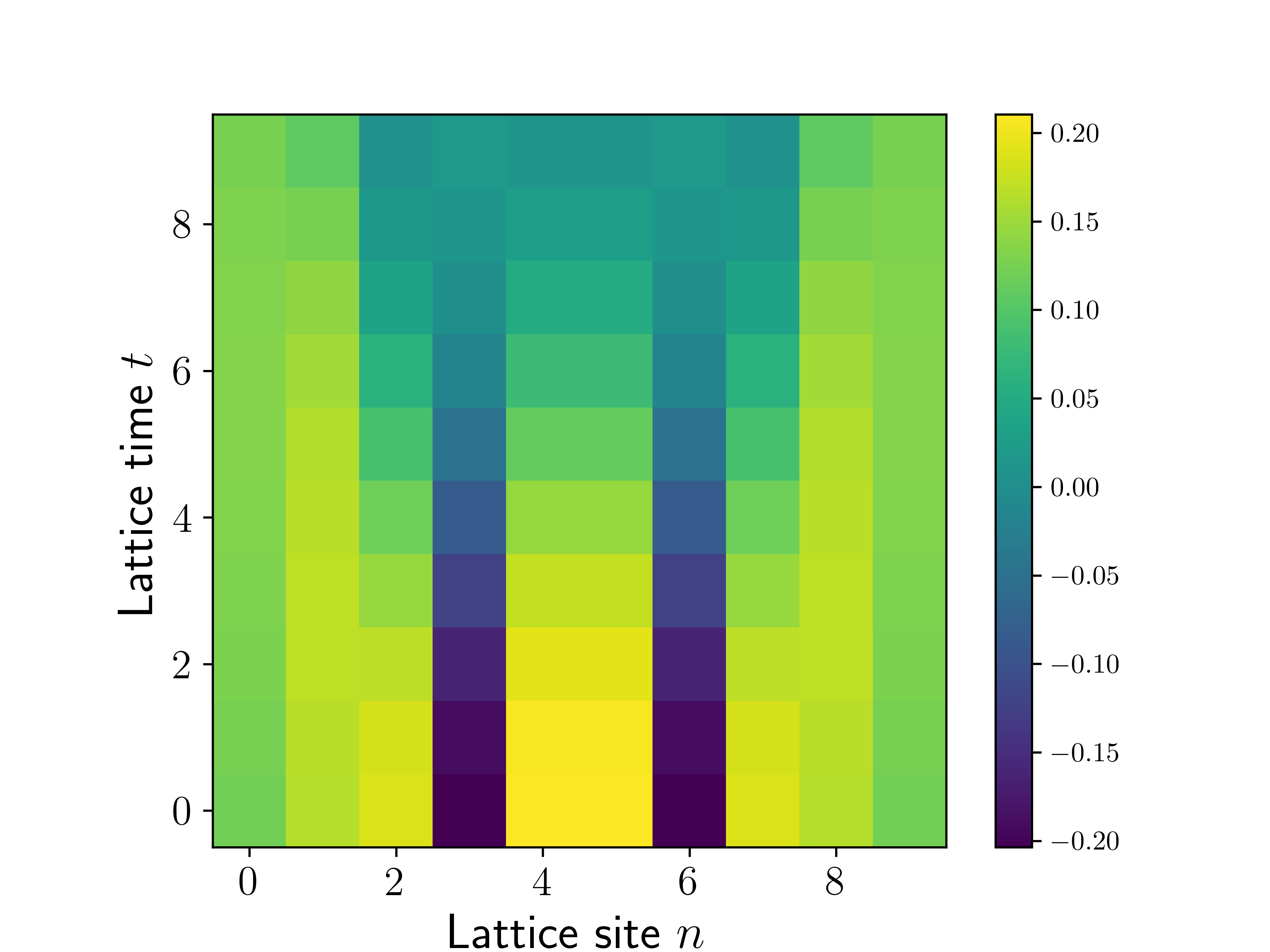}
        \label{fig:TFIM_evolution_qiskit_left}
    \end{subfigure}
    \hfill
    \begin{subfigure}[b]{0.48\linewidth}
        \centering
        \includegraphics[width=\linewidth]{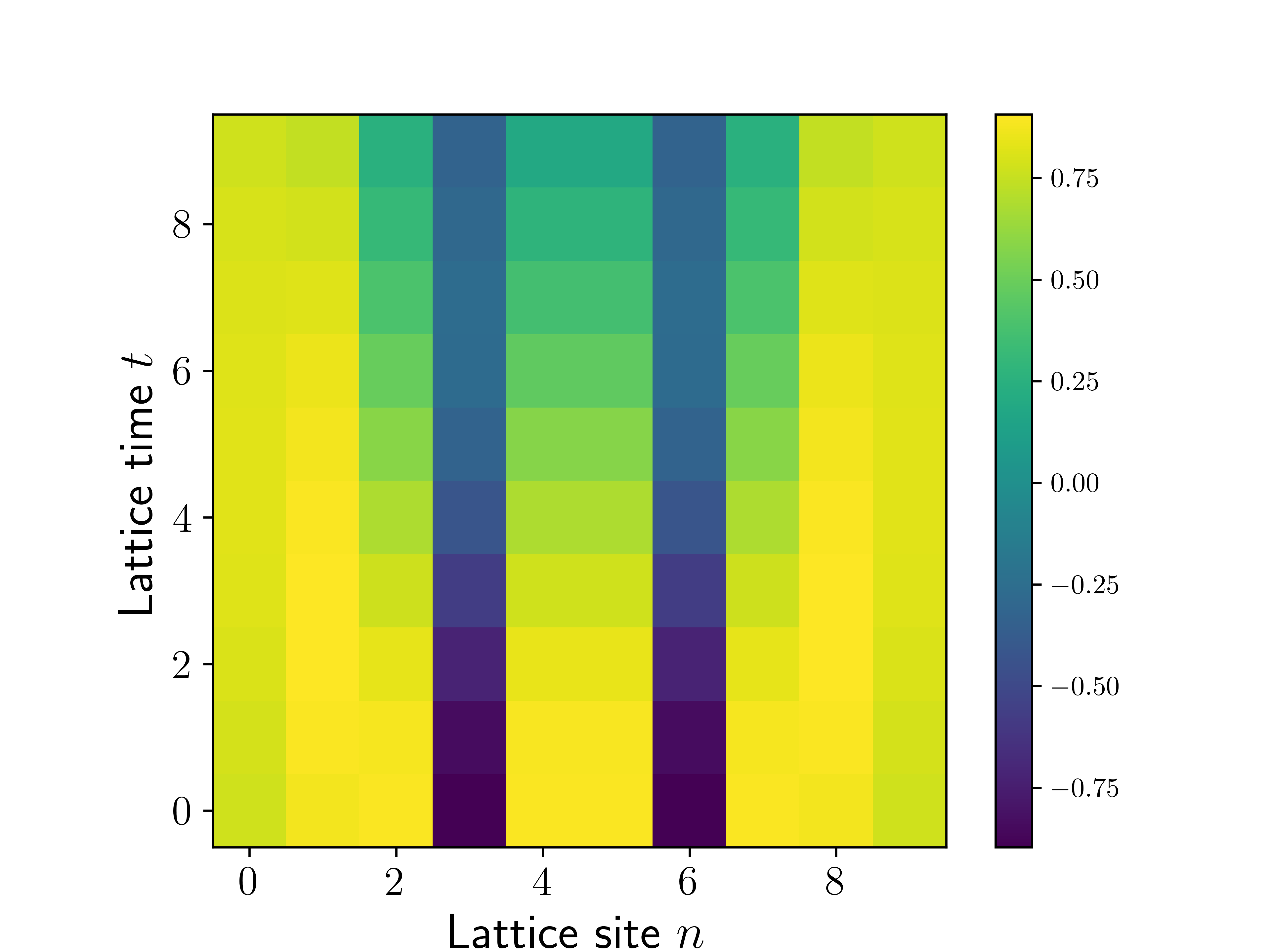}
        \label{fig:TFIM_evolution_qiskit_right}
    \end{subfigure}
    \caption{Time evolution of local disturbances in the transverse field Ising model with \texttt{Qiskit}. The circuit is initialized in the ground state of the Hamiltonian (left: $(h_x, h_z) = (1.06, 0.01)$, right: $(h_x, h_z) = (1.06, 0.5)$), and Pauli-$X$ operators are applied on the third and sixth sites before Trotterized time evolution.}
    \label{fig:TFIM_evolution}
\end{figure}
\begin{figure}[hbt]
    \centering
    \begin{subfigure}[b]{0.48\textwidth}
    \includegraphics[width=\textwidth]{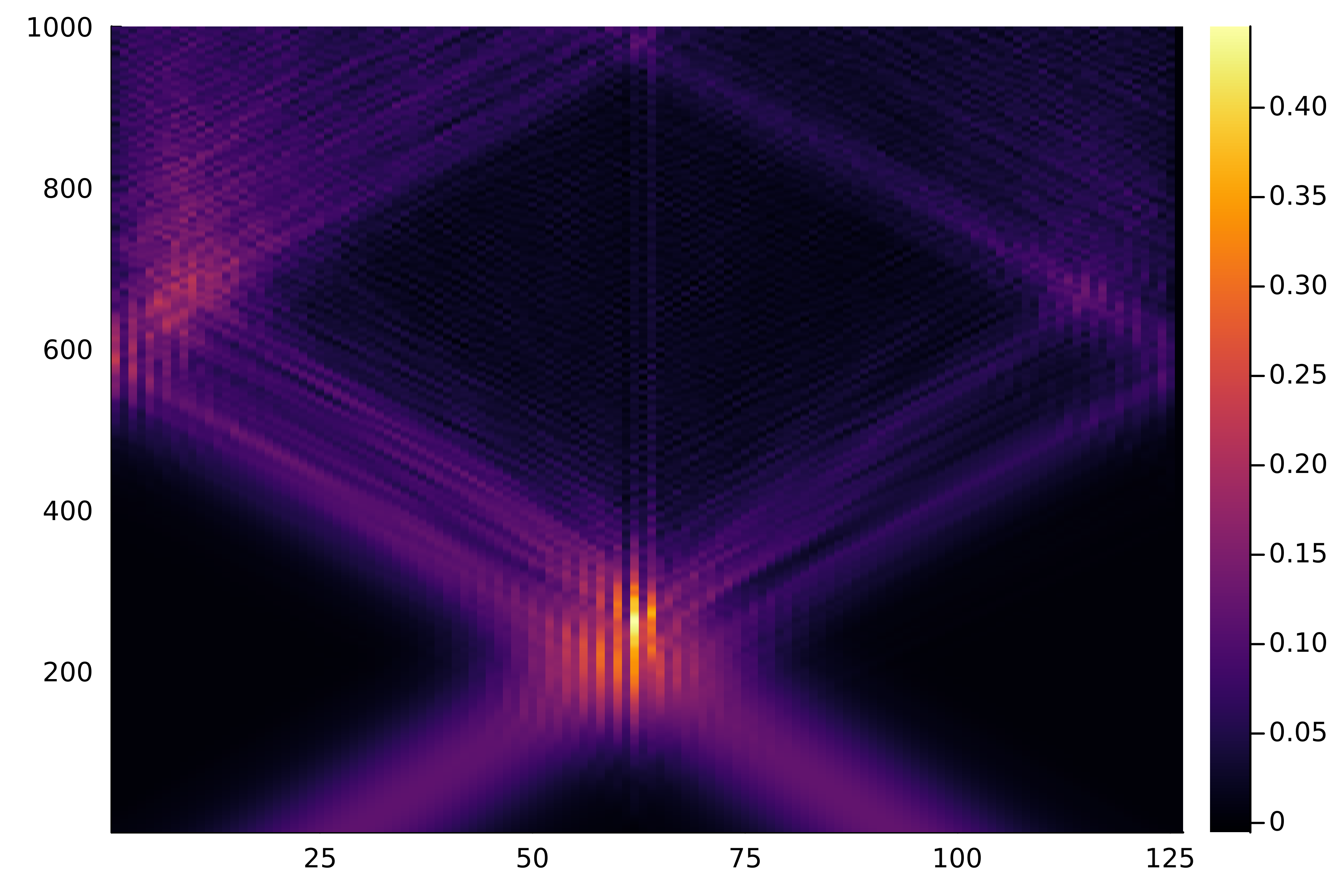}
        \centering
        \label{fig:scattering_bond_dim_5}
    \end{subfigure}
    \hfill
    \begin{subfigure}[b]{0.48\textwidth}
        \centering
        \includegraphics[width=\textwidth]{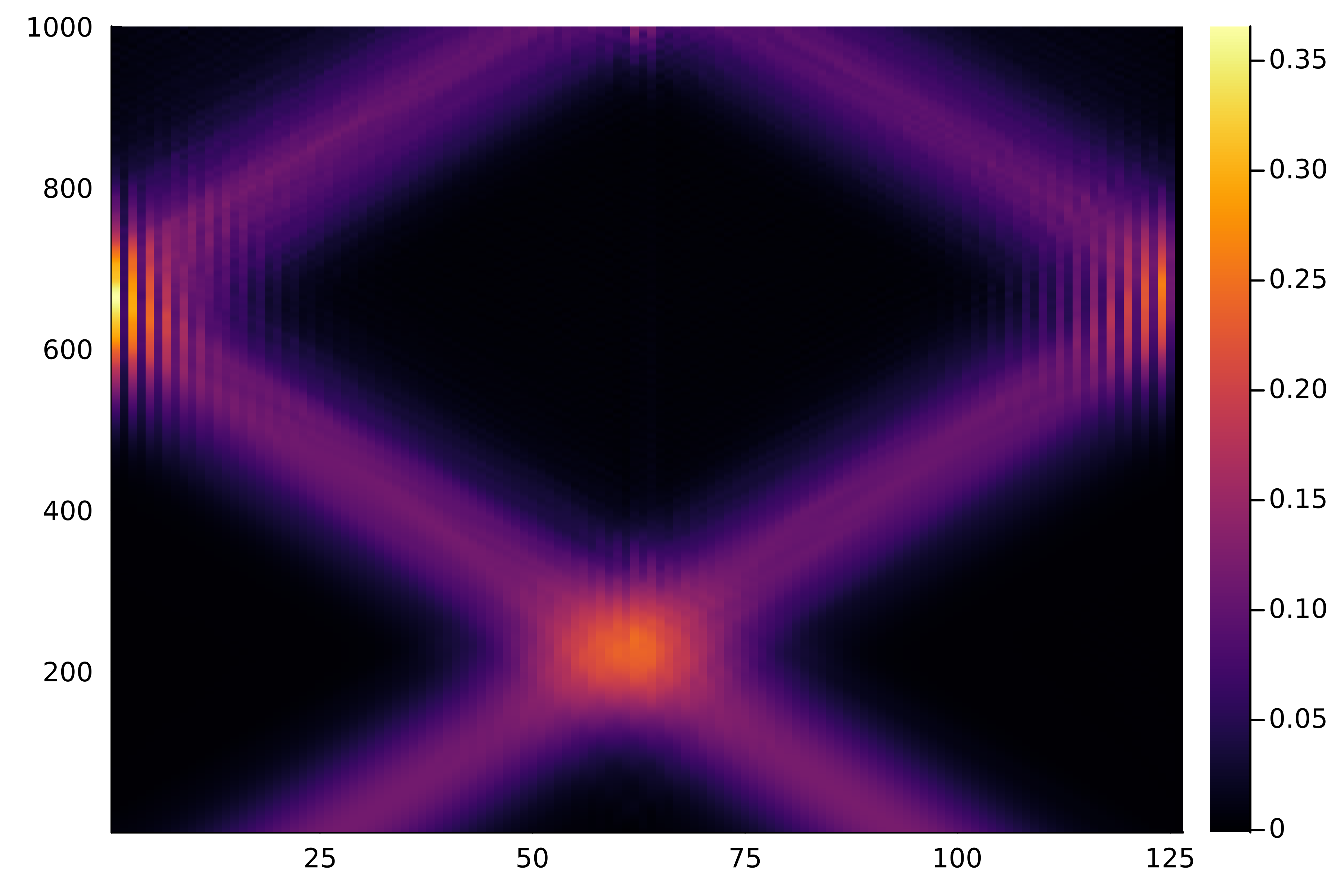}
        \label{fig:scattering_bond_dim_10}
    \end{subfigure}
    \caption{Scattering in the non-integrable regime at $(g_x, g_z) = (1.06, 0.01)$. The plot on left has artificial trails due to the low bond dimension ($D = 5$) unlike the plot on the right ($D = 10$).}
    \label{fig:scattering_wavepackets}
\end{figure}
\chapter{Results: Schwinger model}
By now, hopefully the reader has gathered enough interest to pick up some numerical skills. With this hope, we leave a Mathematica analysis of the energy level difference, small lattice size diagonalization problems for the bosonized Schwinger and Thirring models as exercises. What we will do now is to set up a toy problem in Mathematica to probe string breaking.

\section{A toy bosonized Schwinger model on a small lattice}

In this subsection we set up a very simple lattice toy model of the bosonized
Schwinger theory which is amenable to exact diagonalization and real--time
evolution in \texttt{Mathematica}. The goal is not to faithfully reproduce all
continuum details, but to obtain a qualitative, ``baby'' version of string
breaking and its dependence on the separation between charges and on the
$\theta$--angle.

\subsection{Continuum motivation}

Recall that the bosonized massive Schwinger model with a $\theta$--angle can be
written as
\begin{equation}
  \mathcal{L}
  = \frac{1}{2} (\partial_\mu \phi)^2
    - \frac{1}{2} \frac{e^2}{\pi}\,\phi^2
    - \frac{m}{\pi\alpha}\,\cos\!\big(\sqrt{4\pi}\,\phi - \theta\big)\,.
\end{equation}
In this language the electric field and dynamical charge density are (up to
conventions)
\begin{equation}
  E(x,t) \;=\; \frac{e}{\sqrt{\pi}}\,\phi(x,t) + \frac{e\,\theta}{\pi}\,,\qquad
  j^0(x,t) \;=\; \frac{1}{\sqrt{\pi}}\,\partial_x \phi(x,t)\,,
\end{equation}
so that in the vacuum the electric field is shifted by a constant proportional
to~$\theta$. In the presence of static external charges, the electric field
develops a ``string'' between them; dynamical fermion pairs can then be created
from the vacuum, screening this string and reducing the field in the central
region. Our toy model is designed to mimic this screening process in a very
simple setting.

\subsection{Lattice truncation: oscillators with a cosine potential}

We discretize space into $N$ sites with lattice spacing $a$, and place on each
site a single bosonic mode which is a truncated harmonic oscillator. The exact
continuum field $\phi(x)$ is replaced by site operators $\phi_j$,
$j=1,\dots,N$, obeying canonical commutation relations
$[\phi_j,\Pi_k]={\rm i}\delta_{jk}$, with~$\Pi_j$ the conjugate momenta. In
practice, we represent each $(\phi_j,\Pi_j)$ in a truncated harmonic--oscillator
basis of dimension $d=n_{\max}+1$:
\begin{equation}
  \phi_j \;\equiv\; x_j\,,\qquad
  \Pi_j \;\equiv\; p_j\,,\qquad
  [x_j,p_k]={\rm i}\,\delta_{jk}\,,
\end{equation}
where $(x_j,p_j)$ are obtained from creation and annihilation operators
$a_j,a_j^\dagger$ via the usual relations
\begin{equation}
  x_j = \frac{a_j + a_j^\dagger}{\sqrt{2m_\phi}}\,,\qquad
  p_j = {\rm i}\sqrt{\frac{m_\phi}{2}}\,(a_j^\dagger - a_j)\,,
\end{equation}
and the local Hilbert space is truncated to occupation number
$n=0,1,\dots,n_{\max}$. The full Hilbert space dimension is then
$d^N = (n_{\max}+1)^N$, which for example is $3^5=243$ for $N=5$ and
$n_{\max}=2$.

A simple lattice Hamiltonian that mimics the bosonized Schwinger dynamics is
\begin{equation}
  H_0 = \sum_{j=1}^N \left[
    \frac{1}{2}\,\Pi_j^2
    + \frac{1}{2}\,m_\gamma^2\,\phi_j^2
    + \lambda\,\big(1 - \cos(\beta\,\phi_j - \theta)\big)
  \right]
  + \frac{\kappa}{2} \sum_{j=1}^{N-1} (\phi_{j+1}-\phi_j)^2\,,
  \label{eq:toy-H0}
\end{equation}
where $m_\gamma^2 \sim e^2/\pi$ plays the role of the Schwinger mass,
$\lambda\sim m/(\pi\alpha)$ controls the cosine potential, and
$\beta\sim\sqrt{4\pi}$ sets the normalization of~$\phi$ (in the numerics we
treat $m_\gamma$, $\lambda$, $\beta$ and~$\kappa$ as tunable parameters rather
than fixing them to their exact continuum values). The last term is a discrete
gradient term which regularizes the kinetic energy of~$\phi$.

In this toy--model we take the electric field to be proportional to~$\phi_j$,
with the $\theta$--shift included explicitly:
\begin{equation}
  E_j \;\propto\; \phi_j + \frac{\theta}{\sqrt{\pi}}\,, 
\end{equation}
or, if we wish to emphasize the flux on links rather than at sites, we may also
define a discrete electric field on the link between $j$ and $j+1$ as
\begin{equation}
  E_{j+\frac12} \;\propto\; \phi_{j+1} - \phi_j\,.
\end{equation}
Either choice leads to qualitatively similar behaviour in the small systems we
study below.

\subsection{Mimicking a string with source terms}

We want to study the real time relaxation of a string like configuration. Either we proceed by a) starting with a non-eigenstate and time evolving that or b) starting with the eigenstate of a Hamiltonian and suddenly changing the parameters to a different Hamiltonian. Option b) is called a quench and for simplicity (or variety!) we will consider quenched dynamics. 

To emulate an electric string between static charges we add a source term
linear in~$\phi_j$,
\begin{equation}
  H_{\text{source}}[J] \;=\; \sum_{j=1}^N J_j\,\phi_j\,,
\end{equation}
where the profile $J_j$ encodes the background charges and the resulting
background electric field. In the simplest version we choose $N=5$ and take
$J_j$ to be nonzero only on the central three sites,
\begin{equation}
  J_j^{\text{(init)}} = \big(0,\,J_0,\,J_0,\,J_0,\,0\big)_j\,,
\end{equation}
with $J_0>0$. The initial Hamiltonian is then
\begin{equation}
  H_{\text{init}}(\theta) = H_0(\theta) + H_{\text{source}}[J^{\text{(init)}}]\,.
\end{equation}
The ground state $|\Psi_0^{(\theta)}\rangle$ of $H_{\text{init}}(\theta)$ has a
non--vanishing expectation value of~$\phi_j$ in the central region, and
therefore a nonzero ``electric field'' between the effective charges. This
state plays the role of a stretched string of some effective length~$R$ (here
encoded in the spatial support of~$J_j$).

To study string breaking we perform a quantum quench at $t=0$: we suddenly
change the source profile to a weaker one,
\begin{equation}
  J_j^{\text{(final)}} = \big(0,\,J_1,\,J_1,\,J_1,\,0\big)_j\,,\qquad
  |J_1| < |J_0|\,,
\end{equation}
and evolve with
\begin{equation}
  H_{\text{final}}(\theta) = H_0(\theta) + H_{\text{source}}[J^{\text{(final)}}]\,.
\end{equation}
The initial state at $t=0$ is still $|\Psi_0^{(\theta)}\rangle$, but the time
evolution is now governed by $H_{\text{final}}(\theta)$:
\begin{equation}
  |\Psi^{(\theta)}(t)\rangle = e^{-{\rm i} H_{\text{final}}(\theta)\,t}\,
                               |\Psi_0^{(\theta)}\rangle\,.
\end{equation}
Intuitively, the strong initial string configuration has suddenly been embedded
in a Hamiltonian which prefers a weaker string; the excess field energy is
converted into excitations of the $\phi_j$ modes, mimicking pair production and
screening.

\subsection{String--breaking diagnostics and the role of \texorpdfstring{$x_{\rm Exp}(j,t)$}{xExp[j,t]}}

In the truncated oscillator implementation each site field operator $\phi_j$
is represented by a finite matrix $x_j$. In the numerics we evolve an initial
state $|\Psi_0^{(\theta)}\rangle$ under the final Hamiltonian
$H_{\text{final}}(\theta)$ and monitor the Heisenberg expectation values
\begin{equation}
  x_{\rm Exp}(j,t)
  \;\equiv\;
  \big\langle \Psi^{(\theta)}(t) \big| x_j \big| \Psi^{(\theta)}(t) \big\rangle,
  \qquad
  |\Psi^{(\theta)}(t)\rangle = e^{-{\rm i}H_{\text{final}}(\theta)t}\,
                               |\Psi_0^{(\theta)}\rangle\,.
\end{equation}
\begin{figure}[htb]

  \centering
  \includegraphics[width=0.5\linewidth]{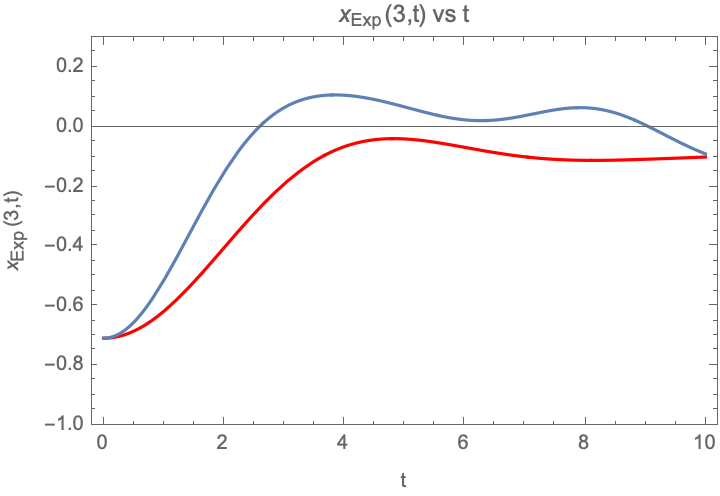}
  \caption{$x_{\rm Exp}(3,t)$ vs $t$. Blue is $\theta=0$ and red is $\theta=\pi$. The magnitude drops as a function of $t$.} \label{fig:string-breaking-mathematica}
\end{figure}

In the \texttt{Mathematica} code (github repository) these expectation values are computed by a
function called \verb|xExp[j_, t_]|, so that \verb|xExp[3, t]| corresponds
precisely to $\langle x_3(t)\rangle = \langle\phi_3(t)\rangle$ on the central
site of a five--site chain. Since in our toy model the electric field (up to
constants and a $\theta$--dependent shift) is taken to be proportional to the
bosonic field,
\begin{equation}
  E_j(t;\theta) \;\propto\; x_{\rm Exp}(j,t) + \frac{\theta}{\sqrt{\pi}}\,,
\end{equation}
the time dependence of $x_{\rm Exp}(3,t)$ is a convenient indicator of how
the electric field in the central ``string'' region relaxes after the quench.

More generally, one can average over several sites in the string region and
define
\begin{equation}
  E_{\text{string}}(t;\theta)
  \;\propto\;
  \frac{1}{N_{\text{string}}}\sum_{j\in\text{string}}
  \left( x_{\rm Exp}(j,t) + \frac{\theta}{\sqrt{\pi}} \right),
\end{equation}
or, if one prefers a link--based diagnostic, use a discrete derivative
$E_{j+1/2}(t)\propto x_{\rm Exp}(j{+}1,t)-x_{\rm Exp}(j,t)$ and average over
links between the charges. In either case, plotting $x_{\rm Exp}(3,t)$ or
$E_{\text{string}}(t;\theta)$ as a function of~$t$ provides a very direct,
``by eye'' visualization of string breaking in this small bosonized Schwinger
toy model.
We will not supply code snippets here. Our short code can be found in the github repository. We show the plot for $x_{\rm Exp}(3,t)$ vs $t$ in Fig.~\ref{fig:string-breaking-mathematica}. The fall in the magnitude of the electric field is apparent. 
\section{MPS analysis}
\begin{figure}[t]
    \centering
    \begin{subfigure}[b]{0.48\textwidth}
        \centering
        \includegraphics[width=\textwidth]{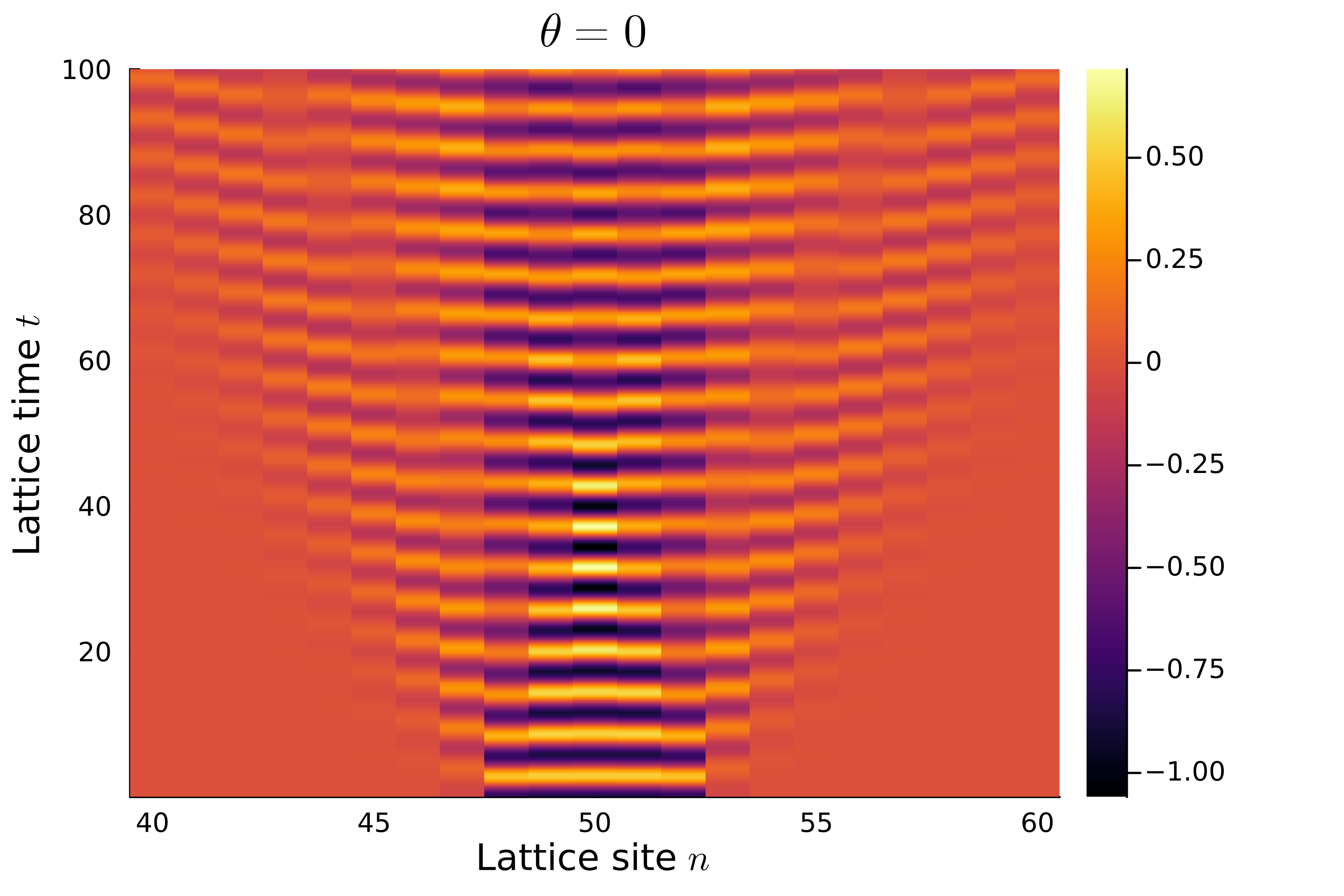}
        \label{fig:oscillatory-string-theta-0}
    \end{subfigure}
    \hfill
    \begin{subfigure}[b]{0.48\textwidth}
        \centering
        \includegraphics[width=\textwidth]{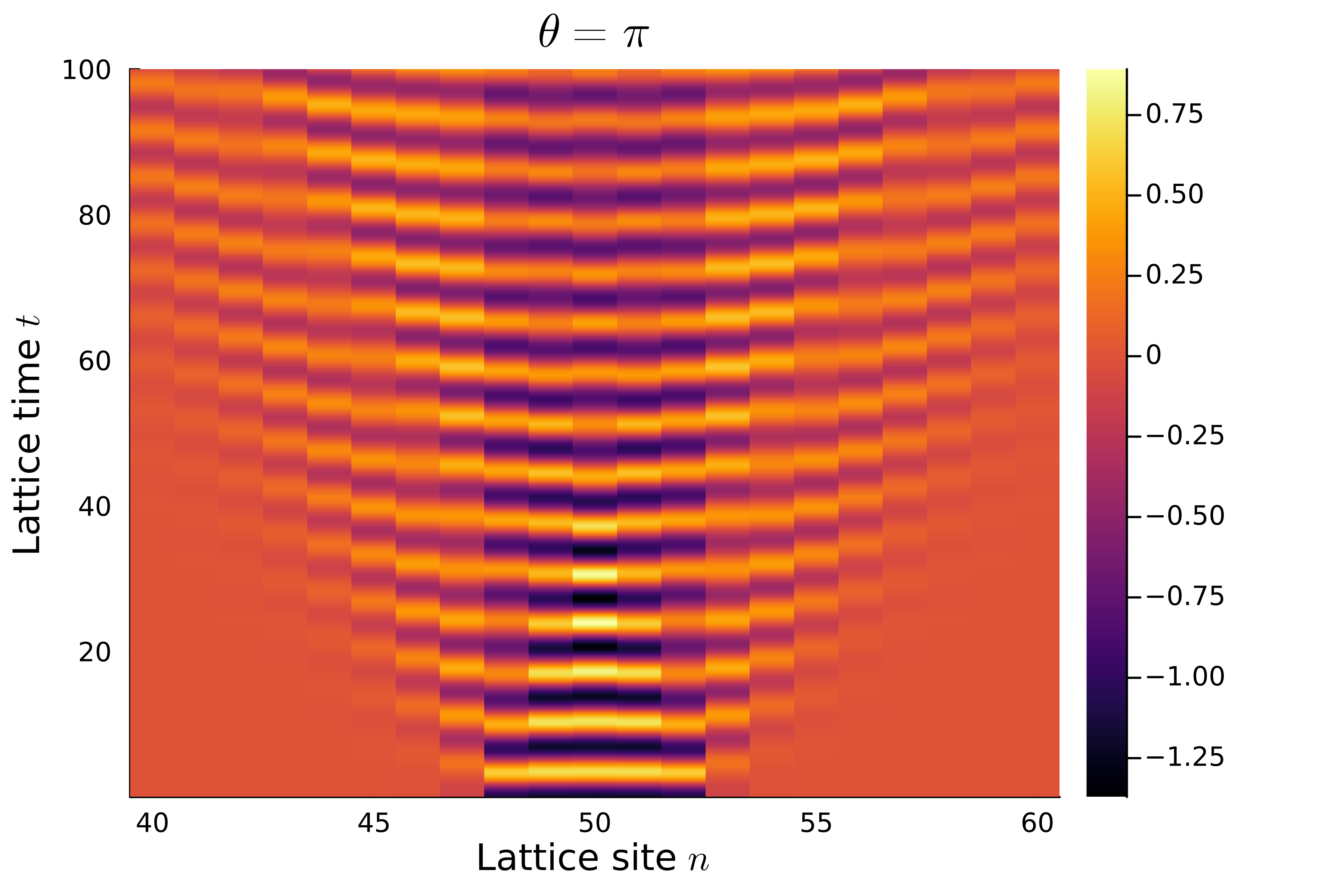}
        \label{fig:oscillatory-string-theta-pi}
    \end{subfigure}
    \caption{Quench from $J_0 = 1$ to $J_2 = 0.2$. The string is initially confined to a window of $5$ (right) sites and exhibits oscillatory dynamics with clear leakage of weight outside the original region. Simulation parameters: $L = 100$, $D = 24$, $d = 12$. We show a plot of the expectation value of the field operator $\expval{\phi_n(t)}$ at the center of the string ($n = 50$ on the lattice) as a function of the lattice time.}
    \label{fig:oscillatory-string}
\end{figure}
\begin{figure}
    \centering
    \includegraphics[width=0.5\linewidth]{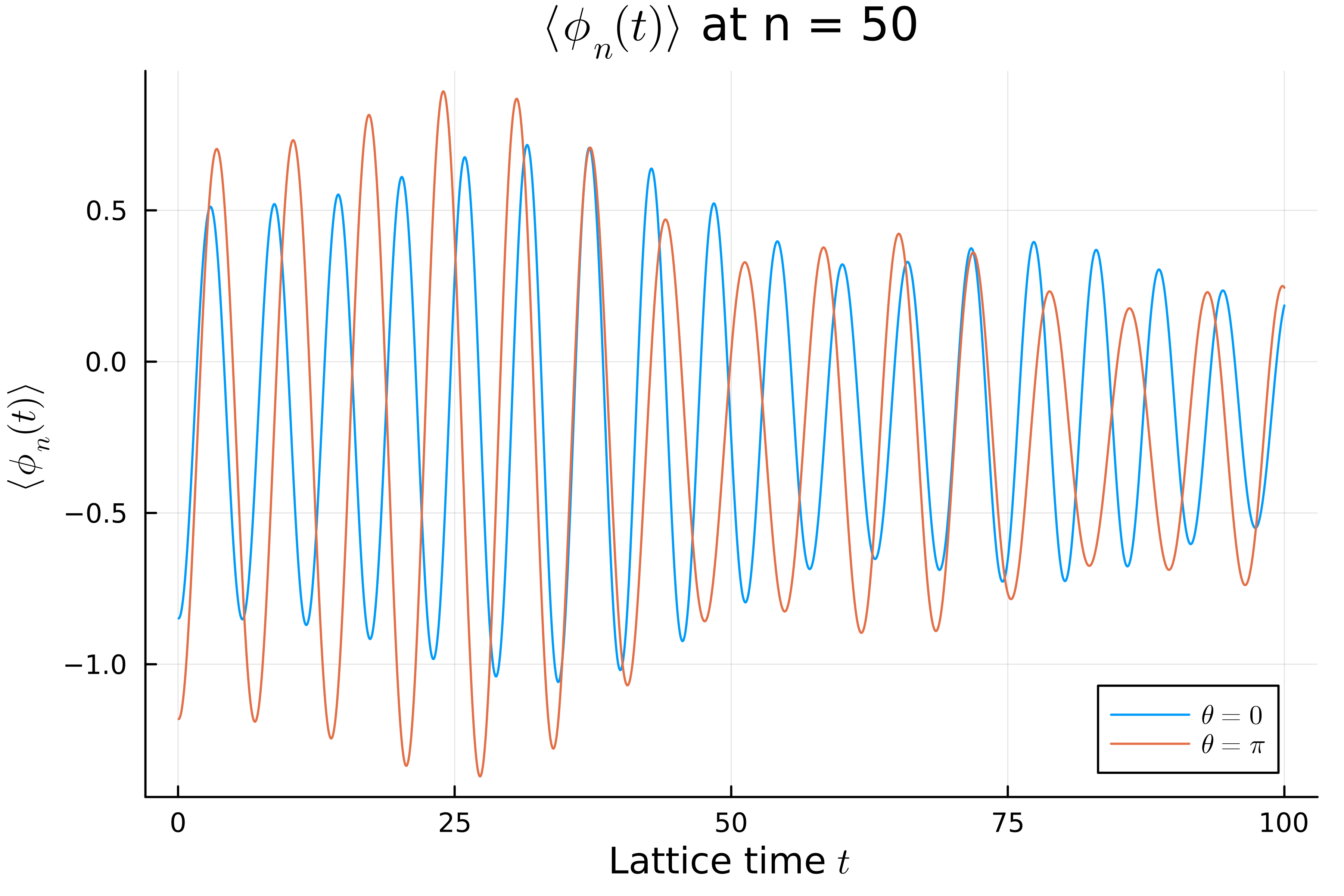}
    \caption{Plot of $\langle \phi_n(t)$ at $n = 50$ (center of the string) as a function of time. For both the cases $\theta = 0$ and $\theta = \pi$, the magnitude of the expectation value of the field operator decreases with time.}
    \label{fig:string_breaking_center}
\end{figure}
In the previous section, our \texttt{Mathematica} exercise was limited to a relatively small truncation of the local Hilbert space for the field operators. To access a substantially larger portion of the full Hilbert space, we now switch to MPS techniques. In close analogy with the earlier setup, we probe real-time quench dynamics by introducing a source term $J^{\text{init}}$ that is non-zero only on a central window of five sites in the middle of a chain of length $L = 100$. The ground state of this inhomogeneous Hamiltonian is prepared using a single-site DMRG algorithm, and the resulting state is then evolved in real time with the TDVP scheme. The results of this exercise are shown in Fig.~\ref{fig:oscillatory-string} and Fig.~\ref{fig:string_breaking_center}.
\chapter{State-of-the-art}

\section{Actual quantum hardware}
For convenience, let us summarize here the state-of-the-art so far as doing experiments on actual quantum hardware is concerned. We should add a disclaimer: As of now there is no quantum advantage over the tensor-network methods. The present results rely on $O(10^2)$ physical qubits. Estimates in \cite{sakamoto} to reliably compute the vacuum persistence amplitude in the Schwinger model are of the order of $O(10^5)$ physical qubits. We are still a long way off. However, it is fascinating to observe the interesting benchmarking results that have already been obtained on the existing NISQ devices during the last 2 years.

\begin{enumerate}
  \item \textbf{$\phi^4$:} 
  Zemlevskiy has demonstrated scalable digital quantum simulations of two--particle
  wave-packet scattering in a truncated scalar $\phi^4$ theory, extracting interaction
  effects in the post--collision energy density and related observables on an IBM
  Heron device (ibm\_fez) using up to $120$ qubits and circuits with $\mathcal{O}(10^3)$
  two--qubit gates~\cite{Zemlevskiy_2025}.

  \item \textbf{Ising field theory:}
  Farrell et al  have carried out large--scale simulations of
  inelastic scattering in the near--Ising field theory using specially prepared
  $W$ states as initial wavepackets, diagnosing particle production via the
  skewness of the post--collision energy density. Their experiments on IBM Heron
  hardware (ibm\_marrakesh) reach $104$ qubits and circuits with $\sim 5\times 10^3$
  two--qubit gates (about $45$ Trotter steps)~\cite{Farrell2025WStates}.

  \item \textbf{Schwinger model:}
  Farrell et al.\ have used trotterized real--time evolution to study hadron
  wavepacket dynamics in the lattice Schwinger model, measuring the propagation
  of hadronic excitations and local chiral observables on IBM Heron processors
  (ibm\_torino) with up to $112$ qubits and circuits containing $\sim 1.4\times 10^4$
  CNOTs (CNOT depth $\sim 370$ over $14$ Trotter steps)~\cite{FarrellPRD2024}.
  Complementary work by Davoudi and collaborators develops hardware--efficient
  strategies for hadron scattering in confining gauge theories: Davoudi, Hsieh
  and Kadam construct interacting mesonic wavepackets directly in the lattice
  gauge theory and demonstrate their preparation on a trapped--ion quantum
  computer~\cite{DavoudiQuantum2024}, while Belyansky \emph{et al.} study
  high--energy quark and meson collisions in the Schwinger model using uniform
  MPS and a circuit--QED scattering proposal~\cite{BelyanskyPRL2024}. 

    \item \textbf{Analog quantum simulators (trapped ions and optical lattices).}
  \begin{enumerate}
    \item \textbf{Few-qubit Schwinger dynamics in trapped ions:}
    In an early trapped--ion experiment, Martinez \emph{et al.} realized a
    digital simulation of the 1+1D Schwinger model on a four--ion processor,
    observing real--time vacuum decay and particle--antiparticle pair production,
    and providing one of the first laboratory demonstrations of lattice gauge
    dynamics~\cite{Martinez2016}. More recently, Nguyen \emph{et al.} have
    performed a fully digital simulation of the lattice Schwinger model on a
    six--ion quantum computer, using symmetry protection and postselection to
    extend real--time dynamics and observe pair creation over many Trotter
    steps~\cite{NguyenPRXQ2022}. Mueller \emph{et al.} have used a similar
    trapped--ion architecture to observe dynamical quantum phase transitions and
    perform entanglement tomography in a simple lattice gauge theory,
    providing a proof--of--principle for accessing non--equal--time correlators
    and entanglement Hamiltonians on hardware~\cite{MuellerPRXQ2023}.

    \item \textbf{2D lattice QED plaquette on a qudit ion computer:}
    Meth \emph{et al.} have implemented a single 2D lattice QED plaquette with
    dynamical matter on a mixed qubit--qudit trapped--ion device (one qutrit
    gauge link plus four qubit matter sites), using VQE and real--time dynamics
    to probe plaquette expectation values and pair creation, and highlighting the
    efficiency gains of qudit encodings~\cite{Meth2025}.

    \item \textbf{Large-scale U(1) gauge theories in optical lattices:}
    Pan's group has realized U(1) lattice gauge theories with $\mathcal{O}(10^2)$
    sites in optical superlattices, observing gauge invariance across phase
    transitions and quench dynamics~\cite{Yang2020}, and more recently probing
    false--vacuum decay and Schwinger--type pair production in a cold--atom
    gauge--theory simulator~\cite{Zhu2024}.
  \end{enumerate}
\end{enumerate}

\section{Term papers: Student work}
In the course, various options for term papers were assigned. In the GitHub repository, we share the write-up and codes for a few of these.
\begin{enumerate}
\item Scattering and resonances in quantum mechanics: Noise analysis and feasibility study, phase shift, time delay, LSZ, Haag-Ruelle theory, JLP-- Soumyadeep Sarma, Indrayudh Das (IISc).
    \item Quantum Optics: Jaynes-Cummings and Rabi in the strong coupling regime using QISKIT and real quantum hardware IBM(torino)--Ankush Kumar and Suman Dafadar (IISc). Jaynes-Cummings and Rabi in the strong coupling regime with QPE, noise modelling and dissipation studies along with Tavis-Cummings discussion--Mahdi Shamsei and Subhadip Dutta (UCalgary). 
    \item Ising Field Theory: W-state preparation using the Unitary Circuit approach and scattering simulations using MPS methods following \cite{Farrell2025WStates}; the students attempted to reproduce \cite{Farrell2025WStates} for a small number of lattice sites--Ritabrata Ghosh and Abhishek Kundu (IISc).
    \item Schwinger model: Staggered fermion and bosonization studies of string breaking, confinement and spontaneous breaking of CP symmetry--Nikshay Chugh and Aman Goyal (IISc); Staggered fermion approach to particle production and string breaking--Chayanka Kakati and Abhijeet Bhatta (IISc). 
\end{enumerate}

\chapter{What next?}

We are poised at an interesting crossroad. Quantum hardware results for quantum field theory are just beginning to get interesting. There will almost certainly come a time when classical methods can no longer compete with quantum devices for certain real--time, strongly coupled, or finite--density problems---but \emph{when}, and for which observables, remains an open question.

In these lectures, we introduced quantum computing techniques for quantum field theory through certain 1+1 dimensional models. The anharmonic oscillator served as the anchor: it provided a unifying language for spectroscopy, phase structure, adiabatic state preparation, and scattering, and it motivated the lattice formulations of the 1+1D field theories considered here. Even in this modest setting, there is much left to explore. Within 1+1D QFTs one can sharpen and extend what we have done along several directions: mapping out phase diagrams more precisely \cite{Davoudi:2023TPQ}, extracting critical exponents and operator content, studying real--time quenches and thermalization, and pushing further into multiparticle scattering and resonance physics. We have also not touched upon non--Abelian gauge theories~\cite{indrakshi,pufu} or multiflavour theories and nontrivial flavour dynamics~\cite{pufu2}, which are natural next steps even in one spatial dimension.

The main long--term challenge, both conceptually and algorithmically, is to generalize these techniques to higher dimensions, ultimately towards realistic 2+1D and 3+1D gauge theories~\cite{MagnificoEtAl2025}. This raises a host of interconnected issues: the growth of local Hilbert space with truncation, the cost of encoding gauge constraints, the depth and width of circuits needed for Trotterized or LCU-based time evolution, and the design of observables that are both physically interesting and experimentally accessible. Near--term work will likely focus on identifying \emph{minimal} higher--dimensional testbeds where one can already hope to see a quantum advantage in some corners (for example, out--of--equilibrium dynamics or sign--problem--dominated regimes), while still keeping qubit counts and circuit depths compatible with noisy hardware.

Projected entangled--pair states (PEPS) provide the natural tensor--network language for 2+1 dimensional lattice field theories and have already yielded competitive classical results for ground--state properties and phase diagrams in a variety of spin and gauge models; see, e.g., \cite{MagnificoEtAl2025}. By encoding gauge constraints and local symmetries directly at the tensor level, they allow one to access confinement, deconfinement and topological order in regimes that are difficult for traditional Monte Carlo, and thus already serve as a nonperturbative benchmark in 2+1D. From the perspective of quantum computing, PEPS and iPEPS algorithms define a clear classical baseline for quantities such as string tensions, Wilson loops, mass gaps and simple correlators, against which near--term quantum devices will first have to be tested. However, PEPS come with their own bottlenecks: exact contraction is hard and practical schemes rely on approximate environments whose cost grows quickly with bond dimension, limiting accessible system sizes and the sharpness of continuum extrapolations. Real--time dynamics, finite density and high--entanglement quenches in 2+1D remain challenging, so even classically the most interesting observables for quantum advantage are not yet under full control. In 3+1D, the situation is even more severe: fully fledged three--dimensional PEPS for non--Abelian gauge theories are currently beyond reach except in very small volumes or highly truncated settings, and there is no consensus yet on scalable contraction strategies. Thus, although PEPS will probably continue to represent the state of the art for many static problems and for higher-dimensional QFTs with low entanglement, their constraints in terms of spatial dimensionality, accessible bond dimension, and real-time dynamics clearly demarcate the regimes in which truly quantum hardware may eventually gain a definitive advantage.

Even within fixed models, state preparation remains a key bottleneck.In these notes, we have mostly focused on conceptually straightforward, though often costly, approaches such as adiabatic state preparation and basic variational ansätze. Developing more efficient and hardware--aware state preparation techniques---for instance along the lines of problem--tailored variational circuits, measurement--based feedback, or more refined adiabatic/shortcut-to-adiabatic protocols~\cite{sougato}---is likely to be essential, particularly for preparing excited states and carefully engineered wavepackets for scattering experiments \cite{Farrell2025WStates, Davoudi:2024scattering}.

Another promising direction is to move beyond qubits. Many of the field theory encodings we discussed are rather unnatural from the continuum perspective, because they force an intrinsically bosonic or continuous system into a tensor product of two--level systems. Using qudits rather than qubits~\cite{barryqudit} offers a more economical representation of local Hilbert spaces and may substantially reduce circuit depth and truncation errors for a fixed physical problem. At present, such hardware is still in its infancy, but it is reasonable to expect steady progress here. In the same vein, it is completely natural to consider continuous--variable architectures and bosonic modes as fundamental building blocks, along the lines of~\cite{cvqcqft}. For scalar field theories in particular, continuous variables often provide encodings that are conceptually closer to the underlying physics than their purely qubit-based counterparts.

A further theme, only touched upon implicitly in these lectures, is the interplay between quantum devices and advanced classical methods such as tensor networks, Monte Carlo with complex weights, and functional renormalization group techniques. For the foreseeable future, the most fruitful approach may be \emph{hybrid}: using classical methods to prepare ground states or low--entanglement initial data, and then delegating the most challenging real--time or high--entanglement pieces of the computation to quantum hardware. Finding an optimal way to decompose a given physics problem into classical and quantum components is itself an unsolved algorithmic challenge.

A major ingredient we have not talked about in these lectures is \emph{fault tolerance}. All of the algorithms discussed in these notes were implicitly analysed in an idealized setting with perfect gates and measurements, whereas any realistic large--scale quantum simulation of quantum field theory will require logical qubits protected by quantum error--correcting codes. Threshold theorems~\cite{PreskillReliableQC} and concrete code families such as the surface code~\cite{FowlerSurfaceCode} tell us that this is possible in principle, but only at the price of substantial overhead in both qubit-count and circuit-depth. Early resource estimates for fault--tolerant implementations of chemistry and Shor--type algorithms~\cite{GidneyEkera} and for Hamiltonian simulation more generally~\cite{LowChuangHamiltonianSim,BabbushReview} suggest that truly asymptotic advantages in 2+1D and 3+1D QFT will likely require millions of physical qubits and careful architectural co--design. On the other hand, the lattice QFT algorithms we have emphasized here  were explicitly designed with the fault--tolerant regime in mind, and provide a blueprint for what fully error--corrected QFT simulations might look like once such hardware becomes available. In the meantime, systematic consideration of error–corrected resource costs and code compatibility can already serve as realistic objectives for near‑term research and development efforts.

Ultimately, the techniques from these lectures are not the finish line—they’re the launchpad. The anharmonic oscillator and the simple 1+1D theories we have studied form a controlled ``play area'' where ideas can be tested safely. The real challenge---and the real opportunity---is to carry these ideas into more complex theories, higher dimensions, and more demanding observables, in tandem with the rapid evolution of quantum hardware. The coming decade of quantum field theory on quantum computers will be shaped within this conversation among physical questions, computational algorithms, and experimental devices.

It is apt to conclude these lectures quoting a famous poem by the Indian Nobel Laureate, Rabindranath Tagore---the poem talks about a dream that we feel captures the very spirit of true knowledge:

\medskip

{\it \begin{center}Where the mind is without fear and the head is held high;\\
Where knowledge is free;\\
Where the world has not been broken up into fragments;\\
by narrow domestic walls;\\
Where words come out from the depth of truth;\\
Where tireless striving stretches its arms towards perfection;\\
Where the clear stream of reason has not lost its way;\\
into the dreary desert sand of dead habit;
Where the mind is led forward by thee; \\ into ever-widening thought and action—;\\
Into that heaven of freedom, my Father, let my country awake.\end{center}}\flushright{---Rabindranath Tagore, 1912.}

\printbibliography
\end{document}